\documentclass[sort&compress,number,final,3p,times]{elsarticle}

\usepackage{amsmath}
\usepackage{amsfonts}
\usepackage{amssymb}
\usepackage{graphicx}
\usepackage{float}
\journal{Physics Report}

\begin{document}

\begin{frontmatter}

\title{Quantum field theory on toroidal
topology: algebraic structure and applications}

\author[UVic,TRIUMF]{F.C. Khanna}
\ead{khannaf@uvic.ca}
\author[CBPF]{A.P.C. Malbouisson}
\ead{adolfo@cbpf.br}
\author[UFBA]{J.M.C. Malbouisson}
\ead{jmalboui@ufba.br}
\author[UNB]{A.E. Santana}
\ead{asantana@unb.br}

\address[UVic]{Department of Physics and Astronomy, University of Victoria,
Victoria, BC V8P 5C2, Canada}
\address[TRIUMF]{TRIUMF, Vancouver, BC, V6T 2A3, Canada}
\address[CBPF]{Centro Brasileiro de Pesquisas F\'{\i}sicas/MCT,
22290-180, Rio de Janeiro, RJ, Brazil}
\address[UFBA]{Instituto de F\'{\i}sica, Universidade Federal da
Bahia, 40210-340, Salvador, BA, Brazil}
\address[UNB]{ International Center for Condensed
Matter Physics, Instituto de F\'{\i}sica, Universidade de Bras\'{i}lia,
70910-900, Bras\'{i}lia, DF, Brazil}

\begin{abstract}

The development of quantum theory on a torus has a long history,
and can be traced back to the 1920s, with the attempts by
Nordstr\"om, Kaluza and Klein  to define a fourth spatial
dimension with a finite size, being curved in the form of a torus,
such that Einstein and Maxwell equations would be unified. Many
developments were carried out considering cosmological problems in
association with particles physics, leading to methods that are
useful for areas of physics, in which size effects play an
important role. This interest in finite size effect systems has
been increasing rapidly over the last decades, due principally to
experimental improvements. In this review, the foundations of
compactified quantum field theory on a torus are presented in a
unified way, in order to consider applications in particle and
condensed matted physics. The theory on a torus $\Gamma
_{D}^{d}=({\mathbb{S}}^{1})^{d}\times {\mathbb{R}} ^{D-d}$ is
developed from a Lie-group representation and c*-algebra
formalisms. As a first application,  the quantum field theory at
finite temperature, in its real- and imaginary-time versions, is
addressed by focussing on its topological structure, the torus
$\Gamma _{4}^{1}$. The toroidal quantum-field theory provides the
basis for a consistent approach of spontaneous symmetry breaking
driven by both temperature and spatial boundaries. Then the
superconductivity in films, wires and grains are analyzed, leading
to some results that are comparable with experiments. The Casimir
effect is studied taking  the electromagnetic and Dirac field on a
torus. In this case, the method of analysis is based on a
generalized Bogoliubov transformation, that separates the Green
function into two parts: one is associated with the empty
space-time, while the other describes the impact of
compactification. This provides  a natural procedure for
calculating the renormalized energy-momentum tensor. Self
interacting four-fermion systems, described by the Gross-Neveu and
Nambu-Jona-Lasinio models, are considered. Then finite size
effects on the hadronic phase structure are investigated, taking
into account density and temperature. As a final application,
effects of extra spatial dimensions are addressed, by developing a
quantum electrodynamics in a five-dimensional space-time, where
the fifth-dimension is compactified on a torus. The formalism,
initially developed for particle physics, provides results
compatible with other trials of probing the existence  of
extra-dimensions.
\end{abstract}


\end{frontmatter}

\tableofcontents

\section{Introduction}

Over the last decades, due to experimental improvements, there is
an increasing interest in physical systems where finite size
effects play an important role. This is the case of experiments
considering effects of extra spatial dimensions or the dependence
of the critical temperature on the size of superconductors in
films, wires and grains~\cite{Strongin1970,Kodama1983}.
Compactification is a basic reason for the existence of the
Casimir effect, which, for the electromagnetic field, has been
measured with a high accuracy only
recently~\cite{Lamoreaux1997,Lamoreaux1998,Lamoreaux1999,Mohideen1998,Asorey2012}.
Another outstanding example of size effects is in the transition
of the confined to the deconfined state of matter, in particle
physics, giving rise to the phase-transition from hadrons to a
plasma of quarks and gluons~\cite{Muta1987,Bailin1993}. These
results have motivated the development of formalisms to treat size
effects~\cite{Brezin1985,Cardy1988,Zinn-Justin1996,Arkani-Hamed1998,
RandallSundrum1999,Antoniadis1990,Antoniadismunoz1993,%
AntoniadisDimopoulos1999,AntoniadisPRD1999}, such as the quantum
field theory on a torus, with well established perturbative
methods~\cite{KhannamalsaesAnnals2011}. Taking into consideration
practical applications, in this review, we present developments
for the quantum field theory defined on spaces with topologies
$\Gamma _{D}^{d}=({\mathbb{S}}^{1})^{d}\times { \mathbb{R}}
^{D-d}$, where $ D$ is the dimension of the whole space manifold
and $d$ is the number of compactified dimensions on the
hyper-torus $({\mathbb{S}}^{1})^{d}$, such that $d\leq D $; these
can be time (real or imaginary) and spatial dimensions, describing
systems in thermal equilibrium subjected to spatial constraints.

Currently, topologies of this type may be associated with extra
dimensions; however this is quite an old idea. Historically, one
of the first works exploring the notion that our world might have
more than four dimensions was carried out by
Nordstr\"om~\cite{Nordstrom1914}, as a generalization of the
Einstein theory of gravitation to a 5-dimensional space-time.
Later on, Kaluza~\cite{Kaluza1921} and Klein~\cite{Klein1926}
proposed that the fourth spatial dimension would have a finite
size, being curved in the form of a circle.  The result is that
the equations are separated into sets, one of which describes the
Einstein equations and the other is equivalent to Maxwell
equations. Geometrically, the extra fifth dimension can be viewed
as the circle group U(1), which corresponds to the formulation of
the electromagnetic theory as a gauge theory with gauge group
U(1). There are many theoretical studies in these directions, in
particular in string theory which has been reviewed in other
places~\cite{Polchinski1986,AtickWitten1988,AbdalaGadelha2003,%
AbdalaGadelha2001,Polchinski2000}. The existence of such extra
dimensions have not been observed in experiments, although there
are several studies considering extra-dimension effects, such as
the seminal paper by Randal and Sundrum~\cite{RandallSundrum1999}.
This is a way to investigate, for instance, the electroweak
transition and baryogenesis, by taking a 5-dimensional space-time,
using finite-temperature field theory with a compactified extra
dimension~\cite{Panico2005,Panico2006,Antoniadis1990,Dvali2002,Hall2002,%
Kubo2002,Burdman2003,Fairlie1979,Manton1979,Serone2005,Arkani-Hamed2002,%
Arkani-Hamed2002a}. Considering that these extra compactified
dimensions are beyond our four-dimensional world, it is reasonable
to assume that their presence may give rise to effects at
different scales, not just those in the cosmological or high
energy physics realm, but also in low energy
phenomena~\cite{Roy2009,Ttira2010}. With this perspective, effects
on the anomalous magnetic moment of the muon, associated with
extra-dimensional excitations of the photon and of W and Z bosons,
have been studied~\cite{Nath1999}. Since the $g-2$ experiment at
the Brookhaven National Laboratory (USA) in 2004, and subsequent
ones,  the expected value for $g$ obtained by standard theoretical
calculations (that predict $g=2$) could not be confirmed. The
reason is that both theoretical predictions and experimental
results have a large uncertainty. Although a conclusive response
is not available, a value of $g\neq 2$ has  not been
excluded~\cite{Aubert2009}. In the framework of quantum
electrodynamics, a recent experiment for the electron magnetic
moment gives a more precise value for $g$: the claimed uncertainty
is nearly 6 times lower than in the past. However, there is still
a deviation from the value $g=2$~\cite{Odom2006}. In atomic
physics, very accurate measurements of the asymptotic quantum
effects on Rydberg excitations have also been carried
out~\cite{Jentschura2005}. In this case, another interesting
consequence of the possible existence of extra dimensions is that
the electric charge may not be exactly conserved; a subject that
has been under discussion for a long
time~\cite{Dubovsky2000,Ignatev1979,Mohapatra1987,Suzuki1988a}. In
four-dimensional theories, a very small deviation from electric
charge conservation would lead to incompatibilities with
low-energy tests of quantum electrodynamics~\cite{Dubovsky2000}.
These inconsistencies can be cured by the introduction of
hypothetical millicharged particles~\cite{Suzuki1988a}. However,
this artifact would not be necessary if our world were considered
as  a submanifold of a higher-dimensional
space~\cite{Dubovsky2000}. From a theoretical point of view, in
short, many of these studies, with compactified extra-dimensions,
lie in the framework of topologies
$\Gamma_D^d$~\cite{Chatzistavrakidis2012,Floratos2012,Haghighat2011,%
Ferrer2001,Ferrer2001a}.

This concept of compactification is not restricted to the
discussion of extra-dimensions. An important development, which
has its roots in the late fifties, is the first systematic
approach to quantum field theory at finite temperature: the
imaginary-time (Matsubara) formalism~\cite{matsubara1955}. A
fundamental aspect of this method is that it relies on the
periodicity (antiperiodicity) of the correlation functions of
bosons
(fermions)~\cite{threezawas1957,MartinShwingerKMS1959,KuboKMS1957,HaagKMS1967}.
This periodicity in imaginary-time approach is equivalent to
formulating the theory in a compactified torus
${\mathbb{S}}^{1}\times{\mathbb{R}}^{D-1}$, where
${\mathbb{S}}^{1}$ is a circumference of length proportional to
the inverse of the
temperature~\cite{Polchinski1986,AtickWitten1988}. The same
mathematical apparatus is applied to a field theory compactified
in both space and
time~\cite{FordYoshimura1979,BirrellFord1980,Ford1980,FordSvaiter1995,%
TFDCompacPRA2002,JotAdolfo2002}. It should be mentioned the
pioneering contribution by Birrell and
Ford~\cite{BirrellFord1980}, who developed a generalization of the
Matsubara mechanism to include compactification of space
coordinates. These ideas led to numerous other developments that
include spontaneous symmetry breaking for field theories on torus,
establishing solid aspects on previous attempts to bring together
particle physics and cosmology, such as in the concept of
topological mass
generation~\cite{Kibble1980,Toms1980,Toms1980a,Toms1980b,Kennedy1981,%
Goncharov1982,Goncharov1982a,Hosotani1983,Actor1990,Actor1990a}.

The topological framework to study  finite temperature effects and
spatial constraints is developed by considering a $D$-dimensional
manifold with a topology of the type $\Gamma
_{D}^{d}={\mathbb{S}}^{1_{1}}\times {\mathbb{S}}^{1_{2}}\cdots
\times {\mathbb{S}}^{1_{d}}\times {\mathbb{R}}^{D-d}$, with
${\mathbb{S}} ^{1_{1}}$ corresponding to the compactification of
time and $\,{\mathbb{S}} ^{1_{2}},\dots, {\mathbb{S}}^{1_{d}}$
referring to the compactification of $ d-1$ spatial dimensions. A
characteristic of this topological structure is that it leads to
modifications of  boundary conditions imposed on fields and
correlation functions, but it does not modify the local field
equations; $i.e$, the topology plays a role on global properties
of the system, but not on local ones, which are associated with
the invariants of the Lie space-time
symmetries~\cite{BalazChaosinsphere1986,LimaDaniel2007}. These
developments, that also include algebraic and perturbative
analysis, make a toroidal quantum field theory attractive for a
broad range of applications.

For instance, an analysis of the spontaneous symmetry breaking for
scalar field theories at finite temperature with a compactified
spatial dimension~\cite{MalbouissonsaesNPB2002} has been carried
out for
superconductors~\cite{Ourbook2009,AbreuMalsaes2009,LinharesAdolfoExp2006,%
Abreudecalanmalsaes2005}. In this case, the Ginzburg-Landau model
for phase transitions is generalized. The model is defined on a
$3$-dimensional Euclidean space with one, two or three
compactified spatial dimensions; the size of which is respectively
the thickness of a film, the cross section of a wire or the volume
of a grain. The critical temperature is found as a function of
these quantities and minimal sizes, sustaining the transition, are
determined. Such achievements are in reasonable agreement with
experimental results~\cite{LinharesAdolfoExp2006,Abreu2004}.

A similar analysis of the spontaneous symmetry breaking has been
considered for four-fermion interacting fields, a  model for
describing systems in condensed matter physics, such as
superconductivity~\cite{Doniach1974} and aspects of graphene
defined from a Dirac theory in $(2+1)$
dimensions~\cite{Schnetz2006,Herbut2006,Jackiw2007,Jurivcic2009}.
In this case,  the quantized Hall conductivities, which are
identified as the topological (toroidal) TKNN (Thouless, Kohmoto,
Nightingale, den Nijs)~\cite{Thouless1982,Kohmoto1985}  integers,
are also associated with graphene lattice~\cite{Hasegawa2006}.
These are important applications of the toroidal topology that
have been explored and revised
extensively~\cite{Kane2005,Wang2012}.

Four-fermion interaction is also employed in phenomenological
approaches for particle physics. Simple models describing such
interaction are the Nambu-Jona-Lasinio
model~\cite{Nambu1961,Nambu1961a} and the Gross-Neveu
model~\cite{GrossNeveu1974}. Both have been formulated on a torus,
in order to study, as effective theories for quantum
chromodynamics, the phase-transition in the confining/deconfining
region~\cite{MalbouissonJuaes2004,Malbouisson2006,MalbsKhaaesEPL2010,%
AbreuMarcelo2006,Ebert2008,AbreuMalsaes2009,Abreumals2010,Abreu2011,Abreu2011a}.
Beyond a phase diagram structure, this  formalism provides a
critical temperature for the confinement/deconfinement transition
in accordance with lattice calculations.

Another application of the toroidal quantum field theory  is the
Casimir effect. As  first analyzed by Casimir~\cite{Casimir1948},
the vacuum fluctuations of the electromagnetic field confined
between two conducting plates with separation $L$ give rise to an
attractive force between the plates. The effect has been studied
in different geometries, topologies, fields and physical boundary
conditions~\cite{Milonni1993,Elizalde1994,Elizalde2012,Mostepanenko1997,%
Bordag1999,Bordag2001,Bordag1997,Elizalde1998,Elizalde1998a,%
Elizalde2003a,Cognola2005}. The investigations are of interest in
diverse areas, such as nano-devices in condensed matter physics,
the confinement/deconfinement transition in particle physics and
cosmological
models~\cite{Lamoreaux1997,Lamoreaux1998,Lamoreaux1999,Milonni1993,%
Mostepanenko1997,Bordag1999,Bordag2001,Levin1993,Romeo1995,Seife1997,%
Boyer2003,Milton1999,Plunien1986,Saito1991,Gundersen1988,Lutken1988,%
Lutken1984,Ravndal1989,Silva2001,Kenneth2002,
TFDCompacPRA2002,Farina2006,Ozcan2012,
Brevik2010,Ellingsen2010,Milton2009,Tatur2009,
Theo2012,Fialkovsky2011,Elizalde2003,Elizalde2011}. Many aspects
of the Casimir effect have been analyzed by studying fluctuations
of the vacuum of a field theory compactified on a torus. This is,
in particular, due to the association of the toroidal topology
with the Dirichlet or Neumann boundary
conditions~\cite{Ourbook2009,Khoo2011,Zhai2011,Oikonomou2011,
Oikonomou2011a,Oikonomou2010,Buchmueller2009,Bellucci2009,Chopovsky2012,%
Belich2011}. It is important to emphasize the nature of
calculations in the local formulation, initiated by Brown and
Maclay~\cite{Brown1969}, to derive the energy-momentum tensor of
quantum fields: a procedure that leads naturally to standard
renormalization scheme associated with the Casimir effect.

Our goal here is to present the theory of the quantum field theory
on a torus as a representation theory, focussing on the
aforementioned applications.  The presentation is organized in the
following way. In Section 2, we review some aspects of the quantum
field theory at finite temperature, emphasizing its topological
structure, the torus $\Gamma _{4}^{1}$, in imaginary- and
real-time formalism. These results are generalized, in Section 3,
for theories on torus $\Gamma _{D}^{d}$. The algebraic structure
of such a theory is analyzed in Section 4. In Section 5, the
Casimir effect is reviewed for the electromagnetic and Dirac
fields. In Section 6, the notion of spontaneous symmetry breaking
is developed for a quantum theory on a torus. These results are
used, in Section 7, to analyze superconductivity in films, wires
and grains. In Section 8, the Gross-Neveu and the
Nambu-Jona-Lasinio model are studied on toroidal topologies. In
Section 9, a quantum electrodynamics with an extra-compactified
dimension is developed. The final conclusions and additional
comments are presented in Section 10.

\section{Thermal field theory and toroidal topology}

A systematic approach to many-body physics at finite-temperature
was initiated by Matsubara~\cite{matsubara1955}. His method is
equivalent to a time-axis Wick rotation of a quantum mechanical
system, giving rise to what is known as \textit{imaginary time}
formalism. Further developments included the search for
mathematical structures, such as geometrical and topological
features, and the generalization of the theory for relativistic
particles physics, following along the lines of zero temperature
quantum field
theory~\cite{KadanoffBaymbook1962,Abrikosov1963,FetterWalecka1971,%
Kapusta1989,leBellacbook1996,DasbookT1997}.

A first generalization of the Matsubara work was carried out by
Ezawa, Tomozawa and Umezawa~\cite{threezawas1957}, who extended
the imaginary-time formalism to quantum field theories. They
discovered, in particular, the periodicity (anti-periodicity)
condition for correlation functions of boson (fermion) fields; a
concept that later became known as the KMS (Kubo, Martin and
Schwinger) condition~\cite{leBellacbook1996,DasbookT1997}. Kubo
introduced this condition in his paper on linear-response
theory~\cite{KuboKMS1957} in quantum statistical mechanics. Later,
Martin and Schwinger~\cite{MartinShwingerKMS1959}, developing a
real-time formalism for thermal systems in a non-equilibrium
state~\cite{SchwingerJMP1961}, used the KMS condition in quantum
electrodynamics~\cite{MartinShwingerKMS1959}. These
accomplishments were explored further by Kadanoff and
Baym~\cite{KadanoffBaymbook1962}. The KMS condition, an acronym
coined by Haag, Hugenholtz and Winnink~\cite{HaagKMS1967}, was
early identified as a structural mathematical aspect of the
quantum field theory at finite temperature, which was in turn
associated with the
c*-algebra~\cite{Segal1947,Takesakibook1970,Emchbook1972,Bratteli1987}.

The imaginary-time quantum field theory has been applied to
different areas. For instance, numerous studies using quantum
chromodynamics~\cite{SmilgaQCDbook2001} have been carried out in
attempts to understand  the quark-gluon plasma and the
hadronization phase-transition~\cite{BlaizotBoltzQCD1999}. An
important step to this understanding was carried out by Dolan and
Jackiw~\cite{DolanJackiw1974}, who studied the effective potential
of the $\phi^{4}$ theory  at finite temperature, developing the
concept of spontaneous symmetry-breaking/restoration.

The imaginary-time formalism is restricted to equilibrium systems.
However, there are many examples in high energy and condensed
matter physics where time dependance is crucial. This has
motivated, over the decades, the search for a real-time formalism
at finite
temperature~\cite{FloreaniniJackiw1988,EboliJackiwPiPRD1989,Zubarev1991,%
KimsantanakhannaPLA2000,PimentelFainberg2003,Bonin2011,PerezDasFrenkel2005,%
FoscoAdolfo2008,Kita2010}. One of these real-time methods is the
closed-time path formulation initially due to
Schwinger~\cite{SchwingerJMP1961}. The approach is constructed by
following a path in the complex-time plane and important
contributions, which deserve to be mentioned,  were carried out by
Martin and Schwinger~\cite{MartinShwingerKMS1959},
Mahanthappa~\cite{Mahanthappa11962}, Bakshi and
Mahanthappa\cite{Mahantapa21963,Mahantapa31963},
Keldysh~\cite{Keldysh1965}, and Kadanoff and
Baym~\cite{KadanoffBaymbook1962}. From this procedure an effective
doubling of the degrees of freedom emerges, such that Green
functions are represented by $2\times2$ matrices. Actually, such a
doubling has been recognized as an intrinsic characteristic of
real-time finite-temperature field theories, providing a correct
definition for perturbative analysis~\cite{leBellacbook1996}. This
method is called the Schwinger-Keldysh formalism.

The real-time formalism can be derived from a linear-algebra
representation theory. Such a theory was first presented by
Takahashi and
Umezawa~\cite{TakahashiUmezawaTFD1975,Umezawabook1982,Umezawabook1993},
and was called \textit{Thermofield Dynamics} (TFD). As a
consequence of the real-time requirement, a doubling is defined in
the original Hilbert space of the system, through the algebraic
structure called tilde conjugation rules and thermo-algebras. The
temperature is then introduced by a Bogoliubov transformation;
that is, a rotation in the doubled space. Furthermore, the
propagators are $2\times2$ matrices. From this fact, an algebraic
association with the Schwinger-Keldysh method is
established~\cite{Umezawabook1993,ChuHumezawa1994}.

A thermal theory can  be also studied by formulating TFD within
the framework of $c$*-algebras and thermo-Lie groups, a modular
representation of
Lie-algebras~\cite{Ojima1981,SantanaKhanna1995,SantanaNeto1999,%
MontignyAESKhann2000}. This apparatus provides a physical
interpretation for the theory, in particular for the
doubling~\cite{KhannamalsaesAnnals2009,Ourbook2009,KhannamalsaesAnnals2011},
and opens several possibilities for the study of thermal
effects~\cite{Kopfaes1997,Celeghini1998455}. These algebraic
developments include perturbative schemes and classical
representations~\cite{Umezawabook1982,SemenoffUmezawa1983,%
Arimitsumezawa1987,SantanaChuKhann1996,LoriClassicaTFD1997},
with applications to a variety of systems%
~\cite{Umezawabook1993,WhiteheadUmezawa1984,BarnettKnight1985,%
SuzukiSpinTFD1986,KobesSemenoff1985,%
MannRevzen1989,MannRevHume1989,ChaturvediTFDOptcs1990,%
ChaturvediOptcsTFD1999,Malbouissonbaseia2000,Rakhimovkhanna2001,%
AbdalaGadelha2001,AbdalaGadelha2003,SantanaKhannaRev2002,%
KhannaJotaEsdrasFermions2007,BalachandranTFD2010,%
LeinekerAmilcar2011,ChekerkerLandrem2011}. %

A distinguishing feature of thermal theories is that, due to the
KMS condition, the final prescription of the Matsubara or TFD
formalism may be regarded as a compactification of the time
coordinate. This corresponds to a field theory on the topology
$\mathbb{S}^{1}\times\mathbb{R}^{D-1}$, where $\mathbb{S}^{1}$ is
a circumference of length $\beta=1/T$, where $T$ is the
temperature, describing the time compactification, and
$\mathbb{R}^{D-1}$ stands for $D-1$ space-like coordinates. In
this section, we review some basic elements of the thermal quantum
field theory, emphasizing such topological aspects.

\subsection{Imaginary-time formalism}

Let us start with the standard definition of the statistical
average of an observable $A$, i.e.
\begin{equation}
\langle
A\rangle_{\beta}=\text{Tr}[\rho_{\beta}A]=\frac{1}{Z(\beta)}
\text{Tr}\ [e^{-\beta H}A],\label{dec20111}
\end{equation}
where $Z(\beta)$, the partition function, is given by $Z(\beta)=$
Tr$\ (e^{-\beta H})$, where $\beta=1/T$, with $T$ being the
temperature and $H$ the Hamiltonian. We use natural units, such
that $k_{B}=c=\hbar=1$. In this section, for simplicity, we
restrict to the boson field.

An important result regarding the nature of the canonical (or the
grand canonical) ensemble is the Kubo-Matin-Schwinger (KMS)
condition, stating that the statistical average of an operator in
the Heisenberg picture, $A_{H}(t)=e^{itH}A(0)e^{-itH}$, is
periodic in time with a period given by $i\beta$. This result can
be proved directly from the statistical average, i.e.
\begin{equation}
\langle A_{H}(t)\rangle_{\beta}=\frac{1}{Z}\text{Tr}[\text{\
}e^{-\beta H}A_{H}(t)]\equiv\langle
A_{H}(t-i\beta)\rangle_{\beta}. \label{kms12101}
\end{equation}
The change in the argument of $A_{H},$ $t\rightarrow t-i\beta,$ is
called a Wick rotation of the time axis.

Consider the free Klein-Gordon field defined in a D-dimensional
Euclidian space-time, $\mathbb{R}$, with a point given by
$x=(x^{0},\mathbf{z})$, where $x^{0}$ stands for the time and
$\mathbf{z}=(z^{1},z^{2},...,z^{D-1})$ the space coordinate, and
the Green function
\[
G_{0}(x-y;\beta)=i\langle T[\phi(x)\phi(y)]\rangle_{\beta}.
\]
It satisfies the KMS condition, Eq.~(\ref{kms12101}), i.e.,
\begin{equation}
G_{0}(x-y;\beta)=G_{0}(x-y+i\beta {n}_{0};\beta),
\label{cbpf15122}
\end{equation}
where ${n}_{0}=(1,0,0,0,...)$. The function $G_{0}(x-y;\beta)$ is
a solution of the following equation
\begin{equation}
(\square+m^{2})G_{0}(x-y;\beta)=-\delta(\tau_{x}-\tau_{y})\delta
(\mathbf{x}-\mathbf{y}), \label{cbpf15121}
\end{equation}
where $\tau=it,$ such that
$\square+m^{2}=-\partial_{\tau}^{2}-\nabla
^{2}+m^{2}$, and is given by%
\begin{equation}
G_{0}(x-y;\beta)=\frac{-1}{i\beta}\sum_{n}\int\frac{d^{D-1}p}{2\pi^{3}}
\frac{e^{-ik_{n}\cdot x}}{k_{n}^{2}-m^{2}+i\varepsilon},
\label{8propmat1}
\end{equation}
where $k_{n}=(k_{n}^{0},\mathbf{k})$ and $k_{n}^{0}=2\pi n/\beta$
are the Matsubara frequencies. This solution, the thermal Feynman
propagator, is a direct consequence of the KMS condition and is
unique, since the periodic boundary condition and the Feynman
contour are respected.

The normalized generating functions for this Green function is
\begin{equation}
Z_{0}[J\mathbf{;}\beta]=\exp\left[  \frac{i}{2}\int d^{D}xd^{D}y\,J(x)G_{0}%
(x-y;\beta)J(y)\right]  , \label{cbpf22122}%
\end{equation}
where
\[
G_{0}(x-y;\beta)=i\frac{\delta^{2}Z_{0}[J,\beta]}{\delta
J(x)\delta J(y)}|_{J=0}.
\]
reproducing the basic result of the Gibbs formalism.

Using the notion of spectral function, as discussed by Kadanoff
and Baym~\cite{KadanoffBaymbook1962}, Dolan and
Jackiw~\cite{DolanJackiw1974} managed to write the thermal
propagator in a Fourier-integral representation, thus restoring
time as a real quantity. The basic result is that the
Fourier-series representation, given in Eq.~(\ref{8propmat1}), can
be rewritten, using analytical continuation and the
spectral function, as%
\begin{equation}
G_{0}(x-y;\beta)=\int\frac{d^{D}k}{(2\pi)^{D}}e^{-ik(x-y)}\
G_{0}(k;\beta),
\label{bo1012}%
\end{equation}
where%
\[
G_{0}(k;\beta)=G_{0}(k)+\
f_{\beta}(k^{0})[G_{0}(k)-G_{0}^{\ast}(k)]
\]
and%
\[
f_{\beta}(k^{0})=\sum\limits_{l_{0}=1}^{\infty}e^{-\beta\omega_{k}l_{0}}
=\frac{1}{e^{\beta\omega_{k}}-1}\equiv n(k^{0};\beta),
\]
with $n(k^{0};\beta)$ being the boson distribution function at
temperature $T$ with $\omega _{k}=k^{0}$. Then we have
\[
G_{0}(k;\beta)=\frac{-1}{k^{2}-m^{2}+i\varepsilon}+n(k^{0};\beta)2\pi
i\delta(k^{2}-m^{2}).
\]

It is important to emphasize the following aspect. The thermal
Green function for the free-boson field is given in
Eq.~(\ref{8propmat1}), satisfying Eq.~(\ref{cbpf15121}), the
Klein-Gordon equation, and the KMS boundary-condition. Since the
Klein-Gordon equation expresses an isometry locally in space-time,
such results are equivalent to
writing the Klein-Gordon field on a toroidal topology $\Gamma_{4}%
^{1}=\mathbb{S}^{1}\times\mathbb{R}^{D}$, where $\mathbb{S}^{1}$
describes the compactification in the imaginary time coordinate in
a circle of circumference $\beta$. This interpretation is valid
not only for free-fields, but also for interacting fields. Since
the topology maintains the nature of the local interaction, the
Feynman diagrams in  $\Gamma_{D}^{1}$ theory are the same except
by the following redefinition in the expression of the
propagators. In other words, the Feynman rules in the momentum
space are modified by taking, for the time-like compactified
dimension, the integrals in momentum space
replaced by sums as%
\begin{equation}
\int
\frac{dp^{0}}{2\pi}\rightarrow\frac{1}{i\beta}\sum_{n_{0}=-\infty}^{\infty}
\label{portoabril1}%
\end{equation}
with
\begin{equation}
p^{0}\rightarrow p^{n_{0}}=\frac{2\pi n_{0}}{i\beta},
\label{portoabril2}
\end{equation}
for bosons, and
\begin{equation}
p^{0}\rightarrow p^{n_{0}}=\frac{2\pi(n_{0}+\frac{1}{2})}{i\beta},
\label{portoabril3}%
\end{equation}
for fermions.

\subsection{Real-time formalism}

There are two versions for a real-time finite-temperature quantum
field theory. One was formulated by
Schwinger \cite{MartinShwingerKMS1959,Mahanthappa11962,Mahantapa21963,%
Mahantapa31963} and Keldysh~\cite{Keldysh1965}, which is based on
using a path in the complex time plane~\cite{leBellacbook1996}.
The other is the thermofield dynamics (TFD) proposed by Takahashi
and Umezawa~\cite{TakahashiUmezawaTFD1975}. In this case, the
thermal theory is constructed on a Hilbert space and thermal
effects are introduced by a Bogoliubov
transformation~\cite{Ourbook2009}. In equilibrium, these two
real-time formalisms are the same~\cite{ChuHumezawa1994}. We focus
here on basic elements of the TFD approach, emphasizing that a
real-time formalism is a theory on the topology $\Gamma_{4}^{1}$.

\subsubsection{Thermal state in a Hilbert space}

For a system in thermal equilibrium, the ensemble average of an
operator $A$ is given by $\langle A\rangle_{\beta}=Z^{-1}(\beta
){\rm Tr}(e^{-\beta H}A)$. Then considering
$H|n\rangle=E_{n}|n\rangle$, with the set $\{|n\rangle\}$ being an
orthonormal basis of the Hilbert space, we write $\langle
A\rangle_{\beta}=Z^{-1}(\beta)\sum_{n}e^{-\beta E_{n}}\langle
n|A|n\rangle$. In TFD this average is written as
\begin{equation}
\langle A\rangle_{\beta}\equiv\langle0(\beta)|A|0(\beta)\rangle,
\label{tfd11}
\end{equation}
where the thermal state $|0(\beta)\rangle$ is given
by~\cite{Umezawabook1993}
\[
|0(\beta)\rangle=Z(\beta)^{-1/2}\sum_{n}e^{-\beta\varepsilon_{n}/2}
|n,\tilde{n}\rangle,
\]
such that a doubling of the Hilbert space is introduced. A vector
of the basis is given by
$|n,\tilde{m}\rangle=|n\rangle\otimes|\tilde{m}\rangle$, and the
operator $A$ acts on non-tilde vectors, i.e.
\[
\langle n,\tilde{n}|A|m,\tilde{m}\rangle=\langle
n|\otimes\langle\tilde
{n}|A|m\rangle\otimes|\tilde{m}\rangle=\langle
n|A|m\rangle\langle\tilde {n}|\tilde{m}\rangle=A_{nm}\delta_{nm}.
\]
The tilde in a vector $|m,\tilde{m}\rangle$ indicates that
$|\tilde{m}\rangle$ is the replica of $|m\rangle,$ with $m$ and
$\tilde{m}$ standing for the same number: $m=\tilde{m}.$ This is
why we have written $\langle\tilde{n}|\tilde
{m}\rangle=\delta_{nm}$, without reference to the tilde in the
$\delta_{mn}$. In a vector like $|m,\tilde{n}\rangle,$ the tilde
emphasizes the element of the tilde-Hilbert space only. The vector
$|0(\beta)\rangle$ is then a pure state, defined in this doubled
Hilbert space, but equivalent to a mixed state describing the
thermal equilibrium of a system as far as the averages are
concerned.

Let us consider the boson oscillator. Neglecting the zero-point
energy, the Hamiltonian is
\[
H=wa^{\dagger}a.
\]
The creation and destruction operators, $a^{\dagger}$ and $a$
respectively, satisfy the algebra
\begin{equation}
\lbrack a,a^{\dagger}]=1\,;\;\;[a,a]=[a^{\dagger},a^{\dagger}]=0.
\label{c2bos1}%
\end{equation}
The eigenvalues and eigenstates of $H$ are specified by
$H|n\rangle =nw|n\rangle,\ \ \ n=0,1,2,...,$ where $|0\rangle$ is
the vacuum state. These states $|n\rangle$ are orthonormal, i.e.,
$\langle m|n\rangle=\delta_{mn}$, and the number operator,
$N=a^{\dagger}a$, is such that $N|n\rangle =n|n\rangle.$ Since
$a^{\dagger}$ and $a$ describe bosons, $|n\rangle$ is a state with
$n$ bosons. In order to construct the TFD formalism, we have to
double degrees of freedom, giving rise to tilde operators as
$\tilde {a}^{\dagger}$ and $\tilde{a}$, commuting with non-tilde
operators. For the basic algebraic relations, then we have
\begin{equation}
\lbrack\tilde{a},\tilde{a}^{\dagger}]=1\,,\text{ \ \ \
}[\tilde{a},\tilde
{a}]=[\tilde{a}^{\dagger},\tilde{a}^{\dagger}]=0,
\end{equation}
and the algebra for the non-tilde operators, given in
Eqs.~(\ref{c2bos1}). The other commutation relations are null.

The thermal state $|0(\beta)\rangle$ is
\begin{align}
|0(\beta)\rangle &
=\frac{1}{\sqrt{Z(\beta)}}\sum\limits_{n}e^{-n\beta
w/2}|n,\tilde{n}\rangle\nonumber\\
&  =\frac{1}{\sqrt{Z(\beta)}}\sum\limits_{n}e^{-n\beta w/2}\frac{1}%
{(n!)^{1/2}}\frac{1}{(\tilde{n}!)^{1/2}}(a^{\dagger})^{n}(\tilde{a}^{\dagger
})^{n}|0,\tilde{0}\rangle. \label{c2vac1}
\end{align}
It follows that
\[
\langle0(\beta)|0(\beta)\rangle =
\frac{1}{Z(\beta)}\sum\limits_{n,m}\langle
m,\widetilde{m}|e^{-\beta w(n+m)/2}|n,\tilde{n}\rangle =
\frac{1}{Z(\beta)}\sum\limits_{n}e^{-\beta wn}.
\]
Using $\langle0(\beta)|0(\beta)\rangle=1$, we find
\begin{equation}
Z(\beta)=\frac{1}{1-e^{-\beta w}}. \label{c2par1}%
\end{equation}
From Eq.~(\ref{c2vac1}) we have
\begin{equation}
|0(\beta)\rangle=\sqrt{1-e^{-\beta w}}\sum\limits_{n}\frac{e^{-n\beta w/2}%
}{n!}(a^{\dagger})^{n}(\tilde{a}^{\dagger})^{n}|0,\tilde{0}\rangle.
\label{c2vac2}%
\end{equation}
In this way we are able to proceed with calculations in
statistical mechanics using, instead of the canonical density
matrix, the state $|0(\beta)\rangle$. To explore this possibility,
Eq.~(\ref{c2vac2}) is written in the form
$|0(\beta)\rangle=U(\beta)|0,\tilde{0}\rangle,$ where $U(\beta)$
is a unitary operator.

The sum in Eq.~(\ref{c2vac2}) is an exponential, i.e
\begin{equation}
|0(\beta)\rangle=\sqrt{1-e^{-\beta w}} \exp(e^{-\beta
w/2}a^{\dagger}\tilde
{a}^{\dagger})|0,\tilde{0}\rangle. \label{c2vac3}%
\end{equation}
This result is written as an exponential function only, and as
such a unitary operator, by taking into account the operator
relation
\begin{equation}
e^{\alpha(A+B)}=e^{\tanh\alpha B}e^{\ln\cosh\alpha
C}e^{\tanh\alpha A},
\label{c2ope1}%
\end{equation}
where $C=[A,B]$. Defining
\begin{align}
\cosh\theta(\beta)  &  =\frac{1}{\sqrt{1-e^{-\beta w}}}\equiv
u(\beta
),\label{c2bog1}\\
\sinh\theta(\beta)  &  =\frac{e^{-\beta w/2}}{\sqrt{1-e^{-\beta
w}}} \equiv
v(\beta), \label{c2bog2}%
\end{align}
we obtain
\begin{align}
|0(\beta)\rangle &
=\cosh^{-1}\theta(\beta)e^{\tanh\theta(\beta)a^{\dagger}
\tilde{a}^{\dagger}}|0,\tilde{0}\rangle\nonumber\\
&  =\exp\left[  \tanh\theta a^{\dagger}\tilde{a}^{\dagger}\right]
\exp\left[ -\ln\cosh\theta(
\tilde{a}\tilde{a}^{\dagger}+a^{\dagger}a)\right] \exp\left[
\tanh\theta(- \tilde{a}a)\right]  |0,\tilde{0}\rangle,
\label{c2vac4}%
\end{align}
where we have used the commutation relation
$[\tilde{a},\tilde{a}^{\dagger
}]=1$ and $e^{f(\theta)\tilde{a}^{\dagger}\tilde{a}}|0,\tilde{0}%
\rangle=|0,\tilde{0}\rangle, $ where $f(\theta)$ is an arbitrary
function of $\theta.$

Using the identity given in Eq.~(\ref{c2ope1}) with $A =
-\tilde{a}a$, $B=a^{\dagger}\tilde{a}^{\dagger}$, $C = [A,B] =
-\tilde{a}\tilde{a}^{\dagger}-a^{\dagger}a$, and $\alpha = \theta
= \theta(\beta)$, we have
\begin{equation}
|0(\beta)\rangle=e^{-iG(\beta)}|0,\tilde{0}\rangle, \label{c2vac5}%
\end{equation}
where
\begin{equation}
G(\beta)=-i\theta(\beta)(\tilde{a}a-\tilde{a}^{\dagger}a^{\dagger}).
\label{c2vac6}%
\end{equation}
Hence the unitary operator, transforming $|0,\tilde{0}\rangle$
into $|0(\beta)\rangle,$ is given by
\begin{equation}
U(\beta)=e^{-iG(\beta)}. \label{c2vac7}%
\end{equation}
This operator $U(\beta)$ defines a Bogoliubov transformation.

Using $U(\beta)$, let us introduce the following thermal operators
through the relations
\[
a(\beta) = U(\beta)aU^{\dagger}(\beta) \; , \;\;\;\;\;
\tilde{a}(\beta) = U(\beta)\tilde{a}U^{\dagger}(\beta).
\]
The importance of these operators lies in the fact that $a(\beta
)|0(\beta)\rangle=U(\beta)aU^{\dagger}(\beta)U(\beta)|0,\widetilde{0}%
\rangle=U(\beta)a|0,\tilde{0}\rangle=0,$ and
$\tilde{a}(\beta)|0(\beta )\rangle=0.$ Then $|0(\beta)\rangle$ is
a vacuum for $a(\beta)$ and $\tilde {a}(\beta)$, but it is not a
vacuum for $a$ and $\tilde{a}$. In this sense, $|0(\beta)\rangle$
is a pure state for thermal operators, and a thermal state for
non-thermal operators. This is the reason that $|0(\beta)\rangle$
is called a thermal vacuum.

Since $U(\beta)$ is a unitary transformation, the algebra of the
original operators $a$ and $\tilde{a}$ is kept invariant, i.e. the
operators $a(\beta)$ and $\tilde{a}(\beta)$ satisfy the following
commutation relations
\begin{equation}
\lbrack
a(\beta),a^{\dagger}(\beta)]=1;\,\,\,\,[\tilde{a}(\beta),\tilde
{a}^{\dagger}(\beta)]=1, \label{c2bos6}%
\end{equation}
with all the other commutation relations being zero. Let us
consider, as an example, the average of the number operator
$N=a^{\dagger}a$. We find that
\begin{align}
n(\beta)  &  =\langle
N\rangle=\langle0(\beta)|a^{\dagger}a|0(\beta
)\rangle\nonumber\\
&  =\langle0(\beta)|[u(\beta)a^{\dagger}(\beta)+v(\beta)\tilde{a}%
(\beta)][u(\beta)a(\beta)+v(\beta)\tilde{a}^{\dagger}(\beta)]|0(\beta
)\rangle\nonumber\\
&  =v^{2}(\beta)=\frac{1}{e^{\beta w}-1}, \label{c2num1}%
\end{align}
where we have used $a(\beta)|0(\beta)\rangle=0$ and
$\tilde{a}(\beta )|0(\beta)\rangle=0$. This is the boson
distribution function for a system in thermal equilibrium.

Using $u^{2}(\beta) - v^{2}(\beta)=1$, we find
\begin{equation}
a^{\dagger}(\beta)a(\beta) -\tilde{a}^{\dagger}(\beta)\tilde{a}
(\beta )=a^{\dagger}a-\tilde{a}^{\dagger}\tilde{a}.
\end{equation}
This result can be used to determine the form of the generator of
time translation, the Hamiltonian $\widehat{H}$ of the theory.
Demanding the invariance of $\widehat{H}$ under the Bogoliubov
transformation, the simplest form for $\widehat{H}$ is
$\widehat{H} = H-\widetilde{H}$, such that
\[
\widehat{H}(\beta) = H(\beta)-\widetilde{H}(\beta)=\omega[
a^{\dagger }(\beta)a(\beta) -\tilde{a}^{\dagger}(\beta)\tilde{a}
(\beta)] =
\omega[a^{\dagger}a-\tilde{a}^{\dagger}\tilde{a}]=\widehat{H}.
\]

With this expression for $\widehat{H}$, we have all the elements
to fix the tilde mapping, i.e., $\widetilde{\ \ \
}:A\rightarrow\widetilde{A}$. We first observe that,  the
equations of motion, in the Heisenberg picture, for an arbitrary
operator $A$ and its partner $\widetilde{A}$ are
$i\partial_{t}A(t) = [A(t),\widehat{H}]$ and
$i\partial_{t}\widetilde{A}(t) = [\widetilde{A}(t),\widehat{H}]$.
This leads to
\begin{align*}
i\partial_{t}A(t)  &  =[A(t),\widetilde{H}]=A(t)H-HA(t)\\
-i\partial_{t}\widetilde{A}(t)  &  =[\widetilde{A}(t),\widetilde
{H}]=A(t)\widetilde{H}-\widetilde{H}A(t).
\end{align*}
Comparing these two equations, we conclude the following
conditions for the tilde mapping:
\begin{align}
(A_{i}A_{j})\widetilde{}  &  =\widetilde{A}_{i}\widetilde{A}_{j},
\label{c1til2}\\
(cA_{i}+A_{j})\widetilde{}  &  =c^{\ast}\widetilde{A}_{i}+\widetilde{A}%
_{j},\label{c1til3}\\
(A_{i}^{\dagger})\widetilde{}  &  =(\widetilde{A}_{i})^{\dagger},
\label{c1til4}\\
(\widetilde{A}_{i})\widetilde{}  &  =A_{i} ,\label{c1til5}\\
\lbrack A_{i},\widetilde{A}_{j}]  &  =0.
\end{align}
These properties are called  tilde conjugation rules.

The thermal Fock space, $\mathcal{H}_{T}$, is constructed from the
vacuum $|0(\beta)\rangle,$ and is spanned by the set of states
given by
\[
\{\ |0(\beta)\rangle,\ a^{\dagger}(\beta)|0(\beta)\rangle,\ \
\tilde {a}^{\dagger}(\beta)|0(\beta)\rangle,\,\dots\,,\
\frac{1}{\sqrt{n!}}\frac {1}{\sqrt{m!}}\left(
a^{\dagger}(\beta)\right)  ^{n}\left(  \tilde
{a}^{\dagger}(\beta)\right)  ^{m}|0(\beta)\rangle\,\dots\}.
\]
These states and their superpositions are physically identified
since each one corresponds to a density matrix~\cite{Ourbook2009}.
This is the case of the thermal vacuum, $|0(\beta)\rangle$ that
corresponds to the equilibrium density matrix, $\rho_{\beta}$.

A doublet notation is introduced by defining
\begin{equation}
\left(
\begin{array}
[c]{c}%
a(\beta)\\
\tilde{a}^{\dagger}(\beta)
\end{array}
\right)  =B(\beta)\left(
\begin{array}
[c]{c}%
a\\
\tilde{a}^{\dagger}%
\end{array}
\right)  , \label{c2bog4}%
\end{equation}
where
\begin{equation}
B(\beta)=\left(
\begin{array}
[c]{cc}%
u(\beta) & -v(\beta)\\
-v(\beta) & u(\beta)
\end{array}
\right)  . \label{c2bog5}%
\end{equation}
Given two arbitrary (boson) operators $A$ and $\widetilde{A},$ a
doublet notation is given by
\begin{equation}
\left(  A^{a}\right)  =\left(
\begin{array}
[c]{c}%
A^{1}\\
A^{2}%
\end{array}
\right)  =\left(
\begin{array}
[c]{c}%
A\\
\widetilde{A}^{\dagger}%
\end{array}
\right)  , \label{c2bog6}%
\end{equation}
with a tilde transposition given by
\begin{equation}
(\overline{A}^{a})=(A^{\dagger},-\widetilde{A}). \label{c2bog7}
\end{equation}
We now address the problem of a thermal quantum field.

\subsubsection{Thermal quantum field in real-time formalism}

Using the fact that for a boson oscillator the Hamiltonian is
given by $\widehat{H}=H-\widetilde{H}$, we can construct a
Lagrangian formalism given by
$\widehat{\mathcal{L}}=\mathcal{L-}\widetilde{\mathcal{L}}$. The
extension for a field follows along the same lines. Then, for the
Klein-Gordon field with an external source we have
\[
\widehat{\mathcal{L}} = \mathcal{L-}\widetilde{\mathcal{L}} =
\frac{1}{2}\partial_{\alpha}\phi\partial^{\alpha}\phi-\frac{1}{2}m^{2}
\phi^{2} +
J\phi-\frac{1}{2}\partial_{\alpha}\widetilde{\phi}\partial^{\alpha
}\widetilde{\phi}+\frac{1}{2}m^{2}\widetilde{\phi}^{2}-\widetilde{J}
\widetilde{\phi}.
\]
In order for the Hamiltonian formalism to be derived, we define
the canonical momentum density by
\[
\pi(x) =
\frac{\partial\mathcal{L}(\phi,\partial\phi)}{\partial\dot{\phi} }
\; , \;\;\;\;\; \widetilde{\pi}(x) =
\frac{\partial\widetilde{\mathcal{L}}(\widetilde{\phi
},\partial\widetilde{\phi})}{\partial\dot{\widetilde{\phi}}} .
\]
The Hamiltonian is defined by
\begin{equation}
\widehat{H}=\int\widehat{\mathcal{H}}\,d^{3}x=\int[{\mathcal{H}}(\phi
,\pi)-\widetilde{\mathcal{H}}(\widetilde{\phi},\widetilde{\pi})]\,d^{3}x,
\label{ham71}%
\end{equation}
where the Hamiltonian density is
\[
\widehat{\mathcal{H}}=\frac{1}{2}\pi^{2}+\frac{1}{2}(\nabla\phi)^{2}+\frac
{1}{2}m^{2}\phi^{2}-J\phi-\frac{1}{2}\widetilde{\pi}^{2}-\frac{1}{2}
(\nabla\widetilde{\phi})^{2}-\frac{1}{2}m^{2}\widetilde{\phi}^{2}
+\widetilde{J}\widetilde{\phi}.
\]

A quantum field theory is introduced by requiring that the
equal-time non-zero commutation relations are fulfilled,
\begin{align}
\lbrack\phi(t,\mathbf{x),}\pi(t,\mathbf{y)}]  &  =i\delta(\mathbf{x-y}%
),\label{comrel72}\\
\lbrack\widetilde{\phi}(t,\mathbf{x),}\widetilde{\pi}(t,\mathbf{y)}]
&
=-i\delta(\mathbf{x-y}), \label{comrel74}%
\end{align}
The fields $\phi$ and $\pi$ are operators defined to act on a
Hilbert space $\mathcal{H}_{T}$. We use the Bogoliubov
transformation to introduce thermal operators. In this case there
are infinite modes and so a Bogoliubov
transformation is defined for each mode, i.e.%
\[
\phi(x;\beta)=\int\frac{d^{3}k}{(2\pi)^{3}}\frac{1}{2\omega_{k}}%
\ [a(k;\beta)e^{-ikx}+a^{\dag}(k;\beta)e^{ikx}]
\]
and%
\[
\widetilde{\phi}(x;\beta)=\int\frac{d^{3}k}{(2\pi)^{3}}\frac{1}{2\omega_{k}%
}\ [\tilde{a}(k;\beta)e^{ikx}+\tilde{a}^{\dag}(k;\beta)e^{-ikx}],
\]
where $a(k;\beta)$ ($\tilde{a}(k;\beta))$ and $a^{\dag}(k;\beta)$
($\tilde {a}^{\dag}(k;\beta))$ are thermal (tilde) annihilation
and creation operators. For the momenta, $\pi(x;\beta)$ and
$\widetilde{\pi}(x;\beta),$ we have
\[
\pi(x;\beta) = \dot{\phi}(x;\beta) = \int\frac{d^{3}k}{(2\pi)^{3}
}(-i)\ \frac{1}{2}[a(k;\beta)e^{-ikx}-a^{\dag}(k;\beta)e^{ikx}]
\]
and
\[
\widetilde{\pi}(x;\beta) = \dot{\widetilde{\phi}}(x;\beta)=\int
\frac{d^{3}k}{(2\pi)^{3}}\
\frac{i}{2}[\tilde{a}(k;\beta)e^{ikx}-\tilde
{a}^{\dag}(k;\beta)e^{-ikx}],
\]
where we have used the tilde conjugation rules to write
$\widetilde{\phi }(x;\beta)$ and $\widetilde{\pi}(x;\beta)$ from
$\phi(x;\beta)$ and $\pi(x;\beta),$ respectively.

The algebra given by Eqs.~(\ref{comrel72}) and (\ref{comrel74}) is
still valid for the operators $\phi(x;\beta),$
$\widetilde{\phi}(x;\beta),\pi(x;\beta)$ and\
$\widetilde{\pi}(x;\beta).$ Then the commutation relations for the
thermal modes read
\begin{align}
\lbrack a(k;\beta),a^{\dag}(k^{\prime};\beta)]  &  =(2\pi)^{3}2k_{0}%
\delta(\mathbf{k}-\mathbf{k}^{\prime}),\label{com72}\\
\lbrack\tilde{a}(k;\beta),\tilde{a}^{\dag}(k^{\prime};\beta)]  &
=(2\pi
)^{3}2k_{0}\delta(\mathbf{k}-\mathbf{k}^{\prime}), \label{com7311}%
\end{align}
with all the other commutation relations being zero. The general
Bogoliubov transformation applied to all modes is written in the
form
\begin{equation}
U(\beta)  = \exp\left\{
\sum_{k}\theta_{k}(\beta)[a^{\dag}(k)\tilde
{a}^{\dag}(k)-a(k)\tilde{a}(k)]\right\} = \prod_{k}U(k,\beta),
\label{bog73111}
\end{equation}
where
\[
U(k,\beta)=\exp\{\theta_{k}(\beta)[a^{\dag}(k)\tilde{a}^{\dag}(k)-a(k)\tilde
{a}(k)]\},
\]
with $\theta_{k}$ defined by $\cosh\theta_{k}=v(k,\beta),$ in the
limit of the continuum. However, in this limit the unitary nature
of the Bogoliubov transformation is lost, a property that gives
rise to non-equivalent vacua in the theory~\cite{Umezawabook1993}.
Despite the loss of unitarity, the Bogoliubov transformation is
still canonical, in the sense that, the algebraic structure of the
theory is preserved.

The thermal Hilbert space, $\mathcal{H}_{T}$, is constructed from
the thermal vacuum,
$|0(\beta)\rangle=U(\beta)|0,\tilde{0}\rangle,$ where $|0,\tilde
{0}\rangle={\bigotimes}_{k} |0,\tilde{0}\rangle_{k}$ and
$|0,\tilde {0}\rangle_{k}$ \ is the vacuum for the mode $k.$ The
thermal vacuum is such that
$a(k;\beta)|0(\beta)\rangle=\tilde{a}(k;\beta)|0(\beta)=0$ and
$\langle0(\beta)|0(\beta)\rangle=1$.

The thermal and non-thermal operators are related by
\begin{equation}
a(k;\beta)=U(\beta)a(k)U^{-1}(\beta)=u(k,\beta)a(k)-v(k,\beta)\tilde
{a}^{\dagger}(k), \label{bog71}%
\end{equation}
where $v(k,\beta)=1/\sqrt{\exp(\beta\omega_{k})-1}$ and
$u^{2}(k,\beta )-v^{2}(k,\beta)=1$. The other operators,
$a^{\dagger}(k),$ $\tilde{a}(k)$ and $\tilde{a}^{\dagger}(k)$ are
derived by using the Hermitian and the tilde conjugation rules.

The thermal Feynman propagator for the real scalar field is then
defined by
\begin{equation}
G(x-y;\beta)^{ab} = -i\langle0(\beta)|T[{\Phi}(x)^{a}{\Phi}^{+}
(y;\beta)^{b}]|0(\beta)\rangle = \frac{1}{(2\pi)^{4}}\int
d^{4}k\,G(k;\beta)^{ab}e^{-ik(x-y)}, \label{adre3}
\end{equation}
where $a,b=1,2$, with $a=1$ standing for non-tilde operators and
$a=2$ standing for tilde operators; $G(k;\beta)^{ab}=B^{-1}(k_{0}
;\beta)G_{0}(k)^{ab}B(k_{0};\beta),$ with
\begin{equation}
(G_{0}(k)^{ab})=\left(
\begin{array}
[c]{cc}
\frac{1}{k^{2}-m^{2}+i\epsilon} & 0\\
0 & \frac{-1}{k^{2}-m^{2}-i\epsilon}
\end{array}
\right)  , \label{andre2}
\end{equation}
such that the components of $G(k;\beta)^{ab}$ are given by
\begin{align*}
G(k;\beta)^{11}  &  =\frac{1}{k^{2}-m^{2}+i\epsilon}-2\pi i\,n(k_{0}%
)\,\delta(k^{2}-m^{2}),\\
G(k;\beta)^{22}  &  =\frac{-1}{k^{2}-m^{2}-i\epsilon}-2\pi i\,n(k_{0}%
)\,\delta(k^{2}-m^{2}),\\
G(k;\beta)^{12}  &  =G(k;\beta)^{21}=-2\pi i\,[n(k_{0})+n(k_{0})^{2}%
]^{1/2}\,\delta(k^{2}-m^{2}),
\end{align*}
where $n(k_{0})=v_{k}(\beta)^{2}$. The propagator
$G(k;\beta)^{11}$ is the same as in the Fourier integral
representation of the Matsubara method; and the two-by-two Green
function in Eq.~(\ref{adre3}) is similar to the propagator in the
Schwinger-Keldysh approach.

Closing this section, let us emphasize some aspects of the real
time formalism. First, it is important to note that doubling is
not a characteristic of TFD only, but rather an ingredient present
in any thermal theory. In terms of the density matrix, the
doubling is needed when we write $\rho(t)$ as a projector, i.e.
$\rho\simeq|\psi\rangle\langle\psi|$, and
the Liouville von-Neumann equation is written in the form $i\partial_{t}%
\rho(t)=L\rho(t).$ In this later case, the time evolution is
controlled by $L=[H,.\,]$, the Liouvilian, which is an object
associated with, but different from, the Hamiltonian operator,
$H$. In TFD as in the density matrix formalism, this doubling has
been physically identified and plays a central role in the
construction of mixed states, for instance. Another aspect to be
emphasized is that a real-time formalism is also a theory on
$\Gamma_{4}^{1}$ topology, where the state $|0(\beta)\rangle$ is
the counterpart ingredient of the imaginary-time procedure. The
generalization of the thermal theory, in both versions of Fourier
series representation or the integral representation, leads
naturally to a theory on $\Gamma_{D}^{d}$ topology. These aspects
are explored in the next section.

As a final observation, it is to be  noted that the TFD structure
has been described in geometric terms. Indeed, Israel showed that
one can observe a thermal vacuum for
 particle modes limited by a space-time horizon~\cite{IsraelTFD1976}. The procedure was
 a geometric description of TFD, which  generalized the results of Fulling, Gibbons,
 Hawking and Unruh about black
 holes. These ideas were improved in several directions, using Rindler coordinates
 and the notion of entanglement of the left and right Rindler
 quanta~\cite{Gui1992,Gui1992a,RovelliTFDIsrael2012}.

\section{Field theory on the topology $\Gamma_{D}^{d}$}

In this section, we generalize the argument advanced in the
previous section establishing that a quantum field theory at
finite temperature corresponds to a field theory on a topology
$\Gamma_{D}^{1}=(\mathbb{S}^{1})\times \mathbb{R}^{D-1}$, where
the imaginary time is the compactified dimension. We proceed
further with these concepts, to include in the analysis not only
time but also space coordinates, in such way that any set of
dimensions of the manifold $\mathbb{R}^{D}$ can be compactified,
defining a theory on the topology
$\Gamma_{D}^{d}=(\mathbb{S}^{1})^{d}\times\mathbb{R}^{D-d}$, with
$1\leq d\leq D$. This establishes a formalism that suffices to
deal with the general question of compactification on the
$\Gamma_{D}^{d}$ topology at finite-temperature, in the imaginary
or real
time~\cite{BirrellFord1980,Ford1980,KhannamalsaesAnnals2009,Ourbook2009,%
KhannamalsaesAnnals2011}.

Using a generalized Bogoliubov transformation, the effect of
compactification on $\Gamma_{D}^{d}$ is introduced and a Fourier
integral representation for the propagator is derived. Initially,
an analysis is carried out for free boson and fermion fields. An
extension of the formalism for abelian and non-abelian
gauge-fields is derived by functional methods. Exploring the
canonical formalism, the S-matrix is
developed~\cite{KhannamalsaesAnnals2009,Ourbook2009,KhannamalsaesAnnals2011}.

\subsection{Generalized Matsubara procedure}

We consider a D-dimensional space-time, $\mathbb{R}$, with a point
given by $x=(x^{0},\mathbf{z})$, where $x^{0}$ stands for the time
and $\mathbf{z} =(z^{1},z^{2},...,z^{D-1})$ space coordinates.
Since the toroidal topology keeps the nature of the local
interaction, the Feynman rules for a field theory on the
hyper-torus $\Gamma_{D}^{d}$ are a generalization of the Matsubara
prescription, Eqs.~(\ref{portoabril1})-(\ref{portoabril3}).
Explicitly, the Feynman rules in the momentum space are modified
by taking, for each compactified space dimension, the integrals in
momentum space
replaced by sums, in the following way:%
\[
\int
\frac{dp^{i}}{2\pi}\rightarrow\frac{1}{L_{i}}\sum_{l_{i}=-\infty}^{\infty}
\]
with $i=1,...,D-1$,
\[
p_{i}\rightarrow p_{l_{i}}=\frac{2\pi l_{i}}{L_{i}},
\]
for bosons, and
\[
p_{i}\rightarrow p_{l_{i}}=\frac{2\pi(l_{i}+\frac{1}{2})}{L_{i}},
\]
for fermions .

For a free boson field compactified in $d\leq D$ dimensions,
satisfying periodic boundary condition, we have
\begin{equation}
G_{0}(x-x^{\prime};\alpha)=\frac{1}{i^{d}\,\alpha_{0}\cdot\cdot\cdot
\alpha_{d-1}}\sum\limits_{n_{0}\cdot\cdot\cdot n_{d-1}}\int\frac
{d^{D-d}\mathbf{k}}{(2\pi)^{D-d}}\,\frac{e^{-ik_{\alpha}(x-x^{\prime})}%
}{k_{\alpha}^{2}-m^{2}+i\varepsilon}, \label{may10201}%
\end{equation}
where $k_{\alpha}=(k_{n_{0}}^{0},k_{n_{1}}^{1},\dots,k_{n_{d-1}}^{d-1}%
,k^{d},\dots,k^{D-1})$, with
\[
k_{n_{j}}^{j}=\frac{2\pi n_{j}}{\alpha_{j}},\ \ 0\leq j\leq d-1,
\]
$n_{j}\in\mathbb{Z}$ and $d^{D-d}\mathbf{k}=dk^{d}dk^{d+1}\cdots
dk^{D-1}$. The compactification parameters
$\alpha=(\alpha_{0},\alpha_{1},...,\alpha _{D})$ stand for the
effect of temperature ($\alpha_{0}=\beta=T^{-1}$) and for space
compactifion ($\alpha_{j}$). In what follows, we employ the name
Matsubara representation (or prescription) referring to both time
and space compactification. The Green function
$G_{0}(x-x^{\prime};\alpha)$ is a solution of the Klein-Gordon
equation. This means that $G_{0}(x-x^{\prime };\alpha)$ is the
Green function of a boson field defined locally in the Minskowki
space-time. Globally this theory is such that $G_{0}(x-x^{\prime
};\alpha)$ has to satisfy periodic boundary conditions (the
generalization of the KMS condition). These facts assure us that
$G_{0}(x-x^{\prime};\alpha)$   is the Green function of a field
theory defined on a hyper-torus,
$\Gamma_{D}^{d}=(\mathbb{S}^{1})^{d}\times\mathbb{R}^{D-d}$, with
$1\leq d\leq D$, where the circumference of the $j$-th
$\mathbb{S}^{1}$ is specified by $\alpha_{j}$.

We proceed to study representations in terms of the spectral
function, as derived for the case of temperature by Dolan and
Jackiw~\cite{DolanJackiw1974}. For the case of compactified
spatial dimensions, the spectral function is defined in terms of
the momentum, such that the calculation follows in parallel with
the case of temperature. At the end of the calculation a Wick
rotation is performed in order to recover the physical meaning of
the theory~\cite{KhannamalsaesAnnals2011}.  We start, for
simplicity, with one-compactified dimension. Take the topology
$\Gamma_{D}^{1}$ where the imaginary time-axis is compactified. In
this case, we denote, $\alpha =(\beta,0,\dots,0)=\beta n_{0}$, $
n_{0}=(1,0,\dots,0)$, with $T=\beta^{-1}$ being the temperature,
such that the Green function is given by Eq.~(\ref{8propmat1}) and
its Fourier integral representation is given in
Eq.~(\ref{bo1012}). In the case of compactification of the
coordinate $x^{1}$, for the topology $\Gamma_{D}^{1}$, we take
$\alpha=(0,iL_{1},0,\dots ,0)=iL_{1} n_{1}$, with $
n_{1}=(0,1,0,\dots,0)$. The factor $i$ in the parameter
$\alpha_{j}$ corresponding to the compactification of a space
coordinate makes explicit that we are working with the Minkowski
metric; the period in the $x^{1}$ direction is real and equal to
$L_{1}$. The propagator has the Matsubara representation
\[
G_{0}(x-y;L_{1})=\frac{1}{L_{1}}\sum\limits_{l_{1}}\int\frac{d^{D-1}%
\mathbf{k}}{(2\pi)^{D-1}}\frac{e^{-ik_{l_{1}}(x-y)}}{(k_{l_{1}}%
)^{2}-m^{2}+i\varepsilon},
\]
where $k_{l_{1}}=(k^{0},k_{l_{1}}^{1},k^{2},\dots,k^{D-1})$, with $k_{l_{1}%
}^{1}=2\pi l_{1}/L_{1}$. The Fourier-integral representation is
derived along the same way as for
temperature~\cite{KhannamalsaesAnnals2011}, leading to
\[
G_{0}(x-y;L_{1})=\int\frac{d^{D}k}{(2\pi)^{D}}e^{-ik(x-y)}\
G_{0}(k;L_{1}),
\]
where
\[
G_{0}(k;L_{1})=\frac{-1}{k^{2}-m^{2}+i\varepsilon}+f_{L_{1}}(k^{1})2\pi
i\delta(k^{2}-m^{2}),
\]
with
\[
f_{L_{1}}(k^{1})=\sum\limits_{l_{1}=1}^{\infty}e^{-iL_{1}k^{1}l_{1}}.
\]

Now let us consider the topology $\Gamma_{D}^{2}$, accounting for
a double compactification, one being the imaginary time and the
other the $x^{1}$
direction. In this case $\alpha=(\beta,iL_{1},0,\dots,0)=\beta n%
_{0}+iL_{1} n_{1}$. The Fourier-series representation is
\[
G_{0}(x-y;\beta,L_{1})=\frac{1}{i\beta L_{1}}\sum\limits_{l_{0},l_{1}}%
\int\frac{d^{D-2}\mathbf{k}}{(2\pi)^{D-2}}\frac{e^{-ik_{l_{0}l_{1}}%
(x-y)}}{(k_{l_{0}l_{1}})^{2}-m^{2}+i\varepsilon},
\]
where
$k_{l_{0}l_{1}}=(k_{l_{0}}^{0},k_{l_{1}}^{1},k^{2},\dots,k^{D-1})$,
with $k_{l_{0}}^{0}=2\pi l_{0}/\beta$ and $k_{l_{1}}^{1}=2\pi
l_{1}/L_{1}$. The corresponding Fourier-integral representation is
\[
G_{0}(x-y;\beta,L_{1})=\int\frac{d^{D}k}{(2\pi)^{D}}e^{-ik(x-y)}%
\ G_{0}(k;\beta,L_{1}),
\]
where
\[
G_{0}(k;\beta,L_{1})=\frac{-1}{k^{2}-m^{2}+i\varepsilon}+f_{\beta L_{1}}%
(k^{0},k^{1})2\pi i\delta(k^{2}-m^{2}),
\]
with
\[
f_{\beta
L_{1}}(k^{0},k^{1})=f_{\beta}(k^{0})+f_{L_{1}}(k^{1})+2f_{\beta
}(k^{0})f_{L_{1}}(k^{1}).
\]
The generalization of this result for $d$ compactified dimensions
leads to a general structure of the
propagator~\cite{KhannamalsaesAnnals2011}, which is given by
\begin{equation}
G_{0}(x-y;\alpha)=\int\frac{d^{D}k}{(2\pi)^{D}}e^{-ik(x-y)}\
G_{0}(k;\alpha),
\label{bo10121}%
\end{equation}
where%
\begin{equation}
G_{0}(k;\alpha)=G_{0}(k)+\
f_{\alpha}(k_{\alpha})[G_{0}(k)-G_{0}^{\ast}(k)];
\label{CBPF22.12.3}%
\end{equation}
the function $f_{\alpha}(k_{\alpha})$
is~\cite{KhannamalsaesAnnals2011}
\begin{eqnarray}
f_{\alpha }(p^{\alpha }) & = & \sum_{s=1}^{d} \sum_{\{\sigma
_{s}\}} 2^{s-1} \left( \prod_{j=1}^{s} f_{\alpha _{\sigma
_{j}}}(p^{\sigma _{j}})\right)
\notag \\
& = & \sum_{s=1}^{d} \sum_{\{\sigma _{s}\}} 2^{s-1}
\sum_{l_{\sigma _{1}},...,l_{\sigma _{s}}=1}^{\infty }
(-\eta)^{s+\sum_{r=1}^{s}l_{\sigma _{r}}}\,\exp
\{-\sum_{j=1}^{s}\alpha _{\sigma _{j}}l_{\sigma _{j}}p^{\sigma
_{j}}\},  \label{v2gen}
\end{eqnarray}%
where $\eta =1\;(-1)$ for fermions (bosons) and $\{\sigma _{s}\}$
denotes the set of all combinations with $s$ elements, $\{\sigma
_{1},\sigma
_{2},...\sigma _{s}\}$, of the first $d$ natural numbers $\{0,1,2,...,d-1\}$%
, that is all subsets containing $s$ elements; in order to obtain
the physical condition of finite temperature and spatial
confinement, $\alpha
_{0}$ has to be taken as a positive real number, $\beta = T^{-1}$, while $%
\alpha _{n}$, for $n=1,2,...,d-1$, must be pure imaginary of the form $%
iL_{n} $.

In short, we have presented a formalism to consider the quantum
field theory in a flat manifold with topology (${\mathbb
S}^{1})^{d} \times \mathbb{R}^{D-d}$, such that fields and  Green
functions fulfill periodic (bosons) or antiperiodic(fermions)
boundary conditions. The result of the topological analysis is a
generalization of the Matsubara formalism. The representation in
terms of the generalization of TFD can be accomplished as well.
This aspect is developed in part in the rest of this section and
in Section 4, where a representation theory for such compactified
quantum fields on a torus is presented.

\subsection{Quantum fields and the Bogolibov transformation on $\Gamma_{D}
^{d}$}

\subsubsection{Boson fields}

Let us consider a free boson field. The modular conjugation rules
can be applied to any relation among the dynamical variables, in
particular to the equation of motion in the Heisenberg picture.
The set of doubled equations are then derived by writing the
hat-Hamiltonian, the generator of time translation, as
$\widehat{H}=H-\widetilde{H}$. In this case the time evolution
generator is $\widehat{H}.$ Then we have the Lagrangian densities
$\mathcal{L}(x)$ and $\mathcal{L}(x;\alpha)$ given, respectively,
by
\begin{align}
\mathcal{L}(x) &
=\frac{1}{2}\partial_{\mu}\phi(x)\partial^{\mu}\phi
(x)-\frac{m^{2}}{2}\phi(x)^{2},\label{may10202}\\
\mathcal{L}(x;\alpha) &
=\frac{1}{2}\partial_{\mu}\phi(x;\alpha)\partial^{\mu
}\phi(x;\alpha)-\frac{m^{2}}{2}\phi(x;\alpha)^{2},\label{aug091441}%
\end{align}
where the field $\phi(x;\alpha)$ is defined by
\[
\phi(x;\alpha)=U(\alpha)\phi(x)U^{-1}(\alpha).
\]

The mapping $U(\alpha)$ is taken as a Bogoliubov transformation
and is defined, as usual, by a two-mode squeezed operator. For
fields expanded in terms of modes, we define
\begin{equation}
U(\alpha) = \exp\left\{  \sum_{k}\theta(k_{\alpha};\alpha)[a^{\dag
}(k)\widetilde{a}^{\dag}(k)-a(k)\widetilde{a}(k)]\right\} =
\prod_{k}U(k;\alpha),\label{bog731}
\end{equation}
where
\[
U(k_{\alpha};\alpha)=\exp\{\theta(k_{\alpha};\alpha)[a^{\dag}(k)\widetilde
{a}^{\dag}(k)-a(k)\widetilde{a}(k)]\},
\]
with $\theta(k_{\alpha};\alpha)$ being a function of the momentum,
$k_{\alpha }$, and of the parameters $\alpha,$ both to be
specified. The label $k$ in the sum and in the product of the
equations above is to be taken in the continuum limit, for each
mode. Then we have
\begin{equation}
\phi(x;\alpha)=\int\frac{d^{D-1}\mathbf{k}}{(2\pi)^{D-1}}\frac{1}{2k_{0}
}[a(k;\alpha)e^{-ikx}+a^{\dag}(k;\alpha)e^{ikx}].\label{xifield}
\end{equation}
To obtain this expression, we have used the non-zero commutation
relations
\begin{equation}
\lbrack a(k;\alpha),a^{\dag}(k^{\prime};\alpha)]=(2\pi)^{3}2k_{0}
\delta(\mathbf{k}-\mathbf{k}^{\prime}),\label{com73}
\end{equation}
with
\begin{equation}
a(k;\alpha) = U(k_{\alpha};\alpha)a(k)U^{-1}(k_{\alpha};\alpha) =
u(k_{\alpha};\alpha)a(k)-v(k_{\alpha};\alpha)\,\widetilde{a}^{\dagger}(k),
\end{equation}
where $u(k_{\alpha};\alpha)$ and $v(k_{\alpha};\alpha)$ are given
in terms of $\theta(k_{\alpha};\alpha)$ by
\[
u(k_{\alpha};\alpha) = \cosh\theta(k_{\alpha};\alpha) \; ,
\;\;\;\;\; v(k_{\alpha};\alpha) = \sinh\theta(k_{\alpha};\alpha).
\]
The inverse is
\[
a(k)=u(k_{\alpha};\alpha)a(k;\alpha)+v(k_{\alpha};\alpha)\,\widetilde
{a}^{\dagger}(k;\alpha),
\]
such that the other operators $a^{\dag}(k),\widetilde{a}(k)$ and
$\widetilde{a}^{\dagger}(k)$ are obtained by applying the
hermitian conjugation or the tilde conjugation, or both.

It is worth noting that the transformation $U(\alpha)$ can be
mapped into a
$2\times2$ representation of the Bogoliubov transformation, i.e.%
\begin{equation}
B(k_{\alpha};\alpha)=\left(
\begin{array}
[c]{cc}%
u(k_{\alpha};\alpha) & -v(k_{\alpha};\alpha)\\
-v(k_{\alpha};\alpha) & u(k_{\alpha};\alpha)
\end{array}
\right)  ,\label{Bmar101}%
\end{equation}
with $u^{2}(k_{\alpha};\alpha)-v^{2}(k_{\alpha};\alpha)=1$, acting
on the pair of commutant operators as
\[
\left(
\begin{array}
[c]{c}%
a(k;\alpha)\\
\widetilde{a}^{\dagger}(k;\alpha)
\end{array}
\right)  =B(k_{\alpha};\alpha)\left(
\begin{array}
[c]{c}%
a(k)\\
\widetilde{a}^{\dagger}(k)
\end{array}
\right)  .
\]
A Bogoliubov transformation of this type gives rise to a compact
and elegant $2\times2$ representation of the propagator in the
real-time formalism.

%\underset{k}

The Hilbert space is constructed from the $\alpha$-state,
$|0(\alpha
)\rangle=U(\alpha)|0,\widetilde{0}\rangle,$ where $|0,\widetilde{0}%
\rangle = {\bigotimes}_k |0,\widetilde{0}\rangle_{k}$ and
$|0,\widetilde{0}\rangle_{k}$ is the vacuum for the mode $k.$ Then
we have the following:
$a(k;\alpha)|0(\alpha)\rangle=\widetilde{a}(k;\alpha)|0(\alpha)\rangle=0$
and $\langle0(\alpha)|0(\alpha)\rangle=1.$ This shows that
$|0(\alpha)\rangle$ is a vacuum for $\alpha$-operators
$a(k;\alpha)$. However, it is important to note that it is a
condensate for the operators $a(k)$ and $a^{\dagger}(k),$ since
for instance  $a(k)|0(\alpha)\rangle \neq 0$. An arbitrary basis
vector is given in the form
\begin{equation}
|\psi(\alpha);\{m\};\{k\}\rangle =
[a^{\dag}(k_{1};\alpha)]^{m_{1}} \cdots\lbrack
a^{\dag}(k_{M};\alpha)]^{m_{M}}
\lbrack\widetilde{a}^{\dag}(k_{1};\alpha)]^{n_{1}}\cdots
\lbrack\widetilde{a}^{\dag}(k_{N};\alpha)]^{n_{N}}|0(\alpha)\rangle
,\label{therm12345}
\end{equation}
where $n_{i},m_{j}=0,1,2,...,$ with $N$ and $M$ being indices for
an arbitrary mode.

Consider only one field-mode, for simplicity. Then we write
$|0(\alpha )\rangle$ in terms of $u(\alpha)$ and $v(\alpha)$ as
\begin{equation}
|0(\alpha)\rangle =
\frac{1}{u(\alpha)}\exp\left[\frac{v(\alpha)}{u(\alpha
)}a^{\dagger}\widetilde{a}^{\dagger}\right] \left| 0,\widetilde{0}
\right\rangle = \frac{1}{u(\alpha)}\sum\limits_{n}\left(
\frac{v(\alpha)}{u(\alpha
)}\right)^{n}|n,\widetilde{n}\rangle.\label{vacuu12}
\end{equation}
At this point, the physical meaning of an arbitrary $\alpha$-state
given in Eq.~(\ref{therm12345} ) is not established. This aspect
becomes clear by considering the Green function defined by
\[
G_{0}(x-y;\alpha)=-i\langle0(\alpha)|\mathrm{T}[\phi(x)\phi(y)]|0(\alpha
)\rangle.
\]
 Using
$U(\alpha)$ in Eq.~(\ref{bog731}), we find that the $\alpha$-Green
function is written as
\[
G_{0}(x-y;\alpha)=-i\langle\widetilde{0},0|\mathrm{T}[\phi(x;\alpha
)\phi(y;\alpha)]|0,\widetilde{0}\rangle.
\]
Then, using the field expansion, Eq.~(\ref{xifield}), and the
commutation relation Eq.~(\ref{com7311}), we obtain
\begin{equation}
G_{0}(x-y;\alpha)=\int\frac{d^{D}k}{(2\pi)^{D}}e^{-ik(x-y)}\ G_{0}%
(k;\alpha),\label{bo10}%
\end{equation}
where%
\[
G_{0}(k;\alpha)=G_{0}(k)+\
v^{2}(k_{\alpha};\alpha)[G_{0}(k)-G_{0}^{\ast}(k)].
\]

This propagator is formally identical to $G_{0}(x-y;\alpha)$
written in the integral representation given by
Eqs.~(\ref{bo10121}) and (\ref{CBPF22.12.3}). Then the analysis in
terms of representation of Lie-groups and the Bogoliubov
transformation leads to the integral representation by performing
the mapping $v^{2}(k_{\alpha};\alpha)\rightarrow
f_{\alpha}(k_{\alpha})$. It is worthwhile
 to note that this is possible, since $v^{2}(k_{\alpha};\alpha)$ has not been
fully specified up to this point. Considering the specific case of
compactification in time, in
order to describe temperature only, the real quantity $v^{2}(k_{\alpha}%
;\alpha)$ is mapped on the real quantity $f_{\beta}(w)\equiv
n(\beta)$. Including space compactification,
$f_{\alpha}(k_{\alpha})$ is a complex function. In such a case,
$f_{\alpha }(k_{\alpha})$ is an analytical continuation of the
real function $v^{2}(k_{\alpha};\alpha)$; a procedure that is
possible, since, $v^{2}(k_{\alpha};\alpha)$ is arbitrary.
Therefore, for space
compactification, we can also perform the mapping $v^{2}(k_{\alpha}%
;\alpha)\rightarrow f_{\alpha}(k_{\alpha})$ in $G_{0}(k;\alpha)$,
in order to recover the propagator shown in Eqs.~(\ref{bo10121})
and (\ref{CBPF22.12.3}). From now on, we denote the vector
$|\alpha_{w}^{\alpha}\rangle$ by
$|\alpha\rangle$ and the function $f_{\alpha}(k_{\alpha})$ by $v^{2}%
(k_{\alpha};\alpha)$.

\subsubsection{Fermion field}

A similar mathematical structure is introduced for the
compactification of fermion fields. First, we have  to construct
the state $|\alpha\rangle$ explicitly.

The Lagrangian density for the free Dirac field is
\begin{equation}
\mathcal{L}(x)=\frac{1}{2}\overline{\psi}(x)\left[  \gamma\cdot i
\overleftrightarrow{\partial}-m \right]  \psi(x)
\end{equation}
and for the $\alpha$-field we have
\begin{equation}
\mathcal{L}(x;\alpha)=\frac{1}{2}\overline{\psi}(x;\alpha) \left[
\gamma\cdot i \overleftrightarrow{\partial}-m \right]
\psi(x;\alpha).
\end{equation}
The field $\psi(x;\alpha)$ is expanded as
\[
\psi(x;\alpha)=\int\frac{d^{D-1}k}{(2\pi)^{D-1}}\frac{m}{k_{0}}
\sum
\limits_{\zeta=1}^{2}\left[  c_{\zeta}(k;\alpha)u^{(\zeta)}(k)e^{-ikx}%
+d_{\zeta}^{\dagger}(k;\alpha)v^{(\zeta)}(k)e^{ikx}\right]  ,
\]
where $u^{(\zeta)}(k)$ and $v^{(\zeta)}(k)$ are basic spinors. The
fermion $\alpha$-operators $c(k;\alpha)$ and $d(k;\alpha)$ stand
for particle and anti-particle destruction operators,
respectively. The fermion field $\psi(x;\alpha)$ is defined by
\[
\psi(x;\alpha)=U(\alpha)\psi(x)U^{-1}(\alpha),
\]
where $U(\alpha)$ is
\[
U(\alpha) = \exp\left\{ \sum_{k}\{\theta_{c}(k;\alpha)[c^{\dag}(k)
\widetilde{c}^{\dag}(k)-c(k)\widetilde{c}(k)] +
\theta_{d}(k;\alpha)[d^{\dag
}(k)\widetilde{d}^{\dag}(k)-d(k)\widetilde{d}(k)]\}\right\} =
\prod\limits_{k}U_{c}(k;\alpha)U_{d}(k;\alpha),
\]
with
\begin{align}
U_{c}(k;\alpha)  &
=\exp\{\theta_{c}(k;\alpha)[c^{\dag}(k)\widetilde{c}
^{\dag}(k)-c(k)\widetilde{c}(k)]\},\nonumber\\
U_{d}(k;\alpha)  &
=\exp\{\theta_{d}(k;\alpha)[d^{\dag}(k)\widetilde{d}
^{\dag}(k)-d(k)\widetilde{d}(k)]\}.\nonumber
\end{align}
In terms of non $\alpha$-operators, the fermion $\alpha$-operators
$c(k;\alpha)$ and $d(k;\alpha)$ are written as
\begin{align}
c(k;\alpha)  &  = U(k;\alpha)c(k)U^{-1}(k;\alpha) =
u_{c}(k;\alpha)c(k)-v_{c}(k;\alpha)\widetilde{c}^{\dagger}(k), \nonumber\\
d(k;\alpha)  &  = U(k;\alpha)d(k)U^{-1}(k;\alpha) =
u_{d}(k;\alpha)d(k)-v_{d}(k;\alpha)\widetilde{d}^{\dagger}(k).\nonumber
\end{align}
The parameters $\theta_{c}(k;\alpha)$ and $\theta_{d}(k;\alpha)$
are such that $\sin\theta_{c}(k;\alpha)=v_{c}(k_{\alpha};\alpha)$,
and $\sin\theta _{d}(k;\alpha)=v_{d}(k_{\alpha};\alpha)$,
resulting in $v_{c}^{2}(k_{\alpha
};\alpha)+u_{c}^{2}(k_{\alpha};\alpha)=1$ and $v_{d}^{2}(k_{\alpha}%
;\alpha)+u_{d}^{2}(k_{\alpha};\alpha)=1$. The inverse formulas for
the $\alpha$-operators are
\begin{align}
c(k)  &
=u_{c}(k_{\alpha};\alpha)c(k;\alpha)+v_{c}(k_{\alpha};\alpha)
\widetilde{c}^{\dagger}(k;\alpha),\nonumber\\
d(k)  &
=u_{d}(k_{\alpha};\alpha)d(k;\alpha)+v_{d}(k;\alpha)\widetilde{d}
^{\dagger}(k;\alpha),\nonumber
\end{align}
where the operators $c$ and $d$ carry a spin index.

These operators satisfy the anti-commutation relations
\[
\{c_{\zeta}(k,\alpha\mathbf{),}c_{\varkappa}^{\dagger}(k^{\prime}%
,\alpha\mathbf{)}\}=\{d_{\zeta}(k,\alpha\mathbf{),}d_{\varkappa}^{\dagger
}(k^{\prime},\alpha\mathbf{)}\}=(2\pi)^{3}\frac{k_{0}}{m}\delta(\mathbf{k-k}%
^{\prime})\delta_{\zeta\varkappa},
\]
with all   other anti-commutation relations being zero. In order
to be consistent with the Lie algebra, and with the definition of
the $\alpha$
-operators, a fermion operator, $A,$ is such that $\widetilde{\widetilde{A}%
}=-A$ and a tilde-fermion operator anti-commutes with a non-tilde
operator. This is consistent in the following sense. Consider, for
instance,
$c(k;\alpha)=U(k;\alpha)c(k)U^{-1}(k;\alpha)$ and $\widetilde{c}%
(k;\alpha)=U(k;\alpha)\widetilde{c}(k)U^{-1}(k;\alpha)$. In order
to map $c(k;\alpha)\rightarrow\widetilde{c}(k;\alpha)$ by using
the modular conjugation, directly, it leads to
$\widetilde{\widetilde{c}}(k)=-c(k)$.

Let us define the $\alpha$-state
$|0(\alpha)\rangle=U(\alpha)|0,\tilde{0} \rangle,$ where
\[
|0,\tilde{0}\rangle=\bigotimes_{k}|0,\tilde{0}\rangle_{k}%
\]
and $|0,\tilde{0}\rangle_{k}$ is the vacuum for the mode $k$ for
particles and anti-particles. This $\alpha$-state satisfies the
condition $\langle
0(\alpha)|0(\alpha)\rangle=1$. Moreover, we have%
\begin{align}
c(k;\alpha)|0(\alpha)\rangle &  =\widetilde{c}(k;\alpha)|0(\alpha)\rangle=0,\\
d(k;\alpha)|0(\alpha)\rangle &
=\widetilde{d}(k;\alpha)|0(\alpha)\rangle=0.
\end{align}
Then $|0(\alpha)\rangle$ is a vacuum state for the
$\alpha$-operators $c(k;\alpha)$ and $d(k;\alpha)$. Basis vectors
are given in the form
\[
\lbrack c^{\dag}(k_{1};\alpha)]^{r_{1}}\cdots[d^{\dag}(k_{M};\alpha)]^{r_{M}%
}[\widetilde{c}^{\dag}(k_{1};\alpha)]^{s_{1}}\cdots[\widetilde{d}
^{\dag }(k_{N};\alpha)]^{s_{N}}|0(\alpha)\rangle,
\]
where $r_{i},s_{i}=0,1$. A general $\alpha$-state can then be
defined by a linear combinations of such basis vectors.

Let us consider some particular cases, first, the case of
temperature. The topology is $\Gamma_{D}^{1}$, and we take
$\alpha=(\beta,0,\dots,0)$, leading to
\begin{align}
v_{c}^{2}(k^{0};\beta)  &  =\frac{1}{e^{\beta(w_{k}-\mu_{c})}+1},\\
v_{d}^{2}(k^{0};\beta)  &  =\frac{1}{e^{\beta(w_{k}+\mu_{d})}+1},
\end{align}
where $\mu_{c}$ and $\mu_{d}$ are the chemical potential for
particles and antiparticles, respectively. For simplicity, we take
$\mu_{c}=\mu_{d}=0$, and write
$v_{F}(k^{0};\beta)=v_{c}(k^{0};\beta)=v_{d}(k^{0};\beta)$, such
that
\[
v_{F}^{2}(k^{0};\beta)=\frac{1}{e^{\beta
w_{k}}+1}=\sum\limits_{n=1}^{\infty }(-1)^{1+n}e^{-\beta w_{k}n}.
\]
For the case of spatial compactification, we take $\alpha=(0,iL_{1}%
,0,\dots,0)$. By a type of Wick rotation, we derive
$v_{F}^{2}(k^{1};L_{1})$
from $v_{F}^{2}(k^{0};\beta)$, resulting in%
\[
v_{F}^{2}(k^{1};L_{1})=\sum\limits_{n=1}^{\infty}(-1)^{1+n}e^{-iL_{1}k^{1}n}.
\]
For spatial compactification and temperature, we have
\[
v_{F}^{2}(k^{0},k^{1};\beta,L_{1})=v_{F}^{2}(k^{1};\beta)+v_{F}^{2}%
(k^{1};L_{1})+2v_{F}^{2}(k^{1};\beta)v_{F}^{2}(k^{1};L_{1}).
\]

The $\alpha$-Green function is given by
\begin{equation}
S_{0}(x-y;\alpha)=-i\langle0(\alpha)|\mathrm{T}[\psi(x)\overline{\psi}
(y)]|0(\alpha)\rangle. \label{prop78}
\end{equation}
Let us write
\begin{equation}
iS_{0}(x-y;\alpha)=\theta(x^{0}-y^{0})S(x-y;\alpha)-\theta(y^{0}-x^{0})
\overline{S}(y-x;\alpha), \label{feyn79}
\end{equation}
with
$S(x-y;\alpha)=\langle0(\alpha)|\psi(x)\overline{\psi}(y)|0(\alpha
)\rangle$ and
$\overline{S}(x-y;\alpha)=\langle0(\alpha)|\overline{\psi}
(y)\psi(x)|0(\alpha)\rangle$. Calculating $S(x-y;\alpha)$ we
obtain
\begin{equation}
S(x-y;\alpha) =
(i\gamma\cdot\partial+m)\int\frac{d^{D-1}k}{(2\pi)^{D-1}
}\frac{1}{2\omega_{k}} \lbrack e^{-ik(x-y)}-
v_{F}^{2}(k_{\alpha};\alpha)(e^{-ik(x-y)}
-e^{ik(x-y)})],\label{prop713}
\end{equation}
For $\overline{S}(x-y;\alpha)$, we have
\begin{equation}
\overline{S}(x-y;\alpha)
=(i\gamma\cdot\partial+m)\int\frac{d^{D-1}k}{ (2\pi)^{D-1}}
\frac{1}{2\omega_{k}} \lbrack- e^{ik(x-y)} +
v_{F}^{2}(k_{\alpha};\alpha)(e^{-ik(x-y)}
+e^{ik(x-y)})].\label{prop724}
\end{equation}
This leads to%
\begin{equation}
S_{0}(x-y;\alpha)=(i\gamma\cdot\partial+m)G_{0}^{F}(x-y;\alpha),
\label{Sfermions}%
\end{equation}
where%
\begin{equation}
G_{0}^{F}(x-y;\alpha)=\int\frac{d^{D}k}{(2\pi)^{D}}e^{-ik(x-y)} G_{0}%
^{F}(k;\alpha), \label{Gxfermi}%
\end{equation}
and%
\begin{equation}
G_{0}^{F}(k;\alpha)=G_{0}(k)+\ v_{F}^{2}(k_{\alpha};\alpha)[G_{0}%
(k)-G_{0}^{\ast}(k)]. \label{Gkfermi}%
\end{equation}

This Green function is similar to the boson Green function,
Eq.~(\ref{bo10}); the difference is the fermion function
$v_{F}^{2}(k_{\alpha};\alpha)$. Using a the Bogoliubov
transformation, written in the form of a $2\times2$ matrix for
particles (subindex $c$) and anti-particles (subindex $d$), i.e.
\begin{equation}
B_{c,d}(k_{\alpha};\alpha)=\left(
\begin{array}
[c]{cc}%
u_{c,d}(k_{\alpha};\alpha) & v_{c,d}(k_{\alpha};\alpha)\\
-v_{c,d}(k_{\alpha};\alpha) & u_{c,d}(k_{\alpha};\alpha)
\end{array}
\right)  . \label{bog23july}%
\end{equation}
a  $2\times2$ Green function is introduced. This will be explored
later in the application of the Casimir effect.

\subsection{Generating functional}

We now construct the generating functional for interacting fields
in a flat space with topology $\Gamma_{D}^{d}$.

\subsubsection{Bosons}

For a system of free bosons, we consider,  up to normalization
factors, the following generating functional
\begin{equation}
Z_{0} \simeq \int D\phi e^{iS} = \int D\phi\exp\left[i\int
dx\mathcal{L}\right] = \int D \phi\exp \left\{-i\int dx
\left[\frac{1}{2}\phi(\square+m^{2})\phi-J\phi \right]\right\},
\end{equation}
where $J$ is a source. Such a functional is written as
\begin{equation}
Z_{0}\simeq\exp\left\{\frac{i}{2}\int dxdy
\left[J(x)(\square+m^{2}-i\varepsilon
)^{-1}J(y)\right]\right\}, \label{andr1}%
\end{equation}
describing the usual generating functional for bosons. However, we
would like to introduce the topology $\Gamma_{D}^{d}$. This is
possible by finding a solution of the Klein-Gordon equation
\begin{equation}
(\square+m^{2}+i\varepsilon)G_{0}(x-y;\alpha)=-\delta(x-y).
\label{KG11mar1}
\end{equation}
Using this result in Eq.~(\ref{andr1}), we find the normalized
functional
\begin{equation}
Z_{0}[J;\alpha]=\exp\left\{\frac{i}{2}\int dxdy
\left[J(x)G_{0}(x-y;\alpha)J(y)\right]\right\}.
\label{andr4}%
\end{equation}
Then\ we have%
\[
G_{0}(x-y;\alpha) = i \left.
\frac{\delta^{2}Z_{0}[J;\alpha]}{\delta J(y)\delta J(x)}
\right|_{J=0}.
\]
In order to treat interactions, we analyze the usual approach with
the $\alpha$-Green function. The Lagrangian density is
\[
\mathcal{L}(x)=\frac{1}{2}\partial_{\mu}\phi(x)\partial^{\mu}\phi
(x)-\frac{m^{2}}{2}\phi^{2}+{\mathcal{L}}_{int},
\]
where ${\mathcal{L}}_{int}={\mathcal{L}}_{int}(\phi)$ is the
interaction Lagrangian density. The functional $Z[J;\alpha]$
satisfies the equation
\[
(\square+m)\frac{\delta Z[J;\alpha]}{i\delta J(x)}+L_{int}\left(
\frac{1}{i }\frac{\delta}{\delta J}\right)
Z[J;\alpha]=J(x)Z[J;\alpha]
\]
with the normalized solution given by%
\[
Z[J;\alpha]=\frac{\exp\left[  i\int dxL_{int}\left(
\frac{1}{i}\frac{\delta }{\delta J}\right)  \right]
Z_{0}[J;\alpha]}{\left.\exp\left[  i\int dxL_{int} \left(
\frac{1}{i}\frac{\delta}{\delta J}\right)  \right]  Z_{0}
[J;\alpha] \right|_{J=0}}.
\]
Observe that the topology does not change the interaction. This is
a consequence of the isomorphism and the fact that we are
considering a local interaction. Now we turn our attention to
constructing the $\alpha$-generator functional for fermions.

\subsubsection{Fermions}

The Lagrangian density for a free fermion system with sources is
\[
\mathcal{L}=i\overline{\psi}\gamma^{\mu}\partial_{\mu}\psi-m\overline{
\psi }\psi+\overline{\psi}\eta+\overline{\eta}\psi.
\]
The functional $\ Z_{0}\simeq\int D\psi D\overline{\psi}e^{iS}$ is
then reduced to
\begin{equation}
\ Z_{0}[\eta,\overline{\eta};\alpha] = \exp\left\{-i\int dxdy
\left[\overline{\eta} (x)S_{0}(x-y;\alpha)\eta(x)\right]\right\},
\label{KG11mar2}
\end{equation}
where%
\[
S_{0}(x-y;\alpha)^{-1}=\ i\gamma^{\mu}\partial_{\mu}-m.
\]
Since $S_{0}(x-y;\alpha)^{-1}S_{0}(x-y;\alpha)=\delta(x-y)$, and
$G_{0}(x-y;\alpha)$ satisfies Eq.~(\ref{KG11mar1}), we find
\[
S_{0}(x-y;\alpha)=(i\gamma\cdot\partial+m)G_{0}^{F}(x-y;\alpha).
\]
The functional given in Eq.~(\ref{KG11mar2}) provides the same
expression for
the propagator, as derived in the canonical formalism, i.e.%
\[
S_{0}(x-y;\alpha) = i \left.
\frac{\delta^{2}}{\delta\overline{\eta}\delta\eta}
Z_{0}[\eta,\overline{\eta};\alpha]
\right|_{\eta\mathbf{=}\overline{\eta}=0}.
\]

For interacting fields, we obtain
\[
Z[\overline{\eta},\eta;\alpha]=\frac{\exp\left[  i\int
dxL_{int}\left(
\frac{1}{i}\frac{\delta}{\delta\overline{\eta}};\frac{1}{i}\frac{\delta}{
\delta\eta}\right)  \right]
Z_{0}[\eta,\overline{\eta};\alpha]}{\left.\exp\left[ i\int
dxL_{int}\left(  \frac{1}{i}\frac{\delta}{\delta\overline{\eta}}
;\frac{1}{i}\frac{\delta}{\delta\eta}\right)  \right]  Z_{0}[\eta
,\overline{\eta};\alpha]
\right|_{\overline{\mathbf{\eta}}=\mathbf{\eta}=0}}.
\]
It is important to note that, when $\alpha\rightarrow\infty$ we
have to recover the flat space-time field theory, for both \
bosons and fermions.

\subsection{Gauge fields}

The Lagrangian density for quantum chromodynamics is given by
\begin{equation}
\mathcal{L}  =
\overline{\psi}(x)[iD_{\mu}\gamma^{\mu}-m]\psi(x)-\frac
{1}{4}F_{\mu\nu}F^{\mu\nu} -
\frac{1}{2\sigma}(\partial^{\mu}A_{\mu}^{r}(x))^{2} + A_{\mu}^{r}
(x)t^{r}J^{\mu}\left(  x\right)
+\partial^{\mu}\chi^{\ast}(x)D_{\mu}\chi(y),
\end{equation}
where
\[
F_{\mu\nu}^{r}=\partial_{\mu}A_{\nu}^{r}(x)-\partial_{\nu}A_{\mu}%
^{r}(x)+gc^{rsl}A_{\mu}^{s}(x)A_{\nu}^{l}(x)
\]
and $F_{\mu\nu}=\sum\limits_{r}F_{\mu\nu}^{r}t^{r}$ is the field
tensor describing gluons; $t^{r}$ and $c^{rsl}$ are, respectively,
generators and structure constants of the gauge group $SU(3)$; the
covariant derivative is given by
$D_{\mu}=\partial_{\mu}+igA_{\mu}=\partial_{\mu}+igA_{\mu}
^{r}(x)t^{r}$ and $\psi(x)$ is the quark field, including the
flavor and color indices. The ghost field is given by $\chi(x)$.
The quantity $\frac {1}{2\sigma}(\partial^{\mu}A_{\mu}^{r}(x))^{2}
$ is the gauge term, with $\sigma$ being the gauge-fixing
parameter.

The generating functional using the Lagrangian density
$\mathcal{L}$ is
\[
Z[J,\eta,\overline{\eta},\alpha,\alpha^{\ast}] = \int DAD\psi
D\overline {\psi}D\chi D\chi^{\ast} \,\exp\left[  i\int
d^{4}x\left( \mathcal{L}+AJ+\overline{\eta}
\psi+\overline{\psi}\eta+\alpha^{\ast}\chi+\chi^{\ast}\alpha\right)
\right] ,
\]
where $\alpha^{\ast}\ $\ and $\alpha$ are Grassmann variables
describing sources for ghost fields, and $\overline{\eta}$ and
$\eta$ are the Grassman-variable sources for quarks fields, and
$J$ stands for the source of the gluon-field. It is important to
note that we are using non-tilde fields, in such a way that the
propagator is a c-number.

The Lagrangian density is written in terms of interacting and
noninteracting parts as
$\mathcal{L}=\mathcal{L}_{0}+\mathcal{L}_{I}$ with $\mathcal{L}
_{0}=\mathcal{L}_{0}^{G}+\mathcal{L}_{0}^{FP}+\mathcal{L}_{0}^{Q},$
where $\mathcal{L}_{0}^{G}$ is the free gauge field contribution
including a gauge fixing term, i.e.
\[
\mathcal{L}_{0}^{G}=-\frac{1}{4}(\partial_{\mu}A_{\nu}^{r}-\partial_{\nu}%
A\mu^{r})(\partial^{\mu}A^{\nu r}-\partial^{\nu}A^{\mu
r})-\frac{1}{2\sigma }(\partial^{\mu}A_{\mu}^{r})^{2}.
\]
The term $\mathcal{L}_{0}^{FP} $ corresponds to the Faddeev-Popov
field,
\[
\mathcal{L}_{0}^{FP}=(\partial^{\mu}\chi_{\mu}^{r\ast})(\partial^{\mu}%
\chi_{\mu}^{r}),
\]
and $\mathcal{L}_{0}^{Q}$ describes the quark field,%
\[
\mathcal{L}_{0}^{F}=\overline{\psi}(x)[\gamma\cdot
i\partial-m]\psi(x).
\]
The interaction term is
\begin{equation}
\mathcal{L}_{I} = -\frac{g}{2}c^{rsl}(\partial_{\mu}A_{\nu}^{r}
-\partial_{\nu}A_{\mu}^{r})A^{s\mu}A^{l\nu} -
\frac{g^{2}}{2}c^{rst}c^{ult}
A_{\mu}^{r}A_{\nu}^{s}A^{u\mu}A^{l\nu} -
gc^{rsl}(\partial^{\mu}\chi^{r\ast})A_{\mu}^{l}\chi^{s}(y) +
g\overline {\psi}t^{r}\gamma^{\mu}A_{\mu}^{r}\psi.
\end{equation}

Following steps similar to those in the scalar field case, we
write for the gauge field
\[
Z_{0}^{G(rs)}[J]=\exp \left\{ \frac{i}{2}\int dxdy \left[
J^{\mu}(x)D_{0\mu\nu}^{(rs)}(x-y;\alpha)J^{\nu}(y) \right]
\right\},
\]
where
\[
D_{0}^{(rs)\mu\nu}(x;\alpha)= \int\frac{d^{D}k}{(2\pi)^{D}}\,e^{-ikx}%
D_{0}^{(rs)\mu\nu}(k;\alpha)
\]
with
\[
D_{0}^{(rs)\mu\nu}(k;\alpha)=\delta^{rs}d^{\mu\nu}(k)G_{0}(x-y;\alpha),
\]
and%
\[
d^{\mu\nu}(k)=g^{\mu\nu}-(1-\sigma)\frac{k^{\mu}k^{\nu}}{k^{2}}.
\]

For the Fadeev-Popov field we have
\[
Z_{0}^{FP}[\overline{\alpha},\alpha] = \exp \left\{\frac{i}{2}\int
dxdy \left[\overline
{\alpha}(x)G_{0}(x-y;\alpha)\alpha(y)\right]\right\},
\]
where $\overline{\alpha}$ and $\alpha$ are Grassmann variables. It
is important to note that $G_{0}(x-y;\alpha)$ is the propagator
for the scalar field. Then we write the full generating functional
for the non-abelian gauge field as
\[
Z[J,\overline{\xi},\xi\mathbf{,}\overline{\eta},\eta]=\frac{\mathcal{E}%
[\partial_{source}]Z_{0}[J,\overline{\xi},\xi\mathbf{,}\overline{\eta},\eta
]}{\mathcal{E}[\partial_{source}]Z_{0}[0]},
\]
where%
\[
\mathcal{E}[\partial_{source}]=\exp\left[  i\int dxL_{int}\left(
\frac{1} {i}\frac{\delta}{\delta
J},\frac{1}{i}\frac{\delta}{\delta\overline{\xi}
},\frac{1}{i}\frac{\delta}{\delta\xi},\frac{1}{i}\frac{\delta}{\delta
\overline{\eta}},\frac{1}{i}\frac{\delta}{\delta\eta}\right)
\right]
\]
and%
\[
Z_{0}[0] = \left.
Z_{0}^{G(rs)}[J]Z_{0}^{FP(rs)}[\overline{\xi},\xi]Z_{0}
^{F(rs)}[\overline{\eta},\eta]
\right|_{J=\overline{\xi}=\xi=\overline{\eta}=\eta=0}.
\]

As an example, the gluon-quark-quark three point function is
derived to order $g,$
\[
G_{\mu}^{a}(x_{1},x_{2},x_{3};\alpha)  =
-t^{a}\int\frac{d^{D}p_{1}}
{(2\pi)^{D}}\frac{d^{D}p_{2}}{(2\pi)^{D}} e^{-i\{p_{1}\cdot
x_{1}-p_{2}\cdot x_{2}-(p_{1}-p_{2})\cdot x_{3}\}} \,d_{\mu\nu}\,
S_{0}(p_{1};\alpha)\gamma^{v}S_{0}(p_{1};\alpha)D_{0}(p_{1}
-p_{2};\alpha).
\]
We observe that $G_{\mu}^{a}(x_{1},x_{2},x_{3};\alpha)$ has a part
independent
of the topology, i.e. the flat space contribution $G_{\mu}^{a}(x_{1}%
,x_{2},x_{3})$. This is due to the form of the integral
representation of the propagators $S_{0}(p_{1};\alpha)$ and
$D_{0}(p_{1}-p_{2};\alpha)$ and represents a general property of
the theory.

\subsection{$S$-Matrix and reaction rates}

Now the $S$-matrix is developed on the hyper-torus. We use the
canonical formalism for abelian fields and derive the reaction
rate formulas as functions of parameters describing the space-time
compactification, particularizing our discussion to the
$4$-dimensional Minkowski space.

\subsubsection{$S$-Matrix}

Consider a field operator $\phi(x)$\thinspace\ such that%
\begin{align}
\lim_{t\rightarrow-\infty}\phi(x;\alpha)  &  =\phi_{in}(x;\alpha),\\
\lim_{t\rightarrow\infty}\phi(x;\alpha)  &  =\phi_{out}(x;\alpha),
\end{align}
where $\phi_{in}(x;\alpha)$ and $\phi_{out}(x;\alpha)$ stand for
the in- and out-fields before and after interaction takes place,
respectively. These two fields are assumed to be related by a
canonical transformation
\[
\phi_{out}(x;\alpha)=S^{-1}\phi_{in}(x;\alpha)S,
\]
where $S$ is a unitary operator.

We define the evolution operator, $U(t,t^{\prime}),\,$  relating
the interacting field to the incoming field, $i.e$
\begin{equation}
\phi(x;\alpha)=U^{-1}(t,-\infty)\phi_{in}(x;\alpha)U(t,-\infty),
\label{heis14}%
\end{equation}
with $U(-\infty,-\infty)=1.$ The operator $\phi(x;\alpha)$
satisfies the Heisenberg equation
\[
-i\partial_{t}\phi(x;\alpha)=[\widehat{H},\phi(x;\alpha)],
\]
where the generator of time translation, $\widehat{H}$, is written
as $\widehat{H}=\widehat{H}_{0}+\widehat{H}_{I}$, with $H_{0}$ and
$H_{I}$ being the free-particle and interaction Hamiltonians,
respectively. The field $\phi_{in}(x;\alpha)$ satisfies
\begin{equation}
-i\partial_{t}\phi_{in}(x;\alpha)=[\widehat{H}_{0},\phi_{in}(x;\alpha)].
\label{heis24}%
\end{equation}

Requiring unitarity of $U(t,t^{\prime}),$ we have
\[
\partial_{t}(U(t,t^{\prime})U^{-1}(t,t^{\prime}))=0.
\]
In addition, from Eq.~(\ref{heis14}) we have
\[
\partial_{t}\phi_{in}(x;\alpha) = \partial_{t}[U(t,-\infty)\phi
(x;\alpha)U^{-1}(t,-\infty)] =
[U(t,-\infty)\partial_{t}U^{-1}(t,-\infty) +i\widehat{H},\phi_{in}
(x;\alpha)].
\]
Comparing with Eq.~(\ref{heis24}), we obtain%
\[
i\partial_{t}U(t,-\infty)=\widehat{H}_{I}(t)U(t,-\infty).
\]
This equation is written as,
\[
U(t,-\infty)=I-i\int_{-\infty}^{t}dt_{1}\widehat{H}_{I}(t_{1})U(t_{1}
,-\infty),
\]
that is solved by iteration, resulting in%
\begin{align}
U(t,-\infty)  &  =
I-i\int_{-\infty}^{t}dt_{1}\widehat{H}_{I}(t_{1}) + \, \cdots \, +
(-i)^{n}\int_{-\infty}^{t}\cdots\int_{-\infty}^{t_{n-1}}dt_{1}\cdots
dt_{n}
\widehat{H}_{I}(t_{1})\cdots \widehat{H}_{I}(t_{n}) + \, \cdots  \nonumber\\
&  = \mathrm{T}\exp\left[
-i\int_{-\infty}^{t}dt^{\prime}\widehat{H} _{I}(t^{\prime})\right]
,
\end{align}
where $\mathrm{T}$ is the time-ordering operator.

The $S$-matrix is defined by
$S=\lim_{t\rightarrow\infty}U(t,-\infty),$ such
that $S=\sum_{n=0}^{\infty}S^{(n)}$, where%
\[
S^{(n)}=\frac{(-i)^{n}}{n!}\int_{-\infty}^{\infty}\cdots\int_{-\infty}^{\infty
}dt_{1}\cdots dt_{n}T\left[  \widehat{H}_{I}(t_{1})\cdots\widehat{H}_{I}%
(t_{n})\right]  .
\]
Then we have
\[
S=\mathrm{T}\exp\left[  -i\int_{-\infty}^{\infty}dt^{\prime}\widehat{H}%
_{I}(t^{\prime})\right]  .
\]
The transition operator, $\mathcal{T}$ , is defined by
$\mathcal{T}=S-I.$ Observe that
$\widehat{H}(\alpha)\equiv\widehat{H},$ and in the definition of
the $S$-matrix there is no need to introduce a tilde $S$-matrix,
as is the case of TFD~\cite{Ourbook2009,KhannamalsaesAnnals2011}.
Here this is a consequence of the GNS construction.

\subsubsection{Reaction rates}

Consider the scattering process
\[
p_{1}+p_{2}+\cdots +p_{r}\rightarrow
p_{1}^{\prime}+p_{2}^{\prime}+\cdots +p_{r} ^{\prime},
\]
where $p_{i}$ and $p_{i}^{\prime}$ are momenta of the particles in
the initial and final state, respectively. The amplitude for this
process is obtained by the usual Feynman rules as
\[
\left\langle f\right\vert S\left\vert i\right\rangle
=\sum_{n=0}^{\infty }\left\langle f\right\vert S^{(n)}\left\vert
i\right\rangle ,
\]
where $\left\vert i\right\rangle
=a_{p_{1}}^{\dagger}a_{p_{2}}^{\dagger
}...a_{p_{r}}^{\dagger}\left\vert 0\right\rangle $ and $\left\vert
f\right\rangle
=a_{p_{1}^{\prime}}^{\dagger}a_{p_{2}^{\prime}}^{\dagger
}...a_{p_{r}^{\prime}}^{\dagger}\left\vert 0\right\rangle $ with
$\left\vert 0\right\rangle $ being the vacuum state, such that
$a_{p}\left\vert 0\right\rangle =0$. For the topology
$\Gamma_{D}^{d},$ a similar procedure may be used by just
replacing $\left\vert i\right\rangle $ and $\left\vert
f\right\rangle $ states for $\left\vert i;\alpha\right\rangle $
and $\left\vert f;\alpha\right\rangle $ . The amplitude for the
process is then
given as%
\[
\left\langle f;\alpha\right\vert \hat{S}\left\vert
i;\alpha\right\rangle =\sum_{n=0}^{\infty}\left\langle
f;\alpha\right\vert \hat{S}^{n}\left\vert i;\alpha\right\rangle ,
\]
where
\begin{align}
\left\vert i;\alpha\right\rangle  &  =a_{p_{1}}^{\dagger}(\alpha)\,a_{p_{2}%
}^{\dagger}(\alpha)\cdots a_{p_{r}}^{\dagger}(\alpha)\left\vert
0(\alpha
)\right\rangle ,\\
\left\vert f;\alpha\right\rangle  &  =a_{p_{1}^{\prime}}^{\dagger}\,%
(\alpha)a_{p_{2}^{\prime}}^{\dagger}(\alpha)\cdots
a_{p_{r}^{\prime}}^{\dagger }(\alpha)\left\vert
0(\alpha)\right\rangle ,
\end{align}
The vacuum state on the topology $\Gamma_{D}^{d}$ is given by
$\left\vert 0(\alpha)\right\rangle $. As emphasized earlier, the
phase-space factors are not changed by the topology. The meaning
of these states is described in Section IV.

The differential cross-section for the particular process
\[
p_{1}+p_{2}\rightarrow p_{1}^{\prime}+p_{2}^{\prime}+\cdots
+p_{r}^{\prime}
\]
is given by
\begin{equation}
d\sigma =
(2\pi)^{4}\delta^{4}(p_{1}^{\prime}+p_{2}^{\prime}+p_{3}^{\prime
}+\cdots+p_{r}^{\prime}-p_{1}-p_{2})
\frac{1}{4E_{1}E_{2}v_{rel}}\prod_{l}(2m_{l})\prod_{j=1}^{r}\frac{
d^{3}p_{j}^{\prime}}{(2\pi)^{3}2E_{j}^{\prime}}\left\vert M_{fi}
(\alpha)\right\vert ^{2}, \label{5a}
\end{equation}
where
$E_{j}^{\prime}=\sqrt{m_{j}^{^{\prime}2}+\mathbf{p}_{j}^{^{\prime}2}}$
and $v_{rel}$ is the relative velocity of the two initial
particles with 3-momenta $\mathbf{p}_{1}$ and\textbf{\
}$\mathbf{p}_{2}$. The factor $2m_{j} $ appears for each lepton in
the initial and final state. Here $E_{1}$ and $E_{2}$ are the
energies of the two particles with momenta $\mathbf{p}_{1}$
and\textbf{\ }$\mathbf{p}_{2}$, respectively. The amplitude
$M_{fi}$ is related to the $S$-matrix element by
\begin{equation}
\left\langle f;\alpha\right\vert S\left\vert i;\alpha\right\rangle
=i\left( 2\pi\right)  ^{4}M_{fi}(\alpha)\prod_{i}\left(
\frac{1}{2VE_{i}}\right)
^{\frac{1}{2}}\prod_{f}\left[  \frac{1}{2VE_{f}}\right]  ^{\frac{1}{2} }%
\delta^{4}(p_{f}-p_{i}). \label{6}%
\end{equation}
Here $p_{f}$ and $p_{i}$ are the total $4$-momenta in the final
and initial state, respectively. The product extends over all the
external fermions and bosons, with $E_{i}$ and $E_{f}$ being the
energy of particles in the initial and final states, respectively
and $V$ is the volume.

\subsubsection{Decay of particles}

Consider the decay of the boson field $\sigma$ into $2\pi$ , with
an interaction Lagrangian density given by
\begin{equation}
\mathcal{L}_{I}=\lambda\sigma(x)\pi\,(x)\pi(x). \label{7}%
\end{equation}
The initial and final states in $\Gamma_{D}^{d}$ are,
respectively,
\[
\left\vert i;\alpha\right\rangle =a_{k}^{\dag}(\alpha)\left\vert
0(\alpha)\right\rangle ,
\]
and
\[
\left\vert f;\alpha\right\rangle
=b_{k_{1}}^{\dag}(\alpha)b_{k_{2}}^{\dag }(\alpha)\left\vert
0(\alpha)\right\rangle ,
\]
where $a_{k}^{\dag}(\alpha)$ and $b_{k}^{\dag}(\alpha)$ are
creation operators in the topology $\Gamma_{D}^{d}$ for the
$\sigma$- and $\pi$- particles, respectively, with momenta $k$. At
the tree level, the transition matrix element is
\[
\left\langle f;\alpha\right\vert \hat{S}\left\vert
i;\alpha\right\rangle  = i\lambda\int dx\left\langle
0(\alpha)\right\vert b_{k_{2}}(\alpha)b_{k_{1} }(\alpha) \,
\lbrack\sigma(x)\pi(x)\pi(x)-\tilde{\sigma}(x)\tilde{\pi}
(x)\tilde{\pi}(x)]a_{k}^{\dag}(\alpha)\left\vert
0(\alpha)\right\rangle .
\]
Using the expansion of the boson fields, $\sigma(x)$ and $\pi(x)$,
in momentum space, the Bogoliubov transformation and the
commutation relations, the two terms of the matrix element are
calculated. For instance we have
\[
\left\langle 0(\alpha)\right\vert
\sigma(x)a_{k}^{\dag}(\alpha)\left\vert 0(\alpha)\right\rangle
=e^{-ikx}\cosh\theta(k;\alpha),
\]
Combining these factors, the amplitude for the process is
\[
M_{fi}(\beta)  = \lambda\lbrack\cosh(k;\alpha)\cosh\theta(k_{1}
;\alpha)\cosh\theta(k_{2};\alpha) -
\sinh\theta(k;\alpha)\sinh\theta(k_{1};\alpha)\sinh\theta(k_{2}
;\alpha)].
\]
It is to be noted that  the indices
\textquotedblleft1\textquotedblright\ and
\textquotedblleft2\textquotedblright\ in $k_{1}$ and $k_{2}$ are
referring here to two outgoing particles.

The decay rate for the $\sigma$-meson is given as
\begin{align}
\Gamma(w,\alpha)  &
=\frac{1}{2w}\int\frac{d^{3}k_{1}d^{3}k_{2}(2\pi
)^{4}\delta^{4}(k-k_{1}-k_{2})}{(2w_{1})(2w_{2})(2\pi)^{3}(2\pi)^{3}%
}\left\vert M_{fi}(\alpha)\right\vert ^{2}\nonumber\\
&  =\frac{\lambda^{2}}{32w\pi^{2}}\int\frac{d^{3}k_{1}}{w_{1}}\frac{d^{3}%
k_{2}}{w_{2}}\delta^{4}(k-k_{1}-k_{2})W(w;w_{1},w_{2};\alpha),
\label{May24101}%
\end{align}
where%
\[
W(w;w_{1},w_{2};\alpha)   =
|\cosh\theta(k;\alpha)\cosh\theta(k_{1}
;\alpha)\cosh\theta(k_{2};\alpha) -
\sinh\theta(k;\alpha)\sinh\theta(k_{1};\alpha)\sinh\theta(k_{2}
;\alpha)|^{2},\nonumber
\]
with
\[
w_{i}=\sqrt{\kappa_{i}^{2}+m^{2}}\; , \;\;\;
w=\sqrt{\mathbf{k}^{2}+M^{2}}.
\]
Using $\sinh^{2}\theta(k;\alpha)=v^{2}(k_{\alpha};\alpha)\equiv
n(k;\alpha)$
and $\cosh^{2}\theta(k;\alpha)=u^{2}(k;\alpha)\equiv1+n(k;\alpha)$, we have%
\begin{eqnarray}
W(w;w_{1},w_{2};\alpha)  &  = & \left\vert
\sqrt{1+n(k;\alpha)}\right\vert ^{2}\left\vert
\sqrt{1+n(k_{1};\alpha)}\right\vert ^{2}\left\vert
\sqrt{1+n(k_{2};\alpha)}\right\vert ^{2} + \left\vert
\sqrt{n(k;\alpha)}\right\vert ^{2}\left\vert \sqrt
{n(k_{1};\alpha)}\right\vert ^{2}\left\vert
\sqrt{n(k_{2};\alpha)}\right\vert
^{2}\nonumber\\
& &
-\,2\,\mathfrak{Re}\left\{[1+n(k;\alpha)][1+n(k_{1};\alpha)][1+n(k_{2}
;\alpha)]\,
n(k;\alpha)n(k_{1};\alpha)n(k_{2};\alpha)\right\}^{1/2}.
\end{eqnarray}
Considering the rest frame of the decaying particle: $w=M,$
$\mathbf{k}
=0,\,w_{i}=\sqrt{\mathbf{k}_{i}^{2}+m^{2}}=\sqrt{\mathbf{q}^{2}+m^{2}}
=w_{q} $ , and the case of temperature only, i.e.
$\alpha=(\beta,0,0,0)$, we recover the result obtained
early~\cite{Rakhimovkhanna2001}.

One aspect to be emphasized is the notion of quasi-particles. The
energy spectrum of   particles taking place in the reaction has
changed as a consequence of the compactification. This new
spectrum corresponds to the energy of the quasi-particles. The
broken symmetry here is due to a topological specification in the
Minkowski space-time. This interpretation is valid also for
 thermal effects, considered from a topological point of view.

Summarizing, in this section we have developed a theory for
quantum fields
defined in a $D$ -dimensional space-time having a topology $\Gamma_{D}%
^{d}=(\mathbb{S}^{1})^{d}\times\mathbb{M}^{D-d}$, with $1\leq
d\leq D$. This describes both spatial constraints and thermal
effects. The propagator for bosons and fermions is found to be a
generalization of the Fourier integral representation of the
imaginary-time propagator. It is worth emphasizing that, the
Feynman rules follow in parallel as in the Minkowski space-time
theory and that the compactification corresponds to a process of
condensation in the vacuum state, described by a generalization of
the Bogoliubov transformation in TFD. The topology then leads to
the notion of quasi-particles.
%~\cite{ford3,genr8,chodos22,tori1,eliz1,gen6,ito1,gen3,%
%genr9,gen7,gen1,luc1,luc2,luc3,mul1,genr4,ebe1,ebe2,ebe5}%
  In the next section we use the modular construction of c$^{\ast}$-algebra to
derive this theory from representations of linear algebras,
exploring the Poincar\'{e} symmetry. Readers interested in
application can jump to section~5.

\section{Algebraic structure of field theory on $\Gamma_{D}^{d}$}

The $c^{\ast}$-algebra has played a central role in the
development of functional analysis  and has attracted attention
due to its importance in
non-commutative geometry~\cite{ConnesbookNoncommuat1992,ConnesDouglas1998,%
Prosen1999,Krajewski2000}. However, this algebra was earlier
associated with the quantum field theory at finite temperature,
through the imaginary time formalism~\cite{Emchbook1972}.
Actually, the search for the algebraic structure of the Gibbs
ensemble theory leads to the Tomita-Takesaki, or the standard
representation of the
$c^{\ast}$-algebra~\cite{Emchbook1972,Bratteli1987,Takesakibook1970}.
For the real-time formalism, the $c^{\ast }$-algebra approach was
first analyzed by Ojima~\cite{Ojima1981}. Later, the use of the
Tomita-Takesaki Hilbert space, as the carrier space for
representations of groups, was
developed~\cite{SantanaNeto1999,SantanaArthurPA2000,KhannamalsaesAnnals2011},
in order to build thermal theories from symmetry: that is called
the thermo group. A result emerging from this analysis is that the
tilde-conjugation rules, the doubling in TFD, are identified with
the modular conjugation of the standard representation. In
addition, the Bogoliubov transformation, the other basic TFD
ingredient, corresponds to a linear mapping involving the
commutants of the von Neumann algebra. This TFD apparatus has been
developed and applied in a wide range of
problems~\cite{Umezawabook1982,Umezawabook1993,DasbookT1997},
including non-commutative field theories, another formalism
originally associated with the $c^{\ast}$-algebra~\cite{Rivelles2001,%
Szabo2003,
BalachandranTFD2010,Leineker2010fx,LeinekerAmilcar2011}. The
physical interpretation of the doubling in the operators has been
fully identified from a modular representation that has been used
to study representations of the thermal Poincar\'{e}
group~\cite{KhannamalsaesAnnals2011,Ourbook2009}. The modular
conjugation is defined in order to respect the Lie algebra
structure. Then such a procedure provides a consistent way to
define the modular conjugation for fermions, which is usually a
non-simple task due to a lack of criterion~\cite{Ojima1981}. This
representation of thermal theories is generalized for a general
torus $\Gamma_{D}^{d}$~\cite{KhannamalsaesAnnals2011}. In the
following, we discuss the formalism emphasizing the
$^{\ast}$-algebra.

\subsection{ C$^{\ast}$-algebra and compactified propagators}

In order to fix the notation, some aspects of  $c^{\ast}$-algebra
are briefly reviewed. A $c^{\ast}$-algebra  $\mathcal{A}$ is a von
Neumann algebra over the field of complex numbers $\mathbb{C}$
with two different maps, an involutive mapping $^{\ast
}:\mathcal{A\,}\rightarrow$ $\mathcal{A\ }$and the \textit{norm},
which is a mapping defined by
$\Vert\cdot\parallel:\mathcal{A\,\,}\rightarrow \mathbf{R_{+\text{
}}}$~\cite{Emchbook1972,Bratteli1987,Takesakibook1970}. Both
mappings satisfy the following properties,
\begin{align*}
(A^{*})^{*}  &  =A,\\
(A+{\lambda}B)^{*}  &  =A^{*}+{\lambda}^{*}B^{*},\\
(AB)^{*}  &  =B^{*}A^{*},\\
\Vert A+\lambda B\Vert &  =\Vert A||+|\lambda|\,\Vert B,\\
\Vert AB \Vert &  \leq \Vert A\Vert\,\Vert B\Vert\\
\Vert A\Vert &  = 0 \Leftrightarrow A=0\\
\Vert A^{*}\Vert &  =\Vert A\Vert\\
\Vert A^{*}A\Vert &  =\Vert A\Vert^{2}
\end{align*}
where $A,B\in\mathcal{A}$, and $\lambda\in\mathbf{C}$.

The set of normal forms $\omega$ on $\ $ $\mathcal{A}$ is called
\textit{pre-dual} of $\mathcal{A}$ and is denoted by
$\mathcal{A_{*}}$. When $\mathcal{A}$, a $c^{*}$-algebra with
identity, can be identified as the dual of the pre-dual,
$\mathcal{A}$ is called a $w^{*}$-algebra. Let $(\mathcal{H}
_{w},\pi_{w}(\mathcal{A}))$ be a faithful realization of
$\mathcal{A} $,$\mathcal{\,\ }$where $\mathcal{H}_{w}$ is a
Hilbert space;
$\pi_{w}(\mathcal{A}):\mathcal{H}_{w}\rightarrow\mathcal{H}_{w}$
is, then, a $^{*}$-isomorphism of $\mathcal{A\ }$defined by linear
operators in $\mathcal{H}_{w}$. Taking
$\mid$$\xi_{w}\rangle\in\mathcal{H}_{w}$ to be normalized, it
follows that $\langle\xi_{w}\mid\pi_{w}(A)\mid\xi_{w}\rangle,$ for
every $A\in\mathcal{A}$, defines a state over $\mathcal{A\ }$
denoted by
$\omega_{\xi}(A)=$$\langle\xi_{w}\mid\pi_{w}(A)\mid\xi_{w}\rangle$.
Such states are called \textit{vector states}. And, as was
demonstrated by Gel'fand, Naimark and Segal (GNS), the inverse is
also true; i.e. every state $\omega$ of a $w^{*}$-algebra
$\mathcal{A\ }$admits a vector representation $\mid\xi_{w}\rangle$
$\in\mathcal{H}_{w}$ such that $w(A)\equiv$$\langle
\xi_{w}\mid\pi_{w}(A)\mid\,\xi_{w}\rangle$. This realization is
called the \textit{GNS
construction}~\cite{Emchbook1972,Bratteli1987,Takesakibook1970},
which is valid if the dual coincides with the pre-dual.

A Hilbert $h^{\ast}$-algebra is a $c^{\ast}$-algebra such that the
norm is induced by an inner product fulfilling the properties:
$(i)$ $(A|B)=(A^{\ast}|B^{\ast})$; and $(ii)$
$(AB|C)=(B|A^{\ast}C)$. It can be shown that an $h^{\ast}$-algebra
is isomorphic to $\mathcal{H}{\otimes }\mathcal{H}^{\ast}$, where
$\mathcal{H}^{\ast}$ is the dual of the Hilbert space
$\mathcal{H}$. Consider $\sigma:\mathcal{H}_{w}\rightarrow
\mathcal{H}_{w}$ to be a (modular) conjugation in
$\mathcal{H}_{w}$, that is, $\sigma$ is an anti-linear isometry
such that $\sigma^{2}=1$. The set
$(\mathcal{H}_{w},\pi_{w}(\mathcal{A}))$ is a Tomita-Takesaki
(standard) representation of $\mathcal{A}$, if
$\sigma\pi_{w}(\mathcal{A})\sigma
=\widetilde{\pi}_{w}(\mathcal{A})$ defines a
$^{\ast}$-anti-isomorphism on the linear operators. Then
$(\mathcal{H}_{w},\widetilde{\pi}_{w} (\mathcal{A}))$ is a
faithful anti-realization of $\mathcal{A}$. It is to be noted that
$\widetilde{\pi}_{w}(\mathcal{A})$ is the commutant of $\pi
_{w}(\mathcal{A})$; i.,e. $[\pi_{w}(\mathcal{A}),\widetilde{\pi}
_{w}(\mathcal{A})]=0$. In this representation, the state vectors
are invariant under $\sigma$; i.e.,
$\sigma|\xi_{w}\rangle=|\xi_{w}\rangle$. Elements of the set
$\pi_{w}(\mathcal{A})$ will be denoted by $A$ and those of
$\widetilde{\pi}_{w}(\mathcal{A})$ by $\widetilde{A}$.

The tilde and non-tilde operators are mapped to each other by the
$\sigma$ modular conjugation, and fulfill the tilde-conjugation
rules (see last section), with $|\xi_{w}\rangle^{\widetilde{}\
}=|\xi_{w}\rangle
\,\,$and\thinspace\thinspace$\langle\xi_{w}|^{\widetilde{}}=\langle\xi_{w}|$.
These tilde-conjugation rules are derived in TFD, usually, in
association with properties of non-interacting physical systems.
The derivation presented here validates their use for interacting
fields as well~\cite{KhannamalsaesAnnals2011,Ourbook2009}.

An interesting aspect of this construction is that properties of $^{\ast}%
$-automorphisms in $\mathcal{A}$ can be defined through a unitary
operator, say $\Delta(\tau)$, invariant under the modular
conjugation, i.e. $[\Delta(\tau),\sigma]=0$. Then writing
$\Delta(\tau)$ as $\Delta(\tau )=\exp(i\tau\widehat{A})$, where
$\widehat{A}$ is the generator of symmetry, we have
$\sigma\widehat{A}\sigma=-\widehat{A}$. Therefore, the generator
$\widehat{A}$ is an odd polynomial function of $A-\widetilde{A}$,
i.e.
\begin{equation}
\widehat{A}=f(A-\widetilde{A})=\sum_{n=0}^{\infty}c_{n}(A-\widetilde{A}
)^{2n+1}, \label{um}%
\end{equation}
where the coefficients $c_{n}\in\mathbb{R}$.

Consider the simplest case where $c_{0}=1,$ $c_{n}=0,\forall\
n\neq0,$ i.e. $\widehat{A}=A-\widetilde{A}.$ Taking $A$ to be the
Hamiltonian, $H,$ the time-translation generator is given by
$\widehat{H}=H-\widetilde{H}$ . The parameter $\tau$ is related to
a Wick rotation such that $\tau\rightarrow i\beta$; resulting in
$\Delta(\beta)=e^{-\beta\widehat{H}}$, where $\beta=T^{-1}$ , $T$
being the temperature. This is the modular operator in
$c^{\ast}$-algebras. As a consequence, a realization for $w(A)$ as
a Gibbs ensemble average is
introduced~\cite{Emchbook1972,Bratteli1987,Takesakibook1970},
\begin{equation}
w_{A}^{\beta}=\frac{\mathrm{Tr}(e^{-\beta H}A)}{\mathrm{Tr}e^{-\beta H}%
}.\label{ade11}%
\end{equation}

We now proceed with the generalization of this construction for
finite temperature, corresponding to the compactification of the
imaginary time, to accommodate also spatial compactification. To
do so, we replace $H$ by the generator of space-time translations,
$P^{\mu}$, in a $d$-dimensional subspace of a $D$-dimensional
Minkowski space-time, $\mathbb{R}^{D}$, with $d\leq D$. Then we
generalize Eq.~(\ref{ade11}) to the form
\begin{equation}
w_{A}^{\alpha}=\frac{\mathrm{Tr}(e^{-\alpha_{\mu}P^{\mu}}A)}{\mathrm{Tr}%
e^{-\alpha_{\mu}P^{\mu}}},\label{adol11}%
\end{equation}
where $\alpha_{\mu}$ are group parameters. This leads to the
following statement:

\begin{itemize}
\item Theorem 1: For $A(x)$ in the $c^{\ast}$-algebra $\mathcal{A}$, there is
a function $w_{A}^{\alpha}(x)$, $x$ in $\mathbb{R}^{D}\mathbf{,}$
defined by Eq.~(\ref{adol11}) such that
\begin{equation}
w_{A}^{\alpha}(x)=w_{A}^{\alpha}(x+i\alpha),\label{adol13}%
\end{equation}
where $\alpha=(\alpha_{0},\alpha_{1},...,\alpha_{d-1},0,\dots,0)$.
This implies that $w_{A}^{\alpha}(x)$ is periodic, in the
$d$-dimensional subspace, with $\alpha_{0}$ being the period in
the imaginary time, i.e. $\alpha _{0}=\beta$, and
$\alpha_{j}=iL_{j}$, $j=1,...,d-1$, are identified with the
periodicity in spatial coordinates. For fermions we have
anti-periodicity.
\end{itemize}

It is important to note that $w_{A}^{\alpha}(x)$ preserves the
isometry, since it is defined by elements of the isometry group.
Therefore, the theory is defined on the topology
$\Gamma_{D}^{d}=(\mathbb{S}^{1})^{d}\times \mathbb{R}^{D-d}$. For
the particular case of $d=1$, taking $\alpha_{0}=\beta $, we have
to identify Eq.~(\ref{adol13}) as the KMS
condition~\cite{KhannamalsaesAnnals2011}. Then by using the GNS
construction, a quantum theory in thermal equilibrium is
equivalent to taking this theory on a $\Gamma_{D}^{1}$ topology in
the imaginary-time axis, where the circumference of
$\mathbb{S}^{1}$ is $\beta$. The generalization of this result for
space coordinates is given by Eq.~(\ref{adol13}); it corresponds
to extending the KMS condition for field theories on toroidal
topology $(\mathbb{S}^{1})^{d}\times\mathbb{R}^{D-d}$.   As an
example, consider the free propagator for the Klein-Gordon-field.
We take
\[
A(x,x^{\prime})=\mathrm{T}[\phi(x)\phi(x^{\prime})],
\]
where $\mathrm{T}$ is the time-ordering operator. The propagator
for the compactified field is $G_{0}(x-x^{\prime};\alpha)\equiv
w_{A}^{\alpha }(x,x^{\prime})$, such that
\begin{equation}
G_{0}(x-x^{\prime};\alpha)=\frac{\mathrm{Tr}(e^{-\alpha_{\mu}P^{\mu}}%
T\phi(x)\phi(x^{\prime}))}{\mathrm{Tr}e^{-\alpha_{\mu}p^{\mu}}}%
\,.\label{aug09148}%
\end{equation}

The average given in Eq.~(\ref{adol11}) can also be written as%
\begin{equation}
w_{A}^{\alpha}(x)\equiv\langle\xi_{w}^{\alpha}|A(x)|\xi_{w}^{\alpha}%
\rangle,\label{CBPF1}%
\end{equation}
as the second part of the GNS construction. In the following, we
turn our attention to constructing the state
$|\xi_{w}^{\alpha}\rangle$ explicitly. Actually, we will obtain
that the state $|0(\alpha)\rangle$, introduced in the previous
section, provides an  example of states in the GNS construction
for a quantum field theory on a topology $\Gamma_{D}^{d}$. In
order to achieve that, first we study, elements of representation
for Lie algebras, by using the modular representations of a
c*-algebra. Applying the results for the Poincar\'{e} group, we
construct representations describing fields compactified in
space-time.

\subsection{Modular representation of Lie symmetries}

Consider $\ell=\{a_{i},i=1,2,3,...\}$ a Lie algebra over the
(real) field $\mathbb{R}$, of a Lie group $\mathcal{G}$,
characterized by the algebraic relations
$(a_{i},a_{j})=C_{ijk}a_{k}$, where $C_{ijk}\in\mathbb{R}$ are the
structure constants and $(,)$ is the Lie product (summation over
repeated indices is implied). Using the modular conjugation,
$^{\ast}$-representations for $\ell,$ denoted by $^{\ast}\ell$,
are constructed. Let us take $\pi (\ell),$ a representation for
$\ell$ as a von Neumann algebra, and $\widetilde{\pi}(\ell)$ as
the representation for the corresponding commutant. Each element
in $\ell$ is denoted by $\pi(a_{i})=A_{i}$ and $\widetilde{\pi
}(a_{i})=\widetilde{A}_{i}$; thus we
have~\cite{SantanaKhanna1995},
\begin{align}
\lbrack{\tilde{A}}_{i},{\tilde{A}}_{j}] &  =-iC_{ijk}{\tilde{A}}%
_{k},\label{oct1}\\
\lbrack A_{i},A_{j}] &  =iC_{ijk}A_{k},\label{oct2}\\
\lbrack{\tilde{A}}_{i},A_{j}] &  =0.\label{oct3}%
\end{align}
The modular generators of symmetry, as given by Eqs.~(\ref{um}),
take the
form $\widehat{A}=A-\widetilde{A}$. Then we have from Eqs.~(\ref{oct1}%
)-(\ref{oct3}) that the $^{\ast}\ell$ algebra is given by%
\begin{align}
\lbrack\widehat{A}_{i},\widehat{A}_{j}] &  =iC_{ijk}\widehat{A}_{k}%
,\label{aug09141}\\
\lbrack\widehat{A}_{i},A_{j}] &  =iC_{ijk}A_{k},\label{aug09142}\\
\lbrack A_{i},A_{j}] &  =iC_{ijk}A_{k}.\label{aug09143}%
\end{align}
This is a semidirect product of the faithful representation
$\pi(a_{i})=A_{i}$ and the other faithful representation
$\widehat{\pi}(A_{i})=\widehat{A}_{i},$ with $\pi(a_{i})$
providing elements of the invariant subalgebra. This is the proof
of the following statement.

\begin{itemize}
\item Theorem 2. Consider the Tomita-Takesaki representation, where the von
Neumann algebra is a Lie algebra, $\ell$. Then the modular
representation for $\ell$ is given by
Eqs.~(\ref{aug09141})-(\ref{aug09143}), the $^{\ast}\ell
$-algebra, where the invariant subalgebra describes properties of
observables of the theory, that are transformed under the symmetry
defined by the generators of modular transformations.
\end{itemize}

Another aspect to be explored is a set of linear mappings
$U(\xi):\pi
_{w}(\mathcal{A})\times\widetilde{\pi}_{w}(\mathcal{A})\rightarrow\pi
_{w}(\mathcal{A})\times\widetilde{\pi}_{w}(\mathcal{A})$ with the
characteristics of a Bogoliubov transformation, i.e. $U(\xi)$ is
canonical, in the sense of keeping the algebraic relations, and
unitary but only for a finite dimensional basis. Then we have a
group with elements $U(\xi)$ specified by the parameters\ $\xi$. \
This is due to the two commutant sets in the von Neumann algebra.
The characteristic of $U(\xi)$ as a linear mapping is guaranteed
by the canonical invariance of $^{\ast}\ell$-algebra. In terms of
tilde and non-tilde operators, we have
\[
A(\xi) = U(\xi)AU(\xi)^{-1} \; , \;\;\;\;\; \widetilde{A}(\xi) =
U(\xi)\widetilde{A}U(\xi)^{-1},
\]
such that
\[
\lbrack{\tilde{A}(\xi)}_{i},{\tilde{A}(\xi)}_{j}] =
-iC_{ijk}{\tilde{A} (\xi)}_{k} \, , \;\;\;\;\; \lbrack
A(\xi)_{i},A(\xi)_{j}] = iC_{ijk}A(\xi)_{k} \; , \;\;\;\;\;
\lbrack{\tilde{A}(\xi)}_{i},A{(\xi)}_{j}] = 0.
\]

\subsection{Generators of symmetry and observables}

In order to identify physical aspects for this approach, it is
important to note that the set of kinematical variables, say
$\mathcal{V}$, is a vector space of mappings in a Hilbert space
denoted by $\mathcal{H}_{T}$. The set $\mathcal{V}$ is composed of
two subspaces and is written as
$\mathcal{V}=\mathcal{V}_{obs}\oplus\mathcal{V}_{gen}$, where
$\mathcal{V}_{obs}$ $\ $\ and $\mathcal{V}_{gen}$  are set of
kinematical observables and generators of symmetries,
respectively.

In quantum and classical theory, the search for irreducible
representations leads to  $\mathcal{V}_{ob}$ and
$\mathcal{V}_{gen}$ being identical to each other and with
$\mathcal{V}$. Let us discuss this point. Often, to each generator
of symmetry there exists a corresponding observable and both are
described by the same algebraic element. For instance, consider
the generator of rotations $L_{3}=ix_{1}\partial/\partial
x_{2}-ix_{2}\partial/\partial x_{1}$ and the generator of space
translation $P_{1}=-i\ \partial/\partial x_{1}.$ As we know,
$L_{3}$ and $P_{1}$ are also considered as physical observables,
an angular momentum and a linear momentum component, respectively.
The effect of an infinitesimal rotation $\alpha$ around the
$x_{3}$-axis over the observable momentum $P_{1}\ $ is
$\exp(i\alpha L_{3})P_{1}\exp(i\alpha L_{3})\simeq
P_{1}+i\alpha\lbrack L_{3},P_{1}].$ The commutator, expressing the
effect of how much $P_{1}$ has changed, is given by
$[L_{3},P_{1}]=L_{3}P_{1}-P_{1}L_{3}=iP_{2}$. In general, we have
$[L_{i},P_{j}]=i\epsilon_{ijk}P_{k}$. This expression shows that
$P=(P_{1},P_{2},P_{3})$ is transformed as a vector by a rotation .
In other words, the generator $L_{i}$ changes $P_{j}$ through the
commutator operation giving rise to another observable,
$i\epsilon_{ijk}P_{k}.$ In this operation $L_{i}$ has to be
thought as a simple generator (not as an observable) of symmetry.

Note that, although the one-to-one correspondence among
observables and generators of symmetry is based on  physical
ground, there exists no \emph{a priori } dynamical or kinematical
imposition to consider a generator of symmetry and the
corresponding observable as being described by the same
mathematical quantity.  Then here the one-to-one correspondence
among generators and observables is maintained, but
$\mathcal{V}_{ob}$ and $\mathcal{V}_{gen}$ are considered
different mappings in $\mathcal{H} _{T}$. The elements of each set
are denoted in the following way: $\mathcal{V}_{ob}=\{\,A\, \}$
and $\mathcal{V}_{gen}=\{\,\widehat{A}\, \}$.

\subsection{Tilde and non-tilde operators}

A basis vector in Fock space $\mathcal{H}$ is denoted by
$|m\rangle = (m!)^{-1/2}(a^{\dagger})^{m}|0\rangle, $ where
$|0\rangle$ is the vacuum state and $a^{\dagger}$ is the usual
boson creation operator fulfilling the algebra
$[a,a^{\dagger}]=1,$ with the other commutation relations being
zero. We consider then the Hilbert space
$\mathcal{H}_{T}$ as a direct product $\mathcal{H}_{T}=\mathcal{H}%
\otimes{\widetilde{\mathcal{H}}}$. The meaning of the tilde space
${\widetilde{\mathcal{H}}}$ has to be specified by the tilde
conjugation rules regarding the representation space. An arbitrary
basis vector in $\mathcal{H}_{T}$ is obtained by first taking the
tilde conjugation of $\mathcal{H}$, that is
$\sigma\mathcal{H}={\widetilde{\mathcal{H}}}$. For a vector
$|m\rangle$ in $\mathcal{H}$ we have
\[
\sigma|m\rangle=\sigma\frac{1}{(m!)^{1/2}}(a^{\dagger})^{m}|0\rangle=\frac
{1}{(m!)^{1/2}}(\tilde{a}^{\dagger})^{m}\sigma|0\rangle,
\]
where we have used $\sigma^{2}=1$ and $\sigma
a^{\dagger}\sigma=\tilde {a}^{\dagger}$. The conjugation of the
vacuum state, $\sigma|0\rangle$, is given by $
\sigma|0\rangle=|0\rangle^{\:\widetilde{}}=|\tilde{0}\rangle; $
i.e., $|0\rangle^{\:\widetilde{}}=|\tilde{0}\rangle$ is the vacuum
for the tilde operators. Therefore, we have
$\sigma|m\rangle=|\tilde{m}\rangle$ and a general basis vector in
$\mathcal{H}_{T}=\mathcal{H}\otimes{\widetilde {\mathcal{H}}}$ is
\[
|m,\tilde{n}\rangle=\frac{1}{(m!)^{1/2}}(a^{\dagger})^{m}\frac{1}{(n!)^{1/2}%
}(\tilde{a}^{\dagger})^{n}|0,\tilde{0}\rangle,
\]
where $|m,\tilde{n}\rangle=|m\rangle\otimes|\tilde{n}\rangle$ with
$|0,\tilde{0}\rangle=|0\rangle\otimes|\tilde{0}\rangle$. Then we
obtain $ \sigma|m,\tilde{n}\rangle=|n,\tilde{m}\rangle$, such that
$|m,\tilde{m}\rangle^{\:\widetilde{}}=|m,\tilde{m}\rangle$, and
the invariance of the vacuum state under the tilde conjugation,
$|0,\tilde {0}\rangle^{\:\widetilde{}}=|0,\tilde{0}\rangle$.

Given an operator $A$ in $\mathcal{H}$, the corresponding
non-tilde operator, say $\mathcal{A}$, in $\mathcal{H}_{T}$ is
defined by $
\mathcal{A}|m,{\tilde{n}}\rangle=(A|m\rangle)\otimes|\tilde{n}\rangle$.
Similarly, given $\widetilde{A}$ in ${\widetilde{\mathcal{H}}}$,
we define $\widetilde{\mathcal{A}}$ in $\mathcal{H}_{T}$ as $
\widetilde{\mathcal{A}}|m,{\tilde{n}}\rangle=\ |m\rangle\
\otimes\widetilde {A}|\tilde{n}\rangle$.  Using
$1=\sum_{r,s}|r,{\tilde{s}}\rangle\langle\tilde{s},r|$, we have
\begin{equation}
\mathcal{A}|m,{\tilde{n}}\rangle =
\sum_{r,s,t,u}|r,\tilde{s}\rangle
\langle\tilde{s},r|\mathcal{A}|t,\tilde{u}\rangle\langle\tilde{u}
,t|m,\tilde{n}\rangle =
\sum_{r,s}\langle\tilde{s},r|\mathcal{A}|m,\tilde{n}\rangle|r,\tilde
{s}\rangle =
\sum_{r}A_{rm}|r,\tilde{n}\rangle=(A|m\rangle)|{\tilde{n}}
\rangle,\label{3till1}
\end{equation}
where $A_{rm}=\langle r|A|m\rangle$.

The tilde conjugation of Eq.~(\ref{3till1}) leads to $
\widetilde{\mathcal{A}}|m,{\tilde{n}}\rangle^{\:\widetilde{}}=\widetilde
{\mathcal{A}}|n,\tilde{m}\rangle=\sum_{r}A_{rm}^{\ast}|n,\tilde{r}\rangle
$, where  $(A_{rm})^{\:\widetilde{}}=A_{rm}^{\ast}$, since
$A_{rm}$ is a c-number, and
$|m,{\tilde{n}}\rangle^{\:\widetilde{}}=|n,\tilde{m}\rangle$. In
addition, we have
\[
\widetilde{\mathcal{A}}|n,\tilde{m}\rangle =
\sum_{r,s,t,u}|r,\tilde
{s}\rangle\langle\tilde{s},r|\widetilde{\mathcal{A}}|t,\tilde{u}\rangle
\langle\tilde{u},t|n,\tilde{m}\rangle =
\sum_{r,s}\langle\tilde{s},r|\widetilde{\mathcal{A}}|n,\tilde{m}
\rangle|r,\tilde{s}\rangle=\sum_{s}\langle\tilde{s}|\widetilde{A}|\tilde
{m}\rangle|n,\tilde{s}\rangle.
\]
Then we get
\begin{equation}
\langle\tilde{s}|\widetilde{A}|\tilde{m}\rangle=A_{sm}^{\ast}=(A^{T\dagger
})_{sm}\ =(A^{\dagger})_{ms},\label{3paul3335}%
\end{equation}
where $A^{T\dagger}$ is the transpose ($T$) and the Hermitian
conjugate
($\dagger$) of $A$. Writing $\langle\tilde{s},r|\mathcal{A}|m,\tilde{n}%
\rangle=\mathcal{A}_{\tilde{s}rm\tilde{n}}=\mathcal{A}_{rmsn},$ we
have
\begin{equation}
\mathcal{A}_{rmsn}=A_{rm}\delta_{ns}.\label{3paul2333}%
\end{equation}
For the tilde operator, we define $\langle\tilde{s},r|\widetilde{\mathcal{A}%
}|m,\tilde{n}\rangle=\widetilde{\mathcal{A}}_{\tilde{s}rm\tilde{n}}%
=\widetilde{\mathcal{A}}_{snrm},$ resulting in
\begin{equation}
\widetilde{\mathcal{A}}_{snrm}=\delta_{rm}(A^{\dagger})_{ns}\label{3paul2334}%
\end{equation}
From Eqs.~(\ref{3paul2333}) and (\ref{3paul2334}), we write
\[
\mathcal{A}=A\otimes
1\,\,\,\mathrm{and}\,\,\,\widetilde{\mathcal{A}
}=1\otimes A^{\dagger}%
\]

Consider the Pauli matrices,
\begin{equation}
s_{1}=\frac{1}{2}\left(
\begin{array}
[c]{cc}%
0 & 1\\
1 & 0
\end{array}
\right)  ,\ \ s_{2}=\frac{i}{2}\left(
\begin{array}
[c]{cc}%
0 & -1\\
1 & 0
\end{array}
\right)  ,\ \ s_{3}=\frac{1}{2}\left(
\begin{array}
[c]{cc}%
1 & 0\\
0 & -1
\end{array}
\right)  ,\label{paul1}%
\end{equation}
which satisfy the Lie algebra
\begin{equation}
\lbrack s_{i},s_{j}]=i\epsilon_{ij}^{k}s_{k}.\label{3pauli22}%
\end{equation}
The representation for the corresponding operators $\mathcal{A=S}$
and $\widetilde{\mathcal{A}}=\widetilde{\mathcal{S}}$ uses the
Hermitian Pauli matrices. We have $\mathcal{S}_{j}\equiv
s_{j}\otimes1$ and $\widetilde {\mathcal{S}}_{j}\equiv1\otimes
s_{j}$, with $i=1,2,3$. These matrices satisfy the algebra given
by~\cite{KhannamalsaesAnnals2009}
\[
\lbrack\mathcal{S}_{i},S_{j}] = i\epsilon_{ijk}\mathcal{S}_{k}\; ,
\;\;\;\;\;
\lbrack\widetilde{\mathcal{S}}_{i},\widetilde{\mathcal{S}}_{j}] =
-i\epsilon_{ijk}\widetilde{\mathcal{S}}_{k} \; , \;\;\;\;\;
\lbrack\mathcal{S}_{i},\widetilde{\mathcal{S}}_{j}] = 0.
\]

Another representation for $\mathcal{A=S}$ and $\widetilde
{\mathcal{A}}=\widetilde{\mathcal{S}}$ constructed by using
Eq.~(\ref{3paul3335}), i.e.
$\langle\tilde{s}|\widetilde{A}|\tilde{m}\rangle=A_{s,m}^{\ast}$,
with an embedding in the higher dimensional space
$\mathcal{H}_{T}$. In the case of the Pauli matrices, we have
\[
\mathcal{S}_{i}=\left(
\begin{array}
[c]{cc}%
s_{i} & 0\\
0 & \mathbf{1}%
\end{array}
\right)  ,\,\,\,\,\widetilde{\mathcal{S}}_{i}=\frac{1}{2}\left(
\begin{array}
[c]{cc}%
\mathbf{1} & 0\\
0 & s_{i}^{\ast}%
\end{array}
\right)  ,
\]
where $\mathbf{1}$ is a $2\times2$ unit matrix.

A useful result for the tilde and non-tilde operator is derived by
considering the vector
\begin{equation}
|1\rangle=\sum_{m}|m,\tilde{m}\rangle.\label{3till4}
\end{equation}
From Eq.~(\ref{3till1}), we have
\begin{equation}
\mathcal{A}|1\rangle=\sum_{m}\mathcal{A}|m,\tilde{m}\rangle=\sum_{m,r}
A_{rm}|r,\tilde{m}\rangle;\label{3till5}%
\end{equation}
and
\begin{equation}
\widetilde{\mathcal{A}}|1\rangle=\sum_{m}\widetilde{\mathcal{A}}|m,\tilde
{m}\rangle=\sum_{m,r}\widetilde{\mathcal{A}}_{rm}|m,\tilde{r}\rangle
.\label{3till6}%
\end{equation}
Taking the tilde conjugation of Eq.~(\ref{3till5}),  we obtain $
\widetilde{\mathcal{A}}_{r,m}=A_{r,m}^{\ast}=\langle
r|A^{T\dagger} |m\rangle=\langle m|A^{\dagger}|r\rangle $.
Inserting this result
into Eq.~(\ref{3till6}) we get%
 $
\widetilde{\mathcal{A}}|1\rangle=\sum_{m}(A^{\dagger}|m\rangle)|\tilde
{m}\rangle $. %
  Taking $\mathcal{A}=\mathcal{BC},$ such that
$A=BC,$ we obtain
\begin{equation}
\widetilde{\mathcal{B}}\widetilde{\mathcal{C}}|1\rangle=(BC)^{\dagger
}|1\rangle=C^{\dagger}B^{\dagger}|1\rangle.\label{3prod11}%
\end{equation}
The Liouville-von Neumann equation is derived from this
representation. The effect of time transformations of an arbitrary
observable, say $\mathcal{A}(t)$, generated by $\widehat{H}$ is
\begin{equation}
\mathcal{A}(t)=e^{it\widehat{\mathcal{H}}}\mathcal{A}(0)e^{-it\widehat
{\mathcal{H}}},\label{c1gal7}
\end{equation}
that leads to
$i\partial_{t}\mathcal{A}(t)=[\mathcal{A}(t),\widehat{\mathcal{H}}]$.
Assuming the state given by $|\psi(t_{0})\rangle
\in\mathcal{H}_{T}$, the average of an observable $\mathcal{A}(t)$
in  $|\psi(0)\rangle$ is given by
$\langle\mathcal{A}\rangle=\langle
\psi(0)|\mathcal{A}(t)|\psi(0)\rangle,$ with
$\langle\psi(0)|\psi(0)\rangle =1$. Then from Eq.~(\ref{c1gal7}),
we write $\langle\mathcal{A}
\rangle=\langle\psi(t)|\mathcal{A}(0)|\psi(t)\rangle,$ where
$|\psi(t)\rangle$ satisfies the equation
\begin{equation}
i\partial_{t}|\psi(t)\rangle=\widehat{\mathcal{H}}|\psi(t)\rangle
.\label{c1gal9}%
\end{equation}
Despite the appearance, this equation is no longer the
Schr\"{o}dinger equation, due to the structure of
$\mathcal{H}_{T}$, that provides reducible representations.
However, Eq.~(\ref{c1gal9}) is in the Schr\"{o}dinger picture.

Consider a time-dependent operator $F(t)$ acting in $\mathcal{H}$.
Then in the Hilbert space $\mathcal{H}_{T}$, the vector
$|F(t)\rangle$ is defined by the association
$F(t)\rightarrow|F(t)\rangle\,$, such that $
|F(t)\rangle\,\,\equiv F(t)|1\rangle $. Let us verify the scalar
product in $\mathcal{H}_{T},$ with the vectors constructed in this
way. Introducing $|G(t)\rangle=G(t)|1\rangle$, we have
\[
\langle
G|F\rangle=\sum_{m,n}\langle\tilde{n},n|G^{\dagger}F|m,\tilde
{m}\rangle=\sum_{m}\langle
m|G^{\dagger}F|m\rangle=\mathrm{Tr}(G^{\dagger}F)
\]

As far as a state of a quantum system is concerned, we can take
$|\psi (t)\rangle$ in Eq.~(\ref{c1gal9}) to be
$|\psi(t)\rangle\equiv|F(t)\rangle$. If $|\psi(t)\rangle$ is a
normalized state, then $\mathrm{\ Tr}(F^{\dagger }F)=1$. It is
then convenient to represent $F(t)$ as the square root of another
operator, writing, $F(t)=\rho^{1/2}(t)$. In this case
\[
|\psi(t)\rangle=|\rho^{1/2}(t)\rangle=\rho^{1/2}(t)|1\rangle.
\]
Since $\widehat{\mathcal{H}}=\mathcal{H}-\mathcal{\widetilde{H}}
=H\otimes1-1\otimes H^{\dagger}=H\otimes1-1\otimes H,$ we have
\[
\widehat{\mathcal{H}}|\psi(t)\rangle  = \widehat{\mathcal{H}}\rho
^{1/2}(t)|1\rangle=[H\rho^{1/2}(t)-\rho^{1/2}(t)\widetilde{H}]|1\rangle
= [H\rho^{1/2}(t)-\rho^{1/2}(t)H^{\dag}]|1\rangle=[H,\rho^{1/2}
(t)]|1\rangle;
\]
thus we find $i\partial_{t}|\psi(t)\rangle=i\partial_{t}\rho^{1/2}
(t)|1\rangle=[H,\rho^{1/2}(t)]|1\rangle$, such that
\[
i\partial_{t}\rho^{1/2}(t)=[H,\rho^{1/2}(t)].
\]
Calculating $i\partial_{t}|\rho(t)$, where
$\rho(t)=\rho^{1/2\dagger} (t)\rho^{1/2}(t)$, we obtain
$i\partial_{t}\rho(t)=[H,\rho(t)]$, the Liouville-von Neumann
equation. Since $\rho(t)$ is a Hermitian operator with
$\mathrm{Tr}{\rho}=1$, it can be interpreted as the density
matrix.

Take $\rho^{1/2}$ diagonal in the basis $|n,\tilde{m}\rangle$,
such that the state $|\psi(t)\rangle$ is expanded as
\begin{equation}
|\psi(t)\rangle={\sum_{n}}\rho_{n}^{1/2}(t)|n,\tilde{n}\rangle.\label{c1exp1}
\end{equation}
Hence, the average of an observable $\ \mathcal{A}$ is
\[
\langle\mathcal{A}\rangle =
\langle\psi(t)|\mathcal{A}|\psi(t)\rangle =
{\sum_{n,m}}\,\rho_{m}^{\ast1/2}(t)\rho_{n}^{1/2}(t)\langle
m,\tilde {m}|A|n,\tilde{n}\rangle = {\sum_{n}}\,\rho_{n}(t)\langle
n|A|n\rangle =\mathrm{Tr}(\rho A).
\]
The canonical average is introduced by taking $\rho_{\beta
n}^{1/2} =e^{-\beta\varepsilon_{n}/2}/Z(\beta)^{1/2}$, where
$\rho_{\beta}=\rho_{\beta}^{1/2\dagger}\rho_{\beta}^{1/2}$ is the
density matrix for the canonical ensemble. In this case,
$|\psi(t)\rangle=|\psi(\beta)\rangle ={\sum_{n}}\rho_{\beta
n}^{1/2}|n,\tilde{n}\rangle\equiv$ $|0(\beta)\rangle$, the TFD
thermal vacuum.

In short, considering general aspects of symmetries, the
Liouville-von Neumann equation has been derived, but with an
additional ingredient: Eq.~(\ref{c1gal9}) is an amplitude density
matrix equation, such that $|\psi\rangle=\rho^{1/2}|1\rangle$ is
associated with the square root of the density
matrix\cite{Bohm1981,Oliveira2004,Amorim2007}. In this sense, the
representation theory of Lie groups, so often used in the case of
$T=0$ theories, can be useful for thermal physics. This makes
statistical mechanics a self-contained theoretical structure
starting from group theory since, from the Liouville-von Neuman
equation and the density matrix, the entropy is defined in the
standard way, following with the proper connection to
thermodynamics. The self-contained elements are reflected in the
fact that no mention to the Schr\"{o}dinger equation or even the
notion of ensemble has been necessary to build statistical
mechanics. Some algebraic aspects that we have presented here were
earlier found but implicitly presented in the axiomatic structure
of the quantum statistical mechanics based on
$c^{\ast}$-algebra~\cite{Bratteli1987,Takesakibook1970}. The
concept of thermal Lie group is a way to bring part of the
$c^{\ast}$-algebra formalism for the language of Lie
algebras~\cite{Ojima1981,SantanaArthurPA2000,KhannamalsaesAnnals2011}.
In the following we study the Poincar\'e group. Since there is no
risk of confusion, from now on we simplify the notation by using
$\mathcal{A}\equiv A$ and
$\widetilde{\mathcal{A}}\equiv\widetilde{A}$.

\subsection{$^{*}$Poincar\'e Lie algebra}

We use $U(\xi)$ to construct explicitly the states
$w_{A}^{\xi}(x)$ introduced in Eq.~(\ref{adol13}), describing
fields on a $\Gamma_{D}^{d}$ topology. For the Poincar\'{e}
algebra, for instance, we have the $^{\ast}$-Poincar\'{e} Lie
algebra ($^{\ast}\mathfrak{p}$) given by the thermal Poincar\'{e}
Lie algebra as~\cite{SantanaKhanna1995}
\begin{align}
\lbrack M_{\mu\nu},P_{\sigma}]  &  =
i(g_{\nu\sigma}P_{\mu}-g_{\sigma\mu
}P_{\nu}),\label{poin1}\\
\lbrack P_{\mu},P_{\nu}]  &  =0,\label{poin2}\\
\lbrack M_{\mu\nu},M_{\sigma\rho}]  &
=-i(g_{\mu\rho}M_{\nu\sigma}-g_{\nu \rho}M_{\mu\sigma}+
g_{\mu\sigma}M_{\rho\nu}-g_{\nu\sigma}M_{\rho\mu
}),\label{poin3}\\
\lbrack\widehat{M}_{\mu\nu},P_{\sigma}]  &
=[M_{\mu\nu},\widehat{P} _{\sigma
}]=i(g_{\nu\sigma}P_{\mu}-g_{\sigma\mu}P_{\nu}),\label{poin4}\\
\lbrack\widehat{P}_{\mu},P_{\nu}]  &  =0,\label{poin5}\\
\lbrack\widehat{M}_{\mu\nu},M_{\sigma\rho}]  &  =
-i(g_{\mu\rho}M_{\nu\sigma
}-g_{\nu\rho}M_{\mu\sigma}+g_{\mu\sigma}M_{\rho\nu}-g_{\nu\sigma}M_{\rho\mu
}),\label{poin6}\\
\lbrack\widehat{M}_{\mu\nu},\widehat{P}_{\sigma}]  &
=i(g_{\nu\sigma}
\widehat{P}_{\mu}-g_{\sigma\mu}\widehat{P}_{\nu}),\label{poin7}\\
\lbrack\widehat{P}_{\mu},\widehat{P}_{\nu}]  &  =0,\label{poin8}\\
\lbrack\widehat{M}_{\mu\nu},\widehat{M}_{\sigma\rho}]  &
=-i(g_{\mu\rho}
\widehat{M}_{\nu\sigma}-g_{\nu\rho}\widehat{M}_{\mu\sigma}+g_{\mu\sigma
}\widehat{M}_{\rho\nu}-g_{\nu\sigma}\widehat{M}_{\rho\mu}), \label{poin9}%
\end{align}
where $\widehat{M}_{\mu\nu} = \widehat{M}_{\mu\nu}(\theta)$ stands
for the generator of rotations in the Minkowski space and
$\widehat{P}_{\mu} = \widehat{P}_{\mu}(\theta)$ for translations.
This algebra is written in a short notation by
\[%
\begin{tabular}
[c]{ll}%
$\lbrack\mathbf{M,P]}=i\mathbf{P},$ &
$[\widehat{\mathbf{M}},\mathbf{M}]=i
\mathbf{M}$\\
$\lbrack\mathbf{P,P]}=0,$ & $[\widehat{\mathbf{M}},\widehat{ \mathbf{P}%
}]=i\widehat{\mathbf{P}},$\\
$\lbrack\mathbf{M,M]}=i\mathbf{M},$ &
$[\widehat{\mathbf{M}},\widehat{
\mathbf{M}}]=i\widehat{\mathbf{M}}.$\\
$\lbrack\widehat{\mathbf{M}},\mathbf{P]}=i\mathbf{P},$ & $[
\widehat
{\mathbf{P}},\widehat{\mathbf{P}}]=0,$\\
$\lbrack\widehat{\mathbf{P}},\mathbf{P]}=0$. &
\end{tabular}
\
\]

The set of Casimir invariants is
\begin{align}
w^{2}  &  =w_{\mu}w^{\mu},\label{inv1}\\
P^{2}  &  =P_{\mu}P^{\mu},\label{inv2}\\
\widehat{w^{2}}  &
=2\widehat{w}_{\mu}w^{\mu}-\widehat{w}_{\mu}\widehat{w}
^{\mu},\label{inv3}\\
\widehat{P^{2}}  &
=2\widehat{P}_{\mu}P^{\mu}-\widehat{P}_{\mu}\widehat{P}
^{\mu}; \label{inv4}%
\end{align}
where
\[
\widehat{w}_{\mu}=\frac{1}{2}\varepsilon_{\mu\nu\rho\sigma}\widehat{M}
\
^{\nu\sigma}P^{\rho}+\frac{1}{2}\varepsilon_{\mu\nu\rho\sigma}M^{\nu\sigma
}\widehat{P}^{\rho}-\frac{1}{2}\varepsilon_{\mu\nu\rho\sigma}\widehat
{M}\ ^{\nu\sigma}\widehat{P}^{\rho}.
\]
The vector
\[
\overline{w}_{\mu}=\frac{1}{2}\varepsilon_{\mu\nu\rho\sigma}\widehat{M}
\ ^{\nu\sigma}\widehat{P}^{\rho}%
\]
is used to define the scalar $\overline{w}^{2}=\overline{w}_{\mu}
\overline{w}^{\mu}$ , which is not an invariant of the thermal
Poincar\'{e}
algebra but rather of the subalgebra given by Eqs.~(\ref{poin7}%
)-(\ref{poin9}). Using the definition of the hat variables, we get
\begin{equation}
\widehat{w^{2}} =(w_{\mu}w^{\mu})^{\,\widehat{}} = w_{\mu}w^{\mu}
- (w_{\mu }w^{\mu})^{\:\widetilde{}} =
w_{\mu}w^{\mu}-\widetilde{w}_{\mu}\widetilde
{w}^{\mu}, \label{inv5}%
\end{equation}
and, in the same way,
\begin{equation}
\widehat{P^{2}}= \left(  P_{\mu}P^{\mu}\right)  ^{\,\widehat{}} =
P_{\mu
}P^{\mu}-\widetilde{P}_{\mu}\widetilde{P}^{\mu}. \label{inv6}%
\end{equation}

Representations for the thermo-Poincar\'{e} Lie algebra is built
from the Casimir invariants $\widehat{w^{2}}$ and
$\widehat{P^{2}}$. From the definition of tilde variables,
$\widetilde{\mathbf{P}}=\mathbf{P-} \widehat{\mathbf{P}}$ and
$\widetilde{\mathbf{M}}=\mathbf{M-}\widehat {\mathbf{M}}$, we have
for the non-null commutation relations
\begin{align*}
\lbrack\mathbf{M,P]} &  =i\mathbf{P},\\
\lbrack\mathbf{M,M]} &  =i\mathbf{M},\\
\lbrack\widetilde{\mathbf{M}}\mathbf{,}\widetilde{\mathbf{P}}\mathbf{]}
&
=-i\widetilde{\mathbf{P}},\\
\lbrack\widetilde{\mathbf{M}}\mathbf{,}\widetilde{\mathbf{M}}\mathbf{]}
& =-i\widetilde{\mathbf{M}}.
\end{align*}

A quantum field theory can be built following steps similar to the
standard representations. For the scalar representation, using the
invariant
$\widehat{P^{2}}=P_{\mu}P^{\mu}-\widetilde{P}_{\mu}\widetilde{P}^{\mu}$,
we can construct a Lagrangian associated with it. For the
Klein-Gordon field we have the set of equations
\[
(\square^{2}+m^{2}+i\varepsilon)\phi(x)  = 0 \; , \;\;\;\;\;
(\square^{2}+m^{2}-i\varepsilon)\widetilde{\phi}(x) = 0
\]
which are derived from the Lagrangian density%
\[
\widehat{\mathcal{L}}  = \mathcal{L-}\widetilde{\mathcal{L}} =
\frac{1}{2}\partial_{\alpha}\phi\partial^{\alpha}\phi-\frac{1}{2}m^{2}
\phi^{2}-\frac{1}{2}\partial_{\alpha}\widetilde{\phi}\partial^{\alpha
}\widetilde{\phi}-\frac{1}{2}m^{2}\widetilde{\phi}^{2}.
\]
For the Dirac field we have
\[
\widehat{\mathcal{L}}   =\frac{1}{2}\overline{\psi}(x)\gamma\cdot
i\overleftrightarrow{\partial}\psi(x)-m\overline{\psi}(x)\psi(x)\\
 +\frac{1}{2}\widetilde{\overline{\psi}}(x)\gamma\cdot i\overleftrightarrow
{\partial}\widetilde{\psi}(x)+m\widetilde{\overline{\psi}}(x)\widetilde{\psi
}(x)
\]
The $\gamma$-matrices in this equation can be taken in the
representation given by
$\widetilde{\gamma}=(\gamma^{T})^{\dagger}=\gamma^{\ast}$. Both
representations for the tilde matrices are compatible with the
algebra of the $\gamma$-matrices, i.e. $\{\gamma^{\mu},\gamma^{\nu
}\}=2g^{\mu\nu}$ or
$\{\gamma^{\ast\mu},\gamma^{\ast\nu}\}=2g^{\mu\nu}$. In any case,
the Hamiltonian is given by $\widehat{\mathcal{H}}
=\mathcal{H-}\widetilde{\mathcal{H}}$, which is consistent with
the modular-representation.

\section{Casimir effect for the electromagnetic and Dirac fields
on $\Gamma^d_D$\label{sec5}}

The Casimir effect results from the vacuum fluctuations of a
quantum field, when topological or geometrical conditions of the
free space-time are changed. For the case of the electromagnetic
field, confined between two conducting plates with separation $a$,
there is an attractive force between the plates derived from the
negative pressure $P=-\pi^{2}/240a^{4}$~\cite{Casimir1948}, with a
manifestation at the level of mesoscopic scales. Its
generalization has been carried out for different fields in
space-time manifolds with non-trivial topologies and
geometries~\cite{Milonni1993,Mostepanenko1997,Bordag1999,Plunien1986,
Bordag2001,Milton1999,Milton2009,Caruso1991,Svaiter1991,Caruso1999,Farina2006,Cougo-Pinto1999,
Alves2000,Mohideen1998,Revzen1997,Revzen1997a,Feinberg2001,Silva2001,
Robaschik1987,Tadaki1982,Saito1991,Gundersen1988,Lutken1988,LutkenRavdal1988,Lutken1984,
Ravndal1989,RandallSundrum1999,Lukosz1971,Ambjorn1983a,Ambjorn1983,Levin1993,Romeo1995,Seife1997,
QueirozAP2005,TFDCompacPRA2002,Boyer2003,Boyer1974,Elizalde2003,Elizalde2011,%
Seyedzahedi2010,Dodonov1998,Dodonov1998a,Plunien2000,Brevik2000,Dodonov2001,Dodonov2006,Dodonov2012}.
In the late 1990's, after almost fifty years of the Casimir
discovery,  the effect was measured with a precision of few
percent~\cite{Lamoreaux1997,Lamoreaux1998,Lamoreaux1999,Mohideen1998};
a fact that raises interest in relation to microelectronics and
nanotechnology as a practical tool for switching
devices~\cite{Kenneth2002a,Mostepanenko1997}. However, the effect
of temperature, emerging from the positive Stefan-Boltzmann
pressure of a boson or fermion field within a compactified region,
has to be included, since it is significant for separations
$a>1\mu
m$~\cite{Lifshitz1956,Dzyaloshinskii1961,Mehra1967,Brown1969}. The
Casimir effect may be also important for the process of
confinement/deconfinement of the hadronic matter, a phase
transition that may occur at an estimated temperature of
$200\,$MeV, in a region with typical size of 1fm, the order of
magnitude of the proton radius~\cite{Plunien1986}. This has led to
studies of the Casimir effect in the realm of quantum
chromodynamics. In this case, the use of toroidal
topologies~\cite{SaharianBellucci2009,Bellucci2009,Buchmueller2009,
Ellingsen2010,Elizalde2011,Belich2011,QueirozAP2005,Bellucci2011,%
Toldin2010,Mello2010,Khoo2011,
Zhai2011,Oikonomou2011,Oikonomou2011a,Oikonomou2010,Laredo2011,Theo2012,
TomazelliLucio2006}, as an effective scheme for compactification
of fields, is interesting for considering the confinement at the
level of a hadron. Indeed, the confinement is physically
acceptable when it is described by a topological property of the
space-time, that is consistent with the bag model boundary
conditions. In addition,  the boundary condition imposed by the
torus on a field may be associated with the Neumann- or Dirichlet-
or mixed- boundary condition. Then in this case, the toroidal
approach can be used for the electromagnetic field at low
energies, as in condensed matter
physics~\cite{Bellucci2009,Tatur2009,Rudnick2010,Hucht2011}. In
fact, it provides an alternate way for calculations with the
Dirichlet or Neumann boundary condition~\cite{Ourbook2009}.

In this section  some details of the Casimir effect are addressed,
by using the method of toroidal compactified fields based on the
generalized Bogoliubov transformation; although certain aspects of
the  sum of modes techniques are discussed. The basic advantage of
the Bogoliubov  method is the ease of the renormalization scheme,
recovering in particular the basic results of Brown and
Maclay~\cite{Brown1969}, who introduced the local formalism. We
first describe the Casimir effect for the electromagnetic field;
and later, for fermion fields.

\subsection{Casimir effect: local formulation}

Here the objective is to discuss vacuum effects of free fields
under prescribed boundary conditions. The local formulation is
used, by taking advantage of the Green function
method~\cite{Brown1969,Deutsch1979,Plunien1986}. This proceeds
with the following steps. First, the point-splitting technique
and the Lorentz-covariant limit are used to write the
energy-momentum tensor of a free field in the form
\begin{equation}
T^{\mu\nu} = \lim_{x'\rightarrow x} \left[ {\cal
O}^{\mu\nu;ab}_{x,x'}\, {\rm T} \{\phi_{a}(x)\phi_{b}(x')\} +
C^{\mu\nu} \delta(x-x') \right] , \label{Tmini}
\end{equation}
where $a,b$ are are internal indices (like spinor indices for
Dirac fields), ${\cal O}^{\mu\nu;ab}_{x,x'}$ is a tensor
differential operator, $C^{\mu\nu}$ is a $c$-number tensor and
${\rm T}\{ \phi_{a}(x)\phi_{b}(x')\}$ represents the time-ordered
product of field operators. This is  possible since the
energy-momentum tensor for a free field is bilinear in the field
operators and first-order derivatives of the field. Second, the
``renormalized" vacuum expectation value of the energy-momentum
tensor is defined as the difference between the expectation values
of $T^{\mu\nu}$ on the constrained and unconstrained vacuum
configurations,
\begin{equation}
{\cal T}^{\mu\nu} = \left\langle 0 | T^{\mu\nu} | 0
\right\rangle_{\cal S} - \left\langle 0 | T^{\mu\nu} | 0
\right\rangle_{0} ,
\end{equation}
where the labels ``${\cal S}$" and ``0" refer to bounded and
unbounded configurations. We refer to ${\cal T}^{\mu\nu}$ simply
as the energy-momentum tensor of the vacuum. Using the definition
of the propagator,
\begin{equation}
i D_{ab}^{\cal S}(x-x') = \left\langle 0 | {\rm T}\{
\phi_{a}(x)\phi_{b}(x')\} | 0 \right\rangle_{\cal S} ,
\end{equation}
and the corresponding expression for the free case, we have
\begin{equation}
{\cal T}^{\mu\nu} = \left\{ i \left. {\cal O}^{\mu\nu;ab}_{x,x'}
\left[ D_{ab}^{\cal S}(x-x') - D_{ab}^{0}(x-x') \right] \right\}
\right|_{x'\rightarrow x} . \label{Tvac}
\end{equation}

Some essential aspects in deriving Eq.~(\ref{Tvac}) need an
explanation. The local differential operator ${\cal
O}^{\mu\nu;ab}_{x,x'}$ does not change in the presence of boundary
constraints. The term  $C^{\mu\nu} \delta(x-x')$, which leads to a
divergent contribution to its expectation value, is also local and
is canceled out. The divergences emerging  from the local
character of the propagators, calculated at the same point, would
be eliminated. Therefore, the tensor ${\mathcal T}^{\mu\nu}$ would
provide finite  results for the vacuum energy and pressure. The
physical consequences of the non vanishing of ${\mathcal
T}^{\mu\nu}$ are referred to as Casimir effects.

In order to illustrate the procedure described above, consider a
scalar field with Lagrangian density given by
\begin{equation}
{\cal L} = \frac{1}{2} \partial_{\sigma} \phi\, \partial^{\sigma}
\phi - \frac{1}{2} m^2 \phi^2 .
\end{equation}
The canonical energy-momentum tensor, using the point-splitting
technique, is written as
\begin{eqnarray}
T^{\mu\nu} & = & \partial^{\mu} \phi\, \partial^{\nu} \phi -
g^{\mu
\nu} {\cal L} \nonumber \\
& = & \lim_{x'\rightarrow x} {\rm T} \left[ ( \partial^{\mu}
\partial^{\prime \nu} - \frac12 g^{\mu \nu} [ \partial_{\sigma}
\partial^{\prime \sigma} - m^2 ] ) \phi(x)
\phi(x^{\prime}) \right] .
\end{eqnarray}
Now, writing the time-order product of two field-operators in the
form
\[{\rm T}\{ \phi(x) \phi(x^{\prime}) \} = \phi(x)
\phi(x^{\prime}) + \theta(x^{\prime 0} - x^0)
[\phi(x^{\prime}),\phi(x)] ,
\]
where $[\cdot,\cdot]$ represents the commutator, and using the
commutation relations to get the identity
\[
\delta(x^0 - x^{\prime 0}) \left[ \phi(x), \partial^{\prime \nu}
\phi(x^{\prime}) \right] =  n_{0}^{\nu} i\, \delta(x - x^{\prime})
,
\]
where $n_{0} = (1,0,0,0)$ denotes a time-like unit vector, then
the energy-momentum tensor is cast in the form of
Eq.~(\ref{Tmini}), with
\begin{eqnarray*}
{\cal O}^{\mu\nu}_{x,x'} & = & \partial^{\mu}
 \partial^{\prime \nu} - \frac12 g^{\mu \nu} [ \partial_{\sigma}
  \partial^{\prime \sigma} - m^2 ] , \\
C^{\mu \nu} & = & - i (  n_{0}^{\mu}  n_{0}^{\nu} - \frac12 g^{\mu
\nu}) .
\end{eqnarray*}
Therefore,  we have
\begin{equation}
{\cal T}^{\mu\nu}_{scalar} = \left. \left\{ - i  \left[
\partial^{\mu}
\partial'^{\nu} -\frac{1}{2} g^{\mu\nu} \left( \partial_{\sigma}
\partial'^{\sigma} - m^2 \right) \right]  \left[ G_{\cal S}(x-x') -
G_{0}(x-x') \right] \right\} \right|_{x'\rightarrow x} ,
\label{Tscalar}
\end{equation}
where $G_{0}(x-x')$  is the propagator in free
space~\cite{Bogoliubov1960},
\begin{equation}
G_{0}(x) = \frac{1}{4\pi}\delta(x^2) + \frac{i m}{4\pi^2}
\frac{K_1(m\sqrt{-x^2+i\epsilon})}{\sqrt{-x^2+i\epsilon}} ,
\end{equation}
and $G_{\cal S}(x-x')$ denotes the propagator in the presence of
boundaries. The main task in determining ${\cal T}^{\mu\nu}$ for
the massive scalar field is to calculate the constrained
propagator.

As another example, consider the Lagrangian density of the Dirac
field,
\begin{equation}
{\cal L} = \frac{i}{2} {\bar \psi} \gamma^{\mu} {\partial_{\mu}}
\psi - \frac{i}{2} {\partial_{\mu}} {\bar \psi} \gamma^{\mu} \psi
- M {\bar \psi} \psi ,
\end{equation}
with the equations of motion,
\begin{equation}
i\gamma^{\mu}\partial_{\mu}\psi - M\psi = 0\; , \;\;\;\;
i\partial_{\mu}{\bar \psi}\gamma^{\mu} + M{\bar \psi} = 0 .
\label{Deom}
\end{equation}
Then the canonical energy-momentum tensor in the symmetric form is
\begin{equation}
T^{\mu\nu} = \frac{i}{4} \left( {\bar \psi} \gamma^{\mu}
{\partial^{\nu}} \psi + {\bar \psi} \gamma^{\nu} {\partial^{\mu}}
\psi - {\partial^{\mu}} {\bar \psi} \gamma^{\nu} \psi -
{\partial^{\nu}} {\bar \psi} \gamma^{\mu} \psi \right) .
\label{TDirac}
\end{equation}
The tensor  $T^{\mu\nu}$ does not depend on the mass $M$,
explicitly. This is due to the fact that the Lagrangian density,
which is linear in the first derivative of the fields, vanishes
identically for the field configurations that satisfy the
equations of motion; but, naturally, solutions of Eq.~(\ref{Deom})
do depend on the mass, as manifested in the propagator. The spin
indices are summed  in  products involving   ${\bar \psi}$, the
$\gamma$-matrices and $\psi$.

The time-ordered product of fermion operators is written as
\begin{equation*} {\rm T}\{ \psi_a(x) {\bar \psi}_b(x^{\prime})\}
=  \psi_a(x) {\bar \psi}_b(x^{\prime}) -  \theta(x^{\prime 0} -
x^0) \left[\psi_a(x),{\bar \psi}_b(x^{\prime})\right]_+ ,
\end{equation*}
where $[\cdot,\cdot]_+$ represents the anti-commutator; also, from
the anti-commutation relations, one gets the identity
\[
\partial^{\nu} \theta(x^{\prime 0} - x^0) \left[ \psi_a(x),
{\bar \psi}_b(x^{\prime}) \right]_+ = -  n_{0}^{\nu}
\gamma_{ab}^{0}\, \delta({x} - {x}^{\prime}) .
\]
Using these relations, the symmetric energy-momentum tensor is
written in the form of Eq.~(\ref{Tmini}), with
\begin{eqnarray}
{\cal O}^{\mu\nu;ab}_{x,x'} & = & \frac{i}{4} \left\{
\gamma_{ab}^{\mu} ( \partial^{\prime \nu} - \partial^{\nu} ) +
\gamma_{ab}^{\nu} ( \partial^{\prime \mu} - \partial^{\mu} )
\right\} , \nonumber \\
C^{\mu\nu} & = & 4 i  n_{0}^{\mu}  n_{0}^{\nu} ; \nonumber
\end{eqnarray}
and, consequently, we obtain
\[
{\cal T}^{\mu\nu}_{spinor} = \left\{ i \left. {\cal
O}^{\mu\nu;ab}_{x,x'} \left[ S_{ba}^{\cal S}(x-x') -
S_{ba}^{0}(x-x') \right] \right\} \right|_{x'\rightarrow x} ,
\]
where $S_{ab}(x-x')$ is the fermion propagator, written in terms
of the scalar field propagator as
\begin{equation}
S_{ba}(x-x') = \left( i \gamma_{ba}^{\sigma} \partial_{\sigma} + M
\delta_{ba} \right) G_0(x - x') , \label{PropDirac}
\end{equation}
and using that ${\rm Tr}(\gamma^{\mu}) = 0$ and ${\rm
Tr}(\gamma^{\mu} \gamma^{\nu}) = 4 g^{\mu \nu}$, we find
\begin{equation}
{\cal T}^{\mu\nu}_{spinor} = \left. \left\{ 4 i
\partial^{\mu} \partial'^{\nu}  \left[ G_{\cal S}(x-x') -
G_{0}(x-x') \right] \right\} \right|_{x'\rightarrow x} ,
\label{Tspinor}
\end{equation}

\subsection{The Casimir effect for the electromagnetic field}

The Hamiltonian for the free electromagnetic field is
\begin{equation}
H=\sum_{\mathbf{k},\lambda}\omega_{k}\left(n_{k}^{\lambda}+\frac{1}{2}
\right),\label{cascom2}
\end{equation}
where $\lambda=1,2$ stands for the two polarization states of the
photon and $n_{k}
^{\lambda}=a_{\mathbf{k}}^{(\lambda)\dag}a_{\mathbf{k}}^{(\lambda)}$
is the number operator, where $a_{\mathbf{k}}^{(\lambda)}(t)$ and
$a_{\mathbf{k} }^{(\lambda)\dag}$ are the annihilation and the
creation operators of photons with momentum $\mathbf{k}$ and
polarization $\lambda$, satisfying the commutations relations
$[a_{\mathbf{k}}^{(\lambda)},a_{\mathbf{k}^{\prime}
}^{(\lambda^{\prime})\dag}]=\delta_{\lambda\lambda^{\prime}}\delta
_{\mathbf{k},\mathbf{k}^{\prime}}$. The term
$H_{vac}=\sum_{\mathbf{k}}\omega_{k}$  is the vacuum energy, that
is infinite. This is removed by imposing a normal ordering in the
Hamiltonian; which is equivalent to the Takahashi theorem for
massive particles~\cite{Takahashi1969}; i.e. the term $H_{vac}$ is
subtracted out by imposition of the Lorentz symmetry. A similar
procedure is assumed to be valid for non-massive particles, as
photons. Then in the flat space-time, the vacuum state energy is
zero. The situation changes, however, if we impose boundary
conditions on the field, abandoning the flatness condition of the
space-time, at least in global terms. In this case, there are
implications in the energy spectrum, as modifications in the
ground state. This is so due to changes in the modes  $\mathbf{k}$
for each case. The analysis of this change is possible if we
compare the non-trivial case with the field in the flat
space-time, leading to the Casimir effect.  Here we develop this
analysis starting with the energy-momentum tensor for the
electromagnetic field on the topology $\Gamma_{D}^{d}$. In this
case,  the local formulation is interesting for being handled.

Considering the flat space-time, $d=0$, the energy-momentum tensor
operator
for the electromagnetic field is%
\[
T^{\mu\lambda}(x)=-F^{\mu\alpha}(x)F_{\,\,\,\,\alpha}^{\lambda}(x)+\frac{1}%
{4}g^{\mu\lambda}F_{\,\alpha\beta}(x)F^{\alpha\beta}(x),
\]
where $F_{\mu\nu}=\partial_{\mu}A_{\nu}-\partial_{\nu}A_{\mu}$,
with the vector potential, $A_{\mu}$, satisfying the equation
$(g^{\mu\nu}\square -\partial^{\mu}\partial\nu)A_{\nu}(x)=0$. The
tensor $T^{\mu\lambda}(x)$ is
written as  %
\begin{equation}
T^{\mu\lambda}(x)=-\lim_{x\rightarrow x^{\prime}}\left\{
\Delta^{\mu
\nu,\alpha\beta}(x,x^{\prime})T[A_{\alpha}(x),A_{\beta}(x^{\prime}%
)]+2i(n_{0}^{\mu}n_{0}^{\lambda}\
-\frac{1}{4}g^{\mu\nu})\delta(x-x^{\prime
})\right\}  ,\label{caset12}%
\end{equation}
where $\Delta^{\mu\nu,\alpha\beta}=\Gamma_{\ \ \ \ \ \nu}^{\mu\nu
,\lambda\ \ ,\alpha\beta}-\frac{1}{4}g^{\mu\lambda}\Gamma_{\ \ \ \
\nu\rho }^{\nu\rho,\ \ \ \ ,\alpha\beta}$, with
\[
\Gamma^{\mu\nu,\lambda\rho,\alpha\beta}(x,x^{\prime})=(g^{\nu\alpha}%
\partial^{\mu}-g^{\mu\alpha}\partial^{\nu})(g^{\rho\beta}\partial
^{\prime\lambda}-g^{\lambda\beta}\partial^{\prime\rho}).
\]

The vacuum expectation value of $T^{\mu\lambda}(x)$ is
\[
\langle
T^{\mu\nu}(x)\rangle=\langle0|T^{\mu\nu}(x)|0\rangle=-i\lim
_{x\rightarrow
x^{\prime}}\{\Gamma^{\mu\nu}(x,x^{\prime})G_{0}(x-x^{\prime })+2(
n_{0}^{\mu}
n_{0}^{\nu}-\frac{1}{4}g^{\mu\nu})\delta(x-x^{\prime})\},
\]
where $n_{0}^{\mu}$ is the $\mu$-component of the time-like vector
$n_{0}=(1,0,0,0)$ and $\Gamma^{\mu\nu}(x,x^{\prime})=\Gamma_{\ \ \
\ \rho \lambda}^{\mu\nu,\ \ \ \
,\rho\lambda}=2(\partial^{\mu}\partial^{\prime\nu
}-\frac{1}{4}g^{\mu\nu}\partial^{\rho}\partial_{\rho}^{\prime})$.
We have used
$iD_{\alpha\beta}(x-x^{\prime})=\langle0|T[A_{\alpha}(x)A_{\beta}%
(x^{\prime})]|0\rangle=g_{\alpha\beta}G_{0}(x-x^{\prime})$; with
\[
G_{0}(x-x^{\prime})=\frac{1}{4\pi^{2}i}\frac{1}{(x-x^{\prime})^{2}%
-i\varepsilon}.
\]

Now we turn our attention to calculating the energy-momentum
tensor for $\alpha $-dependent fields, in order to take into
account the effects of the toroidal topology. Following the tilde
conjugation rules, as discussed in Sections 3 and 4, the doubled
operator describing the energy-momentum tensor of the
electromagnetic field is~\cite{TFDCompacPRA2002}
\begin{equation}
T^{\mu\lambda(ab)}(x)=-F^{\mu\alpha(a)}(x)F_{\,\,\,\,\alpha}^{\lambda
(b)}(x)+\frac{1}{4}g^{\mu\lambda}F_{\,\beta\alpha}^{(a)}(x)F^{\alpha\beta
(b)}(x),\label{ten1}%
\end{equation}
where the indices $a,b=1,2$ \ are defined according to the doubled
notation, and
$F_{\mu\nu}^{(a)}=\partial_{\mu}A_{\nu}^{a}-\partial_{\nu}A_{\mu}^{a}.$
The vacuum average of the energy-momentum tensor $\langle
T^{\mu\nu
(ab)}\rangle=\langle0,\widetilde{0}|T^{\mu\nu(ab)}|0,\widetilde{0}\rangle$
reads
\[
\langle T^{\mu\nu(ab)}\rangle=-i\lim_{x\rightarrow
x^{\prime}}\{\Gamma^{\mu
\nu}(x,x^{\prime})G_{0}^{(ab)}(x-x^{\prime})+2(n_{0}^{\mu}n_{0}^{\nu}-\frac
{1}{4}g^{\mu\nu})\delta(x-x^{\prime})\delta^{ab}\},
\]
where  the doubled free-photon propagator
\begin{equation}
iD_{\alpha\beta}^{(ab)}(x-x^{\prime})=\langle0,\widetilde{0}|T[A_{\alpha}%
^{a}(x)A_{\beta}^{b}(x)]|0,\widetilde{0}\rangle=g_{\alpha\beta}G_{0}%
^{(ab)}(x-x^{\prime})\label{gree1}%
\end{equation}
is used. The components of $G_{0}^{(ab)}(x-x^{\prime})$ are given
by
\[
G_{0}^{(ab)}(x-x^{\prime})=\frac{1}{(2\pi)^{4}}\int
d^{4}k\,e^{ik(x-x^{\prime })}G_{0}^{(ab)}(k),
\]
with%
\[
\left(  G_{0}^{(ab)}(k)\right)  =\left(
\begin{array}
[c]{cc}%
G_{0}(k) & 0\\
0 & -G_{0}^{\ast}(k)
\end{array}
\right)  =\left(
\begin{array}
[c]{cc}%
\frac{-1}{k^{2}+i\epsilon} & 0\\
0 & \frac{1}{k^{2}-i\epsilon}%
\end{array}
\right)  .
\]
such that
\[
\left(  G_{0}^{(ab)}(x-x^{\prime})\right)  =\left(
\begin{array}
[c]{cc}%
G_{0}(x-x^{\prime}) & 0\\
0 & -G_{0}^{\ast}(x-x^{\prime})
\end{array}
\right)  .
\]

For the $\alpha$-dependent fields, we have
\[
\langle T^{\mu\lambda(ab)}(x;\alpha)\rangle=\langle0,\widetilde{0}%
|T^{\mu\lambda(ab)}(x;\alpha)|0,\widetilde{0}\rangle=\langle0(\alpha
)|T^{\mu\lambda(ab)}(x)|0(\alpha)\rangle,
\]
which is given by
\[
\langle T^{\mu\lambda(ab)}(x;\alpha)\rangle=-i\lim_{x\rightarrow x^{\prime}%
}\{\Gamma^{\mu\lambda}(x,x^{\prime})G^{(ab)}(x-x^{\prime};\alpha)+2(n_{0}%
^{\mu}n_{0}^{\lambda}-\frac{1}{4}g^{\mu\lambda})\delta(x-x^{\prime}%
)\delta^{ab}\}.
\]

The effect of the topology, characterized by the set of parameters
$\alpha,$ is included with some regularization procedure and the
energy-momentum tensor is
\begin{equation}
\mathcal{T}^{\mu\lambda(ab)}(x;\alpha)=\langle T^{\mu\lambda(ab)}%
(x;\alpha)\rangle-\langle T^{\mu\lambda(ab)}(x)\rangle.\label{casad1}%
\end{equation}
This is a finite (renormalized) expression describing measurable
physical quantities. Then we have
\begin{equation}
\mathcal{T}^{\mu\lambda(ab)}(x;\alpha)=-i\lim_{x\rightarrow x^{\prime}%
}\{\Gamma^{\mu\lambda}(x,x^{\prime})\overline{G}^{(ab)}(x-x^{\prime}%
;\alpha)\},\label{emtenso1}%
\end{equation}
where $\overline{G}^{(ab)}(x-x^{\prime};\alpha)=G^{(ab)}(x-x^{\prime}%
;\alpha)-G_{0}^{(ab)}(x-x^{\prime})$. In the Fourier representation we get%
\[
G^{(ab)}(x-x^{\prime};\alpha)=\frac{1}{(2\pi)^{4}}\int d^{4}%
k\,e^{ik(x-x^{\prime})}G^{(ab)}(k;\alpha),
\]
where $G^{(ab)}(k;\alpha)=B_{k}^{-1(ac)}(\alpha)G_{0}^{(cd)}(k)B_{k}%
^{(db)}(\alpha)$; the components of $G_{0}^{(ab)}(k;\alpha)$ are
\begin{align*}
\overline{G}^{(11)}(k;\alpha) &  =\overline{G}^{(22)}(k;\alpha)=v_{k}%
^{2}(\alpha)[G_{0}(k)-G_{0}^{\ast}(k)],\\
\overline{G}^{(12)}(k;\alpha) &
=G^{(21)}(k;\alpha)=v_{B}(k,\alpha
)[1+v_{B}^{2}(k,\alpha)]^{1/2}[G_{0}^{\ast}(k)-G_{0}(k)].
\end{align*}
The generalized Bogoliubov transformation is
\begin{equation}
v^{2}(k_{\alpha};\alpha)=\sum_{s=1}^{N+1}\,2^{s-1}\sum_{\{\sigma_{s}\}}\left(
\prod_{n=1}^{s}f(\alpha_{\sigma_{n}})\right)  \sum_{l_{\sigma_{1}%
},...,l_{\sigma_{s}}=1}^{\infty}\exp\{-\sum_{j=1}^{s}\alpha_{\sigma_{j}%
}l_{\sigma_{j}}k_{\sigma_{j}}\},
\end{equation}
where $\alpha=(\alpha_{0},\alpha_{1},\alpha_{2},...,\alpha_{N})$
and $f(\alpha_{j})=0$ for $\alpha_{j}=0$, $f(\alpha_{j})=1$
otherwise. We denote, without risk of confusion,
$v_{k}^{2}(\alpha)=v^{2}(k_{\alpha};\alpha).$ This
leads to the general case of ($N$+1)-dimensions, considering $v_{k}^{2}%
(\alpha)$, we obtain
\begin{eqnarray}
\overline{{G}}_{0}^{11}(x-x^{\prime};\alpha) & = &
\sum_{s=1}^{N+1}2^{s-1}\sum_{\{\sigma_{s}\}}\left( \prod_{n=1}^{s}
f(\alpha_{\sigma_{n}})\right)
\sum_{l_{\sigma_{1}},...,l_{\sigma_{s}}
=1}^{\infty} \nonumber\\
& & \times \left[
G_{0}^{\ast}(x^{\prime}-x-i\sum_{j=1}^{s}\eta_{\sigma_{j}
}\alpha_{\sigma_{j}}l_{\sigma_{j}}n_{\sigma_{j}})-G_{0}(x-x^{\prime}
-i\sum_{j=1}^{s}\eta_{\sigma_{j}}\alpha_{\sigma_{j}}l_{\sigma_{j}}
n_{\sigma_{j}})\right]  ,
\end{eqnarray}
where $\eta_{\sigma_{j}}=+1$, if $\sigma_{j}=0$, and
$\eta_{\sigma_{j}}=-1$ for $\sigma_{j}=1,2,...,N$. To get the
physical case of finite temperature and spatial confinement,
$\alpha_{0}$ has to be taken as a positive real number while
$\alpha_{n}$, for $n=1,2,...,N$, must be pure imaginary of the
form $iL_{n}$; in these cases, one finds that
$\alpha_{j}^{\ast2}=\alpha_{j}^{2}$.

In a 4-dimensional space-time (corresponding to $N=3$), using the
explicit form of $\overline{{G}}_{0}^{11}(x-x^{\prime};\alpha)$,
we obtain the renormalized $\alpha$-dependent energy-momentum
tensor
\begin{eqnarray}
\mathcal{T}^{\mu\nu(11)}(\alpha) &  = & -i\lim_{x\rightarrow
x^{\prime}}\left\{
\Gamma^{\mu\nu}(x,x^{\prime})\overline{{G}}_{0}^{11}(x-x^{\prime}
;\alpha)\right\}  \nonumber\\
& = &
-\frac{2}{\pi^{2}}\,\sum_{s=1}^{4}2^{s-1}\,\sum_{\{\sigma_{s}\}}\left(
\prod_{n=1}^{s}f(\alpha_{\sigma_{n}})\right) \sum_{l_{\sigma_{1}
},...,l_{\sigma_{s}}=1}^{\infty}\left[
\frac{g^{\mu\nu}}{[\sum_{j=1}^{s}
\eta_{\sigma_{j}}(\alpha_{\sigma_{j}}l_{\sigma_{j}})^{2}]^{2}}\right.
\nonumber\\
& & - \left.
\frac{2\sum_{j,r=1}^{s}(1+\eta_{\sigma_{j}}\eta_{\sigma_{r}
})(\alpha_{\sigma_{j}}l_{\sigma_{j}})(\alpha_{\sigma_{r}}l_{\sigma_{r}
})n_{\sigma_{j}}^{\mu}n_{\sigma_{r}}^{\nu}}{[\sum_{j=1}^{s}\eta_{\sigma_{j}
}(\alpha_{\sigma_{j}}l_{\sigma_{j}})^{2}]^{3}}\right]
.\label{tensaug08}
\end{eqnarray}

Let us analyze the Casimir effect for $d=1$ dimension
compactification. Taking $\alpha=(0,0,0,iL)$, corresponding to
confinement along the $z$-axis, we have
\begin{equation}
v_{k}^{2}(L)=\sum_{l=1}^{\infty}e^{-iLk_{3}l}\,.
\end{equation}
Using this $v_{k}^{2}$, and $\overline{{G}}_{0}^{11}(x-x^{\prime}
;L)=G_{0}^{11}(x-x^{\prime};L)-G_{0}(x-x^{\prime}),$ we get
\begin{equation}
\overline{{G}}_{0}^{11}(x-x^{\prime};L)=\sum_{l=1}^{\infty}[G_{0}^{\ast
}(x^{\prime}-x-Lln_{3})-G_{0}(x-x^{\prime}-Lln_{3})]\hspace{2cm}
\end{equation}
where $n_{3}=(n_{3}^{\mu})=(0,0,0,1)$. Then the energy-momentum
tensor is
\begin{equation}
\mathcal{T}^{\mu\nu(11)}(L)=-\frac{2}{\pi^{2}}\sum_{l=1}^{\infty}\frac
{g^{\mu\nu}+4n_{3}^{\mu}n_{3}^{\nu}}{(Ll)^{4}}=-\frac{\pi^{2}}{45L^{4}}
(g^{\mu\nu}+4n_{3}^{\mu}n_{3}^{\nu}),
\end{equation}
The Casimir energy, $E(L)$, and pressure, $P(L)$, for the
electromagnetic field under periodic boundary conditions are,
respectively,
\[
E(L)=\mathcal{T}^{00(11)}(L)=-\frac{\pi^{2}}{45L^{4}}\,\,\mathrm{\ \ and}%
\,\,\,\,P(L)=\mathcal{T}^{33(11)}(L)=-\frac{\pi^{2}}{15L^{4}}\,.
\]

These expressions are consequences of the periodic conditions
introduced by the torus
$\Gamma_{4}^{1}=\mathbb{S}^{1}\times\mathbb{R}^{3}$, where
$\mathbb{S}^{1}$ stands for the compactification of $x^{3}$-axis
in a circumference of length $L$. It is worth mentioning that if
we take $L=2a$ in the Green functions, this is equivalent to the
contributions of even images used by Brown and Maclay, for
Dirichlet boundary condition, to calculate the Casimir
effect~\cite{Brown1969}. Even images are defined by the even
number of reflections of the electromagnetic field in a region
limited by two conducting plates separated by a distance $a$; i.e.
the field propagates from $\ x^{\prime}$ to $x$ reflecting on the
walls an even number of times. These even images correspond also
to a periodic function with period $L=2a$. Odd images, i.e. the
images generated by an odd number of reflections, do not
contribute to the energy-momentum tensor. This shows that we can
use the toroidal topology method for calculating the Casimir
effect for Dirichlet boundary condition. The method can also be
extended to deal with Neumann and mixed boundary conditions. Using
then $L=2a$, the Casimir energy and pressure for the
electromagnetic field between two parallel conducting plates,
separated by a distance $a$, are given by
\[
E(a)=\mathcal{T}^{00(11)}(a)=-\frac{\pi^{2}}{720a^{4}}\,\,\ \
\mathrm{and}
\,\,\,\,P(a)=\mathcal{T}^{33(11)}(a)=-\frac{\pi^{2}}{240a^{4}}\,.
\]
The negative sign shows that the Casimir force between the plates
is attractive.

Now let us consider the temperature effect. Assuming that the
electromagnetic field satisfies Dirichlet boundary conditions
on parallel conducting plates, normal to the $x^{3}%
$-direction, at finite temperature. This case is defined by the
choice $\alpha=(\beta,0,0,i2a)$, leading to
\begin{eqnarray}
\mathcal{T}^{\mu\nu(11)}(\beta,a) & = & - \frac{2}{\pi^{2}}\left\{
\sum_{l_{0}=1}^{\infty}\frac{g^{\mu\nu}-4n_{0}^{\mu}n_{0}^{\nu}}{(\beta
l_{0})^{4}} +
\sum_{l_{3}=1}^{\infty}\frac{g^{\mu\nu}+4n_{3}^{\mu}n_{3}^{\nu}}
{(2al_{3})^{4}}\right. \nonumber  \\
& & + \left.  4\sum_{l_{0},l_{3}=1}^{\infty}\frac{(\beta l_{0}
)^{2}[g^{\mu\nu}-4n_{0}^{\mu}n_{0}^{\nu}]+(2Ll_{3})^{2}[g^{\mu\nu}+4n_{3}
^{\mu}n_{3}^{\nu}]}{[(\beta l_{0})^{2}+(2al_{3})^{2}]^{3}}\right\}
.
\end{eqnarray}
The Casimir energy, $E(\beta,a)=\mathcal{T}^{00(11)}$,  and
pressure $P(\beta,a)=\mathcal{T}^{33(11)}$) are given   by
\begin{equation}
E(\beta,a)=\frac{\pi^{2}}{15\beta^{4}}-\frac{\pi^{2}}{720a^{4}}+\,\frac{8}
{\pi^{2}}\sum_{l_{0},l_{3}=1}^{\infty}\frac{3(\beta
l_{0})^{2}-(2al_{3})^{2} }{[(\beta l_{0})^{2}+(2al_{3})^{2}]^{3}},
\end{equation}
\begin{equation}
P(\beta,a)=\frac{\pi^{2}}{45\beta^{4}}-\frac{\pi^{2}}{240a^{4}}+\,\frac{8}
{\pi^{2}}\sum_{l_{0},l_{3}=1}^{\infty}\frac{(\beta
l_{0})^{2}-3(2al_{3})^{2} }{[(\beta
l_{0})^{2}+(2al_{3})^{2}]^{3}}.
\end{equation}
The first two terms of these expressions reproduce the black-body
radiation and the Casimir contributions for the energy and the
pressure, separately. The last term represents the interplay
between the two effects~\cite{Plunien1986}. With $L=2a$ we have an
equivalent Stefan-Boltzmann law in the direction $x^{3}$ , i.e. we
have a symmetry by the change $L=2a\leftrightarrow\beta$,
basically as a result of the topology $\mathbb{S}^{1}$, where the
theory is written for each case. This symmetry has been analyzed
in different ways and the result is that when we consider
symmetric boundary conditions on the two plates, these imply
periodicity in the direction $x^{3}$, normal to the plates, with
period $L=2a.$ The fields move unconstrained in the two transverse
directions, obeying the symmetry $L=2a\leftrightarrow\beta$.

The positive black-body contributions for $E$ and $P$ dominate in
the high-temperature limit, while the energy and the pressure are
negative for low $T$. The critical curve $\beta_{c}=\chi_{0}a$,
for the transition from negative to positive values of $P$, is
determined by searching for the value of the ratio $\chi=\beta/a$
for which the pressure vanishes; this value, $\chi_{0}$, is the
solution of the transcendental equation
\[
\frac{\pi^{2}}{45}\frac{1}{\chi^{4}}-\frac{\pi^{2}}{240}+\frac{8}{\pi^{2}}
\sum_{l,n=1}^{\infty}\frac{(\chi l)^{2}-3(2n)^{2}}{[(\chi
l)^{2}+(2n)^{2} ]^{3}}=0,
\]
given, numerically, to be $\chi_{0}\simeq1.15$.

Defining the functions $f(\xi)$ and $s(\xi),$ with
$\xi=\chi^{-1}=a/\beta$ \cite{Brown1969,Plunien1986},
\[
f(\xi)=-\frac{1}{4\pi^{2}}\sum_{j,l=1}^{\infty}\frac{(2\xi)^{4}}{[(2l\xi
)^{2}+(j)^{2}]^{2}},
\]
and
\[
s(\xi)=-\frac{d}{d\xi}f(\xi)=\frac{2^{4}}{\pi^{2}}\sum_{j,l=1}^{\infty}%
\frac{\xi^{3}j^{2}}{[(2l\xi)^{2}+(j)^{2}]^{3}},
\]
the renormalized energy-momentum tensor reads
\[
\mathcal{T}^{\mu\nu(11)}(\beta,a)=\frac{1}{a^{4}}f(\xi)(g^{\mu\nu}+4n_{3}
^{\mu}n_{3}^{\nu})+\frac{1}{\beta
a^{3}}(n^{\mu}n^{\nu}+n_{3}^{\mu}n_{3}^{\nu })s(\xi).
\]
Then the energy density,
$E(\beta,a)=\mathcal{T}^{00(11)}(\beta,a)$, is now written as
\[
E(\beta,a)=\frac{1}{a^{4}}[f(\xi)+\xi s(\xi)].
\]
As this is a thermodynamical expression, the function $f(\xi)$
describes the free-energy density for photons and $s(\xi)$ is the
entropy density.

For the case of a cubic box of edge $a$ at finite temperature,
using Eq.~(\ref{tensaug08}) with $\alpha=(\beta,i2a,i2a,i2a)$, we
have for the pressure
\begin{equation}
P(\beta,a)=\mathcal{T}_{\ }^{33(11)}(\beta,a)=\ h(\chi)\,\frac{1}{a^{4}},\label{boxemf1}%
\end{equation}
where $\chi= \beta/a $ and the function $h(\chi)$ is given by
\begin{eqnarray}
h(\chi) & = &\frac{1}{\pi^{2}}\left\{
\mathcal{C}_{e}+\frac{\pi^{4}}{45} \frac{1}{\chi^{4}} +
8\sum_{l,n=1}^{\infty}\frac{1}{[\chi^{2}l^{2}+4n^{2}]^{2}
}+4\sum_{l,n=1}^{\infty}\frac{\chi^{2}l^{2}-12n^{2}}{[\chi^{2}l^{2}
+4n^{2}]^{3}}\right. \nonumber \\
& & + \, 8
\sum_{l,n,r=1}^{\infty}\frac{1}{[\chi^{2}l^{2}+4(n^{2}+r^{2})]^{2}
} +
16\sum_{l,n,r=1}^{\infty}\frac{\chi^{2}l^{2}+4(n^{2}-3r^{2})}{[\chi^{2}
l^{2}+4(n^{2}+r^{2})]^{3}} \nonumber \\
& & + \left.
16\sum_{l,n,r,q=1}^{\infty}\frac{\chi^{2}l^{2}+4(n^{2}
+r^{2}-3q^{2})}{[\chi^{2}l^{2}+4(n^{2}+r^{2}+q^{2})]^{3}}\right\}
, \nonumber
\end{eqnarray}
with
\begin{equation}
\mathcal{C}_{e} = -
\frac{1}{4}\sum_{l,n=1}^{\infty}\frac{1}{[l^{2}+n^{2}]^{2}
}+\frac{1}{6}\sum_{l,n,r=1}^{\infty}\frac{1}{[l^{2}+n^{2}+r^{2}]^{2}}
-\frac{\pi^{4}}{720} \approx -0.1394 \,.\label{box16}
\end{equation}
We find that the pressure is negative for $T\rightarrow0$,
\[
\mathcal{T}^{33(11)}(a) =
\frac{\mathcal{C}_{e}}{\pi^{2}}\,\frac{1}{a^{4}} \approx
-0.01412\,\frac{1}{a^{4}} \, .
\]
For $T\rightarrow\infty$, the pressure in Eq.~(\ref{boxemf1}) is
dominated by the Stefan-Boltzmann term $\simeq T^{4}$. Therefore,
as the temperature is raised, there is a transition from negative
to positive pressure. We return to this point later.

Another interesting system of parallel plates is given by the
Casimir-Boyer model~\cite{Boyer1974,Boyer2003}, that considers one
conductor plate and the other, non-conductor. In this case, the
free-energy density, $\bar{f}(\xi)$, and
 the entropy density, $\bar{s}(\xi)$, are~\cite{TFDCompacPRA2002}
\[
\bar{f}(\xi)=-\frac{1}{4\pi^{2}}\sum_{l_{0},l_{3}=0^{\prime}}^{\infty
}(-1)^{l_{3}}\frac{(2\xi)^{4}}{[(2l_{3}\xi)^{2}+(l_{0})^{2}]^{2}},
\]
and
\begin{equation*}
\bar{s}(\xi)  =-\frac{d}{d\xi}f(\xi)
=\frac{2^{4}}{\pi^{2}}\sum_{l_{0},l_{3}=0^{\prime}}^{\infty}(-1)^{l_{3}
}\frac{\xi^{3}j^{2}}{[(2l_{3}\xi)^{2}+(l_{0})^{2}]^{3}},
\end{equation*}
where the notation $0^{\prime}$ is to indicate that the term
$l_{0}=l_{3}=0$ is excluded from the sum. Then the energy
momentum-tensor reads
\[
\mathcal{T}^{\mu\nu(11)}(\beta,a)=\frac{1}{a^{4}}\bar{f}(\xi)(g^{\mu\nu
}+4n_{3}^{\mu}n_{3}^{\nu})+\frac{1}{\beta a^{3}}(n_{0}^{\mu}n_{0}^{\nu}%
+n_{3}^{\mu}n_{3}^{\nu})\bar{s}(\xi).
\]
The energy density $E(\beta,a)=\mathcal{T}^{00(11)}(\beta,a)\ $is
written as
\[
E(\beta,a)=\frac{1}{a^{4}}[\bar{f}(\xi)+\xi\bar{s}(\xi)].
\]
For zero temperature the Casimir energy is given by
$E(a)=\frac{7}{8}\frac{\pi^{2}}{720a^{4}}$. This result is $-7/8$
of the Casimir energy for plates of the same material. The force,
in this case, is repulsive.

\subsection{Casimir effect for fermions}

The Casimir effect for a fermion field is of interest in
considering, for instance, the structure of proton in particle
physics. In particular, in the phenomenological MIT bag model
\cite{ChodosJohnson1974}, quarks are\ assumed to be confined in a
small space region, with radius $\sim 1.0fm$, in such a way that
there is no fermionic current outside that region. The fermion
field fulfills the bag model boundary conditions. The Casimir
effect in such a small region is important in order to define the
process of deconfinement. This may appear in heavy ion collisions
at Relativistic Heavy Ion Collider (RHIC) or at Large Hadron
Collider (LHC), giving rise to the quark-gluon plasma. For the
quark field, the problem of the Casimir effect has been quite
often addressed by considering the case of two parallel
plates~\cite{Saito1991,Gundersen1988,Lutken1988,Lutken1984,Ravndal1989,Paola1999,Elizalde2003,QueirozAP2005},
although there are some calculations involving spherical
geometries~\cite{Plunien1986}.

As first demonstrated by Johnson~\cite{Johnson1975}, for plates,
the fermionic Casimir force is attractive as in the case of the
electromagnetic field. On the other hand, depending on the
geometry of confinement, the nature of the Casimir force can
change. This is the case, for instance, for a spherical cavity and
for the Casimir-Boyer model, using mixed boundary conditions for
the electromagnetic field, such that the force is repulsive.
\cite{Lukosz1971,Ambjorn1983a,Ambjorn1983,TFDCompacPRA2002,Boyer1974,Boyer2003}.
Therefore, the analysis considering fermions in a topology of type
$\Gamma_{D}^{d}$ is of interest. We analyze the energy-momentum
tensor for the Casimir effect of a fermion field in a
$d$-dimensional box at finite temperature. As a particular case
the Casimir energy and pressure for the field confined in a
3-dimensional parallelepiped box are calculated. It is found that
the attractive or repulsive nature of the Casimir pressure on
opposite faces changes depending on the relative magnitude of the
length of edges. We also determine the temperature at which the
Casimir pressure in a cubic box changes sign and estimate its
value when the edge of the cube is of the order of the confining
length for baryons. Finally, these results are used to estimate
calculations for estimating the Casimir energy for a
non-interacting massless QCD model.

The energy-momentum tensor for a massless fermionic field is
\begin{align}
T^{\mu\nu}(x) &  =\lim_{x^{\prime}\rightarrow
x}\langle0|i\overline{\psi
}(x^{\prime})\gamma^{\mu}\partial^{\nu}\psi(x)|0\rangle\nonumber
\\
&  =\lim_{x^{\prime}\rightarrow
x}\gamma^{\mu}\partial^{\nu}S(x-x^{\prime
})\nonumber \\
&  =-4i\lim_{x^{\prime}\rightarrow x}\partial^{\mu}\partial^{\nu}
G_{0}(x-x^{\prime}),\label{adee3}
\end{align}
where
\[
S(x-x^{\prime})=-i\langle0|T[\psi(x)\overline{\psi}(x^{\prime})]|0\rangle
\]
and $G_{0}(x-x^{\prime})$ is the propagator of the free massless
bosonic field. With $T^{\mu\nu}(x)$, we introduce the confined
$\alpha$-dependent energy-momentum tensor
$\mathcal{T}^{\mu\nu(ab)}(x;\alpha)$ defined by
\begin{equation}
\mathcal{T}^{\mu\nu(ab)}(x;\alpha)=\langle T^{\mu\nu(ab)}(x;\alpha
)\rangle-\langle T^{\mu\nu(ab)}(x)\rangle,\label{ade122}%
\end{equation}
where $T^{\mu\nu(ab)}(x;\alpha)$ is a function of the field
operators $\psi(x;\alpha)$,$\ \widetilde{\psi}(x;\alpha)$.

Using
\[
S^{(ab)}(x-x^{\prime})=\left(
\begin{array}
[c]{cc}%
S(x-x^{\prime}) & 0\\
0 & \widetilde{S}(x-x^{\prime})
\end{array}
\right)  ,
\]
with $\widetilde{S}(x-x^{\prime})=-S^{\ast}(x^{\prime}-x)$, \ we
have
\begin{equation}
\mathcal{T}^{\mu\nu(ab)}(x;\alpha)=-4i\lim_{x^{\prime}\rightarrow x}%
\partial^{\mu}\partial^{\nu}[G_{0f}^{(ab)}(x-x^{\prime};\alpha)-G_{0f}%
^{(ab)}(x-x^{\prime})].\label{jura2}%
\end{equation}
 The\ $2\times2$ Green functions $G_{0f}^{(ab)}(x-x^{\prime
};\alpha)$ and $G_{0f}^{(ab)}(x-x^{\prime})$ are
\[
G_{0f}^{(ab)}(x-x^{\prime})=\frac{1}{(2\pi)^{4}}\int d^{4}k\,\,G_{0f}%
^{(ab)}(k)\,\,e^{-ik\cdot(x-x^{\prime})},
\]
where
\[
G_{0f}^{(ab)}(k)=\left(
\begin{array}
[c]{cc}%
G_{0}(k) & 0\\
0 & G_{0}^{\ast}(k)
\end{array}
\right)  .
\]
The Green function for fermions, $G_{0f}^{(ab)}(k)$, is different
from that for bosons,$G_{0}^{(ab)}(k)$,  by a sign in the
component $G_{0}^{(22)}(k).$  The $\alpha $-counterpart is
\begin{equation}
G_{0f}^{(ab)}(x-x^{\prime};\alpha)=\frac{1}{(2\pi)^{4}}\int d^{4}%
k\,\,G_{0f}^{(ab)}(k;\alpha)\,\,e^{-ik\cdot(x-x^{\prime})},\label{jurah5}
\end{equation}
with
\[
G_{0f}^{(ab)}(k;\alpha)=B^{-1(ac)}(k;\alpha)G_{0f}^{(cd)}(k)B^{(db)}
(k;\alpha),
\]
where $B^{(ab)}(k;\alpha)$ is the Bogoliubov transformation for
fermions
\[
  B(k;\alpha)  =\left(
\begin{array}
[c]{cc}%
u_{k}(\alpha) & -v_{k}(\alpha)\\
v_{k}(\alpha) & u_{k}(\alpha)
\end{array}
\right)  .
\]
The components of $G_{0}^{(ab)}(k;\alpha)$ are given by
\begin{align*}
G_{0f}^{11}(k;\alpha) &  =G_{0}(k)+v_{k}^{2}(\alpha)[G_{0}^{\ast}%
(k)-G_{0}(k)],\\
G_{0f}^{12}(k;\alpha) &  =G_{0}^{21}(k;\alpha)=v_{k}(\alpha)[1-v_{k}%
^{2}(\alpha)]^{1/2}[G_{0}^{\ast}(k)-G_{0}(k)],\\
G_{0f}^{22}(k;\alpha) &  =G_{0}^{\ast}(k)+v_{k}^{2}(\alpha)[G_{0}%
(k)-G_{0}^{\ast}(k)].
\end{align*}
The physical $\alpha$-tensor  component is given by
$\mathcal{T}^{\mu\nu(11)}(x;\alpha).$

As a first application, the Casimir effect at zero temperature is
derived. For parallel plates perpendicular to the
$x^{3}$-direction and separated by a distance $a$,  we take
$\alpha=i2a$, such that
\begin{equation}
v_{k}^{2}(a)=\sum_{l=1}^{\infty}(-1)^{l+1}e^{-i2ak_{3}l}.\label{vL}
\end{equation}
Using $n_{3}=(n_{3}^{\mu})=(0,0,0,1)$, the energy-momentum tensor
is
\begin{equation}
\mathcal{T}^{\mu\nu(11)}(a)=\frac{4}{\pi^{2}}\sum_{l=1}^{\infty}%
(-1)^{l}\left[
\frac{g^{\mu\nu}+4n_{3}^{\mu}n_{3}^{\nu}}{(2al)^{4}}\right]
,\label{jurah10}%
\end{equation}
leading to the Casimir energy and pressure, that are given,
respectively, by
\begin{equation}
E(a)=\mathcal{T}^{00(11)}(a)=-\frac{7\pi^{2}}{2880}\,\frac{1}{a^{4}%
}\ \ \mathrm{and\ \ }P(a)=\mathcal{T}^{33(11)}(a)=-\frac{7\pi^{2}}{960}%
\,\frac{1}{a^{4}}.\label{sb248}%
\end{equation}
The choice of $\alpha$ as a pure imaginary number is required in
order to obtain the spatial confinement, while the factor $2$ is
needed to ensure antiperiodic boundary conditions on the
propagator and the bag model boundary conditions.

Now the Casimir effect for massless fermions is calculated on a
topology $\Gamma_{N+1}^{d};$ i.e. within a $d$-dimensional box at
finite temperature. The $(1+N)$-dimensional Minkowski space is
considered with $v(\alpha)$ given by
\begin{eqnarray}
v_{k}^{2}(\alpha) & = &
\sum_{j=0}^{N}\sum_{l_{j}=1}^{\infty}(-1)^{1+l_{j}
}\,f(\alpha_{j})\,\exp\{-\alpha_{j}l_{j}k_{j}\}\nonumber\\
& & + \,
\sum_{j<r=0}^{N}2\,f(\alpha_{j})f(\alpha_{r})\sum_{l_{j},l_{r}=1}^{\infty
}(-1)^{2+l_{j}+l_{r}}\,\exp\{-\alpha_{j}l_{j}k_{j}+i\alpha_{r}l_{r}
k_{r}\} + \cdots \nonumber \\
& & +\,2^{N}\,f(\alpha_{0})f(\alpha_{1}) \cdots
f(\alpha_{N})\sum_{l_{0}
,l_{1},...,l_{N}=1}^{\infty}(-1)^{N+1\sum_{r=1}^{N}l_{r}}\,\exp\{-\sum
_{i=0}^{N}\alpha_{i}l_{i}k_{i}\},\label{bboogg1}
\end{eqnarray}
where $\alpha=(\alpha_{0},\alpha_{1},\alpha_{2},...,\alpha_{N}),$
$f(\alpha_{j})=0$ for $\alpha_{j}=0$ and $f(\alpha_{j})=1$
otherwise. This expression leads to the simultaneous
compactification of any $d$ ($1\leq d\leq N+1$) dimensions
corresponding to the non-null parameters $\alpha_{j}$, with
$\alpha_{0}$ corresponding to the time coordinate and $\alpha_{n}$
($n=1,...,N$) referring to the spatial ones. A compact expression
for $v_{k}^{2}(\alpha)$ is
\begin{equation}
v_{k}^{2}(\alpha) =
\sum_{s=1}^{N+1}\,2^{s-1}\,\sum_{\{\sigma_{s}\}}\left(
\prod_{n=1}^{s}f(\alpha_{\sigma_{n}})\right)
\,\sum_{l_{\sigma_{1}
},...,l_{\sigma_{s}}=1}^{\infty}(-1)^{s+\sum_{r=1}^{s}l_{\sigma_{r}}}
\,\exp\{-\sum_{j=1}^{s}\alpha_{\sigma_{j}}l_{\sigma_{j}}k_{\sigma_{j}
}\}.\label{boggg2}
\end{equation}
where $\{\sigma_{s}\}$ denotes the set of all combinations with
$s$ elements, $\{\sigma_{1},\sigma_{2},...\sigma_{s}\}$, of the
first $N+1$ natural numbers $\{0,1,2,...,N\}$, i.e. all subsets
containing $s$ elements, which are expressed  in an ordered form
with $\sigma_{1}<\sigma_{2}<\cdots <\sigma_{s}$.

Using this $v_{k}^{2}(\alpha)$ the $(1\,1)$-component of the
$\alpha$-dependent Green function in the momentum space becomes
\begin{eqnarray}
G_{0}^{11}(k;\alpha) & = &
G_{0}(k)\,+\,\sum_{s=1}^{N+1}\,2^{s-1}\,\sum
_{\{\sigma_{s}\}}\left(
\prod_{n=1}^{s}f(\alpha_{\sigma_{n}})\right)
\nonumber\\
& & \times \,
\sum_{l_{\sigma_{1}},...,l_{\sigma_{s}}=1}^{\infty}(-1)^{s+\sum
_{r=1}^{s}l_{\sigma_{r}}}\,\exp\{-\sum_{j=1}^{s}\alpha_{\sigma_{j}}%
l_{\sigma_{j}}k_{\sigma_{j}}\}\,[G_{0}^{\ast}(k)-G_{0}(k)].
\nonumber
\end{eqnarray}
Taking the inverse Fourier transform of this expression and
defining the vectors $n_{0}=(1,0,0,0,...)$, $n_{1}=(0,1,0,0,...)$,
$...$, $n_{N} =(0,0,0,...,1)$, in the ($1+N)$-dimensional
Minkowski space, written in the contravariant coordinates, the
energy-momentum tensor is
\begin{eqnarray}
\mathcal{T}^{\mu\nu(11)}(\alpha) & = & -
4i\,\sum_{s=1}^{N+1}\,2^{s-1} \,\sum_{\{\sigma_{s}\}}\left(
\prod_{n=1}^{s}f(\alpha_{\sigma_{n}})\right)
\sum_{l_{\sigma_{1}},...,l_{\sigma_{s}}=1}^{\infty}(-1)^{s +
\sum_{r=1}^{s}l_{\sigma_{r}}} \nonumber \\
& & \times\,\lim_{x^{\prime}\rightarrow
x}\,\partial^{\mu}\partial^{\nu}\left[ G_{0}^{\ast}(x^{\prime}-x -
i \sum_{j=1}^{s}\xi_{\sigma_{j}}\alpha_{\sigma_{j}
}l_{\sigma_{j}}n_{\sigma_{j}})-\,G_{0}(x-x^{\prime}-
i\sum_{j=1}^{s}\xi
_{\sigma_{j}}\alpha_{\sigma_{j}}l_{\sigma_{j}}n_{\sigma_{j}})\right]
,\label{TmnGeral}
\end{eqnarray}
where $\xi_{\sigma_{j}}=+1$, if $\sigma_{j}=0$, and
$\xi_{\sigma_{j}}=-1$ for $\sigma_{j}=1,2,...,N$.

In order to fit physical conditions, at finite temperature and
spatial confinement, $\alpha_{0}$ has to be taken as a positive
real number while $\alpha_{n}$, for $n=1,2,...,N$, must be pure
imaginary of the form $i2a_{n}$, such that
$\alpha_{j}^{\ast2}=\alpha_{j}^{2}$. Then the energy-momentum
tensor for the 4-dimensional space-time ($N=3$) is
\begin{eqnarray}
\mathcal{T}^{\mu\nu(11)}(\alpha) & = &
-\frac{4}{\pi^{2}}\,\sum_{s=1}
^{4}\,2^{s-1}\,\sum_{\{\sigma_{s}\}}\left(
\prod_{n=1}^{s}f(\alpha _{\sigma_{n}})\right)
\sum_{l_{\sigma_{1}},...,l_{\sigma_{s}}=1}^{\infty}(-1)^{s+\sum
_{r=1}^{s}l_{\sigma_{r}}}\frac{1}{[\sum_{j=1}^{s}\xi_{\sigma_{j}}
(\alpha_{\sigma_{j}}l_{\sigma_{j}})^{2}]^{2}} \nonumber\\
& & \times\,  \left[ g^{\mu\nu}-\frac
{2\sum_{j,r=1}^{s}(1+\xi_{\sigma_{j}}\xi_{\sigma_{r}})(\alpha_{\sigma_{j}
}l_{\sigma_{j}})(\alpha_{\sigma_{r}}l_{\sigma_{r}})n_{\sigma_{j}}^{\mu
}n_{\sigma_{r}}^{\nu}}{\sum_{j=1}^{s}\xi_{\sigma_{j}}(\alpha_{\sigma_{j}
}l_{\sigma_{j}})^{2}}\right]. \label{tan1}
\end{eqnarray}
An important aspect to note  is that
$\mathcal{T}^{\mu\nu(11)}(\alpha)$ is traceless, as it should be.
To obtain the physical meaning of $\mathcal{T}
^{\mu\nu(11)}(\alpha)$, we have to analyze particular cases. Thus
we rederive, first, some known results for $N=3$. Let us emphasize
that Eq.~(\ref{boggg2}) is the generalization of the Bogoliubov
transformation, compatible with the generalizations of the
Matsubara formalism, for the case of fermions.

The particular case of two parallel plates at zero temperature has
already been analyzed. For this case, taking $\alpha=(0,0,0,i2a)$,
Eq.~(\ref{tan1}) reduces to Eq.~(\ref{jurah10}) and the standard
Casimir effect is recovered. Let us then consider two parallel
plates at finite temperature. Then both time and space
compactification need to be considered; this is carried out by
taking $\alpha=(\beta,0,0,i2a)$ in Eq.~(\ref{tan1}), where
$\beta^{-1}=T$ is the temperature and $a$ is the distance between
plates perpendicular to the $x^{3}$-axis. Then the result is
\begin{eqnarray}
\mathcal{T}^{\mu\nu(11)}(\beta,a) & = & \frac{4}{\pi^{2}}\left\{
\sum_{l_{0}
=1}^{\infty}(-1)^{l_{0}}\frac{[g^{\mu\nu}-4n_{0}^{\mu}n_{0}^{\nu}]}{(\beta
l_{0})^{4}} +
\sum_{l_{3}=1}^{\infty}(-1)^{l_{3}}\frac{[g^{\mu\nu}+4n_{3}^{\mu
}n_{3}^{\nu}]}{(2al_{3})^{4}}\right. \nonumber \\
& & \left.
-\,2\,\sum_{l_{0},l_{3}=1}^{\infty}(-1)^{l_{0}+l_{3}}\left[
\frac{(\beta
l_{0})^{2}[g^{\mu\nu}-4n_{0}^{\mu}n_{0}^{\nu}]}{[(\beta
l_{0})^{2}+(2al_{3})^{2}]^{3}}+\frac{(2al_{3})^{2}[g^{\mu\nu}+4n_{3}^{\mu
}n_{3}^{\nu}]}{[(\beta l_{0})^{2}+(2al_{3})^{2}]^{3}}\right]
\right\}  . \nonumber
\end{eqnarray}

The Casimir energy, $E(\beta ,a)=\mathcal{T}^{00(11)}(\beta,a)$,
is given by
\[
E(\beta,a)=\frac{7\pi^{2}}{60}\,\frac{1}{\beta^{4}}-\frac{7\pi^{2}}%
{2880}\,\frac{1}{a^{4}}-\,\frac{8}{\pi^{2}}\sum_{l_{0},l_{3}=1}^{\infty
}(-1)^{l_{0}+l_{3}}\frac{3(\beta l_{0})^{2}-(2al_{3})^{2}}{[(\beta l_{0}%
)^{2}+(2al_{3})^{2}]^{3}}.
\]
Taking the limit $a\rightarrow\infty$, this energy reduces to the
Stefan-Boltzmann term, while making $\beta\rightarrow\infty$ one
regains the Casimir effect for two plates at zero temperature
presented in Eq.~(\ref{sb248}). The third term, which stands for
the correction due to temperature and spatial compactification,
remains finite as $\beta \rightarrow0$ and, as expected, the high
temperature limit is dominated by the positive contribution of the
Stefan-Boltzmann term.

The Casimir pressure, $P(\beta,a)=\mathcal{T}^{33(11)}(\beta,a)$,
is similarly obtained as
\begin{equation}
P(\beta,a)=\frac{7\pi^{2}}{180}\,\frac{1}{\beta^{4}}-\frac{7\pi^{2}}{960}
\,\frac{1}{a^{4}}+\,\frac{8}{\pi^{2}}\sum_{l_{0},l_{3}=1}^{\infty}%
(-1)^{l_{0}+l_{3}}\frac{(\beta l_{0})^{2}-3(2al_{3})^{2}}{[(\beta l_{0}%
)^{2}+(2al_{3})^{2}]^{3}}. \label{Pcplacas}%
\end{equation}
It is to be noted that for low temperatures (large $\beta$) the
pressure is negative but, as the temperature increases, a
transition to positive values happens. It is possible to determine
the critical curve of this transition, $\beta_{c}=\chi_{0}a$, by
searching for a value of the ratio $\chi=\beta/a$ for which the
pressure vanishes; this value, $\chi_{0}$, is the solution of the
transcendental equation
\[
\frac{7\pi^{2}}{180}\frac{1}{\chi^{4}}-\frac{7\pi^{2}}{960}+\frac{8}{\pi^{2}%
}\sum_{l,n=1}^{\infty}(-1)^{l+n}\frac{(\chi
l)^{2}-3(2n)^{2}}{[(\chi l)^{2}+(2n)^{2}]^{3}}=0,
\]
given, numerically, by $\chi_{0}\simeq1.3818$.

Now we consider the fermion field confined in a $3$-dimensional
closed box having the form of a rectangular parallelepiped with
faces $a_{1}$, $a_{2}$ and $a_{3}$ . At zero temperature, the
physical energy-momentum tensor is obtained from Eq.~(\ref{tan1})
by taking $\alpha=(0,i2a_{1},i2a_{2},i2a_{3})$. The Casimir energy
is then given by
\begin{eqnarray}
E(a_{1},a_{2},a_{3}) & = & -\,\frac{7\pi^{2}}{2880}\left(
\frac{1}{a_{1}^{4}
}+\frac{1}{a_{2}^{4}}+\frac{1}{a_{3}^{4}}\right)
-\,\frac{1}{2\pi^{2}}
\sum_{l_{1},l_{2}=1}^{\infty}\frac{(-1)^{l_{1}+l_{2}}}{[(a_{1}l_{1}
)^{2}+(a_{2}l_{2})^{2}]^{2}}\nonumber\\
& &
-\,\frac{1}{2\pi^{2}}\sum_{l_{1},l_{3}=1}^{\infty}\frac{(-1)^{l_{1}+l_{3}}
}{[(a_{1}l_{1})^{2}+(a_{3}l_{3})^{2}]^{2}}-\frac{1}{2\pi^{2}}\sum_{l_{2}
,l_{3}=1}^{\infty}\frac{(-1)^{l_{2}+l_{3}}}{[(a_{2}l_{2})^{2}+(a_{3}l_{3}
)^{2}]^{2}}\nonumber\\
& &
+\,\frac{1}{\pi^{2}}\sum_{l_{1},l_{2},l_{3}=1}^{\infty}\frac{(-1)^{l_{1}
+l_{2}+l_{3}}}{[(a_{1}l_{1})^{2}+(a_{2}l_{2})^{2}+(a_{3}l_{3})^{2}]^{2}
},\label{Ebox}
\end{eqnarray}
and the Casimir pressure, $P=\mathcal{T}^{33(11)}$, reads
\begin{eqnarray}
P(a_{1},a_{2},a_{3}) & = & -\,\frac{7\pi^{2}}{2880}\left(
\frac{3}{a_{3}^{4}
}-\frac{1}{a_{1}^{4}}-\frac{1}{a_{2}^{4}}\right)
+\,\frac{1}{2\pi^{2}}
\sum_{l_{1},l_{3}=1}^{\infty}(-1)^{l_{1}+l_{3}}\frac{(a_{1}l_{1})^{2}
-3(a_{3}l_{3})^{2}}{[(a_{1}l_{1})^{2}+(a_{3}l_{3})^{2}]^{3}} \nonumber\\
& &
+\,\frac{1}{2\pi^{2}}\sum_{l_{1},l_{2}=1}^{\infty}\frac{(-1)^{l_{1}+l_{2}}
}{[(a_{1}l_{1})^{2}+(a_{2}l_{2})^{2}]^{2}}+\,\frac{1}{2\pi^{2}}\sum
_{l_{2},l_{3}=1}^{\infty}(-1)^{l_{2}+l_{3}}\frac{(a_{2}l_{2})^{2}-3(a_{3}
l_{3})^{2}}{[(a_{2}l_{2})^{2}+(a_{3}l_{3})^{2}]^{3}}\nonumber\\
& &
-\,\frac{1}{\pi^{2}}\sum_{l_{1},l_{2},l_{3}=1}^{\infty}(-1)^{l_{1}
+l_{2}+l_{3}}\frac{(a_{1}l_{1})^{2}+(a_{2}l_{2})^{2}-3(a_{3}l_{3})^{2}
}{[(a_{1}l_{1})^{2}+(a_{2}l_{2})^{2}+(a_{3}l_{3})^{2}]^{3}}.\label{Pbox}
\end{eqnarray}

Different effects arise for different values of the relative
magnitude of the edges of the parallelepiped. Here we address the
symmetric case of a cubic box, $a_{1}=a_{2}=a_{3}=a$. Then the
Casimir energy and pressure, respectively, become
\[
E(a)=-\left(  \frac{7\pi^{2}}{960}+\frac{\mathcal{C}
}{2\pi^{2}}\right) \frac{1}{a^{4}}\ \ \ \mathrm{and\ \
}P(a)=-\left(
\frac{7\pi^{2}}{2880}+\frac{\mathcal{C}}{6\pi^{2}%
}\right)  \frac{1}{a^{4}},
\]
where the constant ${\mathcal{C}}$ is given by
\begin{equation}
{\mathcal{C}}=3\sum_{l,n=1}^{\infty
}\frac{(-1)^{l+n}}{(l^{2}+n^{2})^{2}}-2\sum_{l,n,r=1}^{\infty}\frac
{(-1)^{l+n+r}}{(l^{2}+n^{2}+r^{2})^{2}}\,\simeq\,0.707\,.\label{Const}
\end{equation}
In this case one has
$\mathcal{T}^{33(11)}=\mathcal{T}^{22(11)}=\mathcal{T}^{11(11)}$.
It is clear that both energy and pressure in a cubic box behave
similarly to the case of two parallel plates.

In order to treat the effect of temperature in a box, all four
coordinates in the Minkowski space have to be compactified by
taking $\alpha =(\beta,i2a_{1},i2a_{2},i2a_{3})$ in
Eq.~(\ref{tan1}). This amounts to adding terms involving $\beta$
and the distances to Eqs.~(\ref{Ebox}) and (\ref{Pbox}). In the
simpler case of a cubic box at finite temperature, the expressions
for the Casimir energy and pressure are
\begin{eqnarray}
E(\beta,a) & = & \frac{1}{a^{4}}\left\{
\frac{7\pi^{2}}{60}\frac{1}{\chi^{4}}-\left( \frac{7\pi^{2} }{960}
+ \frac{{\mathcal{C}}}{2\pi^{2}}\right) + \frac{24}
{\pi^{2}}\sum_{l,n=1}^{\infty}(-1)^{l+n}\frac{3\chi^{2}l^{2}-4n^{2}
}{[\chi^{2}l^{2}+4n^{2}]^{3}} \right. \nonumber\\
& & - \left.
\frac{48}{\pi^{2}}\sum_{l,n,r=1}^{\infty}(-1)^{l+n+r}\frac{3\chi^{2}
l^{2}-4(n^{2}+r^{2})}{[\chi^{2}l^{2}+4(n^{2}+r^{2})]^{3}}+\frac
{32}{\pi^{2}}\sum_{l,n,r,q=1}^{\infty}(-1)^{l+n+r+q}\frac{3\chi^{2}
l^{2}-4(n^{2}+r^{2}+q^{2})}{[\chi^{2}l^{2}+4(n^{2}+r^{2}
+q^{2})]^{3}} \right\} ,
\end{eqnarray}
\begin{eqnarray}
P(\chi,a) & = & \frac{1}{a^{4}}\left\{
\frac{7\pi^{2}}{180}\frac{1}{\chi^{4} }-\left(
\frac{7\pi^{2}}{2880}+\frac{{\mathcal{C}}}{6\pi^{2}}\right)
+\frac{16}{\pi^{2}}\sum_{l,n=1}^{\infty}\frac{(-1)^{l+n}}{[\chi^{2}
l^{2}+4n^{2}]^{2}}\right.  \nonumber\\
& &
+\,\frac{8}{\pi^{2}}\sum_{l,n=1}^{\infty}(-1)^{l+n}\frac{\chi^{2}
l^{2}-12n^{2}}{[\chi^{2}l^{2}+4n^{2}]^{3}}-\frac{16}{\pi^{2}}\sum
_{l,n,r=1}^{\infty}\frac{(-1)^{l+n+r}}{[\chi^{2}l^{2}+4(n^{2}+r^{2})]^{2}
}\nonumber\\
& & - \left.
\frac{32}{\pi^{2}}\sum_{l,n,r=1}^{\infty}(-1)^{l+n+r}\frac
{\chi^{2}l^{2}+4(n^{2}-3r^{2})}{[\chi^{2}l^{2}+4(n^{2}+r^{2})]^{3}}+\frac
{32}{\pi^{2}}\sum_{l,n,r,q=1}^{\infty}(-1)^{l+n+r+q}\frac{\chi^{2}
l^{2}+4(n^{2}+r^{2}-3q^{2})}{[\chi^{2}l^{2}+4(n^{2}+r^{2}+q^{2})]^{3}
}\right\}  , \; \label{PboxT}%
\end{eqnarray}
where $\chi=\beta/a$.

The Casimir pressure changes sign from negative to positive values
when the ratio $\chi=\beta/a$ passes through the value
$\chi_{0}\simeq2.00$ . The critical curve is
\begin{equation}
T_{c}=\frac{1}{\chi_{0}\,a} \, .
\end{equation}
Similar result exists, although for a different value of
$\chi_{0}$, for the case of parallel plates, where $T_{c}$ scales,
with the inverse of the length $a$.

Let us present an estimate of the critical temperature
$T_{c}=(\chi _{0}a)^{-1}.$ Taking $a=1 \mathrm{fm}$, a length of
the order of hadron radius, then
$T_{c}=(\chi_{0}a)^{-1}\approx100\,\mathrm{MeV.}$ This temperature
is of the same order of magnitude as the temperature for the
deconfinement transition for hadrons. Let us analyze this
transition in more details, bring together gauge bosons and
fermions.

\subsection{Casimir effect for quarks and gluons in $\Gamma_{4}^{4}$}

A useful and simplified approximation, to discuss the influence of
Casimir effect in the deconfinement of hadron, is to consider
quarks and gluons as massless non-interacting particles. This is
the case of a baryon-free massless quark-gluon plasma, at high
temperature, in a zero-order approximation, where the interaction
and the quark mass can be discarded. This type of approximation
has been used, by considering the quark and gluon field in slabs
and sphere. Here we discuss this matter, assuming that the
massless quark-gluon plasma is confined in a space-time with
topology $\Gamma_{4}^{4}$, where the circumferences are specified
by the set of parameters $\alpha$.

The energy-momentum tensor for the quark-gluon system, in this
approximation, is given as
\[
\mathcal{T}_{qg}^{\mu\nu(11)}(\alpha)=\mathcal{T}_{q}^{\mu\nu(11)}
(\alpha)+\mathcal{T}_{g}^{\mu\nu(11)}(\alpha).
\]
where the quark contribution, $\mathcal{T}_{q}^{\mu\nu(11)}
(\alpha)$, is given by the renormalized energy-momentum tensor for
fermions, Eq.~(\ref{tan1}), multiplied by the factors $n_c$ and
$n_f$ (the numbers of color and flavors respectively); and the
gluon contribution, $\mathcal{T}_{g}^{\mu\nu(11)} (\alpha)$, is
given by the renormalized energy-momentum tensor for the
electromagnetic field, Eq.~(\ref{tensaug08}), multiplied by $n_g$
(the number of gluons). Using   $\Gamma ^4_4$ topology, with equal
compactification length, $L$, in all three spatial directions,
corresponding to a cubic box of edge $L$ at finite temperature, we
fix $\alpha=(\beta,iL,iL,iL)$, such the gluon contribution for the
Casimir pressure is
\[
P_{g}(\beta,L)=\mathcal{T}_{g}^{33(11)}(\beta,L)=n_{g}\,g(\chi)
\,\frac{1}{L^{4}}
\]
where
\begin{eqnarray}
g(\chi) & = & \frac{2}{\pi^{2}}\left\{
\mathcal{C}_{g}+\frac{\pi^{4}}{90} \frac{1}{\chi^{4}} +
4\sum_{l,n=1}^{\infty}\frac{1}{[\chi^{2}l^{2}+n^{2}]^{2} } +
2\sum_{l,n=1}^{\infty}
\frac{\chi^{2}l^{2}-3n^{2}}{[\chi^{2}l^{2}+n^{2}]^{3}
}\right. \nonumber \\
& & \left.
+\,4\sum_{l,n,r=1}^{\infty}\frac{1}{[\chi^{2}l^{2}+n^{2}+r^{2}
]^{2}}+8\sum_{l,n,r=1}^{\infty}\frac{\chi^{2}l^{2}+n^{2}-3r^{2}}{[\chi
^{2}l^{2}+n^{2}+r^{2}]^{3}}+8\sum_{l,n,r,q=1}^{\infty}\frac{\chi^{2}
l^{2}+n^{2}+r^{2}-3q^{2}}{[\chi^{2}l^{2}+n^{2}+r^{2}+q^{2}]^{3}}\right\}
, \nonumber
\end{eqnarray}
with $\chi=\beta/L$ and
\[
\mathcal{C}_{g} = -2\sum_{l,n=1}^{\infty}
\frac{1}{[l^{2}+n^{2}]^{2}} + \frac{4} {3} \sum_{l,n,r=1}^{\infty}
\frac{1}{[l^{2}+n^{2}+r^{2}]^{2}} - \frac{\pi^{4}} {90} \approx
-1.1154 \,.
\]

For the quark field we have
\[
P_{q}(\beta,L)=\mathcal{T}_{q}^{33(11)}(\beta,L) =
n_{c}n_{f}\,f(\chi)\,\frac{1}{L^{4}}
\]
where
\begin{eqnarray}
f(\chi) & = & \frac{1}{\pi^{2}}\left\{
\mathcal{C}_{f}+\frac{7\pi^{4}}
{180}\frac{1}{\chi^{4}}+16\sum_{l,n=1}^{\infty}\frac{(-1)^{l+n}}{[\chi
^{2}l^{2}+n^{2}]^{2}}+8\sum_{l,n=1}^{\infty}(-1)^{l+n}\frac{\chi^{2}
l^{2}-3n^{2}}{[\chi^{2}l^{2}+n^{2}]^{3}}\right. \nonumber \\
& &
-\,16\sum_{l,n,r=1}^{\infty}\frac{(-1)^{l+n+r}}{[\chi^{2}l^{2}+n^{2}
+r^{2}]^{2}}-32\sum_{l,n,r=1}^{\infty}(-1)^{l+n+r}\frac{\chi^{2}l^{2}
+n^{2}-3r^{2}}{[\chi^{2}l^{2}+n^{2}+r^{2}]^{3}} \nonumber \\
& & +\left.
32\sum_{l,n,r,q=1}^{\infty}(-1)^{l+n+r+q}\frac{\chi^{2}l^{2}
+n^{2}+r^{2}-3q^{2}}{[\chi^{2}l^{2}+n^{2}+r^{2}+q^{2}]^{3}}\right\}
\end{eqnarray}
with
\[
\mathcal{C}_{f} =
-8\sum_{l,n=1}^{\infty}\frac{(-1)^{l+n}}{[l^{2}+n^{2}]^{2} } +
\frac{16}{3}\sum_{l,n,r=1}^{\infty}
\frac{(-1)^{l+n+r}}{[l^{2}+n^{2} +r^{2}]^{2}}-\frac{7\pi^{4}}{180}
\approx -5.67
\]

The total Casimir pressure for the system of free, massless,
quarks and gluons is given by
\begin{equation}
P_{qg}(\beta,L) = \left[  n_{c}n_{f}\, f(\chi) + n_{g} \, g(\chi)
\right] \,\frac{1}{L^{4}} \, .
\end{equation}
For high temperatures, both quark and gluons give positive
contributions to the Casimir pressure. However for low
temperatures, the pressure is negative. Then a transition to
positive pressure appears by raising the temperature. The value of
$\chi= \beta/L$ at which the pressure vanishes, in the case of a
cubic box, is the root of the equation $n_{c}n_{f} f(\chi) +
n_{g}g(\chi) = 0$. This leads to the critical curve
\[
T_{c} = \chi_{c}^{-1}\, \frac{1}{L} \, .
\]
Considering a hadron specified by two flavors, $u$ and $d$, each
with 3 colors and an octet of gluons, we have $n_{g}=8$, $n_{c}=3$
and $n_{f}=2$. Then we obtain, numerically, as $\chi_{c} \approx
1.725$. If we take $L\approx1 \mathrm{fm}$, a length of the order
of a hadron radius, we get $T_{c} \approx 115 \, \mathrm{MeV}$.
Such an estimate provides a rough idea of the importance of
Casimir effect in the deconfinement transition for hadrons. This
is to be compared with the temperature, $175\; \mathrm{MeV}$, for
the deconfinement of quarks and gluons from nucleons in lattice
QCD.  These results point to that the Casimir energy may be
important for the production of the quark-gluon plasma.

The method based on the Bogoliubov transformation for compactified
space-time regions provides an effective way to study the Casimir
effect on toroidal topologies.  In general, to derive a finite and
measurable result for the Casimir effect, a renormalization
procedure has to be introduced. Here we take the difference of the
energy-momentum tensor on the topology $\Gamma_{D}^{d}$ and the
energy-momentum tensor written in the empty space-time, leading
directly to a finite  physical tensor
$\mathcal{T}^{\mu\nu}(\alpha)$.   The basic feature supporting
this behavior is that the generalized Bogoliubov transformation
separates the Green function into two parts: one is associated
with the empty space-time, the other describes the impact of
compactification. This represents a natural ease in the
calculation of $\mathcal{T}^{\mu\nu}(\alpha)$, the renormalized
energy-momentum tensor; and hence the Casimir effect.

\section{Restoration of symmetry on toroidal spaces: $\protect\phi^4$ theory}

Spontaneous symmetry breaking is a fundamental concept in modern
physics. In particle physics, this notion is the basis for
understanding fundamental interactions, in the standard model.
There is a vast bibliography on the subject, including historical
and review
articles~\cite{Nambu2009,Gross1996,Guralnik2011,Hill2003,Quigg2007};
while other ones are more important for specific aspects of
particle
physics~\cite{Weinberg1976,Weinberg1979,Dawson1979,Dimopoulos1979}.
There are also contributions considering  general elements of
field theory~\cite{Coleman1973,Coleman1974,Bardeen1983}. In
condensed matter physics, spontaneous symmetry breaking is a
cornerstone to a quantum field theoretical approach of phase
transitions, as is the case of the Ginzburg-Landau
theory~\cite{Landau1937,Ginzburg1950,Affleck1985,Lawrie1994,Lawrie1997}.
General aspects of spontaneous symmetry breaking in
superconductors have been analyzed using a field theoretical
approach~\cite{Weinberg1986}.

In both domains, particle and condensed matter physics, field
theories are defined usually on flat spaces. However, field
theories defined on spaces with some, or all, of its dimensions
compactified on a torus are of interest in several branches of
theoretical physics, since spontaneous symmetry restoration can be
induced by both temperature and spatial
boundaries~\cite{BirrellFord1980,Kibble1980,Toms1980,Toms1980a,Toms1980b,
Kennedy1981,Goncharov1982,Goncharov1982a,Hosotani1983,Actor1990,Actor1990a,
FordSvaiter1995,Ourbook2009}. In this section, the $N$-component
$\varphi ^{4}$ model within the large-$N $ approximation is
considered in order to study finite-size effects, including in
particular the influence of a chemical potential.

On the topology $\Gamma _{D}^{d}$, Feynman rules are modified by
introducing the generalized Matsubara prescription defined for
Euclidean spaces by the following
replacement~\cite{KhannamalsaesAnnals2009,JotAdolfo2002,MalbouissonsaesNPB2002,KhannamalsaesAnnals2011},
\begin{eqnarray}
\int \frac{dk_{0}}{2\pi }\, \rightarrow\,  \frac{1}{\beta
}\sum_{n_{0} = -\infty }^{+\infty }\, & , & \;\;\;\;
k_{0}\rightarrow
\frac{2(n_{0}+c_{0})\pi }{\beta } \notag \\
\int \frac{dk_{i}}{2\pi } \, \rightarrow \, \frac{1}{L_{i}}
\sum_{n_{i}=-\infty }^{+\infty } \, & , & \;\;\;\;
k_{i}\rightarrow \frac{2(n_{i}+c_{i})\pi }{L_{i}}\;,
\label{Matsubara1}
\end{eqnarray}
where $\beta $ is the inverse temperature,   $L_{1},L_{2},\dots
,L_{d-1}$ are the sizes of the compactified spatial dimensions and
$c_{j}=0$ or $c_{j}=1/2$, for bosons or fermions, respectively .

Initially, we review some results for the one-component scalar
field, at the one-loop level, with one-compactified spatial
dimension. The interaction Lagrangian is given by
\begin{equation*}
L_{int}(\phi )=-\frac{1}{4!}\lambda \phi ^{4},
\end{equation*}
with the space-time metric being $diag(1,-1,-1,-1)$, such that the
points $ x^{1}=0$ and $x^{1}=L$ are identified. The topology
affects the boundary conditions on the field operators and the
Green function, but local properties, such as the dynamical
structure, are not altered. The free Feynman propagator on $\Gamma
_{D}^{1}$ satisfies the equation $ (\square
+m^{2})G_{0}(x-y,L)=-\delta (x-y),$ and periodic  conditions $
G_{0}(x-y,L)=G_{0}(x-y+iL{n}_{1},L)$, where ${n} _{1}=(0,1,0,0)$.
Then $G_{0}(x-y,L)$ is given by
\begin{equation*}
G_{0}(x-y,L)=\frac{-1}{L}\sum_{n}\int d^{D-1}k \,
\frac{e^{-ik_{n}\cdot x}}{ k_{n}^{2}-m^{2}+i\varepsilon },
\end{equation*}
where $k_{n}=(k^{0},k_{n}^{1},k^{2},k^{3})\,$, with
$k_{n}^{1}=2\pi n/L$. The self-energy to the first-order in the
coupling constant is $\Sigma =\Sigma _{M}+\Sigma _{L}$, where
\begin{equation*}
\Sigma _{M}=\frac{m^{2}\lambda }{16\pi
^{2}(D-4)}+\frac{m^{2}\lambda }{32\pi ^{2}}(\gamma
-1)+\frac{m^{2}\lambda }{32\pi ^{2}}\ln \left( \frac{m^{2}}{ 4\pi
\mu ^{2}}\right) +O(D-4)
\end{equation*}
and
\begin{equation*}
\Sigma _{L}=\frac{\lambda }{L^{2}}f(\omega )+O(D-4),
\end{equation*}
where $\omega =Lm/2\pi $ and
\begin{equation*}
f(\omega )=\int\limits_{-\infty }^{\infty }dx\,\frac{(x^{2}-\omega
^{2})^{-1/2} }{e^{2\pi x}-1}.
\end{equation*}
In these expressions, $m$ and $\lambda $ are the renormalized mass
and coupling constant, respectively. For $D=4$ (i.e. $\Gamma
_{4}^{1}$), $\Sigma _{M}$ has a simple pole. On the other hand,
$\Sigma _{L}$ is finite, being regarded as an $L$-correction to
the renormalized mass. In  the case of zero bare mass, the boson
acquires mass as the result of the self interaction and the
compactification effect. The free-space result is recovered for $
L\rightarrow \infty $, such that $\Sigma _{L}\rightarrow 0$. These
ideas were developed in different directions, initially for
applications in cosmological
problems,~\cite{Kibble1980,Toms1980,Toms1980a,Toms1980b,Kennedy1981,
Goncharov1982,Goncharov1982a,Hosotani1983,Actor1990,Actor1990a}.
In view of applications to superconductivity and an extension for
fermions in contact interaction, in the following we present some
details of the symmetry breaking formalism for a multi-component
scalar field in a topology $\Gamma _{D}^{d}$.

\subsection{Effective potential in $\Gamma _{D}^{d}$}

Let us now consider the $N$-component,  $\varphi ^{4}$ theory
described by the Lagrangian density,
\begin{equation}
\mathcal{L}=\frac{1}{2}\partial _{\mu }\varphi _{a}\partial ^{\mu
}\varphi _{a}+\frac{1}{2}m^{2}\varphi _{a}\varphi
_{a}+\frac{u}{4!}(\varphi _{a}\varphi _{a})^{2},
\label{Lagrangeanab}
\end{equation}
in $D$-dimensional Euclidian space-time, where $u=\lambda /N$ is
the coupling constant, $m$ is the mass and summation over repeated
indices $a$ is assumed; to simplify, this index will be
suppressed. The system is considered in thermal equilibrium  at
temperature $\beta ^{-1}$, with compactification of $(d-1)$
spatial coordinates with compactification lengths $L_{j}$,
$j=2,3,...,d$. Cartesian coordinates $\mathbf{r}
=(x_{1},...,x_{d},\mathbf{z})$ are used, where $\mathbf{z}$ is a
$(D-d)$
-dimensional vector, with corresponding momentum $\mathbf{k}%
=(k_{1},...,k_{d},\mathbf{q})$. The generalized Matsubara
prescription, as given in Eq.~(\ref{Matsubara1}), is used with
$c_{i}=0$. The large-N limit is considered, such that
$N\rightarrow \infty $, $u\rightarrow 0$, with $ Nu=\lambda $
fixed.

We start with the one-loop contribution to the effective potential
for the non-compactified theory~\cite{Coleman1973,Coleman1974},
\begin{equation}
U_{1}(\varphi _{0})=\sum_{s=1}^{\infty
}\frac{(-1)^{s+1}}{2s}\left[ \frac{ u\varphi _{0}^{2}}{2}\right]
^{s}\int \frac{d^{D}k}{(2\pi )^{D}}\frac{1}{ (k^{2}+m^{2})^{s}},
\label{potefet0b}
\end{equation}
where $m$ is the physical mass and $\varphi _{0}$ is the
normalized vacuum expectation value of the field (the classical
field). For the \textit{Wick-ordered} model, to order $1/N$ in the
one-loop approximation,  it is unnecessary to perform a mass
renormalization. The parameter $m$ in Eq.~(\ref{Lagrangeanab})
plays in this case the role of the physical mass.

Introducing the parameters $c=1/(2\pi)$, $a_j = 1/ (mL_{j})^{2}$,
$g=(u/8\pi ^{2})$, $q_j = k_j/(2 m \pi)$ and performing the
Matsubara replacements, the one-loop contribution to the effective
potential is written as,
\begin{equation*}
U_{1}(\phi _{0},a_{1},...,a_{d})=\sqrt{a_{1}\cdots
a_{d}}\sum_{s=1}^{\infty } \frac{(-1)^{s+1}}{2s}g^{s}\varphi
_{0}^{2s} m^{D-2s} \sum_{n_{1},...,n_{d}=-\infty }^{+\infty
}\,\int \frac{d^{D-d}q}{\left(a_{1}n_{1}^{2}+\cdots
+a_{d}n_{d}^{2}+c^{2}+\mathbf{q}^{2}\right)^{s}}.
\end{equation*}
The integral over the $D-d$ noncompactified momentum variables is
carried out using the well-known dimensional regularization
formula,
\begin{equation}
\int \frac{d^{l}p}{(2\pi )^{l}}\frac{1}{(\mathbf{p}^{2}+M)^{s}} =
\frac{\Gamma \left(s-\frac{l}{2}\right)}{(4\pi )^{l/2}\Gamma
(s)M^{s-l/2}}; \label{dimreg}
\end{equation}
leading to,
\begin{equation}
U_{1}(\varphi _{0},a_{1},...,a_{d})=\sqrt{a_{1}\cdots a_{d}}
\sum_{s=1}^{\infty }f(D,d,s)g^{s}\varphi
_{0}^{2s}Z_{d}^{c^{2}}\left(s-\frac{D-d}{
2};a_{1},...,a_{d}\right), \label{potefet2b}
\end{equation}
where
\begin{equation}
f(D,d,s)=\pi ^{(D-d)/2}\frac{(-1)^{s+1}}{2s\Gamma (s)}\Gamma
\left(s-\frac{D-d}{2} \right)
\end{equation}
and $Z_{d}^{c^{2}}$ is the homogeneous Epstein-Hurwitz
multivariable $zeta$-function defined by,
\begin{equation}
Z_{d}^{c^{2}}(\nu ;a_{1},...,a_{d})=\sum_{n_{1},...,n_{d}=-\infty
}^{\infty }(an_{1}^{2}+\cdots +a{n_{d}}^{2} +c^2)^{-\nu }.
\label{zetab}
\end{equation}

\begin{figure}[tph]
\begin{center}
\scalebox{0.40}{\includegraphics {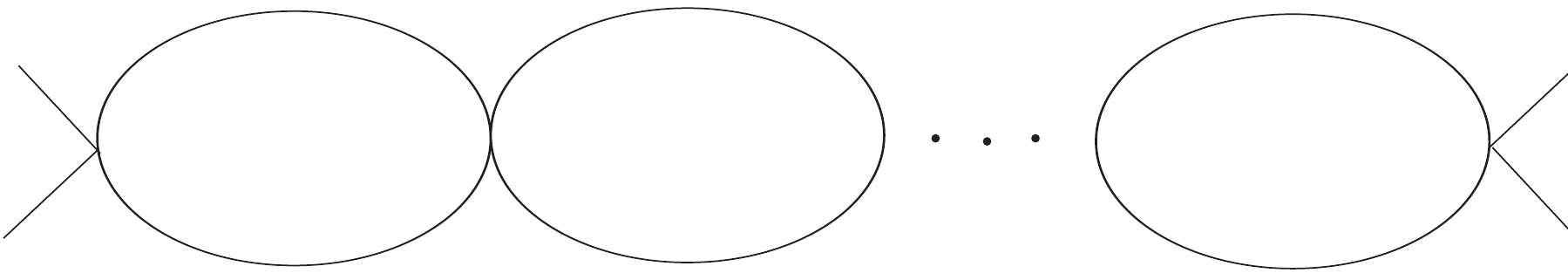} }
\end{center}
\caption{Typical diagram contributing to the four-point function
at leading order in $\frac{1}{N}$. To each vertex there is a
factor $\frac{\protect \lambda }{N}$ and for each single bubble a
color circulation factor of $N$.} \label{figXVIII1}
\end{figure}

The Epstein-Hurwitz function can be extended to the whole complex
$\nu $-plane, with the result~\cite{Kirsten1994,Elizalde2001},
\begin{eqnarray}
Z_{d}^{c^{2}}(\nu;a_1,a_{2},...,a_{d}) & = & \frac{\pi
^{d/2}}{\sqrt{a_{1}\cdots a_{d}}\,\Gamma (\nu )} \left[
\frac{1}{c^{2\nu-d}}\Gamma \left(\nu -\frac{d}{2}\right) + 2
\sum_{j=1}^{d} \sum_{n_{j}=1}^{\infty } \left(\frac{\pi
n_{j}}{c\sqrt{a_j}}\right)^{\nu -\frac{d}{2}}
K_{\nu -\frac{d}{2}}(mL_{j}n_{j})+\cdots \right. \nonumber \\
&& + \left. 2^{d+1}\sum_{n_{1},...,n_{d}=1}^{\infty }
\left(\frac{\pi}{c} \sqrt{ \frac{n_{1}^{2}}{a_1}+\cdots
+\frac{n_{d}^{2}}{a_d} }\right)^{\nu - \frac{d}{2}} K_{\nu
-\frac{d}{2}} \left(2\pi c \sqrt{ \frac{n_{1}^{2}}{a_1}+\cdots
+\frac{n_{d}^{2}}{a_d} } \right) \right] .\label{zetabbb}
\end{eqnarray}
Taking $\nu =s-(D-d)/2$ in this equation and recovering the
original parameters, the one-loop correction to the effective
potential in $D$ dimensions with a compactified $d$-dimensional
subspace is
\begin{eqnarray}
U_{1}(\phi _{0},\beta ,L_{2},...,L_{d}) & = & \sum_{s=1}^{\infty }
u^{s} \varphi_{0}^{2s} h(D,s) \left[ 2^{s-\frac{D}{2}-2} \Gamma
\left( s-\frac{D}{2} \right) m^{D-2s} +
\sum_{i=1}^{d}\sum_{n_{i}=1}^{\infty } \left( \frac{m}{L_{i}
n_{i}} \right)^{\frac{D}{2}-s} K_{\frac{D}{2}-s}
(mL_{i}n_{i})\right. \nonumber
\label{potefet3b} \\
&& + \left. 2 \sum_{i<j=1}^{d} \sum_{n_{i},n_{j}=1}^{\infty }
\left( \frac{m}{\sqrt{L_{i}^{2}n_{i}^{2} + L_{j}^{2}n_{j}^{2}}}
\right)^{\frac{D}{2}-s} K_{\frac{D}{2}-s}\left(m
\sqrt{L_{i}^{2}n_{i}^{2}+L_{j}^{2}n_{j}^{2}}\right) + \,\cdots \right. \notag \\
&& + \left. 2^{d-1}\sum_{n_{1},\cdots n_{d}=1}^{\infty } \left(
\frac{m}{\sqrt{ L_{1}^{2}n_{1}^{2}+\cdots +L_{d}^{2}n_{d}^{2}}}
\right)^{\frac{D}{2}-s} K_{\frac{D}{2
}-s}\left(m\sqrt{L_{1}^{2}n_{1}^{2}+\cdots
+L_{d}^{2}n_{d}^{2}}\right)\right] ,
\end{eqnarray}
with
\begin{equation}
h(D,s)=\frac{1}{2^{D/2-s-1}\pi ^{D/2-2s}}\frac{(-1)^{s+1}}{s\Gamma
(s)}. \label{hb}
\end{equation}
For defining the coupling constant, now the zero external-momentum
four-point function is analyzed,  . The four-point function to the
leading order in $1/N$ is given by the sum of all diagrams of the
type depicted in Fig.~\ref{figXVIII1}. This
gives~\cite{Zinn-Justin1996,MalbouissonsaesNPB2002}
\begin{equation}
\Gamma _{D}^{(4)}(0,\{L_{i}\})=\;\;\frac{u}{1+Nu\Pi
(D,\{L_{i}\})}, \label{4-point1b}
\end{equation}
where $\Pi (D,\{L_{i}\})$ corresponds to the one-loop (bubble)
subdiagram in Fig.~\ref{figXVIII1}.

The following normalization conditions are used,
\begin{equation}
\left. \frac{\partial ^{2}}{\partial \varphi
^{2}}U(D,\{L_{i}\})\right\vert _{\varphi _{0}=0}=m^{2}
\label{renorm1b}
\end{equation}
and
\begin{equation}
\left. \frac{\partial ^{4}}{\partial \varphi
^{4}}U(D,\{L_{i}\})\right\vert _{\varphi _{0}=0}=u.
\label{renorm2b}
\end{equation}
The single bubble function, $\Pi (D,\{L_{i}\})$, is obtained from
the coefficient of the fourth power of the field ($s=2$) in
Eq.~(\ref{potefet3b} ). Then using Eqs.~(\ref{renorm2b}) and
(\ref{potefet3b}), we can write $\Pi (D,\{L_{i}\})$ in the form,
\begin{equation}
\Pi (D,\beta ,L)=H(D)+R(D,\{L_{i}\}).  \label{Sigmab}
\end{equation}
The first term between brackets in Eq.~(\ref{potefet3b}) gives
\begin{equation}
H(D)\propto \Gamma \left(2-\frac{D}{2}\right)m^{D-4}.  \label{Hb}
\end{equation}
For even dimensions $D\geq 4$, $H(D)$ is divergent due to the pole
of the $ \Gamma $-function. Such a term, which is independent of
$\{L_{i}\}$, should be suppressed to get a finite quantity. To
have an uniform procedure in any dimension, this subtraction is
also performed for odd dimension $D$, where no poles of $\Gamma
$-functions are present. Although this subtraction process is not
a perturbative renormalization, the quantities obtained are
referred as \textit{renormalized} quantities.

Then the $L_{i}$-dependent contribution $R(D,\{L_{i}\})$, arising
from the second term between brackets in Eq.~(\ref{potefet3b}), is
given by
\begin{eqnarray}
R(D,\{L_{i}\}) &=&\frac{3}{2}\frac{1}{(2\pi )^{D/2}}\left[
\sum_{i=1}^{d}\sum_{n_{i}=1}^{\infty }\left( \frac{m}{L_{i}n_{i}}\right) ^{%
\frac{D}{2}-2}K_{\frac{D}{2}-2}(mL_{i}n_{i})+2\sum_{i<j=1}^{d}%
\sum_{n_{i},n_{j}=1}^{\infty }\left( \frac{m}{\sqrt{%
L_{i}^{2}n_{i}^{2}+L_{j}^{2}n_{j}^{2}}}\right)
^{\frac{D}{2}-2}\right.
\notag \\
&&\left. \times K_{\frac{D}{2}-2}\left( m\sqrt{%
L_{i}^{2}n_{i}^{2}+L_{j}^{2}n_{j}^{2}}\right) +\ldots +\right.   \notag \\
&&+\left. 2^{d-1}\sum_{n_{1},\ldots ,n_{d}=1}^{\infty }\left( \frac{m}{\sqrt{%
L_{1}^{2}n_{1}^{2}+\ldots +L_{d}^{2}n_{d}^{2}}}\right) ^{\frac{D}{2}-2}K_{%
\frac{D}{2}-2}\left( m\sqrt{L_{1}^{2}n_{1}^{2}+\ldots +L_{d}^{2}n_{d}^{2}}%
\right) \right] .  \label{Gb}
\end{eqnarray}%
This provides a renormalized single bubble function. Using
properties of Bessel functions, for any dimension $D$,
$R(D,\{L_{i}\})$ satisfies the conditions
\begin{equation}
\lim_{L_{i}\rightarrow \infty
}R(D,\{L_{i}\})=0\;,\;\;\;\;\;\;\;\lim_{L_{i}\rightarrow
0}R(D,\{L_{i}\})\rightarrow \infty ,  \label{GG1b}
\end{equation}%
and $R(D,\{L_{i}\})>0$ for any values of $D$, and $L_{i}$.

The $\{L_{i}\}$-dependent renormalized coupling constant $\lambda
(D,\{L_{i}\})$ to the leading order in $1/N$ is defined by,
\begin{equation}
N\Gamma _{D}^{(4)}(0,\{L_{i}\})\equiv \lambda (D,\{L_{i}\})=\;\;\frac{%
\lambda }{1+\lambda R(D,\{L_{i}\})}.  \label{lambRb}
\end{equation}%
The renormalized coupling constant in the absence of boundaries is
\begin{equation}
\lambda (D)=N\lim_{L_{i}\rightarrow \infty }\Gamma
_{D,R}^{(4)}(0,\{L_{i}\})\,=\lambda ,  \label{lambfreeb}
\end{equation}%
where we have used Eq.~(\ref{GG1b}). Thus we conclude the
renormalization scheme is such that the constant $\lambda =Nu$
introduced in the Lagrangian corresponds to the large-$N$ physical
coupling constant in the unbounded space. Then the
$\{L_{i}\}$-dependent renormalized coupling constant is
\begin{equation}
\lambda (D,\{L_{i}\})=\;\frac{\lambda }{1+\lambda R(D,\{L_{i}\})}.
\label{lambRb1}
\end{equation}

\subsection{Spatially induced symmetry breaking}

In this subsection finite-size effects on the mass and the
coupling constant are studied. For simplicity, only the  case of
compactification of one spatial dimension is considered with and
without Wick-ordering.

\subsubsection{Wick-ordered model}

In the Wick-ordered model no mass renormalization is necessary.
For one compactified spatial dimension Eq.~(\ref{Gb}) yields,
\begin{equation}
R(D,L)=\frac{3}{2}\frac{1}{(2\pi )^{D/2}}\sum_{n=1}^{\infty
}\left[ \frac{m}{ nL}\right] ^{D/2-2}K_{\frac{D}{2}-2}(mnL),
\label{Gc}
\end{equation}%
which gives the $L$ -dependent renormalized coupling constant,
\begin{equation}
\lambda (D,L)=\;\frac{\lambda }{1+\lambda R(D,L)}.
\label{lambdac}
\end{equation}%
For $D=3$, using
\begin{equation}
K_{n+\frac{1}{2}}(z)=K_{-n-\frac{1}{2}}(z),\;\;\;\;\;K_{\frac{1}{2}}(z)=%
\sqrt{\frac{\pi }{2z}}e^{-z},  \label{Abramowitz}
\end{equation}%
the coupling constant is
\begin{equation}
\lambda _{W}(D=3,L)=\frac{8\pi m\lambda (e^{mL}-1)}{8m\pi
(e^{mL}-1)+3\lambda }, \label{lambdaexata}
\end{equation}%
where the subscript $W$ indicates explicitly Wick-ordering.
%A plot of $ \lambda _{W}(D=3,L)$ is given in Fig.~\ref{figXVIII2}.

\subsubsection{The model without Wick-ordering}

The effect of suppression of Wick-ordering is that the
renormalized mass cannot be taken as the coefficient $m$ of the
term $\varphi _{a}\varphi _{a}$ in the Lagrangian. The
$L$-corrected physical mass, which is obtained from
Eq.~(\ref{renorm1b}), has the form
\begin{equation}
m^{2}(L)=m^{2}+\frac{4\lambda (N+2)}{N(2\pi
)^{D/2}}\sum_{n=1}^{\infty } \left[ \frac{m(L)}{Ln}\right]
^{D/2-1}K_{\frac{D}{2}-1}(m(L)nL). \label{mDysonc}
\end{equation}
To obtain the $L$-dependent coupling constant, the constant mass
parameter $m $ should be replaced in Eq.~(\ref{lambdac}) and
Eq.~(\ref{Gc}) by the $L$ -corrected mass $m(L)$ and the resulting
system of equations should be solved with respect to $m(L)$. This
is a hard task which can not be implemented analytically.

\begin{figure}[h]
\begin{center}
\scalebox{1.2}{\includegraphics {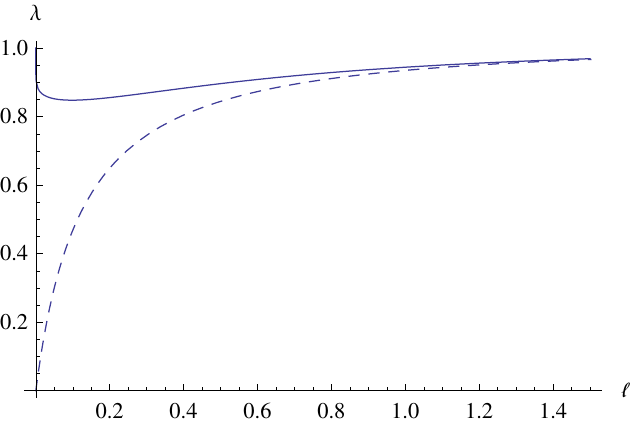} }
\end{center}
\caption{Renormalized coupling constant for the non-Wick ordered
model (full line) and for the Wick ordered model (dashed line) as
a function of the distance between the planes in dimension $D=3$.
The free space coupling constant is fixed as $1.0$ (in units of
$m$) and the reduced length is $\ell = m L$.} \label{figNew}
\end{figure}

However, in dimension $D=3$, a simple expression for the coupling
constant as a function of the renormalized $L$-dependent mass and
$L$ itself can be written; it is
\begin{equation}
\lambda (D=3,L)=\frac{8\pi m(L)\lambda (e^{m(L)L}-1)}{8\pi
m(L)(e^{m(L)L}-1)+3\lambda } ,  \label{lambdaexata1c}
\end{equation}
where
\begin{equation}
m^2(L) = m^2 - \frac{\lambda}{\pi L} \log\left( 1 - e^{-m(L) L}
\right) . \label{massaexata1c}
\end{equation}
Solving numerically the self-consistent Eq.~\ref{massaexata1c} for
$m(L)$ and inserting the results into Eq~(\ref{lambdaexata1c}),
one finds $\lambda (D=3,L)$; this function is plotted in
Fig~\ref{figNew} together with the coupling constant of the
Wick-ordered model.  Comparison of the coupling constant for
Wick-ordered model and without Wick-ordering shows quite different
behaviors. The coupling constant without Wick-ordering slightly
decreases for decreasing values of $L$ until some minimum value
and then starts to increase. In the Wick-ordered model the
coupling constant tends monotonically to zero as $L$ goes to zero.
In the non-Wick ordered model it has a non-vanishing value even
for very small values of $L$. In fact, numerical analysis of the
solution of Eq.~(\ref{massaexata1c}) shows that $m(L)L\rightarrow
0$ and $m^{2}(L)L\rightarrow \infty $ as $L\rightarrow 0$ and,
therefore, the $L$-corrected non-Wick-ordered coupling constant
has a non-vanishing value at $L=0$. This value is equal to the
free space value $\lambda $. As a general conclusion for the
non-Wick-ordered model, the $L$-dependent renormalized coupling
constant departs slightly to lower values, from the free space
coupling constant.

%%%%%%%%%%%%%%
%%%%%%%%%%%%%

For space dimension $D>2$, the correction in $L$ to the squared
mass is positive and the $L$-dependent squared mass is a
monotonically increasing function of $\frac{1}{L}$. Starting in
the ordered phase with a negative squared mass $m^{2}$, the model
exhibits spontaneous symmetry breaking but for a sufficiently
small critical value of $L$ the symmetry is restored. The critical
value of $L$, $L_{c}$, is defined as the value of $L$ for which
the mass in Eq.~(\ref{mDysonc}) vanishes. In the small $L$ regime,
an asymptotic formula for small values of the argument of Bessel
function is
\begin{equation}
K_{\nu }(z)\approx \frac{1}{2}\Gamma (\nu )\left(
\frac{z}{2}\right) ^{-\nu }\;\;\;(z\sim 0\;;\;\;\;Re(\nu )>0),
\label{KFilmb}
\end{equation}%
which can be used in Eq.~(\ref{mDysonc}). Then taking $m(L)=0$ in
the resulting equation, it is not difficult to obtain the
large-$N$ critical value of $L$ in the Euclidian space dimension
$D$ ($D>2$),
\begin{equation}
(L_{c})^{D-2}=-\frac{4\lambda g(D)}{m^{2}},  \label{criticoc}
\end{equation}%
where
\begin{equation}
g(D)=\frac{1}{4\pi ^{\frac{D}{2}}}\Gamma \left( \frac{D}{2}-1
\right)\zeta (D-2),
\end{equation}%
$\zeta (D-2)$ being the Riemann $zeta$-function. For $D=3$ the
$zeta$ -function in $g(D)$ has a pole and a subtraction procedure
is needed: the Laurent expansion of $\zeta (z)$ is used, i.e.,
\begin{equation}
\zeta (z)=\frac{1}{z-1}+\gamma -\gamma _{1}\,(z-1)+\cdots \,,
\label{zetaLaurentc}
\end{equation}%
where $\gamma \simeq 0.577$ and $\gamma _{1}\simeq 0.0728$ are the
Euler-Mascheroni and the first Stieltjes constants, respectively.
The critical value of $L$ in dimension $D=3$ is then,
\begin{equation}
L_{c}=-\frac{\lambda \gamma }{16m^{2}}\pi ^{-\frac{5}{2}}.
\label{criticocd}
\end{equation}%
Previous estimates and numerical simulations for
temperature-driven phase
transitions~\cite{Einhorn1993,Bimonte1997} are then extended to a
phase transition driven by a spatial boundary.

Taking the Wick-ordering, which eliminates all contributions from
the tadpoles, the boundary behavior of the coupling constant and
of the mass are decoupled. Wick-ordering is a useful and
simplifying procedure in applications of the field theory to
particle physics, but the same is not necessarily true in
applications of field theory to investigate critical phenomena,
where the contribution from tadpoles are physically significant.
As a consequence of the suppression of Wick-ordering, the boundary
behavior of the coupling constant is sensibly modified with
respect to the monotonic behavior in the Wick-ordered case.

\subsection{The compactified model at finite temperature: spontaneous
symmetry breaking}

Now the non-Wick-ordered model is considered; such the $\beta $-
and $L$- corrected effective potential is derived from
Eq.~(\ref{potefet3b}) to one-loop approximation, taking $d=2$.
Then the renormalized physical mass is obtained from
Eqs.~(\ref{renorm1b}) and (\ref{potefet3b}), i.e.
\begin{eqnarray}
m^{2}(\beta ,L) & = & m^{2} + \frac{4\lambda }{(2\pi )^{D/2}}
\left[ \sum_{n=1}^{\infty } \left(\frac{m}{n\beta
}\right)^{\frac{D}{2}-1}K_{\frac{D}{2} -1}(n\beta m) +
\sum_{n=1}^{\infty
}\left(\frac{m}{nL}\right)^{\frac{D}{2}-1}K_{\frac{D}{2
}-1}(nLm)\right.  \nonumber \\
&& + \left. 2\sum_{n_{1},n_{2}=1}^{\infty } \left(
\frac{m}{\sqrt{\beta
^{2}n_{1}^{2}+L^{2}n_{2}^{2}}}\right)^{\frac{D}{2}-1}
K_{\frac{D}{2}-1} \left( m\sqrt{ \beta
^{2}n_{1}^{2}+L^{2}n_{2}^{2}} \right) \right] . \label{mDysonb111}
\end{eqnarray}

In the $\beta \times L$ plane, the critical curve is defined by
the vanishing of the corrected mass. In the neighborhood of
criticality, i.e. $ m^{2}(\beta ,L)\approx 0$, the asymptotic
formula for small values of the argument of Bessel functions,
Eq.~(\ref{KFilmb}), is used, such that Eq.~(\ref{mDysonb111})
becomes
\begin{equation}
m^{2}(\beta ,L)\approx m^{2}+\frac{4\lambda }{(\pi )^{D/2}}\Gamma
\left( \frac{D}{2}-1\right) \left[ \left(\beta
^{2-D}+L^{2-D}\right)\zeta (D-2)+2E_{2}\left( \frac{D-2}{2};\beta
,L\right) \right] , \label{mDysoncrb}
\end{equation}
where $\zeta (D-2)$ is the Riemann $zeta$-function, and
$E_{2}((D-2)/2;\beta ,L)$ is the homogeneous two-variable
Epstein-Hurwitz
$zeta$-function~\cite{Kirsten1994,Actor1989,Elizalde2001},
\begin{equation}
E_{2}\left( \frac{D-2}{2};\beta ,L\right)
=\sum_{n_{1},n_{2}=1}^{\infty } \frac{1}{(\beta
^{2}n_{1}^{2}+L^{2}n_{2}^{2})^{\frac{D-2}{2}}}\,.  \label{Zb}
\end{equation}
This function has an analytical continuation to the
complex-plane~\cite{Elizalde1989,Kirsten1994}. However, before
going with this procedure, a symmetrization  has to be performed
in order to maintain the symmetry $\beta \leftrightarrow L$ of
$E_{2}\left( \frac{D-2}{2};\beta ,L\right) $. Then the symmetrized
analytical-continuation Epstein-Hurwitz function
reads~\cite{MalbouissonsaesNPB2002}
\begin{eqnarray}
E_{2}\left( \frac{D-2}{2};\beta ,L\right) & = & -\frac{1}{4}\left(
\beta^{2-D}+L^{2-D}\right) \zeta (D-2)  \nonumber \\
&&+\,\frac{\sqrt{\pi }\Gamma (\frac{D-3}{2})}{4\Gamma
(\frac{D-2}{2})}\left( \frac{1}{\beta L^{D-3}}+\frac{1}{\beta
^{D-3}L}\right) \zeta (D-3) + \frac{\sqrt{\pi }}{\Gamma
(\frac{D-2}{2})}W_{2}(D-2;\beta ,L)\;, \label{Z1b}
\end{eqnarray}%
where
\begin{equation}
W_{2}(D-2;\beta ,L) = \frac{1}{\beta
^{D-2}}\sum_{n_{1},n_{2}=1}^{\infty }\left(\pi \frac{L}{\beta
}n_{1}n_{2}\right)^{\frac{D-3}{2}}K_{\frac{D-3}{2}}\left(2\pi
\frac{L}{\beta }n_{1}n_{2}\right) +
\frac{1}{L^{D-2}}\sum_{n_{1},n_{2}=1}^{\infty }\left(\pi
\frac{\beta }{L}
n_{1}n_{2}\right)^{\frac{D-3}{2}}K_{\frac{D-3}{2}}\left(2\pi
\frac{\beta }{L} n_{1}n_{2}\right).  \label{W2}
\end{equation}

The function $E_{2}((D-2)/2;\beta ,L)$ has two simple poles at
$D=4$ and $D=3 $. The solution of Eq.~(\ref{mDysoncrb}) for
$m(\beta ,L)=0$ and $m^{2}<0 $ defines the critical curve in $D$,
with $D\neq 4$ and $D\neq 3$, as
\begin{equation}
m^{2}+\frac{4\lambda }{(\pi )^{D/2}}\Gamma \left( \frac{D}{2}
-1\right) \left[ \left(\beta _{c}^{2-D}+L_{c}^{2-D}\right)\zeta
(D-2)+2E_{2}\left( \frac{D-2}{2};\beta _{c},L_{c}\right) \right]
=0. \label{mDysoncr1b}
\end{equation}%
For $D=4$ the generalized $zeta$-function $E_{2}$ has a pole and
for $D=3$ both the Riemann $zeta$-function and $E_{2}$ have poles.
We cannot obtain a critical curve in $D=4$ and $D=3$ by a limiting
procedure from Eq.~(\ref{mDysoncr1b}). For $D=4$, which
corresponds to the physically interesting case of the system at
$T\neq 0$ confined between two parallel planes embedded in a
$3$-dimensional Euclidean space, Eqs.~(\ref{mDysoncrb}) and
(\ref{mDysoncr1b}) are meaningless. To obtain a critical curve in
$D=4$, a regularization procedure is carried out. Then the mass is
redefined as,
\begin{equation}
\lim_{D\rightarrow 4^{-}}\left[ m^{2}+\frac{1}{D-4}\frac{4\lambda
}{\pi ^{2}\beta L}\right] =\bar{m}^{2}\;.  \label{massrenb}
\end{equation}
The critical curve in dimension $D=4$ is obtained from
\begin{equation}
\bar{m}^{2}+\frac{\lambda }{3}\left( \frac{1}{\beta
_{c}^{2}}+\frac{1}{L_{c}^{2}} \right) +\frac{\pi {\gamma }}{\beta
_{c}L_{c}}+4\sqrt{\pi }W_{2}(2;\beta _{c},L_{c})=0\;.
\label{mDysoncr2b}
\end{equation}

\subsection{Finite size and chemical potential effects}

Here for $d=2$, finite temperature and one compactified spatial
dimension, results with a finite chemical potential are
considered. The focus, as before, is on the $N$-component $\varphi
^{4}$ model within the large-$N$ approximation. The symmetry
restoration is obtained by extending to a
toroidal topology the two-particle irreducible (2PI) formalism~\cite%
{Cornwall1974,Amelino-Camelia1993} in the Hartree-Fock
approximation. In this approach, all daisy and super-daisy
diagrams, contributing to the effective potential, are taken into
account. Starting from the broken symmetry region, the behavior of
the renormalized mass and the critical temperature is studied.

\subsubsection{One-loop correction to the mass in a toroidal space at finite
chemical potential}

Using the Lagrangian density given in Eq.~(\ref{Lagrangeanab}) to
study symmetry restoration requires an adaptation of the 2PI
formalism for a toroidal space. For an unbounded space at zero
temperature, the stationary condition for the effective action in
the Hartree-Fock approximation translates into a gap equation,
\begin{equation}
G^{-1}(x,y)=D^{-1}(x,y)+\frac{u}{2}\,G(x,x)\delta ^{4}(x-y).
\label{GAP}
\end{equation}%
The Fourier-transformed propagators, $D(k)$ and $G(k)$, are given
by
\begin{equation}
D(k)=\frac{1}{{k}^{2}+m^{2}+\frac{u}{2}\phi
^{2}}\,;\;\;\;\;G(k)=\frac{1}{{k} ^{2}+M^{2}},  \label{D}
\end{equation}%
where $\phi =\sqrt{\left\langle 0\left\vert \varphi _{a}\varphi
_{a}\right\vert 0\right\rangle }$  and $M$ is a
momentum-independent effective mass.

In the 2PI formalism, the gap equation corresponds to the
stationary condition and as such the effective mass depends on
$\phi $ and conveys all daisy and superdaisy graphs contributing
to $G(k)$. Nevertheless, in order to investigate symmetry
restoration, a particular constant value $M$ may be taken in the
spontaneously broken phase. Renormalization of the mass $m$ and of
the coupling constant $u$ can be
performed~\cite{Amelino-Camelia1993}, leading to the equation
\begin{equation}
{M^{2}=-m_{R}^{2}+\frac{u_{R}}{2}\phi ^{2}+\frac{u_{R}}{2}\,G(M)},
\label{M}
\end{equation}%
where $m_{R}^{2}$ and $u_{R}$ are respectively the squared
renormalized mass and the renormalized coupling constant, both at
zero temperature and zero chemical potential, in the absence of
boundaries, {and $G(M)$ is the finite part of the integral
\begin{equation*}
G(x,x)=\frac{1}{(2\pi )^{D}}\int d^{D}{k}\,G(k).
\end{equation*}
The minus sign of the $m_{R}^{2}$ term ensures the spontaneous
symmetry breaking. Then Eq.~(\ref{M}) gives the renormalized mass
in the broken-symmetry phase and is rewritten as
\begin{equation}
{\bar{m}}^{2}(\phi )=-M^{2}+\frac{u_{R}}{2}\int
\frac{d^{D}\,{k}}{(2\pi )^{D}}\frac{1}{{k}^{2}+M^{2}},
\label{lagrangianamu1}
\end{equation}%
where the \textit{effective renormalized mass} ${\bar{m}}^{2}(\phi
)=-m_{R}^{2}+(u_{R}/2)\phi ^{2}$ has been introduced. This
equations is generalized to include the toroidal topology as well
as the chemical potential. Restoration of the symmetry will occur
at the set of points in the toroidal space where ${\bar{m}}^{2}$
is null. }

The system at temperature $\beta ^{-1}$ and one compactified
spatial coordinate ($x_{2}$) with a compactification length
$L_{2}\equiv L$ is considered at finite chemical potential $\mu $.
Then a suitable generalization of the procedure for the $2PI$
formalism~is carried out, to
take into account finite-size and thermal  effects in Eq.~(\ref%
{lagrangianamu1}). The integral over the $D$-dimensional momentum
becomes a
double sum over $n_{\tau }$ and $n_{2}\equiv n_{x}$ together with a ($D-2$%
)-dimensional momentum integral. The renormalized ($\beta ,L,\mu $%
)-dependent mass in the large-$N$ limit is written, by using
dimensional regularization in
Eq.~(\ref{dimreg})~\cite{MalbouissonsaesNPB2002}, in the
form %\begin{}
\begin{equation*}
\bar{m}^{2}(T,L,\mu )=-M^{2}+\frac{u_{R}M^{D-2}}{2}\frac{\pi ^{(D-2)/2}}{%
4\pi ^{2}}\frac{\Gamma \left( s-\frac{D-2}{2}\right) }{\Gamma
\left( s\right) }\sqrt{a_{\tau }a_{x}}\left. \sum_{n_{\tau
},n_{x}=-\infty }^{\infty }\left[ a_{\tau }\left( n_{\tau
}-\frac{i\beta }{2\pi }\mu \right)
^{2}+a_{x}n_{x}^{2}+c^{2}\right] ^{(D-2)/2-s}\right\vert _{s=1},
\end{equation*}%
%\end{}
where the dimensionless quantities $a_{\tau }=(M\beta )^{-2}$, $
a_{x}=(ML)^{-2}$ and $c=(2\pi )^{-1}$ are introduced. The double
sum is recognized as one of the inhomogeneous Epstein-Hurwitz zeta
functions, $ Z_{2}^{c^{2}}(s-\frac{D-2}{2};a_{\tau }.a_{x};b_{\tau
},b_{x})$, which has an analytical extension to the whole complex
$s$-plane~\cite{Elizalde1989,Kirsten1994}; in general for
$j=1,\,2$,
\begin{eqnarray}
Z_{2}^{c^{2}}(\nu ;\{a_{j}\};\{b_{j}\}) &=&\frac{\pi |c|^{2-2\nu
}\,\Gamma (\nu -1)}{\Gamma (\nu )\sqrt{a_{1}a_{2}}}+\frac{4\pi
^{\nu }|c|^{1-\nu }}{ \Gamma (\nu )\sqrt{a_{1}a_{2}}}\left[
\sum_{j=1}^{2}\sum_{n_{j}=1}^{\infty }\cos (2\pi n_{j}b_{j})\left(
\frac{n_{j}}{\sqrt{a_{j}}}\right) ^{\nu -1}K_{\nu -1}\left(
\frac{2\pi c\,n_{j}}{\sqrt{a_{j}}}\right) \right.
\notag \\
&&\left. +\,2\sum_{n_{1},n_{2}=1 }^{\infty }\cos (2\pi
n_{1}b_{1})\cos (2\pi n_{2}b_{2})\left(
\sqrt{\frac{n_{1}^{2}}{a_{1}}+\frac{n_{2}^{2}}{a_{2}} }\right)
^{\nu -1}K_{\nu -1}\left( 2\pi c\,\sqrt{\frac{n_{1}^{2}}{a_{1}}+
\frac{n_{2}^{2}}{a_{2}}}\right) \right] .  \label{Z2mu}
\end{eqnarray}
Here, $a_{1}=a_{\tau }$, $a_{2}=a_{x}$, $b_{1}=b_{\tau }=i\beta
\mu /2\pi $, $b_{2}=b_{x}=0$, $c=1/2\pi $ and $\nu =s-(D-2)/2$.
Then the thermal and boundary corrected mass is obtained in terms
of the original variables, $ \beta ,L,\mu $ and of the fixed
renormalized zero-temperature coupling constant in the absence of
boundaries, $\lambda _{R}=\lim_{N\rightarrow \infty
,u_{R}\rightarrow 0}(N\,u_{R})$. However, the first term in
Eq.~(\ref{Z2mu}) implies that the first term in the corrected mass
is proportional to $\Gamma (1-D/2)$, which is divergent for even
dimensions $D\geq 2$. This term is suppressed by a minimal
subtraction, leading to a finite effective renormalized mass; for
the sake of uniformity, this polar term is also subtracted for odd
dimensions, where no singularity exists, corresponding to a finite
renormalization~\cite{MalbouissonsaesNPB2002}.

In dimension $D$, the following set of dimensionless parameters is
introduced: $\lambda_{R}^{\prime} = \lambda _{R}\,M^{D-4}$; $t =
T/M$; $\chi =1/LM$; and $\omega =\mu /M$. After subtraction of the
polar term, which does not depend on $\beta $, $L$ and $\mu $, the
corrected mass, $\bar{m}^{2}(D,\beta ,L,\mu )$ is obtained. This
implies that the condition for symmetry restoration, $\bar{m}
^{2}(D,\beta ,L,\mu )=0$, can be written in terms of the
dimensionless parameters, replacing $\lambda _{R}^{\prime }$ by
the corrected coupling constant $\lambda _{R}^{\prime }(D,\beta
,L,\mu )$, in such a way that the critical equation reads
\begin{eqnarray}
-1&+&\frac{\lambda _{R}^{\prime }(D,t,\chi ,\omega )}{(2\pi
)^{D/2}}\left[ \sum_{n=1}^{\infty }\cosh \left( \frac{\omega
n}{t}\right) \left( \frac{t}{n} \right)
^{\frac{D}{2}-1}K_{\frac{D}{2}-1}\left( \frac{n}{t}\right)
+\sum_{l=1}^{\infty }\left( \frac{\chi }{l}\right)
^{\frac{D}{2}-1}K_{\frac{D
}{2}-1}\left( \frac{l}{\chi }\right) \right.   \notag \\
&+&\left. 2\sum_{n,l=1}^{\infty }\cosh \left( \frac{\omega
n}{t}\right) \left(
\frac{1}{\sqrt{\frac{n^{2}}{t^{2}}+\frac{l^{2}}{\chi
^{2}}}}\right) ^{\frac{D }{2}-1}K_{\frac{D}{2}-1}\left(
\sqrt{\frac{n^{2}}{t^{2}}+\frac{l^{2}}{\chi ^{2}}}\right) \right]
\; = \; 0 \,.  \label{massa14}
\end{eqnarray}

\subsubsection{Corrections to the coupling constant}

The four-point function~\cite{MalbouissonsaesNPB2002}, given in
Eq.~(\ref{4-point1b}), is modified to incorporate effects of  the
chemical potential in the coupling constant.  At leading order in
$1/N$ it is given by
\begin{equation}
\Gamma _{D}^{(4)}(D,\beta ,L,\mu )=\frac{u_{R}^{\prime
}}{1+Nu_{R}^{\prime }\Pi (D,\beta ,L,\mu )},  \label{4-point1}
\end{equation}
where $u_{R}^{\prime }=u_{R}\,M^{D-4}$ and the expression for the
one-loop diagram is given by
\begin{equation*}
\Pi (D,\beta ,L,\mu )=\left. \frac{\sqrt{a_{\tau }a_{x}}}{16\pi ^{4}}%
\sum_{n_{\tau },n_{x}=-\infty }^{\infty }\int
\frac{d^{D-2}q}{\left[ \mathbf{q}^{2}+a_{\tau }\left( n_{\tau
}-\frac{i\beta }{2\pi }\mu \right)
^{2}+a_{x}n_{x}^{2}+c^{2}\right] ^{s}}\right\vert _{s=2}.
\end{equation*}
Then $\Pi (D,\beta ,L,\mu )$ is written in the form
\begin{equation*}
\Pi (D,\beta ,L,\mu )=H(D)+[1/(2\pi )^{D/2})]R(D,\beta ,L,\mu ).
\end{equation*}
In terms of the dimensionless quantities, $ R(D,\beta ,L,\mu )$ is
given by,
\begin{eqnarray}
R(D,t,\chi ,\omega ) &=&\sum_{n=1}^{\infty }\cosh \left(
\frac{\omega n}{t} \right) \left( \frac{t}{n}\right)
^{\frac{D}{2}-2}K_{\frac{D}{2}-2}\left( \frac{n}{t}\right)
+\sum_{l=1}^{\infty }\left( \frac{\chi }{l}\right) ^{
\frac{D}{2}-2}  K_{\frac{D}{2}{}-2}\left( \frac{l}{\chi }\right) \notag \\
&& +\, 2\sum_{n,l=1}^{\infty }\cosh \left( \frac{\omega
n}{t}\right) \left( \frac{1
}{\sqrt{\frac{n^{2}}{t^{2}}+\frac{l^{2}}{\chi ^{2}}}}\right)
^{\frac{D}{2} -2}K_{\frac{D}{2}-2}\left(
\sqrt{\frac{n^{2}}{t^{2}}+\frac{l^{2}}{\chi ^{2}}} \right)
\label{PRjan231}
\end{eqnarray}
and $H(D)\propto \Gamma \left( 2-\frac{D}{2}\right) $ is a
\textquotedblleft polar" parcel coming from the first term in the
analytic extension of the $ Zeta$-function in Eq.~(\ref{Z2mu}).
For the same reason as before, this term is subtracted.

The dimensionless $t$-, $\chi $- and $\omega $-dependent
renormalized coupling constant $\lambda ^{\prime }(D,t,\chi
,\omega )$ at the leading order in $1/N$ is given as $\lambda
^{\prime }(D,t,\chi ,\omega )\equiv N\Gamma _{D}^{(4)}(D,t,\chi
,\omega )$, i.e.
\begin{equation}
\lambda _{R}^{\prime }(D,t,\chi ,\omega )=\frac{\lambda
_{R}^{\prime }}{ 1+\lambda _{R}^{\prime }[1/(2\pi
)^{D/2})]{R}(D,t,\chi ,\omega )}. \label{lambdaR11}
\end{equation}
Using properties of the Bessel functions, for any dimension $D$
and finite values of the reduced chemical potential $\omega $,
$R((D,t,\chi ,\omega )$ satisfies the conditions
\begin{equation*}
\lim_{t\rightarrow 0,\,\chi \rightarrow \infty }R((D,t,\chi
,\omega )=0,\,;\;\;\;\lim_{t\rightarrow \infty ,\,\chi \rightarrow
0}R((D,t,\chi ,\omega )\rightarrow \infty .
\end{equation*}%
The limit $t=\frac{T}{M}\rightarrow 0,$ $\chi
=\frac{1}{LM}\rightarrow \infty $ corresponds to a spatial
asymptotic freedom for vanishing small values of $L$. The limit
$t\rightarrow \infty ,$ $\chi \rightarrow 0$ corresponds to a
thermal asymptotic freedom to the system in the bulk form.

It is to be noted that, Eqs.~(\ref{massa14}) and (\ref{PRjan231})
are meaningful for a reduced chemical potential satisfying the
condition $0\leq
\omega <1$.  For $D=3$, it is straightforward to observe the range of $%
\omega $, by using the explicit formulas for $K_{\pm 1/2}(z)$ that
will occur in the critical equation and in the correction to the
coupling constant. For $D=4$, the relevant Bessel functions are
$K_{1}$ and $K_{0}$. In the first case we write $K_{1}(z)$ as,
\begin{equation}
K_{1}(z)=\int_{0}^{\infty }d\xi \,e^{-z\,\cosh \xi }\cosh \xi .
\label{K1(1)}
\end{equation}%
Then the  first term between square brackets in
Eq.~(\ref{massa14}) gives rise to expressions proportional to
\begin{equation}
\int_{0}^{\infty }d\xi \,\sum_{n=1}^{\infty }\,\frac{1}{2}\left(
e^{-n\beta \mu }+e^{n\beta \mu }\right) e^{-n\beta M\,\cosh \xi
}\cosh \xi , \label{cond1(1)}
\end{equation}%
where $n/t=n\beta M$. This integral is finite if the sum over $n$
converges for all values of the domain of integration. This
condition is only satisfied for values of the chemical potential
$\mu $ in the interval $[0,M)$, or for $0\leq \omega <1$. A
similar argument applies for the third term in
Eq.~(\ref{massa14}).

In Eq.~(\ref{PRjan231}), the asymptotic formula for large values
of $z$ in the Bessel function is used, i.e. $K_{0}(z)\approx
\sqrt{\frac{\pi }{2z}}e^{-z},$ with $z=(n/t)$. Then for large
values of $n$, the function in the first sum in
Eq.~(\ref{massa14}) has the asymptotic form
\begin{equation*}
f_{n}(t,\omega )\approx \frac{\sqrt{\pi
t}}{\sqrt{2\,n}}\frac{1}{2}\left\{
\exp \left[ -\frac{n}{t}(1-\omega )\right] +\exp \left[ -\frac{n}{t}%
(1+\omega )\right] \right\} .
\end{equation*}%
The second term in the curly brackets is convergent for r all
values of $ \omega \geq 0$. For the first term, the sum over $n$
is convergent only if $0\leq \omega <1$. We get a similar result
for the third sum in Eq.~(\ref{massa14}).

In the following, the restoration of symmetry is investigated,
taking into account thermal, boundary and finite chemical
potential corrections to the coupling constant.

\subsubsection{Boundary and chemical potential effects on the symmetry
restoration}

Accounting thermal and boundary corrections to the coupling
constant at finite chemical potential, we replace, $u$ by the
corrected coupling constant $\lambda (D,t,\chi ,\omega )$. At a
(reduced) temperature $t$, the bounded system at finite chemical
potential corresponds to an effective potential of the form
\begin{equation}
U(D,t,\chi ,\omega )=\frac{1}{2}\bar{m}^{2}(D,t,\chi ,\omega )\phi
^{2} + \lambda _{R}^{\prime }(D,t,\chi ,\omega )\phi ^{4},
\label{efpotential}
\end{equation}
where  $\phi =\sqrt{\langle 0|\varphi _{a}\varphi _{a}|0\rangle
}$. Starting from the ordered phase, a spontaneous symmetry
restoration occurs for the values of $t$, $\chi $ and $\omega $
that make $\bar{m}^{2}(D,t,\chi ,\omega )$ vanish in the ($t,\chi
,\omega $)-space. Since $\lambda _{R}^{\prime }(D,t,\chi ,\omega
)>0$ for all values of $t$, $\chi $ and $\omega $,   critical
lines are determined by the condition
\begin{equation}
\bar{m}^{2}(D,t,\chi ,\omega )=0,  \label{critica}
\end{equation}
which is equivalent to Eq.~(\ref{massa14}). For numerical
evaluations, we fix the value $\lambda _{R}^{\prime }=0.50$ and
take several values of the dimensionless parameters $t$, $\chi $
and $\omega $, for $D=3$.

\begin{figure}[htbp]
\begin{center}
  \begin{minipage}[b]{0.4\linewidth}
    \centering
    \includegraphics[width=\linewidth]{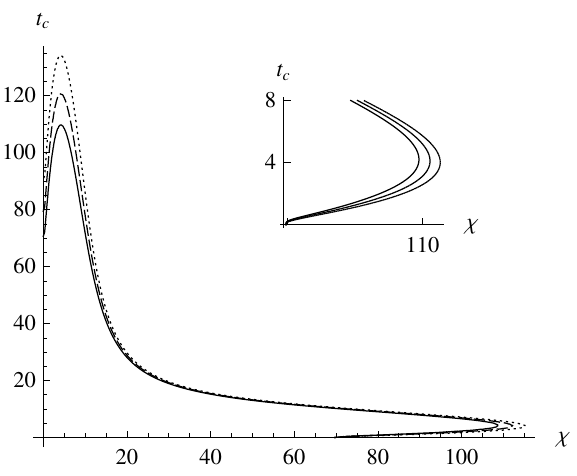}
    \caption{{Reduced critical temperature as a function of the reduced
inverse size of the system for dimension $D=3$ (left plot). We fix
$\protect\lambda _{R}^{\prime }=0.5$ and take the chemical
potential values $\protect\omega =0.1$ (full line), $0.3$ (dashed
line) and $0.4$ (dotted line). The symmetry-breaking regions are
in the \textquotedblleft inner" side of each curve. The
characteristic size of the system is $\protect\chi_0 \approx
69.96$.}} \label{figPhi41}
  \end{minipage}
  \hspace{1.0cm}
  \begin{minipage}[b]{0.4\linewidth}
    \centering
    \includegraphics[width=\linewidth]{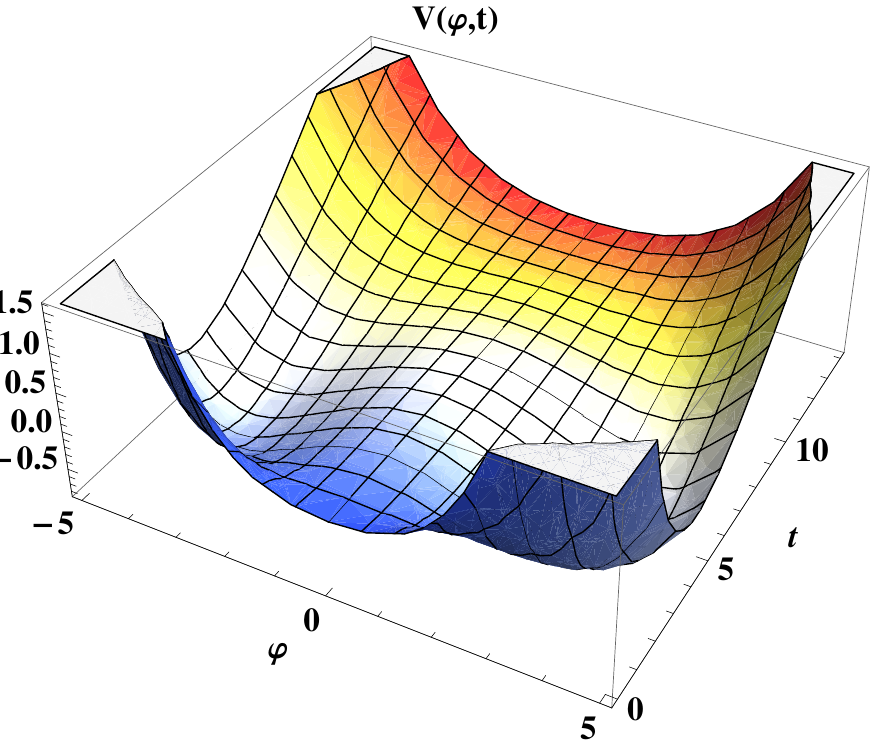}
    \caption{{A tridimensional view of the effective potential
$V=\frac{1}{2}m^{2}\varphi^{2} + \frac{1}{24}\lambda
^{\prime}_{R}\varphi ^{4}$, for  fixed values of the reduced
chemical potential, $\omega=0.30$ and of the reduced inverse size
of the system, $\chi=100$ for $D=3$ }} \vspace{0.5cm}
\label{figPhi441}
  \end{minipage}
  \end{center}
\end{figure}

Fig.~\ref{figPhi41} exhibits the critical temperature as a
function of the reduced inverse size of the system for different
values of the chemical potential. The behavior of the critical
temperature varies by changing the values of the chemical
potential, for small and large values of $\chi $. An interesting
aspect is the existence of a particular size of the system, $
L_{0}$, corresponding to the reduced inverse size $\chi
_{0}\approx 69.96$, where the critical temperature vanishes. This
value for  $\chi _{0}$, which is independent of the chemical
potential, is obtained  by solving Eq.~(\ref{massa14})  for $t=0$.
This is emphasized in the right plot of Fig.~\ref{figPhi41}, which
show in detail the domain around the characteristic value $ \chi
=\chi _{0}$.  For each value of $\omega $, there is a limiting
smallest size of the system, $L_{\mathrm{min}}(\omega )$,
corresponding to a largest reduced inverse size $\chi
_{\mathrm{max}}(\omega )$, such that $\chi _{ \mathrm{max}}(\omega
)>\chi _{0}$, over which the transition ceases to exist. An
interesting aspect is that $L_{\mathrm{min}}(\omega )$ is smaller
for growing values of $\omega $. However, since $\omega $ is in
the range $ 0<\omega <1$, there is as absolute minimum size of the
system.

%%%%%%%%%%%%%%%%
%%%%%%%%%%%%%%%

Moreover, from Fig.~\ref{figPhi41}, $\chi _{0}$ is the border
between two regions: $\chi <\chi _{0}$ and $\chi _{0}<\chi <\chi
_{\mathrm{max}}$. In the first region, the critical temperature is
uniquely defined in terms of the size and the chemical potential
of the system: for each pair $(\chi ,\omega )$ there is only one
critical temperature. In the second region, there are two values
of $t_{c}$  for the same values of $\omega $ and $\chi $ . In the
region $\chi _{0}<\chi <\chi _{\mathrm{max}}$, there are for each
value of $\omega $, two possible critical temperatures, say,
$t_{c}^{(1)}$ and $t_{c}^{(2)}$, with $t_{c}^{(2)}>t_{c}^{(1)}$,
associated respectively to the lower and the upper branches of the
critical curve. This means that in such a region, we have two
possible transitions, a behavior shown in Fig.~\ref{figPhi441},
where we plot the effective potential, for given values of $\omega
$ and $\chi $. It is important to emphasize that  a similar
behavior is found for $D=4$.

These results suggest that finite-size effects with finite
chemical potential are relevant and deeply changes the critical
curves with respect to the ones for the system in bulk form. In
particular, the appearance of a \textquotedblleft doubling" of
critical parameters is an unexpected behavior. This  is to be
contrasted with what happens with the system in bulk form, where
there is always an unique critical temperature, which grows with
increasing chemical potential.

As an overall conclusion, we can say that the concept of
spontaneous symmetry breaking or restoration is  transposed to
spaces endowed with a toroidal topology, being in this case driven
by a set of parameters ($e.g.$, temperature and/or the varying
size of the system). This study extends to systems which include
chemical potential, leading to some non-trivial results.

\section{Ginzburg-Landau model in a topology $\Gamma_D^d$}

The idea of describing thermodynamical phases through classical
(in general complex) fields, the order parameters, was first
introduced by Landau~\cite{Landau1937,Ginzburg1950} and since then
it has been used in a wide range of
applications~\cite{Affleck1985,Lawrie1994,Lawrie1997,Brezin1985,Calan2001,%
Abreu2004}. In the Landau formalism, the free energy  is written
as a functional of the order parameter, $\phi({\bf r})$. First-
and second-order phase transitions are described by choosing an
appropriate expansion of the free energy functional near the
critical point. Important realizations of these ideas have been
implemented over the years, as in superconductivity and in
superfluidity, establishing a connection between the field
theoretical approach and the microscopic point of view.

A first generalization of the Landau approach is the
Ginzburg-Landau (GL) formalism, where the free energy density is
written, in the neighborhood of criticality (natural units, $\hbar
= c = k_B = 1$ are used), as
\begin{equation}
\mathcal{F}(\phi(\mathbf{r}),\mathbf{\nabla}\phi(\mathbf{r})) =
\left| \mathbf{\nabla }\phi(\mathbf{r}) \right|^{2} + a(T) \left|
\phi(\mathbf{r}) \right| ^{2} + \frac{b}{2}\left| \phi(\mathbf{r})
\right|^{4}+c\left| \phi(\mathbf{r}) \right|^{6}\,+\cdots ,
\label{free2}
\end{equation}
such that $a(T) = \alpha (T -T_0)$; $b$ and $c$ are constants
independent of temperature. For a second-order transition, the
expansion up to the term  $\left| \phi(\mathbf{r}) \right|^{4}$ is
used, with $b>0$ and $T_0$ is the critical temperature. For a
first-order transition, the expansion is considered up to the
$\left|\phi(\mathbf{r})\right|^{6}$ term, with $b<0$ and $c>0$,
both independent of temperature. In this case, $T_0$ is not the
critical temperature.

In a field theoretical point of view, the GL free energy density,
in the absence of external fields, is considered as a Hamiltonian
density for the Euclidean self-interacting scalar field theories.
Then we can use methods of quantum field theory to treat
fluctuations of the order parameter in the GL model. Specifically,
for a second-order phase transition, the correction to the mass
term due to compactification of spatial coordinates ($L$) can be
implemented, resulting in $m^{2}(T,L)=\alpha(T-T_c(L))$ close to
criticality; this leads to an $L$-dependent critical temperature.
Such a generalization of the GL model, by including spatial
compactification, is a particular case of field theories on
$\Gamma^{d}_{D}$.

In this section, the quantum field theory formalism on toroidal
topology is applied to address the question of first- and
second-order phase transitions for systems described by the GL
theory. We are concerned with this theory in $D$ Euclidean
dimensions with a $d$-dimensional ($d\leq D$) compactified
subspace. This can be considered as the system constrained to a
region of space delimited by $d$ pairs of parallel planes,
orthogonal to each other, separated by distances $L_i$,
$i=1,2,\cdots$ (a parallelepiped box of edges $L_1,\cdots L_d$).
For practical cases, we will consider $D=3$ with $d=1,\,2,\,3$,
which physically corresponds to a film, a wire of rectangular
cross-section and a parallelepiped grain. Dealing with stationary
field theories, we employ the generalized Matsubara prescription
to implement only spatial compactification. No imaginary-time
compactification is necessary; the temperature is introduced
through the mass-like parameter in the free energy. In this
context, the question of how the critical temperature depends on
the relevant lengths of the system is considered for both second-
and first-order phase transitions. A physical application is found
for the problem of superconducting transitions in films, wires and
grains.

\subsection{Second-order phase transition in the $N$-component model}

Let us consider a system described by an $N$-component bosonic
field, $\varphi _{a}( \mathbf{r})$ with $a=1,2,...,N$, in a
$D$-dimensional Euclidean space, constrained to a $d$-dimensional
($d\leq D$) parallelepiped with length $L_1,L_2,\cdots L_d$,
satisfying periodic boundary conditions on its faces; i.e.
$\forall a$, $\varphi _{a}(x_i = 0, \mathbf{z})=\varphi _{a}(x_i =
L_i, \mathbf{z})$. Cartesian coordinates ${\bf r}=(x_1,...,x_d,
{\bf z})$ are used, where ${\bf z}$ is a $(D-d)$-dimensional
vector, with corresponding momentum ${\bf k}=(k_1,...,k_d,{\bf
q})$, ${\bf q}$ being a $(D-d)$-dimensional vector in momentum
space.

The generating functional  for correlation functions is
\begin{equation}
\mathcal{Z}[\varphi _{a}] =\int \mathcal{D}\varphi _{1}  \cdots
\mathcal{D} \varphi _{N} \,e^{ -\int_{0}^{L_{1}}dx_{1} \cdots
\int_{0}^{L_{d}}dx_{d}\int d^{D-d}\mathbf{z}\;\mathcal{H} (\varphi
(\mathbf{r}),\nabla \varphi (\mathbf{r}))} \,,  \label{ZZZ}
\end{equation}
where $\mathcal{H}(\varphi (\mathbf{r}),\nabla \varphi
(\mathbf{r}))$ is the Hamiltonian density,
$\mathbf{r}=(x_{1},...,x_{d},\mathbf{z})$ with $\mathbf{ z}$ being
a $(D-d)$-dimensional vector. The field has a mixed
series-integral Fourier expansion of the form,
\begin{eqnarray}
 \varphi _a(\{x_i\},{\bf z}) = \frac{1}{L_1 \cdots L_d}
 \sum_{n_1,\cdots ,n_d=-\infty }^{\infty }
 \int d^{D-d}{\bf q}\;
e^{-i\omega _{n_1}x_1-\cdots -i\omega _{n_d}x_d\;-i{\bf q}\cdot
{\bf z}} \tilde{\varphi}_a (\omega _{n_1},\cdots \omega
_{n_d},{\bf q})\,, \label{Fourier}
\end{eqnarray}
where $\omega _{n_i}=2\pi n_i/L_i$, $i=1,\cdots d$. The Feynman
rules are modified according to the particular case of the
generalized Matsubara prescription in Eq.~(\ref{Matsubara1}), for
compactification of spatial coordinates only,
\begin{equation}
\int \frac{dk_{i}}{2\pi }\rightarrow
\frac{1}{L_{i}}\sum_{n_{i}=-\infty }^{+\infty
}\;;\;\;\;\;\;\;k_{i}\rightarrow \frac{2n_{i}\pi }{L_{i}}
\;,\;\;i=1,2...,d.  \label{MatsubaraGeralL}
\end{equation}
In this sense we will refer equivalently in this section to
confinement in a segment of length $L_i$, or to compactification
of the coordinate $x_i$ with a compactification length
$L_i$~\cite{Ourbook2009}.

In the absence of topological restrictions, the $N$-component
vector model is described by the GL Hamiltonian density,
\begin{equation}
{\cal H}{}=\frac{1}{2} {\partial _{\mu} \varphi}_a  {\partial
^{\mu} \varphi}_a + \frac{1}{2}m_{0}^{2}(T)\varphi _{a}\varphi
_{a}+\frac{\lambda}{N}\,(\varphi _{a}\varphi _{a})^{2}\,,
\label{Lagrangeana}
\end{equation}
where $\lambda$ is the coupling constant, $m_{0}^{2}(T)=\alpha
(T-T_{0})$ is the bare mass ($T_{0}$ being the bulk transition
temperature) and summations over repeated indices $a$ are assumed.

Under the spatial constraints described above, the Hamiltonian
becomes,
\begin{equation}
{\cal {H}} = \frac{1}{2} {\partial _{\mu} \varphi}_a  {\partial
^{\mu} \varphi}_a + \frac{1}{2}m_{0}^{2}(T;L_1,...,L_d)\varphi
_{a}\varphi _{a}+ \frac{\lambda }{N}(\varphi_{a}\varphi _{a})^{2},
\end{equation}
where $m_{0}^{2}(T;L_1,...,L_d)$ is a suitably defined
boundary-modified mass parameter such that
\begin{equation}
\lim_{\{L_i\}\rightarrow\infty}m_{0}^{2}(T;L_1,...,L_d) =
m_{0}^{2}(T) \equiv \alpha \left( T - T_0 \right)\,. \label{m0}
\end{equation}
The $\{L_i\}$-corrections entering in the coupling constant
$\lambda$  will be considered in detail later. In the following we
will consider the model described by this Hamiltonian in the large
$N$ limit.

\subsubsection{One-loop effective potential with compactification of a
$d$-dimensional subspace}

The effective potential for systems with spontaneous symmetry
breaking is obtained by following usual
procedures~\cite{Coleman1973}, as an expansion in the number of
loops of Feynman diagrams.  At the 1-loop approximation, we get
the infinite series of 1-loop diagrams with all numbers of
insertions of the $\varphi ^4$ vertex (two external legs in each
vertex). The one-loop contribution to the zero-temperature
effective potential in unbounded space is
\begin{equation}
U_1(\varphi _0) = \sum_{s=1}^\infty \frac{(-1)^{s+1}}{2s}\left[
12(\lambda/N) |\varphi _0|^2\right]^s\int
\frac{d^Dk}{(k^2+m^2)^s}, \label{potefet0(1)}
\end{equation}
where $m$ is the  physical mass.

Then treating the integral concurrently with dimensional and
zeta-function analytic regularizations, as done in the last
section, we get
\begin{eqnarray}
U_{1}(\varphi _{0},\{ L_{i} \}) & =& \sum_{s=1}^{\infty }\,\left[
12 (\lambda /N)\varphi _{0}^{2} \right]^{s} \,h(D,s)\,\left[
2^{s-\frac{D}{2}-2}\Gamma (s-\frac{D}{2})
m^{D-2s}+\sum_{i=1}^{d}\sum_{n_{i}=1}^{\infty
}\left(\frac{m}{L_{i}n_{i}} \right)^{\frac{D}{2}-s}
K_{\frac{D}{2}-s} \left( mL_{i}n_{i}\right)
\right . \nonumber \\
&&+\left.
2\sum_{i<j=1}^{d}\sum_{n_{i},n_{j}=1}^{\infty}\left(\frac{m}{
\sqrt{L_{i}^{2}n_{i}^{2} +
L_{j}^{2}n_{j}^{2}}}\right)^{\frac{D}{2}-s}
K_{\frac{D}{2}-s}\left(
m\sqrt{L_{i}^{2}n_{i}^{2}+L_{j}^{2}n_{j}^{2}}\right) +\cdots
\right.
\nonumber \\
&&+\left. 2^{d-1}\sum_{n_{1},...,n_{d} = 1}^{\infty
}\left(\frac{m}{\sqrt{L_{1}^{2}n_{1}^{2}+\cdots +
L_{d}^{2}n_{d}^{2}}}\right)^{\frac{D}{2}-s}
K_{\frac{D}{2}-s}\left( m\sqrt{L_{1}^{2}n_{1}^{2}+\cdots
+L_{d}^{2}n_{d}^{2}}\right)\right] , \label{potefet3(1)}
\end{eqnarray}
where $$h(D,s) =
\frac{1}{2^{D/2+s-1}\pi^{D/2}}\frac{(-1)^{s+1}}{s\Gamma (s)}$$ and
the $K_{\nu }(z)$ are Bessel functions of the third kind. The mass
and the coupling constant are obtained from the normalization
conditions in Eqs.~(\ref{renorm1b}) and (\ref{renorm2b}).

\subsubsection{Boundary corrections to the coupling constant in the
Large-N limit}

The four-point function at zero external momenta is considered as
the basic object for our definition of the physical coupling
constant. At leading order in $1/N$, it is given by the sum of all
chains of one-loop diagrams, which has the formal expression,
\begin{equation}
\Gamma _{D}^{(4)}({ p}=0,m,\{L_i\})=\frac{\lambda /N}{1+\lambda
\Pi (D,m,\{L_i\})}\,, \label{R1}
\end{equation}
where, after making use of the prescription given in
Eq.~(\ref{Matsubara1}), $\Pi (D,m,\{L_i\})=\Pi({
p}=0,D,m,\{L_i\})$ corresponds to the single bubble four-point
diagram with compactification of a $d$-dimensional subspace. Then
using the normalization condition (\ref{renorm2b}), the single
bubble function $\Pi (D,m,\{L_i\})$ is obtained from the
coefficient of the fourth power of the field ($s=2$) in
Eq.~(\ref{potefet3(1)}). Then we write $\Pi(D,m,\{L_i\})$ as
\begin{equation}
\Pi(D,m,\{L_i\})=H(D,m)+R(D,m,\{L_i\})\,, \label{Sigma}
\end{equation}
where the $\{L_i\}$-dependent term $R(D,m,\{L_i\})$ arises from
the second term between brackets in Eq.~(\ref{potefet3(1)}),
\begin{eqnarray}
R(D,m;\{L_{i}\}) & = &\frac{1}{(2\pi)^{D/2}}\left[\sum_{i=1}^{d}
\sum_{n_{i}=1}^{\infty}\left(\frac{m}{L_{i}n_{i}}\right)^{\frac{D-4}{2}}
K_{\frac{D-4}{2}}(mL_{i}n_{i}) \right. \nonumber\\
&&  + \, 2\sum_{i<j=1}^{d}\sum_{n_{i},n_{j}=1}^{\infty}
\left(\frac{m}{\sqrt{L_{i}^{2}n_{i}^{2}
+L_{j}^{2}n_{j}^{2}}}\right)^{\frac{D-4}{2}}
K_{\frac{D-4}{2}}\left(m\sqrt{L_{i}^{2}n_{i}^{2}+L_{j}^{2}n_{j}^{2}}
\right) \, + \, \cdots  \nonumber \\
&& + \left. 2^{d-1}\sum_{n_{1},\ldots,n_{d}=1}^{\infty}
\left(\frac{m}{\sqrt{L_{1}^{2}n_{1}^{2} + \cdots +
L_{d}^{2}n_{d}^{2}}}\right)^{\frac{D-4}{2}}
K_{\frac{D-4}{2}}\left(m\sqrt{L_{1}^{2}n_{1}^{2} + \cdots +
L_{d}^{2}n_{d}^{2}}\right)\right] \label{G}
\end{eqnarray}
Here $H(D,m)$ is a polar term coming from the first term between
brackets in Eq.~(\ref{potefet3(1)}),
\begin{equation}
H(D,m)\propto \Gamma \left(2-\frac{D}{2}\right) m^{D-4}\,.
\label{H}
\end{equation}
For even dimensions $D\geq 4$, $H(D,m)$ is divergent, due to the
pole of the $\Gamma $-function. Accordingly, this term must be
subtracted to give the physical single bubble function
$R(D,m,\{L_i\})$. In order to have a coherent procedure for a
generic dimension $D$, the subtraction of the term $H(D,m)$ should
be performed even in the case of odd dimensions, where no poles of
$\Gamma $-functions are present. Also, from the properties of
Bessel functions, it follows that, for any dimension $D$,
$R(D,m,\{L_i\})$ is positive, vanishes as $L_i \rightarrow \infty$
and diverges as $L_i \rightarrow 0$.

From the four-point function, we define the $\{L_i\}$-dependent
physical coupling constant $\lambda (m,D,\{L_i\})$, at the leading
order in $1/N$, as
\begin{equation}
\lambda (D,m,\{L_i\}) \equiv N \Gamma _{D,R}^{(4)}({\bf
p}=0,m,\{L_i\}) = \frac{\lambda }{1+\lambda \, R(D,m,\{L_i\})} \,,
\label{lambdaR1}
\end{equation}
where the chosen renormalization scheme ensures that the constant
$\lambda$ corresponds to the physical coupling constant in the
absence of boundaries.

\subsubsection{Critical behavior: size-dependent transition temperature}

For a second-order phase transition, criticality is attained from
the ordered phase, when the inverse squared correlation length,
$\xi ^{-2}(\{L_{i}\},\varphi _{0})$, vanishes in the large-$N$ gap
equation,
\begin{eqnarray}
\xi ^{-2}(\{L_{i}\},\varphi_{0}) & = & m_{0}^{2}+12\lambda
(D,\{L_i\}) \varphi _{0}^{2} + \frac{24\lambda _{R}(D,\{L_i\})
}{L_{1}\cdots L_{d}}   \sum_{\{ n_{j} \}=-\infty }^{\infty }\int
\frac{d^{D-d}q}{(2\pi )^{D-d}}
 \,\frac{1}{{\mathbf q}^{2}+\sum_{j=1}^{d}(\frac{2\pi
n_{j}}{ L_{j} })^{2}+\xi ^{-2}(\{L_{i}\}, \varphi_{0})} \, .\nonumber \\
\label{gap1}
\end{eqnarray}
On the order-disorder border, $\varphi _{0}$ vanishes and the
inverse correlation length equals the physical mass, which is
obtained at the one-loop order from Eqs.~(\ref{potefet3(1)}) and
(\ref{renorm1b}), after suppressing the polar term as before and
replacing $\lambda \rightarrow \lambda _{R}(D,m,\{L_i\})$. Then we
get,
\begin{eqnarray}
m^{2}(D,T,\{L_{i}\}) & = &m_{0}^{2}(T,\{L_{i}\})
+\frac{24\lambda(D,m,\{L_i\})}{(2\pi)^{D/2}}\left[
\sum_{i=1}^{d}\sum_{n_{i}=1}^{\infty }\left(\frac{m}{L_{i}n_{i}}
\right)^{\frac{D}{2}-1}K_{\frac{D}{2}-1}\left(
mL_{i}n_{i}\right)\right.
\nonumber \\
&&+\left. 2\sum_{i<j=1}^{d}\sum_{n_{i},n_{j}=1}^{\infty
}\left(\frac{m}{
\sqrt{L_{i}^{2}n_{i}^{2}+L_{j}^{2}n_{j}^{2}}}\right)^{\frac{D}{2}-1}
K_{\frac{D}{2}-1}\left(
m\sqrt{L_{i}^{2}n_{i}^{2}+L_{j}^{2}n_{j}^{2}}\right) \, + \,
\cdots \right.
\nonumber \\
&&\left. + \, 2^{d-1}\sum_{n_{1},...,n_{d} = 1}^{\infty
}\left(\frac{m}{ \sqrt{L_{1}^{2}n_{1}^{2}+\cdots
+L_{d}^{2}n_{d}^{2}}}\right)^{\frac{D}{2}-1}
K_{\frac{D}{2}-1}\left( m\sqrt{L_{1}^{2}n_{1}^{2}+\cdots
+L_{d}^{2}n_{d}^{2}}\right)\right] ,\label{massa1}
\end{eqnarray}
where $m$ in the right-hand side stands for the thermal and
boundary dependent mass, $m(D,T,\{L_{i}\})$, inclusive in the
factor $\lambda(D,m,\{L_i\})$ which is obtained from
Eqs.~(\ref{lambdaR1}) and (\ref{G}).

Therefore $m(D,T,\{L_i\})$ satisfies an intricate, transcendental
(self-consistent) equation which has no analytical solutions in
general. However, in the neighborhood of criticality
$m^{2}(D,T,\{L_i\})\approx 0$ and we can use the formula for small
values of the argument of the Bessel functions, which leads to the
expression
\begin{equation}
\left( \frac{m}{z} \right)^{\nu} K_{\nu}(m z) \approx 2^{\nu - 1}
\Gamma(\nu) \, z^{-2 \nu} \;\;\;\;\;\;\; (m \approx 0) .
\label{approxim}
\end{equation}
This implies that the mass dependence disappears in all terms of
the right-hand-side of Eqs.~(\ref{G}) and (\ref{massa1}) leading,
close to criticality, to
\begin{eqnarray}
m^{2}(D,T,\{L_{i}\}) & \approx & m_0^{2}(T,\{L_i\})
+\frac{24\lambda (D,\{L_i\})}{(2\pi)^{D/2}}\left[
\sum_{i=1}^{d}\frac{1}{2}\Gamma
\left(\frac{D-2}{2}\right)E_{1}\left(\frac
{D-2}{2};L_{i}\right)\right.
\nonumber \\
& & +\left. 2\sum_{i<j=1}^{d}\frac{1}{2}\Gamma
\left(\frac{D-2}{2}\right)E_{2}\left(\frac
{D-2}{2};L_{i},L_{j}\right) +\cdots + 2^{d-1}\frac{1}{2}\Gamma
\left(\frac{D-2}{2}\right)E_{d} \left(\frac
{D-2}{2};L_{1},...,L_{d}\right)\right] , \nonumber \\
\label{massafront}
\end{eqnarray}
where
\begin{eqnarray}
\lambda (D,\{L_{i}\})&=&\frac{\lambda}{1+\lambda C(D,\{L_i\})},
\label{lambdafront}
\end{eqnarray}
with
\begin{eqnarray}
C_d(D,\{L_i\}) =
\frac{1}{8\pi^{D/2}}\Gamma\left(\frac{D-4}{2}\right)
\left[\sum_{i=1}^{d}L_{i}^{4-D}\zeta(D-4)
+2\sum_{i<j=1}^{d}E_{2}\left(\frac{D-4}{2};L_{i},L_{j}\right)
+\cdots +
2^{d-1}E_{d}\left(\frac{D-4}{2};L_{1},\ldots,L_{d}\right)\right]
. \nonumber \\
\label{C}
\end{eqnarray}

In the above equations, $E_p(\nu;L_1,\dots,L_p)$ is the
multidimensional Epstein function, which can be defined in the
symmetrized form as
\begin{equation}
E_{p}\left(\nu;L_{1},...,L_{p}\right) = \frac{1}{p !}
\sum_{\sigma} \sum_{n_1 = 1}^{\infty} \cdots \sum_{n_p =
1}^{\infty} \left[ \sigma_{1}^{2} n_{1}^{2} + \cdots +
\sigma_{p}^{2} n_{p}^{2} \right]^{\, - \nu} \; , \label{EfuncS}
\end{equation}
where $\sigma_{i}=\sigma(L_{i})$, with $\sigma$ running in the set
of all permutations of the parameters $L_1,...,L_p$, and the
summations over $n_1,...,n_p$ being taken in the given order.
Notice that, for $p=1$, $E_p$ reduces to the Riemann
$zeta$-function, $E_1(\nu;L_j) = L_j^{-2\nu}\zeta(2\nu)$. These
functions satisfy recurrence relations, which permit to write them
in terms of Kelvin and Riemann $zeta$
functions~\cite{Abreudecalanmalsaes2005},
\begin{eqnarray}
E_{p}\left(\nu ;L_{1},...,L_{p} \right) & = & - \,\frac{1}{2\, p}
\sum_{i=1}^{p} E_{p-1}\left( \nu;...,{\widehat {L_{i}}},...\right)
+\, \frac{\sqrt{\pi}}{2\, p\, \Gamma(\nu)} \Gamma\left( \nu -
\frac{1}{2}
 \right) \sum_{i=1}^{p} \frac{1}{L_i} E_{p-1}
 \left( \nu-\frac{1}{2};...,{\widehat {L_{i}}},... \right) \nonumber \\
& & +\, \frac{2\sqrt{\pi}}{p\, \Gamma(\nu)} W_p \left(
\nu-\frac{1}{2},L_1,...,L_p
 \right)\; ,
 \label{Wd}
\end{eqnarray}
where the hat over the parameter $L_{i}$ in the functions
$E_{p-1}$ means that it is excluded from the set
$\{L_{1},...,L_{p}\}$ (the others being the $p-1$ parameters of
$E_{p-1}$), and
\begin{eqnarray}
W_p\left(\eta;L_1,...,L_p\right)& =&\sum_{i=1}^{p} \frac{1}{L_i}
\sum_{n_1,...,n_p=1}^{\infty} \left( \frac{\pi n_i}{L_i \sqrt{(
\cdots + {\widehat {L_i^2 n_i^2}} + \cdots )}} \right)^{\eta}
K_{\eta}\left( \frac{2\pi n_i}{L_i} \sqrt{( \cdots + {\widehat
{L_i^2 n_i^2}} + \cdots )} \right) \; , \label{WWp}
\end{eqnarray}
with $(\cdots + {\widehat {L_i^2 n_i^2}} + \cdots)$ representing
the sum $\sum_{j=1}^{p}L_{j}^{2}n_{j}^{2} \, - \,
L_{i}^{2}n_{i}^{2}$.

Now we analyze the critical equation obtained by setting
$m^2(D,T,\{L_i\})$, given by Eq.~(\ref{massafront}), equal to
zero. It provides curves relating the critical temperature and the
compactification lengths. The cases of physical interest are $D=3$
and $d = 1, 2, 3$ corresponding respectively to a film, a
rectangular wire and a parallelepiped grain.

Taking $d=1$ and making $L_{1}\equiv L$, Eq.~(\ref{massafront})
becomes
\begin{equation}
m^2(D,T,L) = m_0^{2}(T,L)+\frac{6\lambda(D,L)} {\pi^{D/2} L^{D-2}
}\Gamma \left(\frac{D}{2}-1\right)\zeta(D-2) \label{filmecr1}
\end{equation}
where $\zeta (D-2)$ is the Riemann $zeta$-function, defined for
${\rm Re}\{ D-2\} > 1$, and $\lambda(D,L) = \lambda \left[ 1 +
\lambda C_1(D,L) \right]$. For $D = 3$, using that $\Gamma(-1/2) =
-2 \sqrt{\pi}$ and $\zeta(-1) = -1/12$, we find that
$\lambda(3,L)$ is finite,
\begin{equation}
\lambda(3,L)= \frac{48 \pi \lambda}{48 \pi + \lambda L}\,.
\label{lambdaR2}
\end{equation}
However, as $D \rightarrow 3$, $m^2(D,T,L)$ diverges due to the
pole of the function $\zeta(D - 2)$ and a renormalization must be
carried out. Using the Laurent expansion of the zeta function,
Eq.~(\ref{zetaLaurentc}), we define the L-dependent bare mass
$m_0(T,L)$ in such a way that the pole at $D=3$ in
Eq.~(\ref{filmecr1}) is suppressed, that is we take
\begin{equation}
m_0^{2}(T,L) \approx M - \frac{1}{(D-3)}
\frac{6\lambda(D.T,L)}{\pi L}\, , \label{massrenfilme}
\end{equation}
where $M$ is independent of $D$. To fix the finite term $M$, we
make the simplest choice satisfying Eq~(\ref{m0}),
\begin{equation}
M  = \alpha \left( T - T_{0}\right) \, ,
\end{equation}
$T_0$ being the bulk critical temperature. Thus, the renormalized
mass is written in the GL form,
\begin{equation}
m^{2}(T,L) \approx \alpha \left( T - T_{c}(L) \right) \,
\label{MRF}
\end{equation}
where the modified, $L$-dependent, transition temperature is
\begin{equation}
T_{c}(L) = T_{0}-\frac{48\pi C_1 \lambda }{48\pi \alpha L+ \alpha
\lambda L^{2}}\,,  \label{critica1}
\end{equation}
where
\begin{equation}
C_1 = \frac{6\gamma}{\pi} \approx 1.1024.
\end{equation}
From Eq.~(\ref{critica1}), we find that the critical temperature
decreases as $L$ diminishes, vanishing at the value corresponding
to the minimal allowed film thickness for the existence of the
transition,
\begin{equation}
L'_{min}=\frac{24\pi}{\lambda} \left[\sqrt{1+\frac{\lambda
L_{min}}{12\pi}}-1 \right] , \label{Lmin}
\end{equation}
where $L_{min} = C_1 \lambda / \alpha T_0$; the length $L_{min}$
is the minimal film thickness when no boundary corrections to the
coupling constant are taken into
account~\cite{Abreudecalanmalsaes2005,Malbouisson2009}.

We now focus on the case where two spatial dimensions are
compactified, corresponding to a wire with rectangular transversal
section. From Eqs.~(\ref{massafront}-\ref{C}), taking $d=2$ and
using the analytical extension of the Epstein function $E_2$, we
get
\begin{eqnarray}
{m}^{2}(T,D,L_1,L_2) & \approx & {m}^{2}(T,L_1,L_2) +
\frac{3\lambda (D,L_1, L_2)}{\pi ^{D/2}}
\left[\left(\frac{1}{L_{1}^{D-2}} + \frac{1}{L_{2}^{D-2}}\right)
\Gamma \left(\frac{D-2}{2} \right)
\zeta (D-2) \right. \nonumber \\
& & + \left.  \sqrt{\pi} \left( \frac{1}{L_{1}L_{2}^{D-3}} +
  \frac{1}{L_{1}^{D-3}L_{2}} \right) \Gamma \left(
  \frac{D-3}{2} \right) \zeta(D-3) +\, 2 \sqrt{\pi}W_{2}
  \left(\frac{D-3}{2};L_{1},L_{2}\right)
\right]  , \label{EQC1}
\end{eqnarray}
where $\lambda(D,L_{1},L_{2})$ and $W_2((D-3)/2;L_1,L_2)$, given
by the appropriate analytical extension of Eq.~(\ref{C}) and by
Eq~(\ref{WWp}), are finite quantities as $D \rightarrow 3$. On the
other hand, the mass diverges in this limit due to the poles of
the functions $\zeta(D-2)$ and $\Gamma((D-3)/2)$, and a mass
renormalization would be required; however, if we consider a wire
with a square transversal section, $L_{1}=L_{2}=L=\sqrt{A}$, the
two divergent terms cancel out exactly and we find
\begin{equation}
T_{c}(A) = T_{0}-\frac{48\pi C_{2}\lambda}{48\pi\alpha\sqrt{A} +
{\cal E}_2 \alpha\lambda(\sqrt{A})^{2}} , \label{Tcwire}
\end{equation}
where
\begin{equation}
C_2 = \frac{9\gamma}{\pi} + \frac{24}{\pi} \sum_{n_1,n_2=1}^
{\infty}K_{0}(2\pi n_{1}n_{2}) \approx 1.661 \, ,
\end{equation}
and
\begin{equation}
{\cal E}_2 = 1 + \frac{3 \zeta(3)}{\pi^2} +
\frac{24}{\pi}\sum_{n_{1},n_{2}=1}^{\infty}
\frac{n_{1}}{n_{2}}K_{1}(2\pi n_{1}n_{2}) \approx 1.372 .
\label{rho}
\end{equation}
As in the case of films, there is a minimal transverse section
sustaining the transition,
\begin{equation}
A'_{min} = \left[\frac{24\pi}{{\cal E}_2\lambda}
\left(\sqrt{1+\frac{{\cal E}_2\lambda(A_{min})^{1/2}}{12\pi}}-1
\right) \right]^{2} \label{Amin}
\end{equation}
where $A_{min} = (C_2 \lambda / \alpha T_0)^2$ is the minimal area
when no boundary corrections to the coupling constant are
considered~\cite{Abreudecalanmalsaes2005,Malbouisson2009}.

The case where all the spatial dimensions are compactified is
similar to the situation for wires but the expressions are more
complicated. However, if we fix $L_1=L_2=L_3=L$, corresponding to
a cubic grain of volume $V = L^3$, the mass term given by
Eq.~(\ref{massafront}) is finite and the boundary dependent
critical temperature takes the same form as that for films and
square wires, i.e.
\begin{equation}
T_{c}(V) = T_{0}-\frac{48\pi C_{3}\lambda}{48\pi\alpha\ V^{1/3}+
{\cal E}_3 \alpha\lambda(V^{1/3})^{2}} , \label{Tcgrao}
\end{equation}
where
\begin{equation}
C_3 = 1 + \frac{9\gamma}{\pi} + \frac{12}{\pi} \sum_{n_1,n_2
=1}^{\infty}\frac{e^{-2 \pi n_{1}n_{2}}}{n_{1}} +\, \frac{48}{\pi}
\sum_{n_1,n_2 =1}^{\infty} K_{0}(2\pi n_{1}n_{2}) + \frac{48}{\pi}
\sum_{n_1,n_2,n_3=1}^{\infty} K_{0}\left( 2\pi n_{1}
\sqrt{n_{2}^{2}+n_{3}^{2}} \right)\, \approx \, 2.676  \nonumber
\end{equation}
and
\begin{eqnarray} {\cal E}_3 & = & 1 +
\frac{\pi}{15} + \frac{3 \zeta(3)}{\pi^2} + \frac{24}{\pi}
\sum_{n_1,n_2 =1}^{\infty}\left(
\frac{n_{1}}{n_{2}}\right)^{3/2}K_{3/2}(2\pi n_{1}n_{2})  +\,
\frac{48}{\pi} \sum_{n_1,n_2
=1}^{\infty}\frac{n_{1}}{n_{2}}K_{1}(2\pi n_{1}n_{2})
\nonumber \\
 & &  +
\frac{48}{\pi} \sum_{n_1,n_2,n_3=1}^{\infty}
\frac{\sqrt{n_{1}^{2}+n_{2}^{2}}}{n_{3}} K_{1}\left(2\pi
n_{3}\sqrt{n_{1}^{2}+n_{2}^{2}} \right) \,
\approx \, 1.60\,.\nonumber \\
\label{sigma}
\end{eqnarray}
The minimal volume of the grain, allowing the existence of the
phase transition, is
\begin{equation}
V'_{min}=\left[\frac{24\pi}{{\cal E}_3\lambda}
\left(\sqrt{1+\frac{{\cal E}_3\lambda (V_{min})^{1/3}}{12\pi}}-1
\right) \right]^{3},
\end{equation}
where $V_{min} = (C_3 \lambda / \alpha T_0)^3$ corresponds to the
minimal volume for the situation where boundary corrections to the
coupling constant are
ignored~\cite{Abreudecalanmalsaes2005,Malbouisson2009}.

\begin{figure}[ht]
\begin{center}
\vspace{0.5cm} \scalebox{0.90}{{\includegraphics{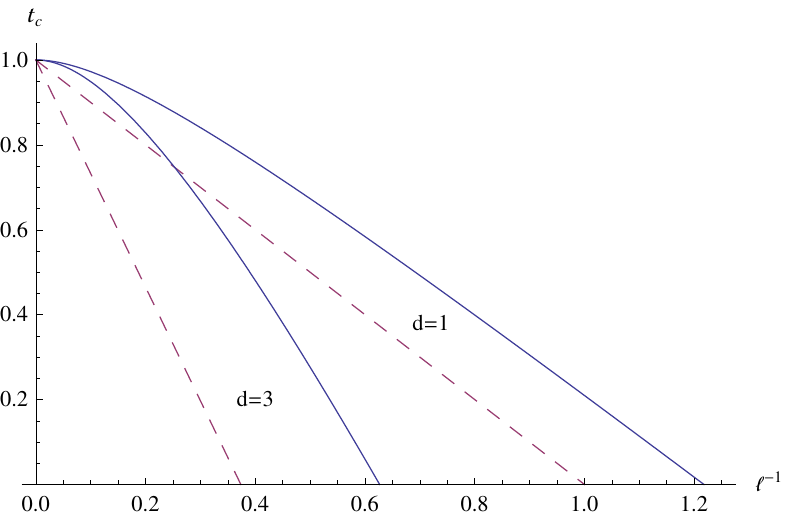}}}
\end{center}
\caption{Reduced transition temperature ($t_c$) as a function of
the inverse of the reduced compactification length ($\ell^{-1}$),
for films ($d = 1$) and cubic grains ($d = 3$), taking $\xi  =
40$. Full and dashed lines correspond to results with and without
correction of the coupling constant, respectively.} \label{figum}
\end{figure}

The above results can be summarized in terms of the {\it reduced
critical temperature} $t_c$ and the {\it reduced length} $\ell$,
defined respectively by
\begin{equation}
t_c=\frac{T_c}{T_0}\,,\;\;\;\;\; \ell = \frac{L}{L_{min}}\, .
\label{reduzidos1}
\end{equation}
Note that the minimal side of the transversal section of the
square wire and the minimal edge of the cubic grain allowing the
existence of the condensed phase without corrections to the
coupling constant are given by $L_{{min}}^{{wire}} =
\sqrt{A_{min}} = {\mathcal C}^{(2)} L_{{min}}$ and
$L_{{min}}^{{grain}} = (V_{min})^{1/3} = {\mathcal C}^{(3)}
L_{{min}}$, where ${\mathcal C}^{(2)} =
{\mathcal{C}}_2/{\mathcal{C}}_1 \approx 1.51$ and ${\mathcal
C}^{(3)} = {\mathcal{C}}_3/{\mathcal{C}}_1 \approx 2.43$. Thus,
for the case where no corrections of the coupling constant are
considered, we have~\cite{Abreudecalanmalsaes2005}
\begin{equation}
t_{c}^{(d)}(\ell)=1-\frac{{\mathcal{C}}^{(d)}}{\ell} \, ,
\label{tcd1}
\end{equation}
where ${\mathcal C}^{(1)} = 1$ and $d = 1, 2, 3$ refer to films,
square wires and cubic grains, respectively. Such a linear
relation between the reduced temperature and the inverse of the
reduced length, $1 - t_c \sim \ell^{-1}$ whatever the number of
compactified dimensions, can also be obtained from scaling
arguments~\cite{Zinn-Justin1996,Cardy1988}. Considering
corrections to the coupling constant, the reduced transition
temperature is written as,
\begin{equation}
t_{c}^{(d)}(\ell)=1-\frac{48\pi{\mathcal{C}}^{(d)}}{48\pi
\ell+{\mathcal{E}}_{d}\xi \ell^2} \label{tcd2}
\end{equation}
where ${\mathcal{E}}_1 = 1$ and $\xi = \lambda L_{min}$. In
Fig.~\ref{figum}, we plot the reduced transition temperature as a
function of the inverse of reduced length for the cases of films
and cubic grains, taking a fixed value of $\xi$. For comparison,
we also plot the straight lines corresponding to the cases where
no corrections to the coupling constant are considered.
%%%%%%%%%%%%%%%%%%%%%%%
%%%%%%%%%%%%%%%%%%%%%%

\subsubsection{Fixed point structure of the compactified
GL model}

We now examine whether there are infrared stable fixed points in
the Euclidean large-$N$ GL model defined in the topology
$\Gamma^{d}_{D}$, i.e. in $D$ dimensions with $d$ ($\leq D$) of
them being compactified. For the bulk system, the fixing-point
structure of a second-order transition was established for a
large-$N$ theory in arbitrary dimension
$D$~\cite{Cardy1988,Zinn-Justin1996,Calan2001,Radzihovsky1995}.
Here, we are concerned with the compactified model, considered as
a mean field theory, and we neglect the minimal coupling with the
vector potential corresponding to the intrinsic gauge
fluctuations.

We consider the $N$-component vector model described by the GL
Hamiltonian density, Eq.~(\ref{Lagrangeana}), in Euclidean
$D$-dimensional space and take the large-$N$ limit with $\lambda$
fixed. The coupling constant is defined in terms of the four-point
one-loop function for small external momenta which, at leading
order in $1/N$, is given by the sum of all chains of one-loop
diagrams. It is given in momentum space, before compactification,
and at the critical point by,
\begin{equation}
\Gamma _D^{(4)}(p,m=0)=\frac{\lambda/N}{1+\lambda\Pi (p,m=0)}\,,
\label{R1c}
\end{equation}
where $\Pi (p,m=0)$ is the single one-loop integral at the
critical point. It is written as,
\begin{equation}
\Pi \left( p,m=0\right) =\int \frac{d^Dk}{(2\pi )^D}\frac 1{\left[
k^2\left( p-k\right) ^2\right] }  =\int_0^1dx\int
\frac{d^Dk}{(2\pi )^D}\frac 1{\left[ k^2+p^2x(1-x)\right] ^2},
\label{PPi}
\end{equation}
where a Feynman parameter $x$ was introduced.

Performing the appropriate generalized Matsubara replacements
(\ref{Matsubara1}) for $d$ dimensions, Eq.~(\ref{PPi}) becomes
\begin{equation}
\Pi (p,D,\{L_i\},m=0) = \frac 1{L_1\cdots
L_d}\sum_{i=1}^d\sum_{\{n_i=-\infty \}}^\infty \int_0^1dx\int
\frac{d^{D-d}q}{(2\pi )^{D-d}}\frac 1{\left[ {\bf q}^2+\omega
_{n_1}^2+\cdots +\omega _{n_d}^2+p^2x(1-x)\right] ^2} \label{Pi2}
\end{equation}
and we define the effective $\{L_i\}$-dependent coupling constant
at the critical point in the large-$N$ limit as,
\begin{equation}
\lambda (p,D,\{L_i\})\equiv \lim_{u\rightarrow
0\,;\;\,N\rightarrow \infty }N\Gamma _D^{(4)}(p,\{L_i\},m=0)=\frac
\lambda {1+\lambda \,\Pi (p,D,\{L_i\},m=0)}. \label{lambda}
\end{equation}

The sum over the $n_i$ and the integral over ${\bf q}$ above
concerns the study of expressions of the form
\begin{equation}
I(s)=\sum_{i=1}^d\sum_{n_i=-\infty }^{+\infty }\int
\frac{d^{D-d}q}{({\bf q} ^2+a_1n_1^2+\cdots +a_dn_d^2+c^2)^s}.
\label{integral1}
\end{equation}
In our case, for the computation of $\Pi (p,D,\{L_i\},m=0)$, we
have $s=2$, $a_i=1/L_i^2$, $ \omega _i^2=(2\pi )^2a_in_i^2$ and
$c^2=p^2x(1-x)/(2\pi )^2$; also, a redefinition of the integration
variables, ${ q}\rightarrow { q}/2\pi $ , has been performed. Such
integral over the $D-d$ noncompactified momentum variables is
performed using the dimensional regularization formula,
Eq.~(\ref{dimreg}), which leads to
\begin{equation}
\Pi(s)=f(D,d,s)Z_d^{c^2}\left( s-\frac{D-d}2;a_1,\ldots
,a_d\right), \label{integral2}
\end{equation}
where
\begin{equation}
f(D,d,s)=\pi ^{(D-d)/2}\frac{\Gamma \left( s-\frac{D-d}2\right)
}{\Gamma (s)}
\end{equation}
and $Z_d^{c^2}(\nu ;a_1,\ldots ,a_d)$ are Epstein-Hurwitz zeta
functions, for $\nu =s-(D-d)/2$, already defined in
Eq.~(\ref{zetab}) and subsequent ones. It can be extended to the
whole complex $\nu$-plane,  leading to the result
\begin{equation}
\Pi (p,D,\{L_i\},m=0)= A(D)|p|^{D-4}+B_d(D,\{L_i\}),
\label{formageral}
\end{equation}
with the coefficient of the $|p|$-term being
\begin{equation}
A(D)=(2\pi)^{-D/2}b(D)\Gamma \left( 2-\frac D2%
\right) ,  \label{A(D)1}
\end{equation}
and where we have defined
\begin{equation}
b(D) =\int_0^1dx\,[x(1-x)]^{D/2-2}  =2^{3-D}\sqrt{\pi
}\frac{\Gamma \left( \frac D2-1\right) }{\Gamma \left(
\frac{D-1}2\right) }. \label{def-b}
\end{equation}
The quantity $B_d(D,\{L_i\})$ is given by
\begin{eqnarray}
B_d(D,\left\{ L_i\right\} ) & = & \frac{h(D,2)}{\left( 2\pi
\right) ^4} \int_0^1dx\,\left[ \sum_{i=1}^d\sum_{n_i=1}^\infty
\left( \frac{\sqrt{ p^2x(1-x)}}{2\pi L_in_i}\right) ^{D/2-2}
K_{D/2-2}\left( \frac 1{2\pi }\sqrt{
p^2x(1-x)}L_in_i\right) \right.  \nonumber \\
& & + \, 2\sum_{i<j=1}^d\sum_{n_i,n_j=1}^\infty \left(
\frac{\sqrt{p^2x(1-x) }}{2\pi
\sqrt{L_i^2n_i^2+L_j^2n_j^2}}\right)^{D/2-2} K_{D/2-2}\left( \frac
1{ 2\pi }\sqrt{p^2x(1-x)}\sqrt{L_i^2n_i^2+L_j^2n_j^2}\right)
+\cdots
\nonumber \\
& & + \, 2^{d-1}\sum_{n_1,\ldots ,n_d=1}^\infty \left(
\frac{\sqrt{p^2x(1-x) }}{2\pi \sqrt{L_1^2n_1^2+\cdots
+L_d^2n_d^2}}\right) ^{D/2-2} K_{D/2-2}\left(\frac{1}{2\pi}
\sqrt{p^2\,x(1-x)}\sqrt{L_1^2n_1^2+\cdots +L_d^2n_d^2}\right).
\nonumber \\ \label{Bd}
\end{eqnarray}
From Eqs.~(\ref{A(D)1}) and (\ref{def-b}), we see that the model
is finite in the domain $2<D<4$. It is not defined for even
dimensions such that $D\leq 2$ and $D\geq 6$. It is well defined
for all odd dimensions, in particular for the physically important
case in condensed matter, $D=3$.

If an infrared-stable fixed point exists for  the model with $d$
compactified dimensions, it is  determined  by a study of the
infrared behavior of the Callan-Symanzik $\beta $ function, {\em
i.e.}, in the neighborhood of $|p|=0$. Therefore, we investigate
the above equations for $|p|\approx 0$. In this case, a typical
term in Eq.~(\ref{Bd}) has the form,
$$
\sum_{n_1,\ldots ,n_p=1}^\infty \left(
\frac{\sqrt{p^2x(1-x)}}{2\pi \sqrt{ L_1^2n_1^2+\cdots
+L_q^2n_q^2}}\right) ^{D/2-s} K_{D/2-s}\left( \frac 1{2\pi }
\sqrt{p^2x(1-x)}\sqrt{L_1^2n_1^2+\cdots +L_q^2n_q^2}\right) ,
$$
with $s=2$ and $q=1,2,\ldots ,d$. In the $|p|\approx 0$ limit, we
can use the approximation given by Eq.~(\ref{approxim}) and the
expression above reduces to
$$
2^{D/2 - s - 1} \Gamma \left( \frac D2-s\right) E_{q}\left( \frac
D2-s;L_1,\ldots ,L_{\ell}\right), \label{GGG2}
$$
where $E_{q}\left( D/2-s;L_1,\ldots ,L_{\ell}\right)$ is one of
the multidimensional Epstein zeta functions in Eq.~(\ref{EfuncS}).
We see from Eq.~(\ref{Bd}) that in the $|p|\approx 0$ limit,  the
remaining $p^2$-dependence is only that of the first term of
Eq.~(\ref{formageral}), which is the same for all number of
compactified dimensions $d$.

For all $d\leq D$, within the domain of validity of $D$, we have,
by inserting Eq.~(\ref{formageral}) in Eq.~(\ref{lambda}), the
running coupling constant
\begin{equation}
\lambda \left( |p|\approx 0,D,\{L_i\}\right) \approx \frac \lambda
{ 1+\lambda \left[ A(D)|p|^{D-4}+B_d\left( D,\left\{ L_i\right\}
\right) \right] }.  \label{g3}
\end{equation}
Let us take $|p|$ as a running scale, and define the dimensionless
coupling
\begin{equation}
g=\lambda \left( p,D,\{L_i\}\right) |p|^{D-4}.  \label{g1}
\end{equation}
The Callan-Zymanzik $\beta $-function controls the rate of the
renormalization-group flow of the running coupling constant and a
 fixed point of this flow is given by a (nontrivial)
zero of the $\beta $ function. Taking $|p|\approx 0$, it is
obtained straightforwardly from Eq.~(\ref{g1}),
\begin{equation}
\beta (g)=|p|\frac{\partial g}{\partial |p|}\approx (D-4)\left[
g-A(D)g^2\right] ,  \label{beta}
\end{equation}
from which we get the infrared-stable fixed point,
\begin{equation}
g_{*}(D)=\frac 1{A(D)}.  \label{gstar}
\end{equation}
We find that the $L_i$-dependent $B_d$-part of the subdiagram $\Pi
$ does not play any role in this expression and, as remarked
above, $A(D)$ is the same for all number of compactified
dimensions, so is $g_{*}$ only dependent on the whole space
dimension. In other words, we get the result that the existence of
an infrared-stable fixed point does not depend on the number of
compactified dimensions. We find  in particular, for $2<D<4$, an
infrared-stable fixed point, in agreement with previous
renormalization-group calculations for materials in bulk form (all
$ L_i=\infty $). Taking $D=3$, we demonstrate directly that  the
transition for $d=1,\,2,\,3$ (films, wires and grains according to
our interpretation), is a second-order one. Moreover, the fixed
point is independent of the size of the compactified dimensions
or, in other words, the nature of the transition is insensitive to
the geometrical constraints.

\subsection{First-order transitions}
\label{first}

We consider the model described by the GL Hamiltonian density in a
Euclidian $D$-dimensional space,
\begin{equation}
\mathcal{H}=\frac 12\left| \partial _\mu \varphi \right| \left|
\partial
^\mu \varphi \right| +\frac {1}{2}m_0^2\left| \varphi \right|
^2-\frac { \lambda } {4}\left| \varphi \right| ^4+\frac {\eta
}{6}\left| \varphi \right| ^6, \label{lagrangiana1a}
\end{equation}
where $\lambda >0$ and $\eta >0$ are the physical quartic and
sextic self-coupling constants. Here the sign of the quartic term
is opposite to that of the second-order phase transition and the
field, $\varphi (x)$, is a complex field. The bare mass is given
by $ m_0^2=\alpha (T/T_0-1$), with $\alpha >0$ and $T_0$ being a
temperature parameter, which is smaller than the critical
temperature for a first-order phase transition. As before, we
consider the system in $D$ dimensions confined to a $d$
-dimensional subspace, a parallelepiped box with edges $L_1,\cdots
L_d$. To get the physical mass we restrict ourselves to the lowest
order terms in $\lambda $ and $\eta$.

%%%%%%%%%%%%%%
\begin{figure}[ht]
\begin{center}
\scalebox{0.45}{\includegraphics {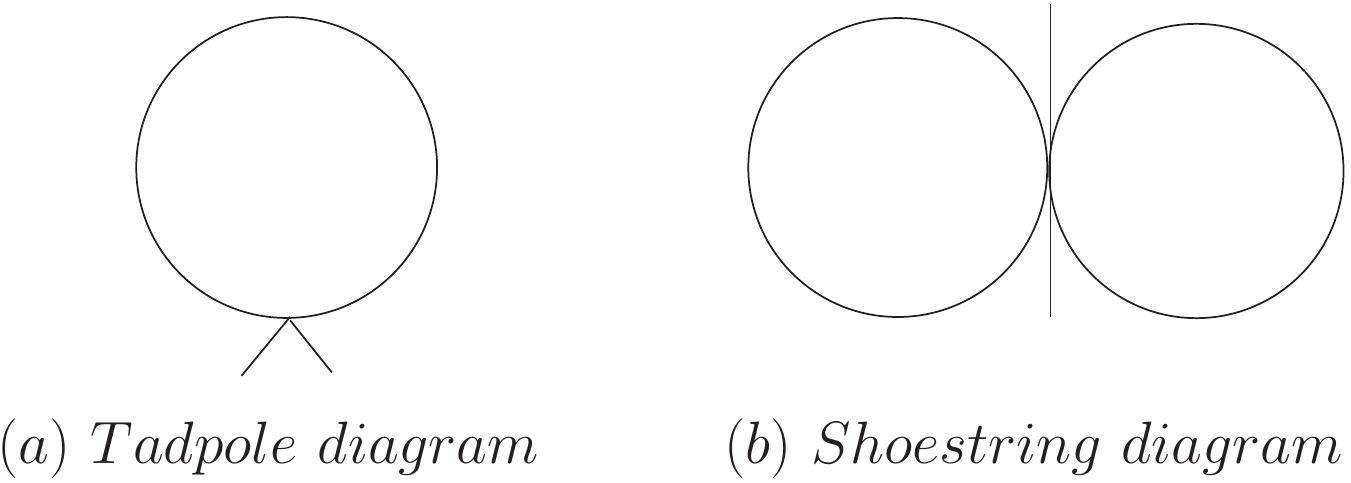} }
\caption{Contributions to the effective potential.}
\label{figXXIII1}
\end{center}
\end{figure}
%%%%%%%%%%%%%

\subsubsection{Finite-size corrections to the mass}

To 1-loop approximation, the procedure follows along the same
lines as in the previous subsection, starting from the expression
for the one-loop contribution to the effective potential in
unbounded space given by Eq.~(\ref {potefet0(1)}). The parameter
$s$ in Eq.~(\ref {potefet0(1)}) counts the number of vertices on
the loop. It follows that only the $s=1$ term contributes to the
mass. It corresponds to the tadpole diagram in
Fig.~\ref{figXXIII1}\rm{a}. It is also clear that all $|\varphi
_0|^6$-vertex and mixed $|\varphi _0|^4$- and $|\varphi _0|^6$
-vertex insertions on the 1-loop diagrams do not contribute when
one computes the second derivative of similar expressions with
respect to the classical field at zero value: only diagrams with
two external legs would survive. This is impossible for a
$|\varphi _0|^6$-vertex insertion at the 1-loop approximation. The
first contribution from the $|\varphi _0|^6$ coupling must come
from a higher-order term in the loop expansion. Two-loop diagrams
with two external legs and only $|\varphi _0|^4$ vertices are of
second order in its coupling constant, as well as all possible
diagrams containing both type of vertices; all these diagrams are
neglected. However, the 2-loop shoestring diagram, in
Fig.~\ref{figXXIII1}\rm{b}, with only one $|\varphi _0|^6$ vertex
and two external legs is a first-order (in $\eta $) contribution
to the effective potential and accordingly it is included in our
approximation. In short, we consider the physical mass as defined
to first-order in both coupling constants, by the contributions of
radiative corrections from only two diagrams: the tadpole and the
shoestring diagram.

The tadpole contribution to the effective potential is given by
\begin{equation}
U_1(\varphi _0,L_1,...,L_d) = \frac{\lambda |\varphi
_0|^{2}}{2\,\left( 2\pi \right) ^{D/2}}\left[
2^{-D/2-1}m^{D-2}\Gamma \left( \frac{2-D} 2\right) +
F_d(D,m,\{L_{i}\}) \right] , \label{potefet3a}
\end{equation}
where
\begin{eqnarray}
F_d(D,m,\{L_{i}\}) & = & \sum_{i=1}^{d} \sum_{n_{i}=1}^{\infty
}\left(
\frac{m}{L_{i}n_{i}}\right)^{D/2-1} K_{D/2-1}(mL_{i}n_{i}) \nonumber \\
& & +\, 2\sum_{i<j=1}^{d}\sum_{n_{i},n_{j}=1}^{\infty }\left(
\frac{m}{\sqrt{ L_{i}^{2}n_{i}^{2}+L_{j}^{2}n_{j}^{2}}}\right)
^{D/2-1} K_{D/2-1}\left(
m\sqrt{L_{i}^{2}n_{i}^{2}+L_{j}^{2}n_{j}^{2}} \right) \,+ \,
\cdots
\nonumber \\
& & +\,2^{d-1}\sum_{n_{1},\ldots ,n_{d}=1}^{\infty }\left(
\frac{m}{\sqrt{ L_{1}^{2}n_{1}^{2}+\cdots
L_{d}^{2}n_{d}^{2}}}\right) ^{D/2-1}K_{D/2-1}\left(
m\sqrt{L_{1}^{2}n_{1}^{2}+\cdots L_{d}^{2}n_{d}^{2} } \right) .
\label{FdDmLs}
\end{eqnarray}

On the other hand, the 2-loop shoestring diagram contribution to
the effective potential in unbounded space ($\{L_i=\infty \}$) is
given by~\cite{Zinn-Justin1996},
\begin{equation}
U_2(\phi _{0})=\frac{\eta |\varphi _0|^{2}}{16}\left[ \int
\frac{d^Dq}{\left( 2\pi \right) ^D}\frac 1{q^2+m^2}\right] ^2,
\label{UU22}
\end{equation}
which is proportional to the square of the tadpole contribution.
Then after compactification of $d$ dimensions with lengths $L_i $,
$ i=1,\ldots ,d$ and integration over the non-compactified
variables, $U_2$ becomes
\begin{equation}
U_2(\varphi _0,L_1,\ldots ,L_d) = \frac{\eta |\varphi
_0|^2}{4\,\left( 2\pi \right) ^D\,}\left[ 2^{-1-D/2}m^{D-2}\Gamma
\left( \frac{2-D}2\right) + \left[ F_d(D,m,\{L_{i}\}) \right]
\right]^2. \label{potefet4aU2}
\end{equation}

In both Eqs.~(\ref{potefet3a}) and (\ref{potefet4aU2}), there is a
term proportional to $\Gamma \left( \frac{2-D}2\right) $ which as
stated before, is divergent for even dimensions $D\geq 2$ and
should be subtracted in order to obtain finite physical
parameters. For odd $D$, the gamma function is finite, but we also
subtract it (corresponding to a finite renormalization) for the
sake of uniformity. After subtraction we get
\begin{equation}
U_1^{\mathrm{(Ren)}}(\varphi _0,L_1,\ldots ,L_d) = \frac{\lambda
|\varphi _0|^{2}}{2\,\left( 2\pi \right) ^{D/2}} \,
F_d(D,m,\{L_{i}\}) \label{u1ren}
\end{equation}
and
\begin{equation}
U_2^{\mathrm{(Ren)}}(\varphi _0,L_1,\ldots ,L_d) = \frac{\eta
|\varphi _0|^{2}}{4\,(2\pi )^D} \, \left[ F_d(D,m,\{L_{i}\})
\right]^2 . \label{U2reno}
\end{equation}

Then the physical mass, to first-order in both coupling constants,
is obtained using Eqs.~(\ref{u1ren}) and (\ref{U2reno}) and also
taking into account the contribution at the tree level; it
satisfies a generalized Dyson-Schwinger equation depending on the
lengths $L_i$ of the confining box,
\begin{equation}
m^{2}(T,\{L_{i}\}) = m_0^{2} - \frac{\lambda }{\left( 2\pi \right)
^{D/2}} \, F_d(D,m,\{L_{i}\}) + \frac{\eta }{2(2\pi )^{D}} \left[
F_d(D,m,\{L_{i}\}) \right]^2 , \label{massren1a}
\end{equation}
where $F_d(D,m,\{L_{i}\})$ is given by Eq.~(\ref{FdDmLs}).

\subsubsection{Phase transition: films, wires and grains}

A first-order transition occurs when all the three minima of the
potential
\begin{equation}
U(\varphi _{0})=\frac{1}{2}m^{2}(T,\{L_{i}\})|\varphi
_{0}|^{2}-\frac{\lambda }{4}|\varphi _{0}|^{4}+\frac{\eta
}{6}|\varphi _{0}|^{6},  \label{potencial1}
\end{equation}%
where $m(T,\{L_{i}\})$ is the physical mass defined above, are
simultaneously on the line $U(\varphi _{0})=0$. This gives the
condition
\begin{equation}
m^{2}(T,\{L_{i}\})=\frac{3\lambda ^{2}}{16\eta }.
\label{condicao1}
\end{equation}%
For $D=3$, the Bessel functions have the explicit form $K_{1/2}(z)
= \sqrt{\pi }e^{-z}/\sqrt{2z}$; then, remembering that
$m_{0}^{2}=\alpha (T/T_{0}-1)$, Eqs.~(\ref{massren1a}) and
(\ref{condicao1}) lead to the critical temperature for any
specific situation.

Having developed the general case of a $d$-dimensional
compactified subspace, it is now easy to obtain the specific
formulas for particular values of $d$. If we choose $d=1$, the
compactification of just one dimension, let us say, along the
$x_1$-axis, we are considering that the system is confined between
two planes, separated by a distance $L_1=L$. Physically, this
corresponds to a film of thickness $L$ and the transition occurs
at the $L$-dependent critical temperature
\begin{eqnarray}
T_c^{\text{film}}(L) & = &T_c\left\{ 1-\left( 1+\frac{3\lambda
^2}{16\eta\alpha } \right)^{-1}\left[ \frac \lambda {8\pi \alpha
L}\ln \left(1-e^{-\sqrt{\frac{ 3\lambda ^2}{16\eta }}L}\right)
+\frac \eta {64\pi ^2\alpha L^2}\left(\ln
(1-e^{-\sqrt{\frac{3\lambda ^2}{16\eta }}L})\right)^2\right]
\right\}, \label{tcfilm}
\end{eqnarray}
where
\begin{equation}
T_{c}=T_{0}\left( 1+\frac{3\lambda ^{2}}{16\eta \alpha } \right)
\label{tc}
\end{equation}
is the bulk ($\{L_{i}\rightarrow \infty \}$) critical temperature
for the first-order phase transition.

In the case $d=2$, the system is confined between two parallel
planes a distance $L_1$ apart from one another normal to the
$x_1$-axis and two other parallel planes, normal to the $x_2$-axis
separated by a distance $L_2$. That is, the material is bounded
within an infinite wire of rectangular cross section $L_1\times
L_2$. To simplify matters, we take $L_1=L_2=L$ and the critical
temperature is written in terms of $L$ as
\begin{equation}
T_c^{\text{wire}}(L) = T_c \left\{ 1-\left( 1+\frac{3\lambda
^2}{16\eta \alpha }\right)^{-1} \left[ \frac \lambda {4\pi \alpha
L}\, F_2(L) + \frac \eta {32\pi ^2\alpha L^2}\, [F_2(L)]^2 \right]
\right\} , \label{tcwire}
\end{equation}
where
\begin{equation}
F_2(L) = 2\ln \left(1-e^{-L\sqrt{\frac{3\lambda ^2}{16\eta
}}}\right)-2\sum_{n_1,n_2=1}^ \infty
\frac{e^{-L\sqrt{\frac{3\lambda ^2}{16\eta}}\sqrt{n_1^2+n_2^2}}}{
\sqrt{n_1^2+n_2^2}} .
\end{equation}

Finally, we may compactify all three dimensions, which leaves us
with a system in the form of a cubic ``grain'' of some material.
The dependence of the critical temperature on its linear
dimension, $L_1=L_2=L_3=L$, is given by
\begin{equation}
T_c^{\text{grain}}(L) = T_c \left\{ 1-\left( 1+\frac{3\lambda
^2}{16\eta \alpha }\right)^{-1} \left[ \frac \lambda {4\pi \alpha
L}\, F_3(L) + \frac \eta {32\pi ^2\alpha L^2}\, [F_3(L)]^2 \right]
\right\} , \label{tcgrain}
\end{equation}
where
\begin{equation}
F_3(L) = 3\ln \left( 1-e^{-L\sqrt{\frac{3\lambda ^2}{16\eta
}}}\right) -2\sum_{j<i=1}^3\sum_{n_i,n_j=1}^\infty \frac{
e^{-L\sqrt{\frac{3\lambda ^2}{16\eta
}}\sqrt{n_i^2+n_j^2}}}{\sqrt{n_i^2+n_j^2}}-4\sum_{n_1,\ldots
,n_3=1}^\infty \frac{e^{-L\sqrt{\frac{3\lambda ^2}{16\eta }}\sqrt{
n_1^2+n_2^2+n_3^2}}}{\sqrt{n_1^2+n_2^2+n_3^2}} .
\end{equation}

Let us observe that the general formalism employed in this section
up to this point, involves extensions to several dimensions of the
one-dimensional mode-sum
regularization~\cite{MalbouissonsaesNPB2002}, which require, in
particular, the definition of symmetrized multidimensional
Epstein-Hurwitz functions with no analog in the one-dimensional
case. It is this kind of mathematical framework that allows us to
obtain general formulas, which may be particularized to films,
wires and grains, thereby implying the peculiar forms of the
critical temperature as a function of the linear dimension $L$,
for these physically interesting cases.

It should be observed the very different form of
Eqs.~(\ref{tcfilm}), (\ref {tcwire}) and (\ref{tcgrain}) when
compared with the corresponding ones for second-order transitions.
In this last case the $L$-dependent transition temperature for
films, wires and grains have the same functional dependence on the
linear dimension, given by Eq.~(\ref{tcd1}). We find that in all
the cases (a film, a wire or a grain), there is a sharp contrast
between the simple inverse linear behavior of $T_c(L)$ for
second-order transitions (without considering corrections to the
coupling constant) and the rather involved dependence on $L$ of
the critical temperature for first-order transitions.

These two types of behavior demand to clarify the subject further;
this can be done by comparing the theoretical curves with
experimental data for superconducting materials. This has been
considered for systems in the form of a film or of a wire
in~\cite{LinharesAdolfoExp2006}. The interested reader will find
there an explicitly comparison between the forms of the $T_c(L)$
curves for both first- and second-order transitions. Also
in~\cite{LinharesAdolfoExp2006}, the degree of agreement between
theoretical expressions for the first-order critical temperature
and some experimental results for a variety of transition-metal
materials~\cite{Raffy1983,Minhaj1994,Pogrebnyakov2003} is
exhibited.

\subsubsection{Effects of finite chemical potential}

In what follows we study effects of the chemical potential on the
size-dependent transition, particularly on the critical
temperature. However, distinctly from the topics studied in the
previous subsection, we take the squared mass $m_0^2$ as a fixed
parameter of the model, not depending on temperature as in the GL
model; now, temperature is introduced by means of methods of
finite temperature field theory. To account for finite chemical
potential effects, we consider the system at a temperature $\beta
^{-1}$ and we compactify one of the spatial coordinates (say $x$)
with a compactification length $L$; in this case, the Matsubara
prescription, Eq.~(\ref{Matsubara1}), has to be changed such that
\begin{equation}
k_0 \equiv k_{\tau} \rightarrow \frac{2 \pi n_{\tau}}{\beta} - i
\mu . \label{Matsubaramu}
\end{equation}

Defining $ a_\tau =1/(m\beta)^2$, $a_x=1/(m L)^2$, with
$c^2=1/4\pi ^2$, the one-loop contribution to the effective
potential at finite chemical potential, can be written as
\begin{eqnarray}
U_1(\varphi _0,a_{\tau}, a_x) =\sum_{s=1}^\infty
\frac{(-1)^{s+1}}{2s}\left[ \lambda|\varphi _0|^2\right]^s
\frac{m^{D-2s}}{\sqrt{a_{\tau} a_x}} \frac 1{\left( 4\pi ^2\right)
^s} \sum_{n_\tau ,n_x=-\infty }^\infty \int
\,\frac{d^{D-2}\,k}{\left[ \mathbf{k}^2+a_\tau \left( n_\tau
-\frac{i\beta }{2\pi }\mu \right) ^2+a_xn_x^2+c^2\right] ^s}.
\end{eqnarray}
The integral in the above equation is calculated using the
dimensional regularization formula in Eq.~(\ref{dimreg}), so that
the tadpole contribution to the effective potential becomes,
\begin{eqnarray}
U_1(\varphi _0,a_{\tau}, a_x) =\sum_{s=1}^\infty
\frac{(-1)^{s+1}}{2s}\left[ \lambda|\varphi _0|^2\right] ^s
\frac{m^{D-2s}}{\sqrt{a_{\tau} a_x}} \frac{\pi ^{(D-2)/2}}{\left(
4\pi ^2\right) ^s} \frac{\Gamma \left( s-\frac{D-2}2\right)
}{\Gamma \left( s\right) } \sum_{n_\tau ,n_x=-\infty }^\infty
\left[ a_\tau \left( n_\tau -\frac{i\beta }{2\pi }\mu \right)
^2+a_xn_x^2+c^2\right] ^{(D-2)/2-s}.  \label{potefet5}
\end{eqnarray}
The double sum in Eq.~(\ref{potefet5}) may be recognized as the
two-variable {\it inhomogeneous} Epstein-Hurwitz zeta function
$Z_{2}^{c^2}(\nu;a_{\tau},a_x;b_{\tau},b_x)$, where $b_\tau
=i\beta \mu /2\pi $, $b_x=0$ and $\nu =s-(D-2)/2$; this function
possesses the analytical continuation given by Eq.~(\ref{Z2mu}).

After suppression of the singular term, using the symmetry
property of Bessel functions, $K_\nu (z)=K_{-\nu }(z)$, and
tanking the value of $ \nu =s-(D-2)/2$ with $s=1$, the tadpole
contribution to the effective potential is obtained as
\begin{equation}
U_1(\varphi _0;\beta ,L,\mu ) = \frac{\lambda |\varphi
_0|^2}{2\,\left( 2\pi \right)^{D/2}} \, F(D,m,\beta,L) ,
\end{equation}
where
\begin{eqnarray}
F(D,m,\beta,L) & = & \sum_{n=1}^\infty \cosh \left( \beta \mu
n\right) \left(\frac m{n\beta }\right)^{D/2-1} K_{D/2-1}\left(
\beta mn\right) + \sum_{n=1}^\infty \left( \frac
m{nL}\right)^{D/2-1} K_{D/2-1}\left( Lmn\right)
\nonumber \\
& & + \, 2\sum_{n_1,n_2=1}^\infty \cosh \left( \beta \mu
n_1\right) \left( \frac m{ \sqrt{\beta ^2n_1^2+L^2n_2^2}}\right)
^{D/2-1} K_{D/2-1}\left( m\sqrt{\beta ^2n_1^2+L^2n_2^2}\right) .
\label{FDmBL}
\end{eqnarray}

We now turn to the 2-loop shoestring diagram contribution to the
effective potential, using again the Matsubara-modified Feynman
rules prescription, Eq.~(\ref{Matsubaramu}), for the compactified
dimensions. In unbounded space ($L\rightarrow\infty $), at zero
temperature and chemical potential, it is given by
Eq.~(\ref{UU22}). Then, at finite temperature and chemical
potential, with one compactified spatial dimension, following
steps analogous to those used to get $U_1$, we obtain
\begin{equation}
U_2(\varphi _0;\beta ,L,\mu) = \frac{\eta |\varphi _0|^{2}
}{4\,(2\pi )^D} \left[ F(D,m,\beta,L) \right] ^2 , \label{U2ren}
\end{equation}
where $F(D,m,\beta,L)$ is given by Eq.~(\ref{FDmBL}).

The mass is obtained from the normalization condition,
Eq.~(\ref{renorm1b}),
\begin{eqnarray}
m^2(\beta ,L,\mu) & = & \left. \frac{\partial ^2}{\partial \varphi
_0^2} U(\varphi _0;\beta ,L,\mu)\right|_{\varphi_{0} =0} =
\left.\frac{\partial ^2}{\partial \varphi _0^2} \left(
U_{\rm{tree}}+U_1(\varphi _0;\beta ,L,\mu ) +
U_2(\varphi _0;\beta ,L,\mu)\right)\right|_{\varphi_{0} =0}. \nonumber \\
\label{cond1}
\end{eqnarray}
We now consider the three-dimensional Euclidean space ($D=3$), in
which case the relevant Bessel functions are $K_{\pm 1/2}(z) =
\sqrt{\pi }e^{-z}/\sqrt{2z}$. Then, after summing the geometric
series, we get
\begin{equation}
m^2(\beta ,L,\mu) = m_0^2 + \frac {\lambda}{4\pi} \, F(\beta,L) +
\frac{ \eta}{16\pi^2} [F(\beta,L)]^2 , \label{m3}
\end{equation}
where
\begin{equation}
F(\beta,L) = - \frac{1}{2 \beta} \left[ \ln \left( 1-e^{-(m-\mu
)\beta }\right) + \ln \left( 1-e^{-(m+\mu )\beta}\right) \right]
 -\frac 1L\ln \left( 1-e^{-mL}\right) +2
\sum_{n_1,n_2=1}^\infty \cosh \left( \beta \mu n_1\right)
\frac{e^{-m\sqrt{ \beta ^2n_1^2+L^2n_2^2}}}{\sqrt{\beta
^2n_1^2+L^2n_2^2}} \, .
\end{equation}

As in the previous subsection, a first-order transition occurs
when all the three minima of the potential
\begin{equation}
U(\varphi _0;\beta ,L,\mu)=\frac 12m^2(\beta ,L,\mu)|\varphi
_0|^2-\frac \lambda 4|\varphi _0|^4+\frac \eta 6|\varphi _0|^6,
\label{potencial}
\end{equation}
are simultaneously on the line $U(\varphi _0;\beta ,L,\mu)=0$,
which gives the condition ,
\begin{equation}
m^2(\beta ,L,\mu)=\frac{3\lambda ^2}{16\eta }.  \label{condicao}
\end{equation}
It can be seen that for $0<\mu<m$ the double sum in Eq.~(\ref{m3})
above for the mass converges. Then, using the condition
(\ref{condicao}), $m=\lambda \sqrt{3}/4\sqrt{\eta}$ in
Eq.~(\ref{m3}) we have a well-defined expression. It gives the
critical temperature as a function of the compactification length
$L$, of the chemical potential $\mu$, and of the fixed mass
parameter $m_0$.

%%%%%%%%%%%%%%%%%%%%%%%
\begin{figure}[ht]
\begin{center}
\vspace{0.5cm} \scalebox{1.1}{{\includegraphics{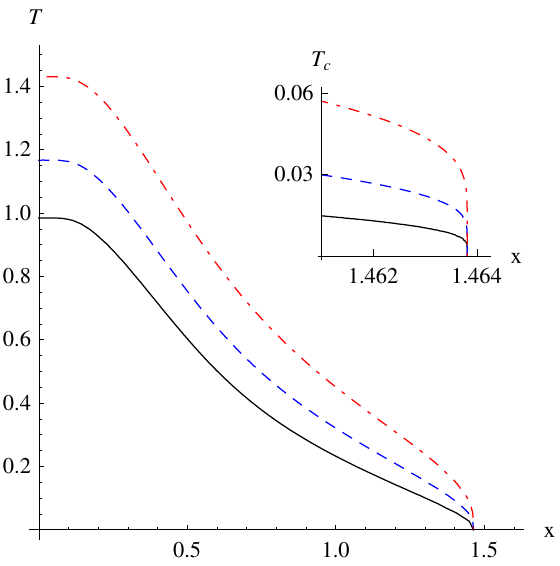}}}
\end{center}
\caption{Critical temperature as a function of the inverse size ,
$x=1/L$, of the system in the horizontal axis. We take $\lambda
=\eta = m_0=1$ in dimensionless units (mass scale $m_0$). From the
lower to the upper curves, we have respectively, $\mu =0.40$ (full
line), $\mu=0.35$ and $\mu=0.25$ (dot-dashed line). The
symmetry-breaking regions are below each curve.} \label{figPhi6}
\end{figure}
%%%%%%%%%%%%%%%%%%%%%%

In order to perform a qualitative analysis of the phase structure
of the model, we take for its parameters  the values
$\lambda=\eta=m_0=1$, in dimensionless units scaled by $m_0$. For
these values, the condition in Eq.~(\ref{condicao}) implies that
the dimensionless chemical potential is to be restricted to the
range $0<\mu\lesssim 0.43$. In Fig.~\ref{figPhi6} we plot the
critical temperature from Eq,~(\ref{m3}) as  a function of the
inverse size of the system, $x=1/L$, for some values of the
chemical potential. We find that for all values of $\mu$, the
critical temperature decreases as the values of $x=1/L$ increase,
in such a way that the symmetry breaking region under the curves
is gradually diminished as the size of the system becomes smaller
and smaller. Moreover, we find that there is a minimal allowed
size for the system, for which this region disappears, a situation
which is similar to the one encountered in the previous section.
We interpret Fig.~\ref{figPhi6} as indicating that there is a
minimal size of the system sustaining the transition; it is of the
order (in the arbitrary units adopted here) of $L\sim 0.68$. This
minimal size is {independent} of the chemical potential, although
the detailed behavior of the critical temperature does depend on
$\mu$. It is worth to remark that, we can reach this conclusion
from an analysis of the critical equation, in an analogous way as
it has been done in the previous section. Also, we find that for a
fixed size of the system, the symmetry breaking region becomes
smaller as the chemical potential increases. As an overall
conclusion in this case, the results suggest that, also for
first-order transitions, as the size of the system is diminished,
finite-size effects start to appear for a given value of the
compactification length, with the critical temperature decreasing
as $L$  is further decreased, making the symmetry breaking region
shrink. For the minimal size  in Fig~\ref{figPhi6}, the transition
ceases to take place.

\section{Phase transitions in four-fermions models on $\Gamma^d_D$}

In this section, the quantum field theory on toroidal topology is
considered for four-fermion interacting systems. The goal is to
analyze the phase structure of such systems, focussing on
applications in particle physics, although general results are
useful in low energy physics.

In particle physics, the strong interaction, that have quarks and
gluons as the basic constituents, presents a rather complicated
structure that is hard to use at normal density and temperature of
hadrons. It is usual then to depend on effective theories, that
are attempts to assume that the gluon fields and the color degrees
of freedom are integrated out, similarly to the Fermi treatment of
the weak interaction. This provides the simplest effective models,
which may be considered as four-fermion contact interaction among
quarks. There are, at least, two schemes to achieve this: the
Nambu-Jona-Lasino Model (NJL)~\cite{Nambu1961,Nambu1961a} and the
Gross-Neveu (GN) model~\cite{GrossNeveu1974}, where the latter is
a sector of the former.

As effective theories for quantum chromodynamics (QCD), NJL and GN
models have been enlightening approaches describing properties of
hadronic matter. Those include asymptotical freedom in the high
energy domain and investigations of the continuous and discrete
chiral symmetry, which are associated with the
confinement/deconfinement phase transition, both in the GN
model~\cite{Barducci1995,Christiansen2000,Schon2000,Hands2001,Hands2002,
Thies2006,Schnetz2006,Jurivcic2009,Zhou1998,Zhou1999,Brzoska2002,
Kneur2007,Boehmer2008,Hofling2002,MalbouissonJuaes2004,
Khanna2005,Kohyama2008,Feinberg2002,Thies2004,Fushchich1989} and
the NJL
model~\cite{Kim1994,He1996,Vshivtsev1998,Kogut2001,Kiriyama2003,
Beneventano2004,Gamayun2005,Kiriyama2006,Ebert2008,
AbreuMarcelo2006,AbreuMalsaes2009,Abreumals2010,Abreu2011,Abreu2011a}.
Rigorous QCD calculations, both at zero and finite temperature,
have been worked out in order to describe such
properties~\cite{Gross1981,Kalashnikov1984}, but mainly treating
the asymptotically free domain at high energies or high
temperatures, where perturbation theory is applicable. Actually,
the intricate mathematical structure of QCD, practically prevents
us from finding analytical results taking into account both
confinement/deconfinement and asymptotic freedom. Four-fermion
interaction models, as effective theory for QCD, are then
practical tools mainly for the search of analytical results.

The four-fermion interaction model is also akin to the interaction
among electrons in condensed matter physics. In particular, the GN
model, is used to describe also properties of graphene, a
honeycomb lattice with two Bravais triangular sublattices. The
Hamiltonian of this structure is mapped in the (2+1) GN
Hamiltonian, such that the $SU(4)$ chiral symmetry, arising from
the arrangement of spins, can be broken into $SU(2)\times SU(2)$,
where parity and time reversal invariance are preserved. This
chiral symmetry breaking is the counterpart of a quantum phase
transition from the semi-metallic phase to a gapped Mott
insulator~\cite{Semenoff1984,
Gusynin2007,Khveshchenko2001,Semenoff1992,Jurivcic2009,
Herbut2006,Herbut2009,Jackiw2007}. Such a breaking of symmetry,
that can be restored by raising the temperature, leads to a gap in
the energy spectrum, that is of interest for electronic
devices~\cite{Novoselov2007}. Similar four-fermion interaction is
the fundamental aspect of the BCS theory for superconductivity. An
important point to be emphasized here is that the susceptibility
arising from the linear response theory has a divergence at a
finite temperature, indicating the existence of a second-order
phase transition between a disordered and a condensed
phase~\cite{Doniach1974}. As already mentioned in Section 6,
spontaneous symmetry breaking is a common phenomenon in several
domains of physics, and usually leads to underlying phase
transitions~\cite{Weinberg1986}. Then  considering applications in
high energy physics, it may be anticipated in a study of
four-fermion model that a phase transition appears, describing
aspects of particle physics.

This is the case of the one-component massive tridimensional GN
model at finite temperature~\cite{MalbsKhaaesEPL2010}. In a
$D$-dimensional Euclidian manifold, $ {\mathbb{R}}^{D}$, we
consider the Hamiltonian for the massive GN model,
\begin{equation}
H = \int d^{D}x\left\{ \psi ^{\dagger }(x)(i\gamma ^{j}\partial
_{j}-m_{0})\psi (x)+\frac{\lambda _{0}}{2}\left[ \psi ^{\dagger
}(x)\psi (x) \right] ^{2}\right\}   \label{GNPR}
\end{equation}
where $m_{0}$ and $\lambda _{0}$ are respectively the physical
zero-temperature mass and coupling constant, $x\in
{\mathbb{R}}^{D}$ and the $\gamma $-matrices are elements of the
Clifford algebra (natural units $ \hbar =c=k_{B}=1$ are used).
This Hamiltonian is obtained by using conventions for Euclidian
field theories in~\cite{Ramond1981}.

From Eq.~(\ref{GNPR}), introducing the thermally corrected mass,
\begin{equation}
m(T)=m_{0}+\Sigma (T),  \label{cri1PR}
\end{equation}
a generalization of the Ginzburg-Landau free energy density is
\begin{equation}
{\mathcal{F}}=a+b(T)\phi ^{2}(x)+c\,\phi ^{4}(x),
\label{july092PR}
\end{equation}
where $b(T)=-m(T)$ and $c=\lambda _{0}/2$. The minus sign for the
mass in Eq.~(\ref{july092PR}) implies that, in the disordered
phase we have $m(T)<0$ and for the ordered phase $m(T)>0$,
consistently. The second order phase transition occurs at the
temperature where $m(T)$ changes sign from negative to positive,
characterizing a spontaneous symmetry breaking. In this formalism,
the quantity $\phi (x)=\sqrt{\left\langle \psi ^{\dagger }(x)\psi
(x)\right\rangle }$, with $\left\langle \cdot \right\rangle $
meaning a thermal average, plays the role of the order parameter
for the transition. The phase transition is obtained for a
critical temperature, $T_{c}$, that is solution of the equation,
$m(T)=0$. This gives rise to a function $ T_c=T_c(m_{0};\lambda
)$~\cite{MalbsKhaaesEPL2010}.

A similar result is obtained from a non-perturbative analysis of
the four-point function, by summing the chains of loop diagrams.
This is the case of  the aforementioned BCS
approach~\cite{Doniach1974}. The singularity of the leading
contribution to the four-point function is related to the
susceptibility, which diverges at a critical temperature. Then in
general a singularity of the four-point function, of a
four-fermion interacting system, indicates the existence of a
phase transition.

It is worth mentioning that, although the GN model is not
renormalizable for dimensions greater than $D=2$, the Euclidian
model has been shown to exist and has been constructed for $D=3$
in the large-$N$ limit~\cite{DeCalanJarrao1991}. In any case,
within the spirit of effective theories, perturbative
renormalizability is not an absolute requirement for an effective
theory to be a physically meaningful
model~\cite{Parisi1975,Arefva1980,Gawedzki1985a,Rosenstein1989,Weinberg1997}.
This possibility has then been explored largely in literature, for
both, GN and NJL models.

In the following, we apply to the GN and NJL models, the formalism
of field theory on toroidal topology to investigate finite-size,
density and magnetic field effects~\cite{Khanna2012,Abreu2011a}.
Then the behavior of the four-point function of the massive
$N$-component GN model on $\Gamma^d_D$ is discussed. Finally, we
address the NJL model on this topology.

\subsection{Phase transition in the massive Gross-Neveu model}

\subsubsection{Phase transition in one-component massive GN model}

The one-component massive GN model in a $D$-dimensional Euclidian
manifold, $ {\mathbb{R}}^{D}$ is described by the Hamiltonian
given in Eq.~(\ref{GNPR}). Thermal and boundary corrections to
$m_{0}$ and to $\lambda _{0}$ are defined by the temperature and
boundary-dependent mass and coupling constant, $m(T,L)$ and
$g(T,L)$ respectively, that assume the form
\begin{equation}
m(T,L)=m_{0}+\Sigma (T,L)  \label{cri1}
\end{equation}
and
\begin{equation}
g(T,L)=\lambda _{0}[1+\lambda _{0}\Pi (T,L)]\;.
\label{4-point1L0}
\end{equation}
Then from Eq.~(\ref{july092PR}), the generalized Ginzburg-Landau
free energy density is
\begin{equation}
{\mathcal{F}}=a-m(T,L)\phi ^{2}(x)+g(T,L)\,\phi ^{4}(x),
\label{july092}
\end{equation}
The minus sign for the mass in Eq.~(\ref{july092}), as before,
implies that, in the disordered phase we have $m(T,L)<0$ and for
the ordered phase $ m(T,L)>0$, consistently. A second order phase
transition occurs in the region where $m(T,L)=0$, characterizing a
spontaneous symmetry breaking.

At criticality, the leading contribution to the four-point
function with zero-external momenta is given by the sum of all
chains of one-loop diagrams, i.e.
\begin{equation}
\Gamma ^{(4)}(\lambda _{0},T,L) = \frac{\lambda _{0}}{[1-\lambda
_{0}\Pi (T,L)] }. \label{g3L1}
\end{equation}
The first two terms of the expansion in powers of $\lambda _{0}$
of such a function are given in Eq.~(\ref{4-point1L0}). The
existence of a singularity of the four-point function in
Eq.~(\ref{g3L1}) indicates a phase transition; in other words,
values of $T$ and $L$ solving the equation,
\begin{equation}
1-\lambda _{0}\Pi (T,L)=0\   \label{jun21}
\end{equation}%
assure a phase transition. The nature of the transition is
obtained by a study of the free energy, Eq.~(\ref{july092}). From
an analysis of the infrared fixed point structure, the transition
is of second order and that the critical temperature is determined
by the condition $m(T,L)=0$. For a given value of $L$, this leads
to a $L$-dependent critical temperature $ T_{c}(L)$.

In order to evaluate the thermal and boundary dependent
self-energy, $\Sigma (D;T,L)$, the generalized Matsubara formalism
for fermion is used. This corresponds to take $c_{i}=1/2$ in
Eq.~(\ref{Matsubara1}), the modified Feynman rules. The Cartesian
coordinates are specified by $\mathbf{r}
=(x^{0},x^{1},\mathbf{z})$, where $\mathbf{z}$ is a
$(D-2)$-dimensional vector. The conjugate momentum of $\mathbf{r}$
is denoted by $\mathbf{k} =(k_{0},k_{1},\mathbf{q})$.

At the one-loop level, the self-energy is given by $\Sigma
(D;\beta ,L;s)|_{s=1}$, where
\begin{equation}
\Sigma (D;\beta ,L;s) = \lambda _{0}\frac{m_{0}}{\beta
L}\sum_{n_{1},n_{2}=- \infty }^{\infty }\int \frac{d^{D-2}q}{(2\pi
)^{D-2}}\frac{1}{(\mathbf{q} ^{2}+\omega _{n_{1}}^{2}+\omega
_{n_{2}}^{2}+m_{0}^{2})^{s}},
\end{equation}
where $\omega _{n_{1}}$ and $\omega _{n_{2}}$ are Matsubara
frequencies given by
\begin{equation*}
\omega _{n_{1}}=\frac{2\pi (n_{1}+\frac{1}{2})}{\beta },\ \ \ \ \
\omega _{n_{2}}=\frac{2\pi (n_{2}+\frac{1}{2})}{L},
\end{equation*}
and $\beta =T^{-1}$. For using dimensional regularization
procedure, the following dimensionless quantities are introduced:
$a_{1}=(m_{0}\beta )^{-2}$ , $a_{2}=(m_{0}L)^{-2}$,
$q_{j}=k_{j}/2\pi m_{0}$, for $j=3,...,D$, $\omega
_{n_{i}}^{\prime }=\omega _{n_{i}}/2\pi m_{0}$, for $i=1,2$, and
$c=1/2\pi $ . Then we get,
\begin{equation}
\Sigma (D;a_{1},a_{2};s) =\lambda _{0}m_{0}\sqrt{a_{1}a_{2}}
\sum_{n_{1},n_{2}=-\infty }^{\infty } \int \frac{d^{D-2}q}{(2\pi
)^{D-2}}\frac{1}{(\mathbf{q}^{2}+\omega _{n_{1}}^{\prime 2}+\omega
_{n_{2}}^{\prime 2}+c^{2})^{s}}.\nonumber
\end{equation}

After dimensional regularization, we obtain
\begin{equation}
\Sigma (D;a_{1},a_{2};s) =\frac{m_{0}^{1-2\nu }\lambda _{0}\Gamma
(\nu )}{ (4\pi )^{(D-2)/2}\Gamma (s)}\,\sqrt{a_{1}a_{2}}
\sum_{n_{1},n_{2} = -\infty }^{\infty }(\omega _{n_{1}}^{\prime
2}+\omega _{n_{2}}^{\prime 2}+c^{2})^{-\nu }, \label{cri3}
\end{equation}
where $\nu =s-(D-2)/2$. This leads to
\begin{equation}
\Sigma (D;a_{1},a_{2};s)= \frac{m_{0}^{1-2\nu }\lambda _{0}\Gamma
(\nu )4^{\nu }}{(4\pi )^{(D-2)/2}\Gamma
(s)}\,\sqrt{a_{1}a_{2}}\left[ Z_{2}^{4c^{2}}(\nu
,a_{1},a_{2})-Z_{1}^{c^{2}}(\nu ,a_{1})
%\right. \hspace{ 0.5cm} &&\left.
-\,Z_{1}^{c^{2}}(\nu ,a_{2}) + Z_{2}^{c^{2}}(\nu ,4a_{1},4a_{2})
\right] , \label{cri4}
\end{equation}
where $Z_{d}^{b^{2}}(\nu ,a_{1},a_{2})$ is the homogeneous
Epstein-Hurwitz multi-variable \emph{zeta}-function, given in
Eq.~(\ref{zetab}), with analytical extension given in
Eq.~(\ref{zetabbb}). For $s=1$ and $D=3$, such that $\nu
=s-(D-2)/2=1/2$,   performing the subtraction of divergent terms,
following the procedure of Section 6, and using $K_{\pm 1/2}(z)=
\sqrt{ \pi }e^{-z}/\sqrt{2z}$ and $\sum_{n=1}^{\infty } e^{-\xi
n}/n=-\ln (1-e^{-\xi })$, the self-energy reads
\begin{equation}
\Sigma (T,L)=m_{0}\,\frac{\lambda _{0}m_{0}}{2\pi
}\,\mathcal{F}\left( \frac{ m_{0}}{T},m_{0}L\right) ,
\end{equation}
where
\begin{equation}
\mathcal{F}(x,y)=-\frac{\ln (1+e^{-x})}{x}-\frac{\ln
(1+e^{-y})}{y} +\,2F(x,y)-4F(x,2y)-4F(2x,y)+8F(2x,2y),
\label{SigmaR11}
\end{equation}
with
\begin{equation}
F(x,y)=\sum_{n,l=1}^{\infty }\frac{\exp
(-\sqrt{x^{2}n^{2}+y^{2}l^{2}})}{ \sqrt{x^{2}n^{2}+y^{2}l^{2}}}.
\end{equation}

Up to the one-loop level, the four-point function with null
external momenta, which defines the $ (\beta ,L)$-dependent
coupling constant, is
\begin{equation}
\Gamma _{D}^{(4)}(\beta ,L;\lambda _{0})\simeq
\lambda_{0}[1+\lambda _{0}\Pi (D,\beta ,L)]\;,  \label{4-point1L}
\end{equation}%
where $\Pi (D,\beta ,L)$ is the $(\beta ,L)$-dependent one-loop
polarization diagram given by
\begin{equation}
\Pi (D,\beta ,L)=\frac{1}{\beta L}\sum_{n_{1},n_{2}=-\infty
}^{\infty }\int \frac{d^{D-2}k}{(2\pi
)^{D-2}}\frac{m_{0}^{2}-(\mathbf{k}^{2}+\omega _{n_{1}}^{2}+\omega
_{n_{2}}^{2})}{(\mathbf{k}^{2}+\omega _{n_{1}}^{2}+\omega
_{n_{2}}^{2}+m_{0}^{2})^{2}}.  \label{sigma0}
\end{equation}
Using the dimensionless quantities and subtracting the polar
terms, as in the preceding section, the finite polarization reads
\begin{equation}
\Pi (T,L)=\frac{m_{0}}{2\pi }\,\mathcal{G}\left( \frac{m_{0}}{T}
,m_{0}L\right) ,
\end{equation}
where the function $\mathcal{G}(x,y)$ is defined by
\begin{eqnarray}
\mathcal{G}(x,y) &=&\frac{\ln
(1+e^{-x})}{x}-\frac{1}{1+e^{x}}+\frac{\ln
(1+e^{-y})}{y}-\,\frac{1}{1+e^{y}}  \notag \\
&&+2G(x,y)-4G(2x,y)-\,4G(x,2y)+8G(2x,2y),\hspace{1cm}
\label{Sigma32N}
\end{eqnarray}
with
\begin{equation}
G(x,y)=\sum_{n,l=1}^{\infty }\exp \left(
-\sqrt{x^{2}n^{2}+y^{2}l^{2}} \right) -F(x,y).  \label{G2}
\end{equation}
This provides us with the finite thermal and boundary-dependent
coupling constant,
\begin{equation}
g(T,L;\lambda _{0})\equiv \Gamma _{3}^{(4)}(T,L,\lambda
_{0})\simeq \lambda _{0}[1+\lambda _{0}\Pi (T,L))]\;.
\label{4-point2L}
\end{equation}

The phase transition occurring in the GN model with a compactified
spatial dimension is now discussed.  Replacing in
Eq.~(\ref{SigmaR11}) $\lambda _{0}$ by $g(T,L;\lambda _{0})$ given
in Eq.~(\ref{4-point2L}), we obtain a new self-energy,
${\mathcal{S}} (T,L;\lambda _{0})$, which incorporates the thermal
and boundary corrections to the coupling constant. Then the
$(T,L)$-dependent mass is
\begin{eqnarray}
m(T,L) &=&m_{0}+{\mathcal{S}}(T,L;\lambda _{0})  \notag \\
&=&m_{0}\left\{ 1+\frac{\lambda _{0}m_{0}}{2\pi
}\,\mathcal{F}\left( \frac{ m_{0}}{T},m_{0}L\right) \left[
1+\frac{\lambda _{0}m_{0}}{2\pi }\,\mathcal{G} \left(
\frac{m_{0}}{T},m_{0}L\right) \right] \right\}   \label{cri6}
\end{eqnarray}%
The  phase transition condition , $m(T,L)=0,$ provides a critical
surface defined by the critical temperature, the size of the
system, $L$, and the the zero-temperature coupling constant in the
the absence of boundaries, $ \lambda _{0}$. This surface defines
$L$-dependent values of the critical temperature, $T_{c}(L;\lambda
_{0})$.

\begin{figure}[htbp]
\begin{center}
  \begin{minipage}[b]{0.4\linewidth}
    \centering
    \includegraphics[width=\linewidth]{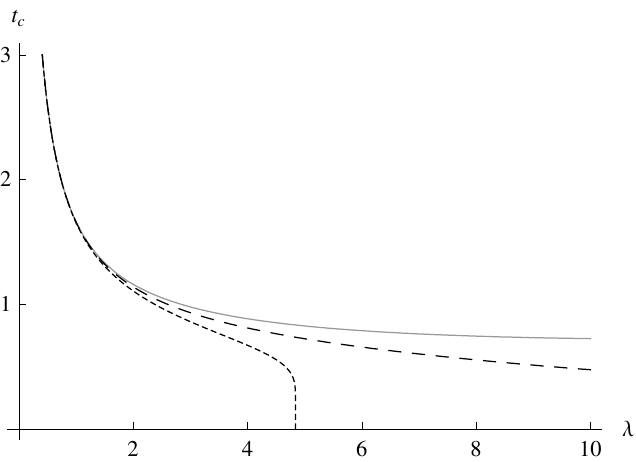}
    \caption{Critical temperature in units of $m_0$, $t_c=T_c/m_0$, as a
    function ofthe free-space coupling constant (in units of $m_0^{-1}$), $
\protect\lambda = m_0\protect\lambda_0 /2\protect\pi$, for some
values of the compactification length (in units of $m_0^{-1}$), $l
= m_0 L$: $\infty$, 1.4 and 1.2 (full, dashed and dotted lines,
respectively).} \label{PRfigura1}
  \end{minipage}
  \hspace{0.5cm}
  \begin{minipage}[b]{0.4\linewidth}
    \centering
    \includegraphics[width=\linewidth]{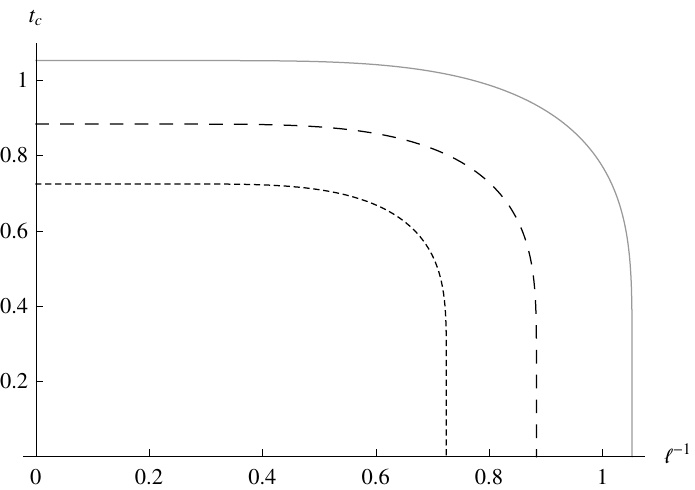}
    \caption{Critical temperature in units of $m_{0}$, $t_{c}=T_{c}/m_{0}$, as a
function of the inverse of the compactification length (in units
of $m_{0}$ ), $l^{-1}=1/m_{0}L$, for some values of
$\protect\lambda =m_{0}\protect \lambda _{0}/2\protect\pi $: 2.5,
4.0 and 10.0 (full, dashed and dotted lines, respectively).}
\label{PRfigura2}
  \end{minipage}
  \end{center}
\end{figure}

The dependence of $T_{c}$ on $\lambda _{0}$ is illustrated in
Fig.~\ref{PRfigura1}, for some values of $L$. In
Fig.~\ref{PRfigura2}, we present the behavior of the critical
temperature as the length of the system diminishes, for some
values of $\lambda _{0}$. From these plots, we find that there
exists a minimal length below which the transition is suppressed.
Moreover, we find that as the size of the system is diminished,
the transition disappears~\cite{MalbsKhaaesEPL2010}. The
suppression of the transition below a minimal size is illustrated
in Fig.~\ref{PRfigura2} for several values of the fixed coupling
constant.

%%%%%%%%%%%%%%%%
%%%%%%%%%%%%%%%

These minimal sizes, $L_{\mathrm{min}} $, are characterized by the
vanishing of the transition temperature and can be estimated. Take
for instance the value $l_{\mathrm{min}}^{-1}\approx 0.826 $
corresponding to $\lambda =5.0$ (the dashed curve in the figure).
This gives $L_{\mathrm{min}}\approx 1.21\,m_{0}^{-1}$. Now, let us
take for $m_{0} $, the mass of the Gross-Neveu fermion, to be the
effective quark mass of the proton~\cite{Maltman1984},
$m_{0}\approx 330\,\mathrm{MeV}$. We get, using the conversion
$1\,\mathrm{MeV}^{-1}=197\,\mathrm{fm}$, $L_{\mathrm{min} }\approx
0.72\,\mathrm{fm}$. This is of the order of magnitude of the
estimated size of a meson, $L_{\mathrm{meson}}$, of $\sim 2/3$ of
the size of a hadron, that is, $L_{\mathrm{meson}}\sim
0.92\,\mathrm{fm}$. Taking for instance the model defined by
$\lambda =5.0$, it can be inferred from Fig.~\ref{PRfigura2} that,
for sizes of the system slightly larger than $L_{
\mathrm{min}}\approx 0.72\,\mathrm{fm}$ (twice or more the minimum
size), of order of magnitude of the estimated size of a meson, the
transition temperature is $\sim 272\,\mathrm{MeV}$, which is
compatible with the deconfining hadronic temperature. Similar
results are obtained for other values of $\lambda $.

For completeness, an indication of a second-order transition is
investigated from a renormalization group argument. In this case,
the existence of an infrared stable fixed point at criticality can
be shown from a study of the infrared behavior of the
beta-function, $i.e.$ in the neighborhood of vanishing external
momentum, $|p|\approx 0$. We consider the thermal
boundary-dependent coupling constant at criticality
$m=m(T_{c}(L))=0$, with an external small momentum $p$, given by
\begin{equation}
g(m=0;|p|\approx 0)=\frac{\lambda _{0}}{[1-\lambda _{0}\Pi
(m=0;p)]}. \label{g3L1p}
\end{equation}

The one-loop polarization $\Pi (D,T,L,p)$ using a Feynman
parameter $x$, and taking the mass parameter as the thermal
boundary-dependent mass $m(T,L)$, which vanishes at criticality.
We have,
\begin{equation}
\Pi (D,T,L,p)=\frac{1}{\beta
L}\int_{0}^{1}dx\sum_{n_{1},n_{2}=-\infty
}^{\infty }\int \frac{d^{D-2}k}{(2\pi )^{D-2}}\frac{m^{2}-(\mathbf{k}%
^{2}+\omega _{n_{1}}^{2}+\omega
_{n_{2}}^{2})}{(\mathbf{k}^{2}+\omega _{n_{1}}^{2}+\omega
_{n_{2}}^{2}+M_{x}^{2})^{2}},  \label{sigma00p}
\end{equation}
where
$M_{x}^{2}=M_{x}^{2}(p,T,L,x)=m^{2}(T,L)+p^{2}x(1-x)$\thinspace .
For $ |p|\approx 0$  and $D=3$, the polarization $\Pi (T,L,p)$
reads
\begin{equation}
\Pi (T,L,p)=\int_{0}^{1}\,dx\,\frac{M_{x}}{2\pi
}\,\mathcal{G}\left( \frac{ M_{x}}{T},M_{x}L\right) .
\end{equation}

At criticality, $m(T,L)=m(T_{c}(L))=0$, and for $|p|\sim 0$,
keeping up to terms linear in $|p|$, from the above equation we
get,
\begin{equation}
\Pi (T_{c},L,|p|\sim 0)\approx A(T_{c},L)|p|+B(T_{c},L)\,,
\label{Pi}
\end{equation}%
with $A(T_{c},L)=A=-\frac{1}{16}$ and
\begin{equation}
B(T_{c},L)=\left( \frac{T_{c}}{2\pi }+\frac{1}{2\pi L}\right) \ln
2-\mathcal{ O}\left( \frac{1}{T_{c}},L\right)
+\,2\mathcal{O}\left( \frac{1}{T_{c}} ,2L\right)
+2\mathcal{O}\left( \frac{2}{T_{c}},L\right) -4\mathcal{O}\left(
\frac{2}{T_{c}},2L\right) ,  \label{B}
\end{equation}%
where
\begin{equation}
\mathcal{O}(x,y) = \sum_{n,l=1}^{\infty
}\frac{1}{\sqrt{x^{2}n^{2}+y^{2}l^{2}}}
\end{equation}%
and  $\int_{0}^{1}\,dx\sqrt{x(1-x)}=\pi /8$ is used.

The coupling constant in Eq.~(\ref{g3L1p}) has dimension of
$|p|^{-1}$. Taking $|p|$ as a running scale, we define a
dimensionless coupling constant
\begin{equation*}
g^{\prime }=|p|g=\frac{|p|\lambda _{0}}{1-\lambda
_{0}[A(T_{c},L)|p|+B(T_{c},L)]}
\end{equation*}%
and the beta-function,
\begin{equation}
\beta (g^{\prime }) =|p|\frac{\partial \,g^{\prime }}{\partial
|p|}; \label{beta}
\end{equation}
we find that the condition of a non-trivial infrared stable fixed
point is fulfilled by the solution
\begin{equation}
g_{\star }^{\prime }=-\frac{1}{A}=16.  \label{fixed}
\end{equation}%
Then the infrared stable fixed point is independent of the length
of the system and of the free-space coupling constant. We conclude
that the phase transition, while it survives with decreasing $L$,
is of second-order. We now discuss the $L$-dependence of the
critical temperature.

\subsubsection{Phase transition in the $N$-component massive GN model}

We have pointed out that the four-point contact interaction of the
GN model is similar to the delta-function interaction in the BCS
theory of superconductivity. In the latter case, as in other
systems of condensed matter, the susceptibility arising from the
linear response theory has a divergence at a finite temperature,
indicating the existence of a second-order phase transition
between a disordered and an ordered i.e. a condensed
phase~\cite{Doniach1974}. As spontaneous symmetry breaking is the
common feature underlying all phase transition
phenomena~\cite{Weinberg1986}, such a divergence would appear in
other domains of physics~\cite{MalbsKhaaesEPL2010}. Here, the
existence of a phase transition in the massive $N$-component  GN
model on $\Gamma^d_D$ is investigated, by analyzing the four-point
function in a non-perturbative way~\cite{Khanna2012}.

Particularly,  the cases $D=2,3,4$ with all spatial dimensions
compactified, initially at zero temperature are addressed, and
then temperature effects are discussed by compactifying the
imaginary time in a length $\beta = T^{-1}$, $T$ being the
temperature. The behavior of the system as a function of its size
and of the temperature are studied, focusing  on the dependence of
the large-$N$ coupling constant on the compactification length and
temperature. Even at $T=0$, a singularity in the $4$-point
function may appear driven by changes in the compactification
length, suggesting the existence of a second-order phase
transition in the system. This can be interpreted as a spatial
confinement transition.

As in previous sections, we use concurrently dimensional and
analytic regularizations and employ a subtraction scheme where the
polar terms  are suppressed. Results obtained with this procedure
have similar structure for all values of $D$, which gives us
confidence that they are meaningful for the $4$-dimensional
space-time.  In all cases, we obtain simultaneously an asymptotic
freedom type of behavior for vanishingly small sizes of the system
and spatial confinement, in the strong coupling regime, for low
temperatures. As the temperature is increased, spatial confinement
disappears, what is interpreted as a deconfining transition. The
values of the confining lengths and the deconfining temperatures
for $D=2,\;3\;{\rm{and}}\;4$ are calculated.

The massive GN model in a $D$-dimensional Euclidean space is
described by the Wick-ordered Lagrangian density
\begin{equation}
\mathcal{L}=:\bar{\psi}(x)(i\,\gamma^{\mu}\partial_{\mu} +m)\psi
(x):+\frac{u}{2}(:\bar{\psi} (x)\psi (x):)^{2}, \label{GN}
\end{equation}
where $m$ is the mass, $u$ is the coupling constant, $x$ is a
point of $\mathbb{ R}^{D}$ and the $\gamma $'s are the Dirac
matrices. The quantity $\psi (x)$ represents a spin $\frac{1}{2}$
field having $N$ (flavor) components, $\psi ^{a}(x)$,
$a=1,2,...,N$, with summations over flavor and spin indices being
understood in Eq.~(\ref{GN}). The large-$N$ limit is considered,
where $N\rightarrow\infty$ and $u \rightarrow 0$ in such way that
$N u = \lambda$ remains finite.

The large-$N$ (effective) coupling constant between the fermions
on $\Gamma^d_D$ is defined in terms of the $4$-point function at
zero external momenta. The $\{ L_i\}$-dependent four-point
function, at leading order in $\frac{1}{N}$, is given by the sum
of chains of one-loop (bubble) diagrams, which can be formally
expressed as
\begin{equation}
\Gamma _{Dd}^{(4)}(0;\{L_i\},u)=\;\frac{u}{1+Nu\Pi_{Dd}(\{L_i\})}
. \label{4-point1}
\end{equation}
The $\{L_i\}$-dependent one-loop Feynman diagram is given by
%\begin
\begin{equation}
\Pi_{Dd}(\{L_i\}) = \frac{1}{L_1\cdots L_d} \sum_{\{n_i\}=-\infty
}^{\infty} \int \frac{d^{D-d} {\bf k}}{(2\pi)^{D-d}} \left[
\frac{m^{2}-{\bf k}^2-\sum_{i=1}^{d}{\nu}_{i}^{2}} {\left({\bf
k}^2 +\sum_{i=1}^{d}{\nu}_{i}^{2}+m^{2}\right)^{2}}\right] ,
\label{sigma0}
\end{equation}
where the Matsubara frequencies $ \nu_{i}$ are given by $  \nu_{i}
= {2\pi (n_i+\frac{1}{2})}/{L_i}\; ,\;\;
 i=1,2,...,d \, ,
$ such that $ \{n _i\} = \{n_1 , \ldots, n _d\}$, with $n_i \in
{\mathbb Z}$, and ${\bf k}$ is a $(D-d)$-dimensional vector in
momentum space.

Introducing the dimensionless quantities $b_i=(m L_i )^{-2}$
($i=1,\dots ,d$) and $q_{j}=k_{j}/2\pi m$ $(j=d+1,\dots ,D)$, it
is found
%\begin
\begin{equation}
\Pi_{Dd}(\{b_i\})  =  \left. \Pi_{Dd}(s;\{b_i\}) \right|_{s=2} =
\frac{m^{D-2}}{4\pi^2}\sqrt{b_1\cdots
b_d}\left.\left\{\frac{1}{2\pi^2} U_{Dd}(s;\{b_i\}) -
U_{Dd}(s-1;\{b_i\}) \right\}\right|_{s=2} , \label{sigma2}
\end{equation}
where
\begin{equation}
U_{Dd}(\mu;\{b_i\}) = \sum_{\{n_i\} = -\infty }^{\infty} \int
\frac{d^{D-d} {\bf q}}{\left[{\bf q}^2
+\sum_{j=1}^{d}b_j(n_{j}+\frac{1}{2})^{2}+(2\pi)^{-2}\right]^{\mu}}
. \label{UDd}
\end{equation}

Using the dimensional regularization formula in Eq.~(\ref{dimreg})
 to perform the integral over
${\bf q}=(q_{d+1},\dots,q_D)$ in Eq.~(\ref{UDd}),  we obtain
\begin{equation}
U_{Dd}(\mu;\{b_i\}) =  \pi^{\frac{D-d}{2}}\,\frac{\Gamma(\mu -
\frac{D-d}{2})}{\Gamma(\mu)} \sum_{\{n_i\}=-\infty }^{\infty}
\left[\sum_{j=1}^{d}b_j\left(n_{j}+\frac{1}{2}\right)^{2}
+(2\pi)^{-2}\right]^{\frac{D-d}{2}-\mu} . \label{UDd2}
\end{equation}
The summations over half-integers in this expression can be
transformed into sums over integers leading to
\begin{eqnarray}
U_{Dd}(\mu;\{b_i\})& = & \pi^{\frac{D-d}{2}} \, \frac{\Gamma
(\mu-\frac{(D-d)}{ 2})}{\Gamma (\mu)}\, 4^{\eta } \nonumber \\
&&  \times \left[ Z_{d}^{h^{2}}(\eta ,b_1,\dots,b_d)
 - \sum_{i=1}^{d}Z_{d}^{h^{2}}(\eta ,\dots,4b_i,\dots)  \right.\nonumber \\
 &&\left.+ \sum_{i<j=1}^{d} Z_{d}^{h^{2}}(\eta ,\dots, 4b_i,\dots,4b_j,
\dots)   - \cdots +  (-1)^{d}\, Z_{d}^{h^{2}}(\eta
,4b_1,\dots,4b_d)\right] ,
 \label{UDd3}
\end{eqnarray}
where $h^2=\pi^{-2}$, $\,\eta =\mu-\frac{D-d}{2}\,$ and
\begin{equation}
Z_{d}^{h^{2}}(\eta,\{a_i\})=\sum_{\{n_i\}=-\infty }^{\infty }
\left[\sum_{j=1}^{d} a_j n_{j}^{2}+h^{2}\right]^{-\eta}
\end{equation}
is the multiple ($d$-dimensional) Epstein-Hurwitz $zeta$-function.

The Epstein-Hurwitz $zeta$-function $Z_{d}^{h^{2}}(\eta,\{a_i\})$
is analytically extended to the whole complex $\eta$-plane
\cite{MalbouissonsaesNPB2002,Kirsten1994,Elizalde2012}; we find
%\begin
\begin{eqnarray}
Z_{d}^{h^{2}}(\eta,\{a_i\})
&=&\frac{\pi^{\frac{d}{2}}}{\sqrt{a_1\cdots a_d}\,\Gamma (\eta)}
\left[ \frac{1}{h^{2(\eta-d)}} \Gamma
\left(\eta-\frac{d}{2}\right)
\right.  \nonumber \\
 & & +\, \sum_{\theta=1}^{d} 2^{\theta+1}
\sum_{\{\sigma_{\theta}\}}
 \sum_{\{n_{\sigma_{\theta}}\}=1}^{\infty}\left( \frac{\pi
}{h}\sqrt{\frac{n_{\sigma_{1}}^{2}}{a_{\sigma_1}} + \cdots +
\frac{n_{\sigma_{\theta}}^{2}}{a_{\sigma_{\theta}}}}\right) ^{\eta
-\frac{d}{2}} \left. K_{\eta -\frac{d}{2}}\left( 2\pi
h\sqrt{\frac{n_{\sigma_{1}}^{2}}{a_{\sigma_1}} + \cdots
+\frac{n_{\sigma_{\theta}}^{2}}{a_{\sigma_{\theta}}}}\right)
\right] , \label{Z2}
\end{eqnarray}
%\end
where $\{\sigma_{\theta}\}$ represents the set of all combinations
of the indices $\{1,2,\dots,d\}$ with $\theta$ elements and
$K_{\alpha }(z)$ is the Bessel function of the third kind.
Consequently, the function $U_{Dd}(\mu;\{b_i\})$ can also be
analytically continued to the whole complex $\mu$-plane.

Taking $Z_{d}^{h^{2}}(\eta,\{a_i\})$ given in Eq.~(\ref{Z2}),
grouping similar terms appearing in the parcels of
Eq.~(\ref{UDd3}) and using the identity
\begin{equation}
\sum_{j=1}^{N}\left( \frac{-1}{2} \right)^j \frac{N!}{j!(N-j)!} =
\frac{1}{2^N}\, ,
\end{equation}
%\begin
\noindent we obtain
\begin{equation}
U_{Dd}(\mu;\{b_i\})  =  \frac{2^{2\mu-D}\pi^{2\mu-\frac{D}{2}}}
{\Gamma(\mu)} \frac{1}{\sqrt{b_1\cdots b_d}} \left[ \Gamma\left(
\mu - \frac{D}{2} \right) + 2^{\frac{D}{2}}\, W_{Dd}(\mu;\{ b_i
\}) \right]  \label{UDd4}
\end{equation}
with $W_{Dd}(\mu;\{ b_i \})$ given by
\begin{equation}
W_{Dd}(\mu;\{ b_i \})  =  2^{1 - \mu} \sum_{j=1}^{d} 2^{2 j}
 \sum_{\{ \rho_j \}} \sum_{\{ c_{\rho_k}=1,4 \}} \left(
\prod_{k=1}^{j} \frac{(-1)^{c_{\rho_k} - 1}}{\sqrt{c_{\rho_k}}}
\right) \,
F_{Dj}(\mu;c_{\rho_1}b_{\rho_1},\dots,c_{\rho_j}b_{\rho_j})  ,
\label{WDd}
\end{equation}
where $\{ \rho_j \}$ stands for the set of all combinations of the
indices $\{1,2,\dots,d\}$ with $j$ elements and the functions
$F_{Dj}(\mu;a_1,\dots,a_j)$, for $j=1,\dots,d$, are defined by
\begin{equation}
F_{Dj}(\mu;a_1,\dots,a_j)  =
\sum_{n_1,\dots,n_j=1}^{\infty}\left( 2
\sqrt{\frac{n_{1}^{2}}{a_{1}} + \cdots +
\frac{n_{j}^{2}}{a_{j}}}\right) ^{\mu - \frac{D}{2}} K_{\mu -
\frac{D}{2}}\left( 2 \sqrt{\frac{n_{1}^{2}}{a_{1}} + \cdots
+\frac{n_{j}^{2}}{a_{j}}}\right)  . \label{FDj}
\end{equation}
%\end
Substituting Eq.~(\ref{UDd4}) into Eq.~(\ref{sigma2}) leads
directly to an analytic extension of $\Pi_{Dd}(s;\{b_i\})$ for
complex values of $s$, in the vicinity of $s=2$. We get,
\begin{eqnarray}
\Pi_{Dd}(s;\{b_i\}) & = & \Pi_{Dd}^{{\rm polar}}(s)
 +  2^{\frac{D}{2}} \left[ 2\, W_{Dd}(s;\{ b_i \}) -  (s-1)\,
 W_{Dd}(s-1;\{ b_i \}) \right]  , \label{SigmaS}
\end{eqnarray}
where
\begin{equation}
\Pi_{Dd}^{{\rm polar}}(s) =
\frac{m^{D-2}\,\pi^{\frac{D}{2}}}{(2\pi)^{D-2s+4}\,\Gamma(s)}
 (s-1-D)\, \Gamma\left( s - 1 - \frac{D}{2} \right)
 \label{SigmaPolar}
\end{equation}
and the functions $W_{Dd}(\mu;\{ b_i \})$ are given by
Eq.~(\ref{WDd}).

It is important to notice that the first term in this expression
for $\Pi_{Dd}(s;\{b_i\})$, $\Pi_{Dd}^{{\rm polar}}(s)$, does not
depend on parameters $b_i$, that is, it is independent of the
compactification lengths $L_i$ ($i=1,\dots,d$). At $s=2$, due to
the poles of the $\Gamma$-function, such a term is divergent for
even dimensions $D\geq 2$. Using again the minimal subtraction
scheme, the finite one-loop diagram reads
\begin{eqnarray}
\Pi_{Dd}(\{b_i\}) &=& \left[ \Pi_{Dd}(s;\{b_i\}) - \Pi_{Dd}^{\rm
polar}(s)
\right]_{s=2} \nonumber\\
&=&\frac{m^{D-2}}{(2\pi)^{\frac{D}{2}}}\, \left[ 2\, W_{Dd}(2;\{
b_i \}) - W_{Dd}(1;\{ b_i \}) \right]  . \label{SigmaR}
\end{eqnarray}
From now on, we shall deal only with finite quantities that are
obtained following this subtraction prescription.

The large-$N$ ($\{b_i\}$-dependent) coupling constant, for $d$
($\leq D$) compactified dimensions, is  obtained by substituting
$\Pi_{Dd}(\{b_i\})$ into Eq.~(\ref{4-point1}) and taking the limit
$N\rightarrow\infty$, $u\rightarrow 0$, with $Nu=\lambda$ fixed;
we get
\begin{equation}
g_{Dd}(\{b_i\},\lambda) = \lim_{N u = \lambda} \left[ N\Gamma
_{Dd}^{(4)}(0,\{b_i\},u)\right] = \frac{\lambda }{1+\lambda \,
\Pi_{Dd}(\{b_i\})} .  \label{ECC}
\end{equation}
It is clear that, while $g_{Dd}(\{b_i\},\lambda)$ depends on the
value of the fixed coupling constant $\lambda$ in a direct way,
its dependence on the compactifiation lengths is dictated by the
behavior of $\Pi_{Dd}$ as $\{b_i\}$ is varied. The dependence of
$g_{Dd}$ on $\{ L_i \}$ and $\lambda$ is the main point to be
discussed in the subsequent analysis.

For all the compactification lengths tending to infinity, i.e. $\{
b_i\rightarrow 0 \}$, thus reducing the problem to the free space
at $T=0$, $\Pi_{Dd}\rightarrow 0$ and we obtain, consistently,
that
\begin{equation}
\lim_{\{L_i\rightarrow\infty\}} g_{Dd}(\{b_i\},\lambda)=\lambda  ,
\end{equation}
where $\lambda$ is the fixed coupling constant in free space at
zero temperature.  In the opposite limit, for any $b_i$ tending to
$\infty$ (i.e. if any compactification length $L_i$ goes to $0$),
the single bubble diagram $\Pi_{Dd}\rightarrow\infty$. This
implies that the effective coupling constant $g_{Dd}$ vanishes,
irrespective of the value of $\lambda$, suggesting that the system
presents an asymptotic-freedom type of behavior for short
distances and/or for high temperatures.

From the extreme limits considered above two situations may
emerge, as one changes the compactification lengths from $0$ to
$\infty$: either $\Pi_{Dd}$ varies from $\infty$ to $0$ through
positive values, or $\Pi_{Dd}$ reaches $0$ before tending to $0$
through negative values. The latter case, which may actually
happen, would lead to an interesting situation where a divergence
of the effective coupling constant would appear at finite values
of the lengths $L_i$. This possibility, and its consequences, will
be investigated explicitly in the following.

For $D=2$ and $d=1$ (two-dimensional space-time with the spatial
coordinate compactified), we put $b_1=(mL)^{-2}$ in
Eqs.~(\ref{WDd}) and (\ref{FDj}). In this case, we have
\begin{equation}
g_{21}(L,\lambda) = \frac{\lambda}{1+\lambda\, \Pi_{21}(L)} ,
\label{G21}
\end{equation}
where
\begin{equation}
\Pi_{21}(L) = 2\, E_1(2 m L) - E_1(m L) , \label{S21}
\end{equation}
with the function $E_1(x)$ being defined by
\begin{equation}
E_1(x) = \frac{1}{\pi} \sum_{n=1}^{\infty} \left\{ - K_{0}(x n) +
(x n)\, K_{1}(x n) \right\} . \label{E1}
\end{equation}

\begin{figure}[htbp]
\begin{center}
  \begin{minipage}[b]{0.4\linewidth}
    \centering
    \includegraphics[width=\linewidth]{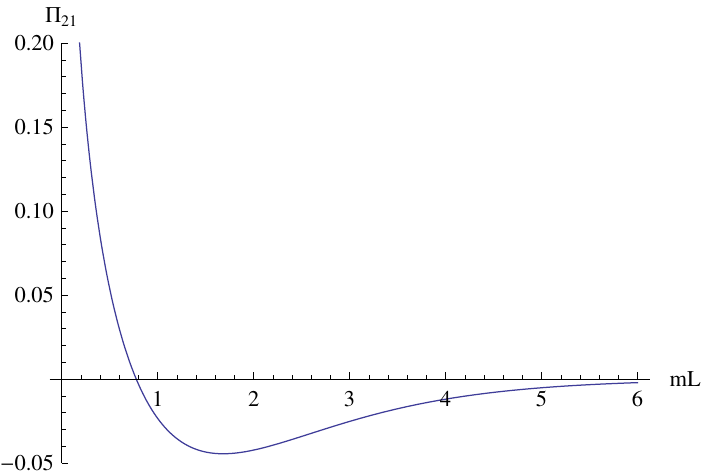}
    \caption{Plot of $\Pi_{21}(L)$ as a function of $mL$.\hspace{1.1cm}}
\label{figS21}\vspace{1.25cm}
  \end{minipage}
  \hspace{1.0cm}
  \begin{minipage}[b]{0.4\linewidth}
    \centering
    \includegraphics[width=\linewidth]{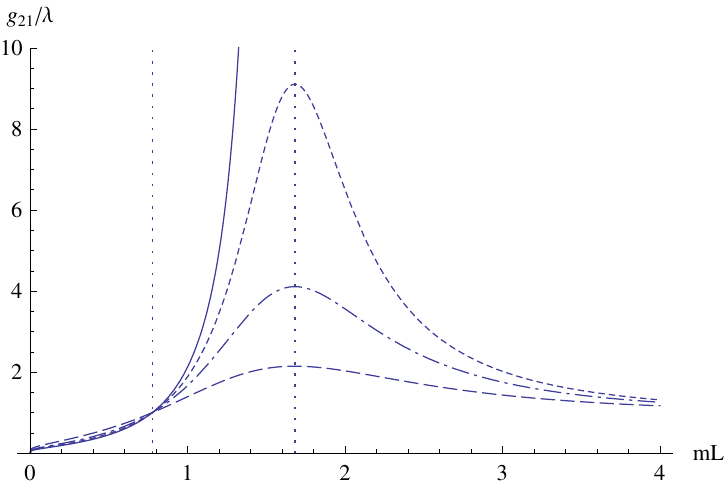}
    \caption{Plots of the relative effective coupling constant,
$g_{21}(L,\lambda)/\lambda$, as a function of $mL$ for some values
of $\lambda$: $20.0$ (dotted line), $17.0$ (dotted-dashed line),
$12.0$ (dashed line) and $22.5$ (full line). The dotted vertical
lines correspond to $L_{\rm min}^{(2)} \simeq 0.78\, m^{-1}$ and
$L_{\rm max}^{(2)}\simeq 1.68\, m^{-1}$.} \label{figG21}
  \end{minipage}
  \end{center}
\end{figure}

The function $\Pi_{21}(L)$ is plotted as a function of $mL$ in
Fig.~\ref{figS21}. From this figure and the numerical treatment of
Eq.~(\ref{S21}), we infer that $\Pi_{21}(L)$ diverges
($\rightarrow +\infty$) when $L\rightarrow 0$ and tends to $0$,
through negative values, as $L\rightarrow\infty$. Also, we find
that $\Pi_{21}(L)$ vanishes for a specific value of $L$, which we
denote by $L_{{\rm min}}^{(2)}$, being negative for all $L
> L_{{\rm min}}^{(2)}$, and assumes a minimum (negative) value
at a value of $L$ denoted by $L_{{\rm max}}^{(2)}$, for reasons
that will be clarified later. Numerically, it is found that
$L_{{\rm min}}^{(2)} \simeq 0.78\, m^{-1}\,$, $L_{{\rm max}}^{(2)}
\simeq 1.68\, m^{-1}\,$ and $\Pi_{21}^{R\,{\rm min}} \simeq
-0.0445$. This behavior of $\Pi_{21}$ as $L$ changes, particularly
the fact that $\Pi_{21}(L) < 0$ for $L > L_{{\rm min}}^{(2)}$,
leads to remarkable properties of the large-$N$ coupling constant
$g_{21}(L,\lambda)$.

The divergence of $\Pi_{21}(L)$ as $L\rightarrow 0$ ensures that,
independently of the value of $\lambda$, $g_{21}(L,\lambda)$
approaches $0$ in this limit and, therefore, the system presents a
kind of asymptotic-freedom behavior for short distances. On the
other hand, since $\Pi_{21}(L)$ assumes negative values for $L
> L_{{\rm min}}^{(2)}$, the denominator of Eq.~(\ref{G21}) will vanish at a
finite value of $L$ if $\lambda$ is sufficiently high. This means
that, starting from a low value of $L$ (within the region of
asymptotic freedom) and increasing the size of the system,
$g_{21}$ will diverge at a finite value of $L$,
$L_{c}^{(2)}(\lambda)$, if $\lambda$ is greater than the
``critical value" $\lambda_c^{(2)} = (- \Pi_{21}^{R\,{\rm
min}})^{-1} \simeq 22.47$. We interpret this result by stating
that, in the strong-coupling regime ($\lambda \geq
\lambda_c^{(2)}$) the system gets spatially confined in a segment
of length $L_{c}^{(2)}(\lambda)$. The behavior of the
$L$-dependent coupling constant as a function of $mL$ is
illustrated in Fig.~\ref{figG21}, for some values of the fixed
coupling constant $\lambda$.

%%%%%%%%%%%%%%%%
%%%%%%%%%%%%%%%

For $\lambda = \lambda_c^{(2)}$, by definition, the divergence of
$g_{21}(L,\lambda)$ is reached as $L$ approaches the value that
makes $\Pi_{21}$ minimal, which we have denoted by $L_{{\rm
max}}^{(2)}$. In the other limit, since
$g_{21}^{-1}(L,\lambda\rightarrow \infty) = \Pi_{21}(L)$,
$L_{c}^{(2)}(\lambda)$ tends to $L_{{\rm min}}^{(2)}$, the zero of
$\Pi_{21}(L)$, as $\lambda\rightarrow \infty$. In other words, the
confining length $L_{c}^{(2)}(\lambda)$ decreases from the maximum
value $L_{{\rm max}}^{(2)}$, when $\lambda = \lambda_c^{(2)}$,
tending to the lower bound $L_{{\rm min}}^{(2)}$ in the limit
$\lambda \rightarrow \infty$.

For the $3$-$D$ model at zero temperature with two compactified
dimensions ($d=2$), denoting the compactification lengths
associated with the two spatial coordinates $x_1$ and $x_2$ by
$L_1$ and $L_2$ ($m^{-1} / \sqrt{b_1}$ and $m^{-1} / \sqrt{b_2}$,
respectively); for simplicity we take $L_1=L_2=L$ and we obtain
\begin{equation}
\Pi_{32}(L_1,L_2) = \frac{m}{\pi} \left[ \frac{1} {L}\log(1 +
e^{-L}) - \frac{1}{1 + e^{L}}+
     G_2(L,L) - 4G_2(L,2L)  + 4 G_2(2L,2L) \right]  ,
  \label{Sigma32N}
\end{equation}
where the function $G_2(x,y)$ is defined by
\begin{eqnarray}
G_2(x,y) & = & \sum_{n,l=1}^{\infty} \exp\left( - \sqrt{x^2 n^2 +
y^2 l^2} \right) \left[ 1 - \frac{1}{\sqrt{x^2 n^2 + y^2 l^2}}
\right] . \label{G2}
\end{eqnarray}

The behavior of  $\Pi_{32}(L)/m$ as a function of $mL$ is similar
to that presented in Fig.~\ref{figS21} for $\Pi_{21}(L)$.  We
find, numerically, that $L_{{\rm min}}^{(3)} \simeq 1.30\,
m^{-1}\,$ and $L_{{\rm max}}^{(3)} \simeq 2.10\, m^{-1}\,$, with
$\Pi_{32}^{R\,{\rm min}} \simeq -0.00986\, m$. This behavior of
$\Pi_{32}(L)$ has profound implications on the effective coupling
constant. Thus for  $D=3$ and $d=2$, Eq.~(\ref{ECC}) is rewritten
as
\begin{equation}
g_{32}(L,\lambda) = \frac{\lambda}{1+\lambda\, \Pi_{32}(L)} .
\label{G32}
\end{equation}
We find that, for $\lambda\geq\lambda_c^{(3)}=(-\Pi_{32}^{R\,{\rm
min}})^{-1}\simeq 101.42\, m^{-1}$, the denominator in
Eq.~(\ref{G32}) vanishes for a finite value of $L$,
$L_c^{(3)}(\lambda)$, leading to a divergence in the effective
coupling constant. The behavior of the effective coupling constant
as a function of $L$, for increasing values of the fixed coupling
constant $\lambda$, is illustrated showing the same pattern as
that of Fig.~\ref{figG21} for the preceding case, with
$L_c^{(3)}(\lambda)$ satisfying $L_{{\rm min}}^{(3)} <
L_c^{(3)}(\lambda) \leq L_{{\rm max}}^{(3)}$.

For the $4$-$D$ GN model with all three spatial coordinates
compactified, taking $L_{i}=L \,\, (i=1,2,3)$  measured in units
of $m^{-1}$, we get
\begin{eqnarray}
\Pi_{43}(L) & = & m^2 \left[ 6 H_1(2L) - 3 H_1(L) + 6
H_2(L,L)  -  24 H_2(L,2L) + 24 H_2(2L,2L)\right. \nonumber \\
 & & -\left. 4 H_3(L,L,L) + 24 H_3(L,L,2L)  - 48 H_3(L,2L,2L) + 32 H_3(2L,2L,2L)
 \right] ,
  \label{Sigma43}
\end{eqnarray}
where the functions $H_j$, $j=1,2,3$, are defined by
\begin{eqnarray}
H_1(x) & = & \frac{1}{2\pi^2} \sum_{n=1}^{\infty} \left[ K_{0}(x
n)
- \frac{K_{1}(x n)}{(x n)} \right] \, , \label{H1} \\
H_2(x,y) & = & \frac{1}{2\pi^2} \sum_{n,l=1}^{\infty} \left[ K_{0}
\left(\sqrt{x^2 n^2 + y^2 l^2} \right) - \frac{K_{1}(\sqrt{x^2 n^2
+ y^2 l^2})}{(\sqrt{x^2 n^2 + y^2 l^2})} \right] \, ,
\label{H2} \\
H_3(x,y,z) & = & \frac{1}{2\pi^2} \sum_{n,l,r=1}^{\infty} \left[
K_{0} \left(\sqrt{x^2 n^2 + y^2 l^2 + z^2 r^2} \right)  -\,
\frac{K_{1}(\sqrt{x^2 n^2 + y^2 l^2})}{(\sqrt{x^2 n^2 + y^2 l^2 +
z^2 r^2})} \right]  . \label{H3}
\end{eqnarray}
Similarly to the cases of $D=2$ and $D=3$, it is found numerically
that $\Pi_{43}(L)$ vanishes for $L = L_{\rm min}^{(4)} \simeq
1.68\, m^{-1}$, being negative for $L > L_{\rm min}^{(4)}$, and
assumes the minimum value, $\Pi_{43}^{R{\rm min}} \simeq
-0.0022751\, m^{2}$, when $L = L_{\rm max}^{(4)} \simeq 2.37\,
m^{-1}$. Therefore, the large-$N$ coupling constant,
\begin{equation}
g_{43}(L,\lambda) = \frac{\lambda}{1 + \lambda \Pi_{43}(L)} \, ,
\label{G43}
\end{equation}
diverges at a finite value of $L$, $L_c^{(4)}(\lambda)$, if
$\lambda \geq \lambda_c^{(4)} = -(\Pi_{43}^{R{\rm min}})^{-1}
\simeq 439.54\, m^{-2}$, meaning that the system gets confined in
a cubic box of edge $L_c^{(4)}(\lambda)$ which is bounded in the
interval between the values $L_{\rm min}^{(4)}$ and $L_{\rm
max}^{(4)}$.

For  $D=2,3,4$, with all spatial coordinates compactified with the
same length, we find that the confining length lies in a finite
interval, $ L_{c}^{(D)}(\lambda) \in \left( L_{{\rm
min}}^{(D)},L_{{\rm max}}^{(D)} \right], $ where the maximum value
corresponds to $\lambda_{c}^{(D)}$, while $L_{{\rm min}}^{(D)}$
sets the bound as $\lambda \rightarrow \infty$. For a given value
of $\lambda$ ($\geq\lambda_{c}^{(D)}$), the confining length
$L_{c}^{(D)}(\lambda)$ is found numerically by determining the
smallest root of the equation
\begin{equation}
g_{D D-1}^{-1}(L,\lambda) = \frac{1}{\lambda} \left[ 1 + \lambda
\Pi_{D D-1}(L) \right] = 0 .
\end{equation}
That is, following the interpretation provided before, starting
from small values of $L$, the first value at which $g_{D
D-1}^{-1}$ vanishes does provide the confining length of the
system, $L_{c}^{(D)}(\lambda)$.

Let us now consider the effect of raising the temperature on the
effective coupling constant for the GN model, with all spatial
dimensions compactified; this is implemented by the Matsubara
procedure for the imaginagy-time. We generally expect that the
dependence of $\Pi_{D D}$ and $g_{D D}$ on $\beta$ should follow
similar patterns as that for the dependence on $L$. Like in the
case where any compactification length $L_i$ tends to zero, we
find that $\Pi_{D D}(L,\beta) \rightarrow \infty$ as $\beta
\rightarrow 0$ ($T \rightarrow \infty$), implying that $g_{D D}
\rightarrow 0$ independently of the value of the fixed coupling
constant $\lambda$. This means that we have an asymptotic-freedom
behavior for very high temperatures. For $\beta\rightarrow\infty$
($T\rightarrow 0$), the behavior of $\Pi_{D D-1}(L)$ has been
described earlier: for sufficiently high values of $\lambda$, the
system is confined in a $(D-1)$-dimensional cube of edge
$L_{c}^{(D)}$. Based on these observations, we expect that,
starting from the compactified model at $T=0$ with $\lambda \geq
\lambda_{c}^{(D)}$, raising the temperature will lead to the
suppression of the divergence of $g_{D D}$ and the consequent
spatial deconfinement of the system, at a specific value of the
temperature, $T_{d}^{(D)}$.

Let us start with the $D=2$ GN model. To account for the effect of
finite temperature, we take the second Euclidean coordinate (the
imaginary time, $x_2$) compactified in a length $L_2 = \beta =
1/T$. In this case, replacing $b_1 = L^{-2}$ and $b_2 =
\beta^{-2}$ ($L$ and $\beta$ measured in units of $m^{-1}$) into
Eqs.~(\ref{WDd}), (\ref{FDj}) and (\ref{SigmaR}), the $L$ and
$\beta$-dependent bubble diagram is written as
\begin{equation}
\Pi_{22}(L,\beta)= 2\, E_1(2 L) - E_1(L) + 2\, E_1(2 \beta) -
E_1(\beta) + 2\, E_2(L,\beta) - 4\, E_2(2 L,\beta) -  4\, E_2(L,2
\beta) + 8\, E_2(2 L,2 \beta) , \label{S22}
\end{equation}
where the function $E_1(x)$ is given by Eq.~(\ref{E1}) and the
function $E_2(x,y)$ is defined by
\begin{equation}
E_2(x,y) = \frac{1}{\pi} \sum_{n,l=1}^{\infty} \left\{ - K_0\left(
\sqrt{x^2 n^2 + y^2 l^2} \right) + \left( \sqrt{x^2 n^2 + y^2 l^2}
\right)\, K_1\left( \sqrt{x^2 n^2 + y^2 l^2} \right)
\right\} . \nonumber \\
\label{E2}
\end{equation}
Notice that, in the limit $\beta\rightarrow\infty$, all
$\beta$-dependent terms vanish and so $\Pi_{22}(L,\beta)$ reduces
to the expression for zero temperature, $\Pi_{21}(L)$. For $\beta
\rightarrow 0$, $\Pi_{22}(L,\beta) \rightarrow \infty$ and,
independently of the value of $\lambda$, the system becomes
asymptotically free. Therefore,  raising the temperature leads to
the suppression of the divergence of $g_{22}$ in the
strong-coupling regime; i.e. for $\lambda \geq \lambda_c^{(2)}$,
there exists a temperature, $T_d^{(2)}(\lambda)$, above which
$g_{22}$ has no divergence and the system is spatially deconfined.
The deconfining temperature $T_d^{(2)}(\lambda)$ is determined by
finding the value of $T$ for which the minimum value of
$g_{22}^{-1}(L,\beta,\lambda)$ changes from negative to positive.

\begin{figure}[ht]
\begin{center}
\scalebox{1.2}{{\includegraphics{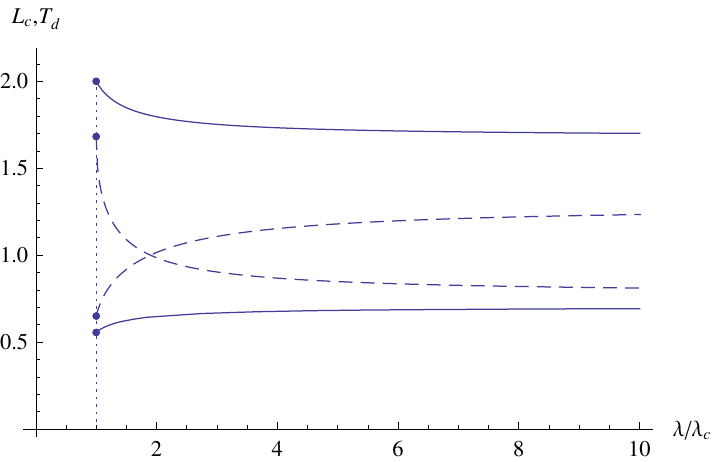}}}
\end{center}
\caption{Deconfining temperature $T_d^{(D)}(\lambda)$ (in units of
$m$), increasing curves, and confining length $L_c^{(D)}(\lambda)$
(in units of $m^{-1}$), decreasing curves, as a function of
$\lambda/\lambda_c^{(D)}$, for $D=2$ (dashed lines) and $D=4$
(full lines).} \label{Tdlamb}
\end{figure}

The GN model in dimensions $D=3$ and $D=4$, with all spatial
dimensions compactified and at finite temperature, can be analyzed
along similar lines as for $D=2$. In all cases, the inverse of the
effective coupling constant in given by
\begin{equation}
g_{DD}^{-1}(L,\beta;\lambda) = \frac{1}{\lambda} \left[ 1 +
\lambda \, \Pi_{DD}(L,\beta) \right] . \label{gLbeta}
\end{equation}
The system is deconfined at a given temperature if the minimal
value of $g_{DD}^{-1}(L,\beta;\lambda)$, with respect to changes
in $L$, is positive. No matter how high the value of $\lambda$ is,
the system becomes deconfined at a temperature $
T_{d}^{(D)}(\lambda) \in \left[ T_{\rm min}^{(D)},T_{\rm
max}^{(D)} \right) $, with the limiting values corresponding to
$\lambda_{c}^{(D)}$ and $\lambda\rightarrow \infty$, respectively.
Notice that, $\,\min_{\{ L \}} g_{DD}^{-1}(L,\beta;\lambda) =
\left[ 1 + \lambda \, M_{D}(\beta) \right] / \lambda$, where
$M_{D}(\beta) = \min_{\{ L \}}\Pi_{DD}(L,\beta) $. In the strong
coupling regime, $\lambda \geq \lambda_c^{(D)}$, the deconfining
temperature is determined by solving the equation $1 + \lambda \,
M_{D}(\beta)=0$; in this regime, this equation always has a
solution, since $M_{D}(\beta\rightarrow\infty) = \Pi_{D
D-1}^{R{\rm min}} = -[ \lambda_{c}^{(D)} ]^{-1}$, while
$M_{D}(\beta) \rightarrow \infty$ as $\beta \rightarrow 0$. For
$D=2$, we find $T_{\rm min}^{(2)} = [\beta_{\rm max}^{(2)}]^{-1}
\simeq 0.65\, m$ and $T_{\rm max}^{(2)} \simeq 1.29\, m$; and, in
the case of $D=3$, we have $T_{\rm min}^{(3)} \simeq 0.54 m$ and
$T_{\rm max}^{(3)} \simeq 0.88 m$. For $D=4$, to avoid an
anomalous behavior, we redefine the strong-coupling regime as
$\lambda \geq \lambda_d^{(4)} \simeq 544.8\,m^{-2}$. The
zero-temperature maximum confining length then becomes $L_{\rm
max}^{(4)} \simeq 2.00\, m^{-1}$ and we find $T_{\rm min}^{(4)}
\simeq 0.55\, m$ and $T_{\rm max}^{(4)} \simeq 0.70\,
m$~\cite{Khanna2012}. The dependence of $T_d^{(D)}$ on $\lambda$
is depicted in Fig.~\ref{Tdlamb}, for $D=2$ and $D=4$; also, in
this figure, the dependence of $L_c^{(D)}$ on $\lambda$ is
presented.

%%%%%%%%%%%%%%%%%%%%%%%
%%%%%%%%%%%%%%%%%%%%%%

The dependencies of $L_{c}^{(D)}$ and $T_{d}^{(D)}$ on the
parameters $\lambda$ and $m$ are intrinsic results of the model.
In all cases the confining length can be written as
$L_{c}^{(D)}(\lambda) = f_D(\lambda)\, m^{-1}$, where the
dimensionless functions $f_D(\lambda)$ are plotted in
Fig.~\ref{Tdlamb}, for $D=2$ and $D=4$. The deconfining
temperature is given by $T_{d}^{(D)}(\lambda) = h_D(\lambda)\, m$,
where $h_D(\lambda)$ are dimensionless functions, also shown in
Fig.~\ref{Tdlamb}. The functions $f_D(\lambda)$ and $h_D(\lambda)$
take on values in finite intervals ($\subset [0.5,2.1]$) and so it
is found that extremely light fermions ($m\rightarrow 0$) are not
confined at all, while extremely heavy ones ($m\rightarrow
\infty$) would be strictly confined in a dot, no matter what the
value of $\lambda$. Also, for all dimensions $D$, the product of
$L_{c}^{(D)}\,T_{d}^{(D)} = f_D(\lambda)\,h_D(\lambda)$ is very
close to the unit in the strong-coupling regime. However, to
estimate values of $L_{c}^{(D)}$ and $T_{d}^{(D)}$ one needs to
fix the parameter $m$, the mass of the  fermions. Taking the GN
model as an effective model for strong interaction, we choose $m
\approx 350 \, {\rm MeV} \simeq 1.75\,{\rm fm}^{-1}$, the
constituent quark mass~\cite{Group2004}. In this case one finds,
for example, $0.96\,{\rm fm} < L_{c}^{(4)} < 1.14\,{\rm fm}$ and
$193\,{\rm MeV} < T_{d}^{(4)} < 245\,{\rm MeV}$; these values
compare amazingly well with the size of
hadrons~\cite{Karshenboim1999} and their estimated deconfining
temperatures~\cite{SmilgaQCDbook2001}.

\subsection{Finite size  effects on the Nambu-Jona-Lasinio model}

The Nambu-Jona-Lasinio (NJL) model~\cite{Nambu1961,Nambu1961a}
provides more results useful for the investigation of dynamical
symmetries when the system is under certain constraints, like
finite temperature, space compactification, finite chemical
potential, or gravitational
field~\cite{Klevansky1991,Hatsuda1994,Buballa2005}. Boundary
effects have been also considered for quark-meson models, in
particular considering a toroidal
topology~\cite{Braun2005,Braun2006,Kim1994,He1996,Kogut2001,Beneventano2004,Gamayun2005,
Kiriyama2006,AbreuMarcelo2006,Ebert2008}. Here, we  discuss
finite-size effects on the dynamical symmetry breaking of the
four-dimensional NJL model at finite temperature. This is carried
out using zeta-function and dimensional regularization methods.
This approach allows to determine analytically the size-dependence
of the effective potential and the gap equation, and to study the
phase structure of the compactified model. We also consider
external magnetic-field effects on the system on the topology
$\Gamma_4^2$, that is with one spatial coordinate compactified and
at $T\neq 0$.

\subsubsection{Effective potential and gap equation}

Consider the massless version of the NJL model, described by the
Lagrangian density,
\begin{equation}
\mathcal{L} = \bar{q}\,  i \gamma{\partial}  q + \frac{G}{2} \sum
^{N^2-1}_{a=0} \left[ \left( \bar{q} \alpha ^a q \right)^2 +
\left( \bar{q} i \gamma _5 \alpha ^a q \right)^2 \right],
\label{NJL}
\end{equation}
where $q$ and $\bar{q}$ are the $N$-component spinors, the
matrices $\alpha ^a$ are the generators of the group $U(N)$, with
$\alpha ^0 = \mathbf{I}/\sqrt{N}$, and $G$ is the coupling
constant.

A bosonization procedure is performed assuming that only one
auxiliary field, associated with the bilinear form $\sigma =
\bar{q} \alpha ^0 q $, takes non-vanishing values and plays the
role of a dynamical fermion mass, in the sense that the system is
in the broken-chiral phase when it is non-zero. We consider the
$D$-dimensional Euclidean space-time, particularizing to $D=4$
latter. We take $\sigma$ uniform so that the effective potential
up to one-loop order, at leading order in $1/N$, is given by
\begin{equation}
\frac{1}{N} U_{eff}  = \frac{ \mathcal{A}_{eff}}{V} = \frac{
\sigma ^2}{2G} + U_{1} (\sigma), \label{effpot}
\end{equation}
where $\mathcal{A}_{eff}$ is the effective action, $V$ is the
volume and
\begin{equation}
U_{1} (\sigma) = - h_{\mathbf{D}} \int \frac{d^D k}{(2\pi)^D}
\ln{\left( \; \frac{k_E^2 + \sigma^2}{\lambda ^2} \right)},
\label{oneloop}
\end{equation}
where $h_{\mathbf{D}}$ is the dimension of the Dirac
representation and $\lambda$ is a scale parameter.

To discuss finite-size effects on the phase structure, we take $d$
($\leq D$) compactified dimensions and denote the Euclidian
coordinate vectors by $x_E = (y,z)$, where $y = \left(
x_E^1,\dots, x_E ^{d}\right)$ has components $y_j \in [0, L_j]$
while $z = \left( x_E ^{d+1},\dots,x_E ^{D} \right)$ refers to the
non-compactified coordinates. Then, accordingly to the generalized
Matsubara rules, the $k_y$-components of the the momentum $k_E$
take discrete values $k_{y j} \rightarrow 2\pi(n_j + c_j)/L_j$,
with $c_j=1/2$ for antiperiodic boundary conditions. For the
system at finite temperature, with $d-1$ compactified spatial
dimensions, we take $L_1 \equiv \beta = 1/T$.

Now, we rewrite Eq.~(\ref{oneloop}) in terms of
zeta-functions~\cite{Elizalde2012} as,
\begin{equation}
U_{1}(\sigma) = \frac{h_{\mathbf{D}}}{2V} \left[
\zeta^{\,\prime}(0) + \ln{\lambda ^2}\zeta (0)\right], \label{U11}
\end{equation}
where $\zeta(s)$ is given by
\begin{equation}
\zeta(s) =  V_{D-d} \sum_{n_{1},...,n_{d} =  -\infty}^{+\infty }
\int \frac{d^{D-d} k_z}{(2\pi)^{D-d}} \left[k_z ^2 + k_y ^2  +
\sigma^{2} \right]^{-s }, \label{zeta1}
\end{equation}
where $k_y ^2 = \sum_{j=1}^{d} (4\pi^2/L_{j}^{2}) \left( n_j + c_j
\right)^2$, with $V_{D-d}$ being the $(D-d)$-dimensional volume.
Performing the $k_z$-integration with dimensional regularization
technique, we obtain
\begin{equation}
\zeta \left(s;\left\{a_j\right\}, \left\{c_j\right\} \right) =
\frac{V_{D-d}}{(4\pi)^{{D-d}/2}} \frac{\Gamma \left( s-
\frac{{D-d}}{2} \right)}{\Gamma \left( s \right) }
Z_{d}^{\sigma^2}\left( s-\frac{{D-d}}{2};\left\{a_j\right\},
\left\{c_j\right\} \right), \label{zeta2}
\end{equation}
where $Z_{d}^{\sigma^2} \left( \nu ; \left\{a_j\right\},
\left\{c_j\right\} \right)$ is the multivariable inhomogeneous
Epstein-Hurwitz zeta function, defined by
\[
Z_{d}^{\sigma^2} \left( \nu ; \left\{a_j\right\},
\left\{c_j\right\} \right) = \sum_{n_{1},...,n_{d} =
-\infty}^{+\infty} \left[ a_1(n_1 + c_1)^2 + \cdots + a_d(n_d +
c_d)^2 + \sigma^2 \right]^{-\nu} .
\]
The function $Z_{d}^{\sigma^2}$ is defined for $\rm{Re} \; \nu >
d/2$, but it can be analytically continued to the whole complex
$\nu$-plane. An analysis of the pole structure of the
zeta-function implies that Eq.~(\ref{U11}) can be written
as~\cite{AbreuMarcelo2006,Elizalde2012},
\begin{equation}
U_{1}\left(\sigma ;\left\{a_j\right\}, \left\{c_j\right\} \right)
= \frac{h_{\mathbf{D}}}{2V_d (4\pi)^{(D-d)/2} } \Gamma \left( -
\frac{D-d}{2} \right) Z_{d}^{\sigma^2}\left(
-\frac{D-d}{2};\left\{a_j\right\}, \left\{c_j\right\} \right),
\label{U1odd}
\end{equation}
for $D-d$ odd, or
\begin{eqnarray}
U_{1}\left(\sigma ;\left\{a_j\right\}, \left\{c_j\right\} \right)
& = & \frac{h_{\mathbf{D}}}{2V_d (4\pi)^{(D-d)/2} }
\frac{(-1)^{\frac{D-d}{2}}}{((D-d)/{2})!} \left\{ Z_{d}^{\sigma^2
\,\prime} \left( -\frac{D-d}{2};\left\{a_j\right\},
\left\{c_j\right\} \right)
\right. \nonumber \\
& & + \left. Z_{d}^{\sigma^2}\left(
-\frac{D-d}{2};\left\{a_j\right\}, \left\{c_j\right\}
\right)\left[ \ln{\lambda^2} - \gamma - \psi \left(\frac{D-d}{2} +
1 \right) \right] \right\}, \label{U1even}
\end{eqnarray}
for $D-d$ even, where $\gamma$ and $\psi(s)$ denote the
Euler-Mascheroni constant and the digamma function, respectively.

To study the phase structure of the model, we analyze the gap
equation, which is obtained by minimizing the effective potential
with respect to $\sigma$,
\begin{equation}
\left.\frac{\partial  }{\partial \sigma}U_{eff} \left(\sigma
;\left\{a_j\right\}, \left\{c_j\right\} \right)\right|_{\sigma =
m} = 0, \label{gap}
\end{equation} where
$m$ is the dynamically generated fermion mass, that is the order
parameter of the chiral phase transition.

Let us initially consider the model in $D=4$ without compactified
coordinates ($d=0$) to set the free space parameters. In this
case, the renormalized effective potential is given by
\begin{equation}
\frac{1}{N} U_{eff} \left(\sigma\right) = \frac{\sigma ^2}{2G_R}+
\frac{h_{\mathbf{D}} (D-1)}{ (4\pi)^{D/2} } \Gamma \left( 1 -
\frac{D}{2} \right) \lambda^{D-2} \sigma ^2 -
\frac{h_{\mathbf{D}}}{D (4\pi)^{D/2} } \Gamma \left( 1 -
\frac{D}{2} \right) \sigma^D , \label{effren}
\end{equation}
where the renormalized coupling constant $G_R$ is defined by
\begin{equation}
\frac{1}{G_R} = \frac{1}{G} - \frac{h_{\mathbf{D}} (D-1)}{
(4\pi)^{D/2} } \Gamma \left( 1 - \frac{D}{2} \right)
\lambda^{D-2}. \label{GR1}
\end{equation}
Note that Eq.~(\ref{effren}) is valid for $ 2 \leq D < 4$, but it
is singular for $D=4$ due to the pole of the gamma-function.
Taking $h_{\mathbf{D}}=4$, in the vicinity of $D=4$, we have
\begin{equation}
\frac{1}{N} \frac{U_{eff} \left(\sigma \right)}{ \lambda ^4} =
\frac{1}{2 G_R} \frac{\sigma ^2}{ \lambda ^4} - \frac{6 \sigma
^2}{(4 \pi)^2 \lambda^2}\left( \frac{1}{4-D} - \gamma + \ln{4 \pi}
+ \frac{1}{3} \right) +   \frac{\sigma ^4}{(4 \pi)^2 \lambda^4}
\left( \frac{1}{4-D} -\gamma + \ln{4 \pi} + \frac{3}{2} -
\ln{\frac{\sigma ^2}{\lambda^2}}\right) . \label{U1d0}
\end{equation}

We can now compare this equation with the corresponding expression
obtained using the cut-off
regularization~\cite{Inagaki1995,Zhou2003},
\begin{equation}
\frac{1}{N} \frac{U_{eff} \left(\sigma  \right)}{ \lambda  ^4} =
\frac{1}{2 G} \frac{\sigma ^2}{ \lambda  ^4} - \frac{6 \sigma
^2}{(4 \pi)^2 \lambda ^2}\left( \ln{\frac{\Lambda ^2}{\lambda
^2}}- \frac{2}{3} \right) + \frac{\sigma ^4}{(4 \pi)^2 \lambda ^4}
\left(\ln{\frac{\Lambda ^2}{\lambda ^2}} + \frac{1}{2} -
\ln{\frac{\sigma ^2}{\lambda ^2}}\right), \label{U1d0cutoff}
\end{equation}
where the cut-off parameter $\Lambda $ must be larger than the
scale $\lambda $. Thus, the zeta-function and cut-off methods are
equivalent through the correspondence $1/(4-D) -\gamma + \ln{4
\pi} + 1 \leftrightarrow \ln{\Lambda ^2/\lambda^2}$. Hence, the
non-trivial solution of the gap equation derived from
Eq.~(\ref{U1d0}) can be written as
\begin{equation}
\frac{1}{G_c} - \frac{1}{G_0} = - \frac{1}{ m \lambda  ^{2}}
\left. \frac{\partial  }{\partial \sigma}U_{1} \left(\sigma
\right)\right|_{\sigma = m} = \frac{4 m ^2}{(4 \pi)^2 \lambda ^2}
\ln{\frac{\Lambda ^2}{m ^2}}, \label{gape1}
\end{equation}
where we have defined the dimensionless coupling constant $G_c =
\lambda^{2} G_R$, and
\begin{equation}
 \frac{1}{G_0} = \left.\frac{\partial  }{\partial \sigma}U_{eff}
 \left( \sigma \right)\right|_{\sigma \rightarrow 0} =
 \frac{12}{(4 \pi)^2} \left( \ln{\frac{\Lambda ^2}{\lambda^2}} -
 \frac{2}{3} \right). \label{G0}
\end{equation}
It is possible to identify the constant $G_0$ in Eq.~(\ref{gape1})
acting as a critical parameter; when $ G_c > G_0 $ we have a
dynamically generated fermion mass. The value for $G_0$ can be
chosen by fixing values for the mass scale $\lambda$ and the
cut-off $\Lambda$ from phenomenological arguments.

To take into account finite-size and temperature effects, we have
to analyze the modified gap equation,
\begin{equation}
\frac{1}{G_c} - \frac{1}{G_0}= -  \frac{1}{ \overline{m} \lambda
^{2}} \left.\frac{\partial }{\partial \sigma}U_{1} \left(\sigma
;\left\{a_j\right\}, \left\{c_j\right\} \right)\right|_{\sigma =
\overline{m}}, \label{gape2}
\end{equation}
where $\overline{m} = \overline{m}(\{ a_j \},\{ c_j \})$ is the
boundary-dependent fermion mass. Then, using Eqs.~(\ref{U1odd})
and (\ref{U1even}), the modified gap equation, Eq.~(\ref{gape2}),
becomes
\begin{equation}
\frac{1}{G_c} = \frac{1}{G_0} + \frac{4}{ \lambda
^{2}V_d(4\pi)^{(D-d)/2} } \Gamma \left( 1 - \frac{D-d}{2} \right)
Z_{d}^{\overline{m}^2 }\left( -\frac{D-d}{2}+1;
\left\{a_j\right\}, \left\{c_j\right\} \right) , \label{gape3}
\end{equation}
for $d=1,3$, while it has the form
\begin{equation}
\frac{1}{G_c} = \frac{1}{G_0} + \frac{4}{ \lambda ^{2} V_2 4\pi }
\left\{ Z_{2}^{\overline{m}^2 \,\prime} \left(
0;\left\{a_j\right\}, \left\{c_j\right\} \right) +
Z_{2}^{\overline{m} ^2}\left( 0 ;\left\{a_j\right\},
\left\{c_j\right\} \right)\left[ \ln{\lambda ^2} - \gamma - \psi
\left( 1 \right) \right] \right\}, \label{gape4}
\end{equation}
for $d=2$. Also, for $D=4$ we find
\begin{equation}
\frac{1}{G_c} = \frac{1}{G_0} - \frac{4}{ \lambda ^{2} V_4 } \,
{\rm FP}  \left[ Z_{4}^{\overline{m}^2 }\left( 1;
\left\{a_j\right\}, \left\{c_j\right\} \right) \right],
\label{gape5}
\end{equation}
where ${\rm FP}[Z_{4}^{\overline{m}^2}]$ means the finite part of
$Z_{4}^{\overline{m}^2 }$. Thus, taking the fermion mass
approaching to zero in Eqs.~(\ref{gape3}), (\ref{gape4}) and
(\ref{gape5}), we obtain the critical values of the coupling
constant $G_c$ with the corrections due to the presence of
boundaries, for the cases $d=1,3$, $d=2$ and $d=4$, respectively.
In this context, $\left. Z_{d}^{\overline{m}^2}
\right|_{\overline{m}^2 \rightarrow 0}$ reduces to a homogeneous
generalized Epstein zeta-function $Z_{d}$.

The construction of the analytical continuation for $Z_{d}$ can be
implemented using the generalized recurrence formula
\begin{equation}
Z_{d} \left( \nu; \left\{a_j\right\}, \left\{c_j\right\} \right) =
\frac{\Gamma\left( \nu - \frac{1}{2} \right)}{ \Gamma(\nu)}
\sqrt{\frac{\pi}{a_d}}\, Z_{d-1} \left(\nu -
\frac{1}{2};\left\{a_{j \neq d}\right\}, \left\{c_{j \neq
d}\right\} \right) +  \frac{4 \pi^s }{\Gamma(\nu)} W_d \left( \nu
- \frac{1}{2}; \left\{a_j\right\}, \left\{c_j\right\} \right),
\label{epstein4(1)}
\end{equation}
where the symbol $\left\{a_{j \neq d}\right\}$ means that the
parameter $a_d$ is excluded from the set $\left\{a_{j}\right\}$,
and
\begin{equation}
W_d \left( \eta ; \left\{a_j\right\}, \left\{c_j\right\} \right) =
\frac{1}{\sqrt{a_d} } \sum_{\{n_{j \neq d} \in \mathbb{Z}\}}
\sum_{n_d=1}^{\infty}\cos{\left(2\pi n_d c_d \right)} \left(
\frac{ n_d}{\sqrt{a_d} X_{d-1}}  \right)^{\eta} K_{\eta}\left(
\frac{2\pi n_d}{\sqrt{a_d} } X_{d-1} \right) \; ; \label{Wd}
\end{equation}
in the above equation
$X_{d-1}=\sqrt{\sum_{k=1}^{d-1}a_{k}\left(n_{k} + c_k \right)^{2}}
$.

\subsubsection{Finite-size effects on the phase structure}

We now examine the gap equation in the limit $\overline{m}
\rightarrow 0$ in order to determine the $L_j$-dependent critical
curves of the phase diagram. First, we study the compactification
of spatial coordinates at zero temperature; for simplicity, we
consider the same compactification length and antiperiodic
boundary conditions for all coordinates, i.e. $L_j=L$ and $c_j
=1/2$ for $j=1,\dots,d$. In this case, from
Eqs.~(\ref{gape3}-\ref{Wd}), we obtain
\begin{equation}
\frac{1}{G_c} = \frac{1}{G_0} - \frac{A_d }{ (L \lambda)^2},
\label{Gc}
\end{equation}
where $A_1 =  1/6$ and
\begin{eqnarray}
A_2 & =&  A_1 - \frac{4}{\pi} \sum_{n_1 = - \infty}^{\infty}
\sum_{n_2 = 1}^{\infty}(-1)^{n_2} \left( \frac{\left| n_1 +
\frac{1}{2}\right| }{n_2} \right) ^{\frac{1}{2}}
K_{-\frac{1}{2}}\left( 2 \pi n_2 \left| n_1 + \frac{1}{2}\right|
\right) \approx 0.22,\nonumber \\
 A_3 & = &  A_2 - \frac{4}{\pi} \sum_{n_1,n_2 = - \infty}^{\infty}
 \sum_{n_3 = 1}^{\infty}(-1)^{n_3}  K_{0}\left( 2 \pi n_3
 \sqrt{\sum_{i=1}^{2}\left( n_i +
 \frac{1}{2}\right)^2  }\right)
 \approx 0.26. \nonumber
\end{eqnarray}

\begin{figure}[htbp]
\begin{center}
  \begin{minipage}[b]{0.4\linewidth}
    \centering
    \includegraphics[width=\linewidth]{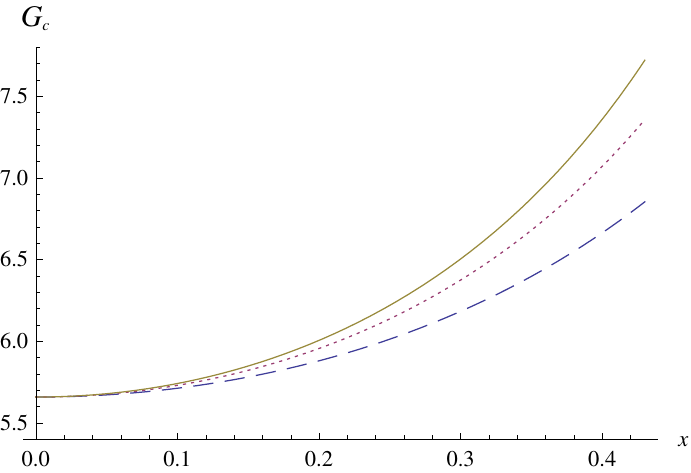}
    \caption{The critical coupling constant $G_c$, Eq.~(\ref{Gc}), as a
function of $x = (L \lambda )^{-1}$.  The dashed, dotted and solid
lines correspond to the cases of $d=1,2,3$ compactified
dimensions. We fixed $G_0 \simeq 5.66$.}
\label{figura1Nova}\vspace{1.5cm}
  \end{minipage}
  \hspace{1.0cm}
  \begin{minipage}[b]{0.4\linewidth}
    \centering
    \includegraphics[width=\linewidth]{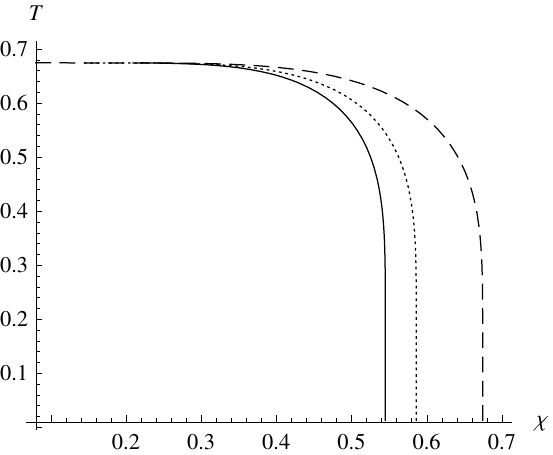}
    \caption{ The phase diagram in the $(\chi,T)$-plane, where $\chi =
(L m )^{-1}$ and $ T = (\beta m )^{-1}$. Dashed, dotted and solid
lines represent $d=2,3,4$ compactified dimensions, respectively.
For each case the non-trivial mass phase corresponds to the region
below the corresponding line. The ratio between the parameters
$\Lambda $ and $m$ is taken as $\Lambda/m= 4.46$.} \label{fig2LN}
  \end{minipage}
  \end{center}
\end{figure}

To illustrate the results above, in Fig.~\ref{figura1Nova} we plot
the critical coupling constant $G_c$, given by Eq.~(\ref{Gc}), as
a function of $x = (L \lambda )^{-1}$; we also fix the value of
$G_0 \simeq 5.66$ by choosing $\Lambda \approx 1.25 \,{\rm GeV}$
and $\lambda \approx 280 \,{\rm MeV}$~\cite{Volkov2005}. For each
of the three cases, the chiral breaking region lies above the
corresponding curve. Since $G_c$ increases as $L$ decreases, we
infer that diminishing the size of the system would require a
stronger interaction to maintain it in the chiral breaking region.

%%%%%%%%%%%%%%%%
%%%%%%%%%%%%%%%

Consider now the system with compactified spatial dimensions at
finite-temperature. We identify the first compactified coordinate
with the Euclidian (imaginary) time, taking $L_1=\beta$. To study
the $(L,T)$-dependent phase diagram, we have to analyze the
critical equation obtained from Eqs.~(\ref{gape1}) and
(\ref{gape2}), that is
\begin{equation}
\frac{4 m^2}{ (4 \pi)^2 \lambda ^2} \ln{\frac{\Lambda ^2}{m^2}} +
\frac{1}{ \overline{m} \lambda^2 } \left.\frac{\partial
}{\partial \sigma}U_{1} \left(\sigma ;\left\{a_j\right\},
\left\{c_j\right\} \right)\right|_{\sigma = \overline{m}} =0,
\label{critical}
\end{equation}
where $a_1=4 \pi^2/ L_1^2 \equiv 4 \pi^2 / \beta ^2$ and, for
simplicity, we take $a_2=\cdots=a_d=4 \pi^2 / L^2$. Proceeding as
before, using again Eqs.~(\ref{gape3}-\ref{Wd}) and taking the
limit $\overline{m}\rightarrow 0$, we get:
\begin{equation}
\frac{4 m^2}{ (4 \pi)^2 } \ln{\frac{\Lambda ^2}{m^2}} -
\frac{A_1}{L^2 } + \frac{4}{\pi L } \sum_{n_1 = - \infty}^{\infty}
\sum_{n_2 = 1}^{\infty}(-1)^{n_2} \left( \frac{\left| n_1 +
\frac{1}{2}\right| }{n_2 \beta L  } \right)^{\frac{1}{2}}
K_{-\frac{1}{2}}\left( 2 \pi n_2 \frac{\beta}{ L}\left| n_1 +
\frac{1}{2}\right| \right)  = 0 \; ,\;\;\;{\rm{for}}\; d=2 ;
\label{ft2}
\end{equation}
\begin{equation}
 \frac{4 m^2}{ (4 \pi)^2 }
\ln{\frac{\Lambda ^2}{m^2}} - \frac{A_2}{L^2 } + \frac{4}{\pi L^2}
\sum_{n_1, n_2 = - \infty}^{\infty} \sum_{n_3 =
1}^{\infty}(-1)^{n_3}  K_{0}\left( 2 \pi n_3 \frac{\beta}{
L}\sqrt{ \left( n_1 + \frac{1}{2} \right)^2 + \left( n_2 +
\frac{1}{2} \right)^2 } \right) = 0 \; , \;\;\;{\rm{for}}\; d=3 ;
\label{ft3}
\end{equation}
\begin{equation}
\frac{4 m^2}{ (4 \pi)^2 } \ln{\frac{\Lambda ^2}{m^2}} -
\frac{A_3}{L^2 } + \frac{4}{\pi L^3} \sum_{n_1, n_2, n_3 = -
\infty}^{\infty} \sum_{n_4 = 1}^{\infty}(-1)^{n_4} \left( \frac{
n_4 \beta L }{\sqrt{\sum_{i=1}^{3} \left( n_j + \frac{1}{2}
\right)^2 }} \right)^{\frac{1}{2}}  K_{\frac{1}{2}}\left( 2 \pi
n_4 \frac{\beta}{ L}\sqrt{\sum_{i=1}^{3} \left( n_j + \frac{1}{2}
\right)^2} \right) = 0 \; , \;\;\;{\rm{for}}\; d=4 . \label{ft4}
\end{equation}

The phase diagrams corresponding to Eqs.~(\ref{ft2}), (\ref{ft3})
and (\ref{ft4}) are plotted in Fig.~\ref{fig2LN}, where $\chi= (L
m )^{-1}$ and $ T = (\beta m )^{-1}$ are the inverse of the
compactification length and the temperature, respectively, in
units of $m$. Chiral breaking regions are below the lines while
above them the system is in the chiral restoration phase. The
critical temperature is maximum ($\approx 0.68 m$) in free space
($ \chi \rightarrow 0$), decreases as $\chi$ increases and
vanishes at a critical value of the size of the system, $L_c$,
corresponding to the minimal size of the system sustaining the
chiral broken phase. For $D=4$, taking $m \approx 300\,{\rm MeV}$,
one finds $T_c \simeq 204 \,{\rm MeV}$ and $L_c \simeq 1.22 \,{\rm
fm}$.

The model can also be analyzed at finite density with or without
compactified spatial dimensions. In both cases, it is found that
nonzero values of the chemical potential can alter the order of
the phase transition~\cite{AbreuMalsaes2009}. Next, we discuss
properties of the finite-size fermion-fermion condensates under
the influence of an external magnetic field.

\subsubsection{Magnetic  effects in the compactified system}

An interesting aspect of the analysis, is to study  changes in the
phase transition induced by the finite size of the system  in a
magnetic background. These effects on four-fermion models, have
been and still are the subject of intense
investigation~\cite{Kim1994,He1996,Vshivtsev1998,Kogut2001,Kiriyama2003,
Beneventano2004,Gamayun2005,Kiriyama2006,Ebert2008,
AbreuMarcelo2006,AbreuMalsaes2009,Abreumals2010,Abreu2011,Abreu2011a}.
The influence of an external magnetic field on the massless
four-dimensional NJL model  is considered. The Lagrangian density
is
\begin{equation}
{\cal L} = \bar{q} \; \left( i \gamma^{\mu}\partial_{\mu} - Q
\gamma^{\mu}A_{\mu} \right) \; q + G \left[ \left( \bar{q} q
\right)^2 + \left( \bar{q} i \gamma _5 \overrightarrow{\tau} \, q
\right)^2 \right], \label{NJL}
\end{equation}
where $q$ and $\bar{q}$ are quark spinors carrying $N_f = 2$
flavors and $N_c = 3$ colors. The components of the vector
$\overrightarrow{\tau}$  are the generators in the flavor space,
$Q$
 electric charge of the quark fields $(Q_u=2e/3, Q_d = - e/3 )$,
and $A^{\mu}$ is the four-potential associated to an external
uniform magnetic field. We choose the gauge $A^{\mu} = (0, -x_{2}
H, 0 , 0)$, with $H$ being constant, which implies that the
external magnetic field, ${\bf H}$, is parallel to the z axis. The
Lagrangian density ${ \cal L}$ is invariant under global chiral
transformations, i.e. $ q \rightarrow \exp{(i \theta \gamma _5
\tau ^3/2 )}\; q$.

We introduce auxiliary fields $\sigma $ and $\pi$, defined by $2 G
\bar{q}  q  \equiv \sigma$ and $-2 G \bar{q}  i \gamma ^5 \tau ^3
\equiv \pi =\pi ^3 $ (we assume $\pi^1 = \pi ^2 = 0$). Then the
Lagrangian density becomes,
\begin{equation}
\widetilde{{\cal L}} = \bar{q} \;  \left( i
\gamma^{\mu}\partial_{\mu} - Q \gamma^{\mu}A_{\mu}  - \sigma - i
\gamma ^5 \tau ^3 \pi \right) \; q - \frac{1}{4 G} \left( \sigma
^2 + \pi ^2 \right),  \label{L2}
\end{equation}
which, after integration over the fermion fields $ q $ and $
\bar{q}$ generates the effective action,
\begin{eqnarray}
\Gamma_{eff} (\sigma, \pi)  =  - \int d ^4 x \frac{1}{4 G} \sigma
^2 - \frac{i}{2} \rm{Tr} \ln{\left( i \gamma^{\mu}\partial_{\mu} -
Q \gamma^{\mu}A_{\mu} - \sigma - i \gamma ^5 \tau ^3 \pi \right)}
, \label{eff_action}
\end{eqnarray}
where $\rm{Tr}$ means  the trace over the color, flavor, Dirac
matrices and coordinate spaces.

We will study the pure chiral sector ($\pi = 0$) and consider the
mean-field approximation, which implies an uniform $\sigma$. Then
the effective potential from Eq.~(\ref{eff_action}) has the form,
\begin{eqnarray}
U (\sigma) & = & \frac{ \sigma ^2}{4 G} + \frac{i}{2 V} \rm{Tr}
\ln{\left( i \gamma^{\mu}\partial_{\mu} - Q \not{\!A} - \sigma
\right)} , \label{pot1}
\end{eqnarray}
where $V$ is the four-dimensional volume. The fermion field,
minimally coupled to the external magnetic field, obeys the Dirac
equation,
\begin{equation}
\left(\gamma^{\mu}\partial_{\mu} - Q \gamma^{\mu}A_{\mu} - \sigma
\right) q = 0. \label{dirac}
\end{equation}
Applying again the Dirac operator,  each component of $ q $
satisfies the equation
\begin{equation}
\left[ \left( i\partial + Q A \right)^2 - \frac{Q}{2} \sigma^{\mu
\nu} F_{\mu \nu} - \sigma ^2 \right] q = 0, \label{dirac2}
\end{equation}
where $\sigma^{\mu \nu} = i [\gamma ^{\mu}, \gamma ^{\nu}] /2 $
and $F_{\mu \nu} = \partial _{\mu} A_{\nu} - \partial _{\nu}
A_{\mu}$.

In the presence of the external magnetic field the natural basis
is the set of the normalized eigenfunctions of the Landau basis.
Then, the  solutions of Eq.~(\ref{dirac2})
are~\cite{Lawrie1994,Lawrie1997,Ritus1972,Leung2006},
\begin{equation}
q(x) = e^{i (p_0 x_0 - p_1 x_1 - p_3 x_3)} u (x_2),
\end{equation}
where $u(x_2)$ satisfies the  equation,
\begin{eqnarray}
\left[ p_2 ^2 + Q^2 H^2 \left( x_2 - \frac{p_1}{Q H} \right)^2
\right] u (x_2) = \left[ p_0 ^2 - p_3 ^2  - \sigma ^2  \mp Q H
\right] u (x_2), \label{dirac3}
\end{eqnarray}
with
\begin{equation}
u_n(x_2) = \frac{1}{\sqrt{2^n n!}} \left(\frac{Q
H}{\pi}\right)^{\frac{1}{4}} H_n \left(\sqrt{Q H}\left[x_2 -
\frac{p_1}{Q H}\right]\right) ;
\end{equation}
here, $H_n$ are the Hermite polynomials and the energy spectrum
provides the dispersion relation, $p_0 ^2 = p_3 ^2 + \sigma ^2 +(2
n + 1 - s ) Q H$, with $n = 0, 1, 2, ...$, corresponding to the
Landau levels, and $s = \pm 1$.

The introduction of the Landau basis implies a change in the
momentum space integrations, as
\begin{equation}
\int \frac{d ^4 p}{(2 \pi)^4} f (p) \rightarrow \frac{|Q| H}{2
\pi} \sum _{s = \pm 1} \sum _{n = 0}^{\infty} \int \frac{d ^2
p}{(2 \pi)^2} f (p_0, p_3, n, s) , \nonumber
\end{equation}
which leads to writing Eq.~(\ref{pot1}) in the form,
\begin{equation}
U (\sigma) = \frac{ \sigma ^2}{4 G} + \frac{i}{2 } {\rm {tr}}
\frac{|Q| H}{2 \pi} \sum_{s = \pm 1} \sum _{n = 0}^{\infty} \int
\frac{d ^2 p}{(2 \pi)^2} \, \ln{\left[ p_0 ^2 - p_3 ^2 - \sigma ^2
-(2 n + 1 - s ) Q H \right]} , \label{pot2}
\end{equation}
where $\rm{tr}$ means the trace over the color and flavor degrees
of freedom.

Magnetic effects are accounted for by the introduction of the
Landau basis, which ``take out" two spatial dimensions.  This
corresponds, using Eq.~(\ref{Matsubara1}), to writing  the
generalized Matsubara prescription in the form,
\begin{equation}
\int \frac{d^2 p_E }{(2 \pi)^2} f (p_0, p_3, n, s) \rightarrow
\frac{1}{\beta L} \sum_{l,m = -\infty}^{\infty}
f(\omega_{l},\omega_{m}, n , s), \label{Matsubaraaqui}
\end{equation}
where  we have performed the replacements,
\begin{eqnarray}
p_0 & \rightarrow & \omega_{l} = \frac{2 \pi }{\beta} \left( l +
\frac{1}{2}\right) - i \mu;\;
l = 0,\pm 1, \pm 2,\cdots, \nonumber \\
p_3 & \rightarrow & \omega_{m} = \frac{2 \pi }{L} \left( m +
\frac{1}{2} \right);\;\;\; m = 0,\pm 1, \pm 2, ..., \nonumber
\end{eqnarray}
with $\mu $ being the chemical potential.

Using Eq.~(\ref{Matsubaraaqui}) in Eq.~(\ref{pot2}), after some
manipulations, the effective potential with magnetic,
finite-temperature and finite-size effects is given as,
\begin{equation}
U (\sigma)   = \frac{ \sigma ^2}{4 G} - \frac{N_c}{2 } \sum _{f}
\frac{|Q_f| H}{2 \pi \beta L} \sum _{s = \pm 1} \sum _{n = 0
}^{\infty} \sum _{l, m = -\infty}^{\infty} \ln{\left[ \omega _l ^2
+ \omega _m ^2 + \sigma ^2 +(2 n + 1 - s ) |Q_f | H \right]},
\label{pot3}
\end{equation}
where $f$ is the sum over the flavor indices.

The effective potential can be written in terms of Epstein-Hurwitz
generalized zeta-functions,  $Z^{c^2}_2(\eta)$,  with $c^2=\sigma
^2 +(2 n + 1 - s ) |Q_f | H$,  giving,
\begin{eqnarray}
U  (\sigma) & = &\frac{\sigma ^2 }{4 G} + \sum _{f} \sum _{s = \pm
1} \frac{N_c |Q_f| H}{4 \pi \beta L} Z^{c^2\,\prime}_2 (0),
\label{pot4}
\end{eqnarray}
where the notation
\[
Z^{c^2\,\prime}_2 (0) = \left.\frac{\partial}{\partial
\eta}Z^{c^2\,\prime}_2 (\eta)\right |_{\eta=0}
\]
is used. An analytical continuation of the  generalized
zeta-function $Z^{c^2}_2(\eta)$, to the whole complex
$\eta$-plane, after use of recurrence formulas and some
manipulations~\cite{AbreuMarcelo2006,AbreuMalsaes2009,Elizalde2012}
leads to,
\begin{eqnarray}
Z^{c^2}_2(\eta) & = & \frac {\beta L}{2 \pi} \frac{\Gamma (\eta
-1)}{ \Gamma (\eta)} F _{ 1}  (\eta - 1) + \frac{\beta L }{\pi }
\frac{1 }{\Gamma (\eta)}  F _{ 2} (\eta - 1)+  \frac{\beta
}{\sqrt{\pi} } \frac{1 }{\Gamma (\eta)}  F _{ 3} \left( \eta -
\frac{1}{2} \right). \label{epstein2}
\end{eqnarray}
The functions $F _{ 1} (\nu)$, $F _{ 2}  (\nu)$ and $F _{ 3}
(\nu)$ are respectively,
\begin{eqnarray}
F _{ 1}  (\nu) & = & (|Q_f | H)^{ -\nu } \left[ \zeta \left( \nu,
\frac{\sigma ^2}{|Q_f | H} \right) - \frac{1}{2}\left(
\frac{\sigma ^2}{|Q_f | H} \right)^{-\nu}
\right],  \nonumber \\
F _{ 2} (\nu) & = & 2 ^{- \nu} \sum _{n = 0 }^{\infty} \sum _{ m =
1}^{\infty} \left( \frac{m L}{\sqrt{(2 n + 1 - s ) |Q_f | H +
\sigma ^2 }  } \right)^{\nu } (-1)^{m} K_{\nu } \left( m L
\sqrt{(2 n + 1 - s ) |Q_f | H + \sigma ^2 } \right) \nonumber
\end{eqnarray}
and
\begin{eqnarray}
F _{ 3}  (\nu) & = & 2 ^{1 - \nu} \sum _{n = 0 }^{\infty} \sum _{
m = -\infty} ^{\infty} \sum _{ l = 1} ^{\infty}  \left( \frac{l
\beta} {\sqrt{ \frac{4 \pi^2}{L^2}\left( m+\frac{1}{2} \right)
^{2}+
(2 n + 1 - s ) |Q_f | H + \sigma ^2 } } \right)^{\nu} (-1)^{l} \nonumber \\
& & \times \, \cosh{(\mu \beta l)} K_{\nu} \left( l \beta
\sqrt{\frac{4 \pi^2}{L^2}\left( m+\frac{1}{2} \right) ^{2}+(2 n +
1 - s ) |Q_f | H + \sigma ^2} \right), \nonumber
\end{eqnarray}
where $\zeta \left( \eta,a \right) $ is the inhomogeneous Riemann
$zeta$-function. The function $Z^{c^2\,\prime}_2 (\eta)$ may be
analyzed using the pole structure of Eq.~(\ref{epstein2}) for
$\eta \rightarrow \epsilon  $ ($\epsilon \ll 1$). We find that
\begin{eqnarray}
\frac{d}{d \eta} \left[ \frac{\Gamma (\eta -1 )}{\Gamma (\eta)} F
_{ 1}(\eta)\right] _{\eta \rightarrow \epsilon} & \approx & F _{
1} (\epsilon - 1 ) - F _{ 1}  ^{\prime} (\epsilon - 1 ) -
\epsilon F_{1} (\epsilon - 1 ), \nonumber \\
\frac{d}{d \eta} \left[ \frac{1}{\Gamma (\eta)} F_{
2}(\eta-1)\right]_{\eta \rightarrow \epsilon}  & \approx &
F_{ 2}(\epsilon - 1), \nonumber \\
\frac{d}{d \eta} \left[ \frac{1}{\Gamma (\eta)} F _{3}
\left(\eta-\frac{1}{2} \right)\right] _{\eta \rightarrow \epsilon}
& \approx & F_{3} \left( \epsilon- \frac{1}{2} \right). \nonumber
\end{eqnarray}
Then, for $ \epsilon \rightarrow 0 $,  the effective potential is
\begin{eqnarray}
U(\sigma) & = & \frac{\sigma ^2 }{4 G} + U _{vac} - \sum _{f}
\frac{N_c (|Q_f| H) ^2}{2 \pi ^2} F_4 \left( \frac{\sigma ^2}
{|Q_f | H} \right)+ \sum _{f}\sum _{s = \pm 1} \frac{N_c |Q_f|
H}{4 \pi ^2} \left[  F_2 \left(-1 \right) + \frac{\pi
^{\frac{1}{2}} }{L} F_3 \left( - \frac{1}{2} \right) \right],
\label{pot5}
\end{eqnarray}
where
\[
F_4 (z) = \zeta ^{\prime} \left(-1, z \right) - \frac{1}{2} \left(
z^2 - z \right) \ln{z} + \frac{z ^2 }{4}
\]
and $U_{vac} $ is the vacuum contribution, given as
\begin{equation}
\frac{1}{N_c} U _{vac} = \frac{N_f }{8 \pi ^2 } \left[ \sigma ^4
\ln{\left( \frac{\Lambda + \sqrt{\Lambda ^2 + \sigma
^2}}{\sigma}\right)}- \Lambda (2 \Lambda ^2 +  \sigma
^2)\sqrt{\Lambda ^2 + \sigma ^2} \right], \label{vac}
\end{equation}
$\Lambda$ being a cutoff parameter.

Then the gap equation is obtained from $\left.  \partial
U(\sigma)/
\partial \sigma\right|_{\sigma = M} = 0,$ where $M = M (T, \mu, L,
H) $ is the $(T, \mu, L, H)$-dependent order parameter of the
chiral symmetry breaking transition; equivalently $M$ plays the
role of a dynamical fermion mass, such that when it has a
non-vanishing value, the system is in the chiral broken phase.
This generates one trivial solution, $M = 0$, and other, non
trivial ones, satisfying the equation,
\begin{equation}
\frac{1}{G} -  \sum _{f} \frac{N_c (|Q_f| H) }{ \pi ^2} \; I
\left( \frac{ M ^2}{|Q_f | H} \right) - \frac{1}{M} \left. \frac{
\partial U_{vac} }{\partial \sigma}\right|_{\sigma = M}  - \sum
_{f}\sum _{s = \pm 1} \frac{N_c |Q_f| H}{ \pi ^2} \left[  F_2
\left(0 \right) + \frac{\pi ^{\frac{1}{2}} }{L} F_3 \left(
\frac{1}{2} \right) \right] = 0 , \label{gap2}
\end{equation}
where
\begin{equation}
I(z) = \ln{\Gamma (z)} - \frac{1}{2} \ln{2 \pi} + z -
\frac{1}{2}(2 z - 1) \ln{z}, \nonumber
\end{equation}
and
\begin{eqnarray}
\left(\frac{\partial U_{vac}}{\partial \sigma}\right)_{\sigma = M}
& = & \frac{N_c N_f \, M}{\pi ^2} \left[ \Lambda \sqrt{\Lambda ^2
+ M ^2} - M ^2 \rm{ln}\left( \frac{\Lambda + \sqrt{\Lambda ^2 + M
^2}}{M} \right) \right].
\end{eqnarray}

We recall that $M$ is the effective quark mass, $x = 1/L$, $T =
1/\beta$, $\omega= e H$ is the cyclotron frequency, $\mu$ is the
chemical potential and $G$  is the coupling constant. It is
interesting to introduce the critical coupling, $G_c$; in the
absence of magnetic field, for vanishing temperature and chemical
potential and in free space, it is given by $G_c = \pi^2/N_c N_f =
\pi ^2 / 6$. The region $G>G_c$ is the region with nontrivial
mass.

In what follows all physical quantities are scaled by the
ultraviolet cutoff parameter $\Lambda$, which has to be determined
by fitting to experimental data; we take dimensionless quantities
by performing the changes: $U/\Lambda ^4 \rightarrow  U$;
$M/\Lambda \rightarrow  M$; $x/\Lambda \rightarrow x$; $T /
\Lambda \rightarrow T$; $\omega / \Lambda ^2 \rightarrow  \omega$;
$\mu / \Lambda \rightarrow \mu$; and $1/ G \Lambda ^2  \rightarrow
1/G$.

%%%%%%%%%%%%%%%%
\begin{figure}[htbp]
\begin{center}
  \begin{minipage}[b]{0.4\linewidth}
    \centering
    \includegraphics[width=\linewidth]{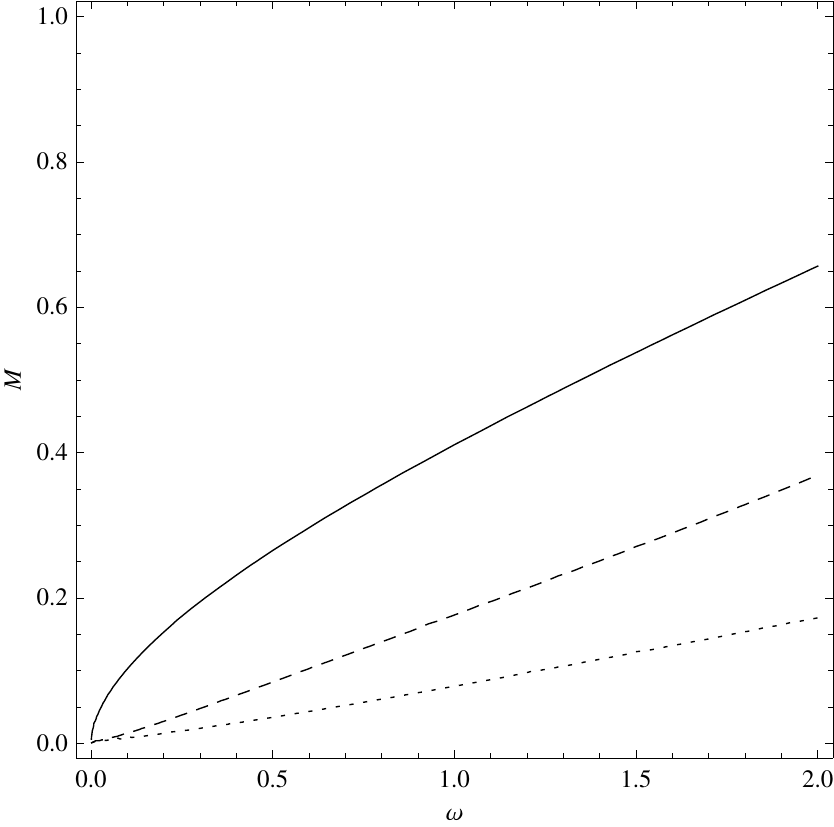}
    \caption{Plot of effective quark mass $M$ versus magnetic field from
Eq.~(\ref{gap2}) at  $T =  0.001$, $\mu =0$, $G = \pi ^2 / 6$.
Solid, Dashed and dotted lines represent $x = 0.01, 0.5$ and $1$,
respectively. }
    \label{Fig7Nova}
  \end{minipage}
  \hspace{0.5cm}
  \begin{minipage}[b]{0.4\linewidth}
    \centering
    \includegraphics[width=\linewidth]{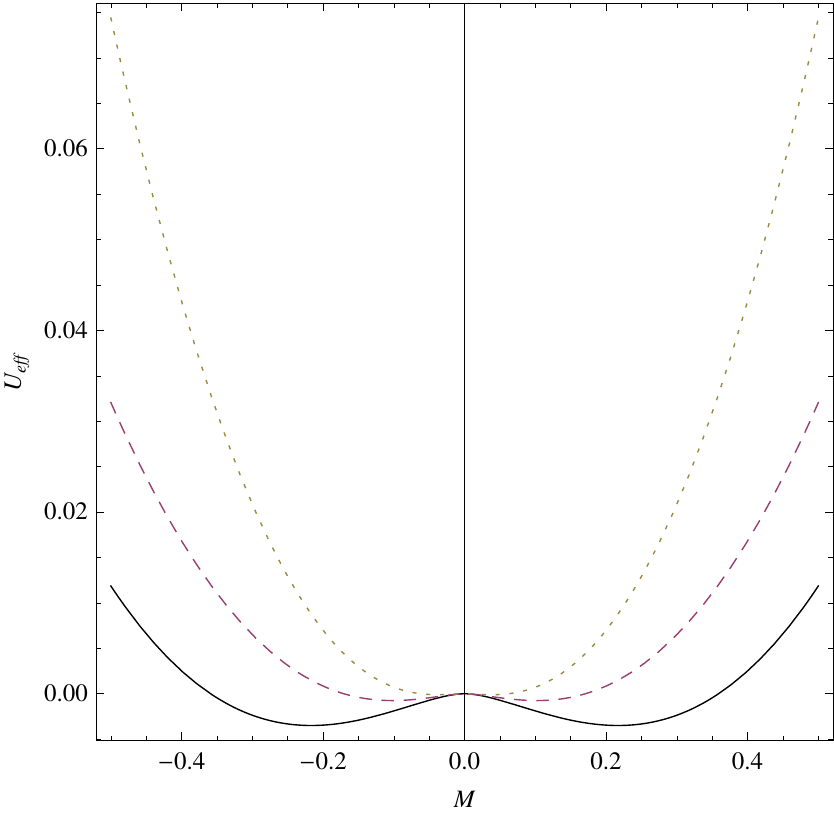}
    \caption{Plot of Effective potential in Eq.~(\ref{pot5}) at   $T =
0.001$, $\mu =0$, $G = \pi ^2 / 12$ and $\omega =2$. Solid, Dashed
and dotted lines represent values of the inverse size of the
system, $ x = 0.01, 0.5$ and $1.5$, respectively.  }
    \label{fig8Nova}
  \end{minipage}
  \end{center}
\end{figure}
%%%%%%%%%%%%%%%

In Fig.~\ref{Fig7Nova} and Fig~\ref{fig8Nova} we plot the
effective quark mass $M$ versus magnetic field  and the effective
potential in Eq.~(\ref{pot5}) in terms  of the inverse size of the
system, $ x $ respectively. From Eq.~(\ref{gap2}) at   $\mu =0$, $
i.e.$  in the absence of boundaries, at zero temperature, chemical
potential and magnetic field. At fixed values of $T$, $\mu$ and
$\omega$, there is a transition from the broken to the unbroken
phase as the size of the system decreases; for fixed values of
$T$, $\mu$ and $x$, there is a transition from the unbroken to the
broken phase as the magnetic field increases. We find that the
increasing of the magnetic field tends to drive the system to an
ordered phase, an effect usually referred as magnetic catalysis.
In addition, in Fig.~\ref{Fig7Nova}, $M$ assumes a vanishing value
at $\omega, \mu, T \approx 0$, an expected result for the coupling
constant at the critical value. In the plot above of the effective
potential, we see that the nature of phase transition is of second
order. The first order transition is not observed, which could be
a combined effect of insufficient  magnetic field and finite-size
effects.

Some points are worthy to be emphasized. For given values of the
chemical potential and magnetic field, the appearance of the
broken phase is inhibited as the size of the system decreases,
$i.e.$, the decreasing of the size of the system makes it
difficult to maintain long-range correlations and thus favors
disorder. There exists a minimal size of the system below which
the transition ceases to exist, in other words,  the
chiral-condensate phase cannot be sustained below this minimal
size. Enhancement of the broken phase for a system with a finite
size, occurs with  increasing  magnetic field, $i.e.$, the
effective mass $M$ increases with $\omega$ at a fixed size of the
system. Also, the dependence of $M$ on the intensity of the
magnetic field is modified by the size of the system.

\section{Compactified extra dimensions in low energy physics:
Electrodynamics in $\Gamma _{5}^{1}$}

Extra dimensions are explored; if they exist, they would manifest
themselves in low energy physics. In particular, some consequences
in quantum electrodynamics ${QED}$ would follow if our world were
5-dimensional. Some effects at the one-loop order are investigated
under the assumption  of a non-trivial vacuum with a non-vanishing
magnetic flux along the extra compactified fifth dimension.
Modifications of the vacuum polarization in extra dimension are
considered. This is a procedure that has been employed in high
energy physics to investigate, for instance, baryogenesis and
electroweak transition, as a generalized version of the standard
model~\cite{Roy2009,Arkani-Hamed1998,Panico2005,Panico2006,Antoniadis1990,%
Dvali2002,Hall2002,Kubo2002,Burdman2003,Fairlie1979,Manton1979,Serone2005,%
Arkani-Hamed2002,Arkani-Hamed2002a}. In particular, Panico and
Serone consider the 5-D non-abelian gauge theory at finite
temperature to study the electroweak phase transition. The
compactified fifth dimension is then taken with a length $L$
having non-vanishing magnetic flux along it. This leads a
first-order phase transition at temperature $T\sim1/L$.

It is reasonable to consider that the presence of  compactified
extra dimensions may have effect not only in particle physics, but
also in gravitation and atomic physics.  This would imply that
signatures of these extra dimensions might appear in phenomena
governed by Newtonian gravitational forces, as it has been
explored
experimentally~\cite{Floratos2012,Floratos2011,Fischbach2001}.
Also, low energy probes for the existence of extra dimensions are
proposed~\cite{McDonald2010}, such as the effects on the anomalous
magnetic moment of the muon, which has been associated with
extra-dimensional excitations of the photon and of the W and Z
bosons~\cite{Nath1999}. In addition,  a value of $g_{\mu}\neq 2$
is not excluded by experiments~\cite{Aubert2009}. Maybe this
discrepancy can be explained as effects due to
extra-dimensions~\cite{Roy2009,Ttira2010}. In particular, Roy and
Bander~\cite{Roy2009} performed a calculation of $g_{\mu}-2$,
assuming a 5-D space-time electroweak theory. They found a
compactification length of the fifth dimension to be $\sim
10^{-16}$m.

Another interesting consequence of the possible existence of extra
dimensions is that the electric charge may not be conserved
exactly~\cite{Dubovsky2000,Ignatev1979,Mohapatra1987,Suzuki1988a}.
In four-dimensional theories, a tiny deviation from electric
charge conservation would lead to contradictions with low-energy
tests of QED~\cite{Dubovsky2000}. These deviations could be fixed
by introducing  millicharged particles~\cite{Suzuki1988a} or by
inclusion of high dimensions~\cite{Dubovsky2000}. In this latter
case, particles initially confined to $4$-dimensional space-time
could, under some circumstances, migrate to  extra dimensions. The
idea is that if the particles are electrically charged, their
migration from 4-dimension into extra dimensions would be
considered as a non-conservation of electric charge. In atomic
physics, measurements of the asymptotic quantum effects on
excitations of Rydberg atoms have also been carried
out~\cite{Jentschura2005,Odom2006}, and non-conservation of charge
can be investigated in similar experiments.

In the following, some details of quantum electrodynamics on
$\Gamma^{1}_{5}$ are reviewed as a possible theory for
extra-dimension physics. Although an abelian theory is considered,
the primary focus is on physical aspects that can be extended to
non-abelian fields.

\subsection{Quantum Electrodynamics in a compactified space}

Consider the case of compactification of one spatial dimension in
an ${\mathbb{R}}^D$ Euclidean spacetime, such that the topology of
the resulting manifold is $\Gamma
_{D}^{1}=({\mathbb{S}}^{1})\times { \mathbb{R}} ^{D-1}$. For one
compactified spatial dimension, from Eqs.~(\ref{Matsubara1}),
modified Feynman rules are
\begin{equation}
\int \frac{dk_s}{2\pi }\rightarrow \frac 1{L}\sum_{n=-\infty
}^{+\infty }\;;\;\;\;k_s\rightarrow \frac{2(n+c)\pi }{L}\;,
\label{Matsubara2}
\end{equation}
where $k_s$ stands for the momentum component corresponding to the
compactified dimension, while $L$  is the size of the compactified
spatial dimension.  The quantity $c$ equals $0$ or $\frac{1}{2}$
for periodic or antiperiodic boundary condition, respectively .

One-loop effects for $QED_{3+1}$ are investigated with an extra
compactified dimension, i.e. a 5-dimensional theory, in a
non-trivial vacuum for the gauge field, defined by a non-vanishing
component along the extra compactified dimension. The system is
defined in terms of an Euclidean action, $S$, of the form,
\begin{equation}\label{eq:defs0}
S({\mathcal A};{\bar\Psi},\Psi) \;=\; S_g({\mathcal A}) \,+\,
S_f({\mathcal A};{\bar\Psi},\Psi)\;,
\end{equation}
where $S_g$ and $S_f$ denote the $U(1)$ gauge field and fermion
actions, respectively. The gauge action is assumed to be
\begin{equation}\label{eq:defsg}
S_g({\mathcal A}) \;=\; \frac{1}{4} \, \int d^5x \, {\mathcal
F}_{\alpha\beta} {\mathcal F}_{\alpha\beta} \;,
\end{equation}
with ${\mathcal F}_{\alpha\beta}\equiv\partial_\alpha {\mathcal
A}_\beta -
\partial_\beta {\mathcal A}_\alpha$;  we adopt the convention
\begin{equation}
\alpha, \beta, \dots  \,=\, 0,\,1,\,2,\,3,\,4 \;,\;\;\; \mu,\nu,
\dots \,=\, 0,\,1,\,2,\,3\;,\;\;\;
 d^5 x \;\equiv\; d^{3+1}x \,dx_4 \,\,\, {\mathrm{and}}\,\,\,
 dx_4=ds,
\end{equation}
with $x$ denoting the coordinates $x_\mu$ in ($3+1$) dimensions,
unless the contrary is explicitly stated.  The extra dimension is
compactified with a radius $R$, so that $L = 2 \pi R\,.$ The Dirac
action, $S_f$, is given by
\begin{equation}
 S_f({\bar\Psi},\Psi;{\mathcal A}) \;=\; \int d^{3+1}x \, ds \; {\bar\Psi}(x,s)
\big( {\mathcal D} + m \big) \Psi(x,s)
\end{equation}
where ${\mathcal D}$ is the $4+1$ dimensional Dirac operator,
\mbox{${\mathcal D}= \gamma_\alpha D_\alpha$}. The covariant
derivative \mbox{$D_\alpha \equiv
\partial_\alpha + i g {\mathcal A}_\alpha$} includes a coupling constant $g$
with dimensions of $({ mass})^{-\frac{1}{2}}$. For Dirac's
$\gamma$-matrices, we assume that $\gamma_s \equiv \gamma_5$,
where the latter is the $\gamma_5$ matrix for the $3+1$ world.

The gauge field configuration ${\mathcal A}_\alpha(x,s)$ is
decomposed into its zero ($A_\alpha$) and non-zero ($Q_\alpha$)
mode components\cite{Panico2005},
\begin{equation}
{\mathcal A}_\alpha(x,s) \;=\; L^{-\frac{1}{2}} \, A_\alpha(x)
\,+\, Q_\alpha(x,s) \;,
\end{equation}
where the two terms are
\begin{equation}
A_\alpha(x) \;=\;  L^{-\frac{1}{2}} \, \int_0^L ds \, {\mathcal
A}_\alpha (x,s) \;, \label{dec1}
\end{equation}
and
\begin{equation}
Q_\alpha(x,s) \;=\; {\mathcal A}_\alpha(x,s) \,-\,
L^{-\frac{1}{2}} \, A_{\alpha}(x) . \label{dec2}
\end{equation}
The factor $L^{-\frac{1}{2}}$ is included  in the zero mode term
for this field to have the usual mass dimension in $3+1$
space-time. The Fourier expansion of the gauge field along the
extra dimension is
\begin{equation}
{\mathcal A}_\alpha (x,s) \;=\; L^{-\frac{1}{2}} \,
\sum_{n=-\infty}^{\infty}\, e^{ i \omega_n s}
\,\widetilde{\mathcal A}_\alpha(x,n) \;,
\end{equation}
with $\omega_n \equiv 2 \pi n/L$, where
\begin{equation}
A_\alpha(x) \,=\, \widetilde{\mathcal A}_\alpha(x,0) \;\;,\;\;\;\;
Q_\alpha(x,s) \,=\, L^{-\frac{1}{2}} \, \sum_{n \neq 0}\, e^{ i
\omega_n s} \,\widetilde{\mathcal A}_\alpha(x,n) \;.
\end{equation}

The concept of dimensional reduction  is defined by the limit of a
compactified theory, where the size of the compactified dimension
goes to zero. At this limit, with finite energy, zero is the only
mode that survives along the compactified dimension. To
dimensionally reduce a theory taking into account the Matsubara
prescription, amounts to considering a compactified theory with
small compactification lengths. Then only the zero mode component
of the gauge field is kept. Therefore, we have
\begin{equation}
S_g({\mathcal A}) \; \to \; S_g (A) \;=\; S_g(A_\mu, \,A_s) \;,
\end{equation}
where
\begin{equation}
S_g(A_\mu, A_s)=\int d^{3+1}x \, \Big[ \frac{1}{2} \partial_\mu
A_s
\partial_\mu A_s +\frac{1}{4} F_{\mu\nu}(A) F_{\mu\nu}(A) \Big],
\end{equation}
with $F_{\mu\nu}(A) \equiv \partial_\nu A_\nu - \partial_\nu
A_\mu$.

For the fermion action $S_f$, the reduction leads to
\begin{equation}
S_f({\mathcal A}; {\bar\Psi}, \Psi ) \; \to \;
S_f(A_\mu,A_s;{\bar\Psi},\Psi) \;.
\end{equation}
Dimensional reduction of the fermion field is performed taking
into account that, in the calculation of the effective gauge field
action, its only contribution comes from the fermion loop. This
loop may be represented as a series of $3+1$ dimensional  loops,
each one with a different  mass  (the different Matsubara
frequencies). Contributions of heavier modes  may be relatively
suppressed, but the fact that there is an infinite number of them
does not allow truncating that series.

After dimensional reduction, the fermion action becomes
\begin{equation}
S_f \;=\; \int d^{3+1} x \, \int_0^L ds \; {\bar\Psi}(x,s) \big(
\not \!\! D + \gamma_s D_s  + m \big)  \Psi(x,s),
\end{equation}
with
\begin{equation}
\not \!\! D \,=\, \gamma_\mu (\partial_\mu + i e A_\mu ) \;\; D_s
\,=\,
\partial_s + i e A_s  ,
\end{equation}
where we have introduced a new, dimensionless coupling constant $e
\equiv g L^{-\frac{1}{2}}$, which plays the role of the electric
charge in $3+1$ dimensions.

Considering the form of the gauge transformations in terms of the
decomposition into zero and non-zero modes,  $A_\mu$ transforms as
a standard gauge field in $3+1$ dimensions~\cite{Zinn-Justin1996},
\begin{equation}
\delta A_\mu (x) \,=\, \partial_\mu \alpha (x);
\end{equation}
the extra dimensional component $A_s$  is a scalar in
$3+1$-dimensions and is shifted by a constant, i.e.
\begin{equation}
\delta A_s (x) \,=\,\Omega \;.
\end{equation}
The gauge field is coupled to the charged fermion field, whose
transformation law under simultaneous action of the previous gauge
transformation is
\begin{equation}
\Psi (x,s) \to  e^{-i e [\alpha(x) + \Omega s ]} \, \Psi (x,s)
\,,\;\;\;\; \bar{\Psi} (x,s) \to  e^{ i e [\alpha(x) + \Omega s ]}
\, \bar{\Psi} (x,s) \;.
\end{equation}
This implies that the constant $\Omega$  must be of the
form~\mbox{$\Omega = 2 \pi k/L e$}, where $k$ is an integer.

\subsection{Effective action and parity conserving term}

Let us first define a part of the effective action, $\Gamma (A)$,
that depends only on the dimensionally reduced gauge field,
\begin{equation}
\Gamma (A) \; \equiv \; \Gamma (A; {\bar\Psi}, \Psi)
\Big|_{{\bar\Psi} = \Psi = 0} \;,
\end{equation}
where $\Gamma (A; {\bar\Psi}, \Psi)$ is the full effective action.
From the functional $\Gamma(A)$, one-particle-irreducible
functions are derived, having as external lines only components of
the gauge field, $A_\mu$, $A_s$. The components $A_\mu$ have a
clear interpretation in $(3+1)$-dimensional space-time. However,
$A_s$, that has no direct meaning, is assumed to have a constant
value, which is determined by an external condition  to the model.
The fermion one-loop gives the only non-trivial contribution to
$\Gamma(A)$, i.e.
\begin{equation}
\Gamma(A) \;=\; \Gamma^{(0)}(A) \,+\,\Gamma^{(1)}(A) \;+\;\cdots,
\end{equation}
where $\Gamma^{(0)}(A) = S_g(A)$ and
\begin{equation}
e^{-\Gamma^{(1)}(A)} \; = \; \int {\mathcal D}\Psi {\mathcal
D}{\bar\Psi} e^{- S_f(A; {\bar\Psi},\Psi) } \;.
\end{equation}

The focus here will be on the effective action for the gauge field
components $A_\mu$, that have a  physical meaning for the
$3+1$-dimensional theory. Following Panico and
Serone~\cite{Panico2005}, $A_s$ is assumed to be such that it
 yields a non-vanishing flux,
\begin{equation}
e \int_0^L ds A_s \;=\; \theta,
\end{equation}
where $\theta$ is a constant.  A simple solution is to take $A_s$
as a constant given by
\begin{equation}
A_s \;=\; \frac{\theta}{e L} \;,
\end{equation}
which fixes the gauge. Since the gauge transformations shift $A_s$
by an integer multiple of $2\pi/e L$,  the value of $\theta$  may
be fixed in the domain $ 0 \leq  \theta < 2 \pi  $. This will be
assumed in what follows. Such a gauge field configuration may be
interpreted as a topological effect, in the sense that it
corresponds locally, although  not globally, to a pure gauge
field configuration~\cite{Ttira2010}.

Performing a Fourier expansion of the fermion field along the $s$
coordinate,
\begin{equation}
\Psi(x,s) = L^{-\frac{1}{2}} \,\sum_{n=-\infty}^\infty e^{i
\omega_n s} \psi_n(x)\,,\;\;\; {\bar\Psi}(x,s) = L^{-\frac{1}{2}}
\,\sum_{n=-\infty}^\infty e^{ - i \omega_n s} {\bar\psi}_n(x) \;,
\end{equation}
and inserting   into the functional expression for
$\Gamma^{(1)}(A)$, we get,
\begin{equation}\label{eq:sfexpand}
S_f \;=\; \sum_{n=-\infty}^{n=+\infty} \int
d^{3+1}x\,\bar{\psi}_n(x) \left(\not \!\! D + i \gamma_s
\left(\omega_n + \frac{\theta}{L} \right) + m \right) \psi_n(x)
\;.
\end{equation}
Using the same expansion, the fermion measure factorizes in the
form,
\begin{equation}
\mathcal{D}\Psi \mathcal{D}\bar{\Psi}\;=\;
\prod_{n=-\infty}^{n=+\infty} \mathcal{D}\psi_n(x)
\mathcal{D}\bar{\psi}_n(x);
\end{equation}
then the Euclidean action corresponding to each mode $n$ may be
written as
\begin{equation}
\int d^{3+1}x\,\bar{\psi}_n(x) \left( \not \!\! D + i \gamma_s
\left( \omega_n + \frac{\theta}{L} \right) + m \right)
\psi_n(x)=\; \int d^{3+1}x \,\,\bar{\psi}_n(x)(\not \!\! D + M_n
\, e^{-i \varphi_n \gamma_5})\psi_n(x),
\end{equation}
where
\begin{equation}
 M_n \equiv \sqrt{m^2+(\omega_n+\theta/L)^2}\;,\;\;\;
\varphi_n = {\rm arctan}(\frac{\omega_n+\theta/L}{m}) \;.
\end{equation}
The  $\gamma_5$ term   implies  that parity symmetry can be
broken. Performing  a change in the fermion variables such that
the explicit dependence on $\gamma_5$ is suppressed, then we have
\begin{equation}
\psi_n(x)\to e^{-i\gamma_5
\varphi_n/2}\psi_n(x)\,,\,\,\,\,\bar{\psi}_n(x) \to
\bar{\psi}_n(x)e^{i\gamma_5 \varphi_n/2} .
\end{equation}
Then  the mode labeled by $n$ has the action,
\begin{equation}
\int d^{3+1}x \,\,\bar{\psi}_n(x)(\not \!\! D + M_n \, e^{-i
\varphi_n \gamma_5})\psi_n(x) \;.
\end{equation}

Now $\Gamma^{(1)}$ may be written as follows,
\begin{equation}
e^{-\Gamma^{(1)}(A)} \;=\; \prod_{n=-\infty}^{+\infty}\Big[
{\mathcal J}_n \; e^{-\Gamma^{(1)}_{3+1}(A,M_n)} \Big]\;,
\end{equation}
where we have the Jacobian,
\begin{equation}\label{eq:defj}
{\mathcal J}_n \;=\; \exp \big(\frac{i e^2}{16 \pi^2} \, \varphi_n
\int d^{3+1}x {\tilde F}_{\mu\nu} F_{\mu\nu} \big)  \;,
\end{equation}
with ${\tilde F}_{\mu\nu}=\frac{1}{2} \epsilon_{\mu\nu\rho\lambda}
F_{\rho\lambda}$. The quantity $\Gamma^{(1)}_{3+1}(A,M_n)$ is the
one-loop fermion contribution to the effective action, for a
fermion whose mass is $M_n$, in $3+1$ dimensions. It may be
expressed as a fermion determinant,
\begin{equation}
e^{-\Gamma^{(1)}_{3+1}(A,M_n)} \;=\; \det(\not \!\! D + M_n) \;.
\end{equation}
A general expression for the one loop effective action is
\begin{equation}
\Gamma^{(1)}(A) \;=\;\Gamma^{(1)}_e (A) \,+\, \Gamma^{(1)}_o (A) ,
\end{equation}
where
\begin{equation} \Gamma^{(1)}_e(A) \;=\;
\sum_{n=-\infty}^{\infty} \Gamma^{(1)}_{3+1}(A,M_n) ,
\end{equation}
\begin{equation}\label{eq:sumj}
\Gamma^{(1)}_o(A) \;=\; - \sum_{n=-\infty}^{\infty} \ln {\mathcal
J}_n \;,
\end{equation}
with  $e$ and $o$ subscripts stand, respectively, for the even and
odd components of the one loop effective action with respect to
the parity transformation.

The parity conserving part of the effective action may be obtained
by performing the sum of the required $QED_{3+1}$ object, with an
$n$-dependent mass-like term, $M_n$.  We consider the part of
$\Gamma^{(1)}_e$ that contributes to the vacuum polarization
tensor for the $A_\mu$ gauge field components. Since we are not
interested in response functions which involve the $s$ component
of the currents, it is useful to define,
\begin{equation}
\Gamma^{(1)}_e (A_\mu) \;\equiv\;\Gamma^{(1)}_e (A_\mu,A_s) \,-\,
\Gamma^{(1)}_e (0,A_s) \;.
\end{equation}
The term $\Gamma^{(1)}_e (0,A_s)\equiv \Gamma_s(A_s)$ does not
contribute to response functions involving $A_\mu$, although it
can be used to study the fermion-loop corrections to an $A_s$
effective potential. The explicit form of this function
is~\cite{Zinn-Justin1996,Fosco1997},
\begin{equation}
 \Gamma_s(A_s) \,=\,- 2 L \int d^{3+1}x \, \int \frac{d^4k}{(2\pi)^4}
\ln \big[ \cosh(L k) \,+\, \cos \theta \big] \;.
\end{equation}

The vacuum polarization tensor $\Pi_{\mu\nu}$ is obtained from the
quadratic term in a functional expansion of
$\Gamma^{(1)}_e(A_\mu)$ in the gauge field,
\begin{equation}
\Gamma^{(1)}_e(A_\mu) = \frac{1}{2} \int d^{3+1}x \int d^{3+1}y
A_\mu(x) \Pi_{\mu \nu}(x,y) A_\nu(y) + \cdots
\end{equation}
It is then sufficient to resort to the analogous expansion for the
$3+1$ dimensional effective action,
\begin{equation}
\Gamma_{3+1}^{(1)}(A,M_n)= \frac{1}{2} \int d^{3+1}x \int d^{3+1}y
\,\left[A_\mu(x) \Pi_{\mu \nu}^{(n)}(x,y) A_\nu(y)\right] + \cdots
\label{Gamma1(3+1)}
\end{equation}
(which is even) so that the vacuum polarization receives
contributions from all the modes $n$,
\begin{equation}\label{eq:pisum}
\Pi^e_{\mu \nu} \;=\;\sum_n \Pi_{\mu \nu}^{(n)}\,,
\end{equation}
where the contribution to each mode is given by $\Pi_{\mu
\nu}^{(n)}=\Pi^{(n)}(k^2) \; \delta^T_{\mu \nu}(k)$, with,
\begin{equation}\label{eq:tamed}
\Pi^{(n)}(k^2) \;=\; \frac{2\,e^2}{\pi} \int_0^1
d\beta\,\,\beta(1-\beta)\,\,\ln
\left[1+\beta(1-\beta)\frac{k^2}{M_n^2}\right]\;.
\end{equation}

Considering the corresponding modification in the photon's
effective action due to the extra dimension, it implies  a
correction to electrostatic potentials. For the Hydrogen atom, the
corrected electrostatic potential is,
\begin{equation}
V_{eff}(r,L) \;=\; - \frac{e^2}{4 \pi r} \,-\,\frac{e^4}{120 \pi^2
m^2} \left[\frac{m L \,\sinh(m L)}{\cosh(m L) - \cos\theta}\right]
\; \delta^{(3)}({\mathbf r})\;.
\end{equation}
The usual correction is obtained when $\theta \to 0$ and $m L \to
0$,
\begin{equation}
V_{eff} (r) \;\to\; - \frac{e^2}{4 \pi r} \,-\,\frac{e^4}{60 \pi^2
m^2} \; \delta^{(3)}({\mathbf r}) \;.
\end{equation}
The relation between the $L$-corrected and the usual potential is
governed by the quantity,
\begin{equation}
\xi(m L, \theta) \;\equiv\;\frac{(m L/2)\,\sinh(m L)}{\cosh(m L) -
\cos \theta}\;. \label{U1}
\end{equation}
The case of a vanishing flux yields simply
$$\xi(m L, 0) \,=\,\frac{m L}{2\tanh\left(\frac{m L}{2}\right)},$$
which for small values of $m L$,  $\xi(m L, 0) \rightarrow 1$, and
grows linearly with $m L$ when \mbox{$m L >> 1$}.

The opposite regime, when the effect of the flux is maximum,
corresponds to $\theta = \pi/2$, i.e.
\begin{equation}
\xi \left(m L, \frac{\pi}{2} \right) \;=\; \frac{m L} {2} \,
\tanh(m L)\;. \label{U2}
\end{equation}
The behavior in this case is quite different; it tends to zero
quadratically for small $m L$, and also grows linearly in the $m L
\rightarrow \infty$, but with a different slope.

It is interesting to note that, from Eq.~(\ref{U1}),  an estimate
for the compactification length $L$ can be obtained for the
vanishing flux approximation. The typical contribution of the
vacuum polarization term for the energy shift in muonic atoms is
of the order of $0.5\,\%$~\cite{Eides2001}; we may then take
$\xi(m L, \,0)\approx 1.0001$. Such a choice implies that a
correction due to an extra dimension does not significantly
changes the values from present data. With this reasoning we have
$L\,\approx 0.03 \,[m]^{-1}$ which in MKS units reads $L \approx
10^{-16}\,{\rm{m}}$. This estimate for the size of the extra
dimension is of the same order as  the one obtained by Roy and
Bander~\cite{Roy2009}.

Summarizing, the formalism presented here has been employed to
study effects from compactified extra dimensions in quantum
electrodynamics, using a model, previously developed for the
electroweak transitions. At the one-loop level, the Hydrogen atom
potential is modified by a factor depending on the size of the
compactified dimension, L, in such a way that present day data are
compatible with $L\sim 10^{-16}{\rm{m}}$. An interesting point is
that this value is of the same order as those obtained from
studies of the anomalous magnetic moment of the muon in the
context of eletroweak interactions. The results will not change
significantly by including higher order effects. This raises a
question concerning the invisibility of extra dimensions if they
were so large. A possible interpretation is that we are strictly
in four-dimensions and  our world is just a brane of a
higher-dimensional space. In such a case, nothing would be seen
directly in fifth dimension. This would imply  that, no  direct
experiment would be able to detect its presence. Only indirect
evidences would be reliable, such as experiments on
non-conservation of electric charge. In this case, the outcoming
of such experiments could be interpreted as particles initially
confined in our $4$-dimensional subspace migrating to other extra
dimensions.

\section{Concluding remarks}

In this article we review applications of the quantum field theory
on toroidal topology, in particle and in condensed matter physics.
In order to unify the presentation, the theory on a torus $\Gamma
_{D}^{d}=({\mathbb{S}}^{1})^{d}\times {\mathbb{R}} ^{D-d}$ is
developed from a Lie-group representation formalism. As a first
application, we revisit, with emphasis on topological aspects, the
quantum field theory at finite temperature: a prototype of a
quantum field theory on a torus $\Gamma _{4}^{1}$, in its real-
and imaginary-time versions.

The toroidal quantum-field theory  provides the basis for a
consistent approach of spontaneous symmetry breaking driven by
both temperature and spatial boundaries. In this framework, we
study superconductivity in films, wires and grains, where some
results are rather successfully comparable with experiments. Other
applications include the Casimir effect and self interacting
four-fermion systems: the Gross-Neveu and Nambu-Jona-Lasinio
models, considered as effective theories for quantum
chromodynamics. In this latter case, finite size effects on the
hadronic phase structure are investigated.

In addition, effects of extra spatial dimensions are addressed. We
proceed by considering quantum electrodynamics in a
five-dimensional space-time, where the fifth-dimension is
compactified on a torus. The formalism, initially developed for
particle physics, provides results compatible with other trials of
probing the existence  of extra-dimensions.

Many aspects of quantum fields on $\Gamma _{D}^{d}$ have been
originally developed in diverse works  analyzing cosmological
problems, such as the topological mass generation. Regarding these
results, we have reviewed only their field-theory elements. Other
applications not detailed here include toroidal topologies in
string theory and quantum Hall effect in Condensed matter physics.
These choices reflect two facts. First, these studies have been
considered in detail elsewhere; second, we were foreseeing
applications specifically in particles and condensed matter
physics, having quantum fields as a starting point.

Due to its nature, the toroidal topology is such that the local
properties of the space-time are kept invariant,  as the Casimir
invariants of the Lie-algebra. The effect of the topology arises
in non-local properties of a system, as in correlation functions.
This implies in the success of the representation theory; which is
not the case of others topologies. This is why we restricted
ourselves to presenting a general theory for quantum fields on
$\Gamma _{D}^{d}$. However, it would be interesting if similar
developments, including analysis of symmetries, were carried out
for other topological spaces, as for example, the spherical one.
To our knowledge, this remains an open
problem.\\
\\

\noindent {\bf Acknowledgements}\\

{We thank CAPES, CNPq and FAPERJ (Brazil) and NSERC (Canada) for
partial support.}

\bibliographystyle{elsarticle-num}

\end{document}